\documentclass[11pt,papersize=8.5in,11in]{report}

\usepackage{comment}

\usepackage[dvips,letterpaper,top=1.0in,bottom=1.25in,left=1.5in,right=1.0in]{geometry}
\usepackage{graphicx}
\usepackage{titlesec}
\usepackage{epsfig}
\usepackage{dcolumn}

\usepackage{amsmath}
\usepackage{amsfonts}
\usepackage{amssymb}
\usepackage{amsthm}
\usepackage[adobe-utopia]{mathdesign}
\usepackage{bm}
\usepackage{float}

\usepackage[usenames,dvipsnames]{xcolor}

\usepackage{subcaption}

\usepackage{slashed}
\usepackage[vcentermath]{youngtab}

\usepackage[font = singlespacing]{caption} 
\usepackage{setspace} 
\usepackage{enumitem} 
\usepackage{verbatim} 
\usepackage{hyperref} 
\usepackage{appendix} 
\usepackage{multirow} 
\usepackage{tocloft}  
\usepackage{CJK}      
\usepackage{hyperref}
\usepackage{color}
\usepackage{grffile}
\usepackage{fancybox}
\newcommand{\sh}[1]{#1\hskip -6pt  / }

\usepackage{cite}

\hypersetup{colorlinks=true, citecolor=blue, linkcolor=blue, urlcolor=blue}
\setlist{nolistsep}
\titleformat{\chapter}[display]{\normalfont\large\bfseries}{\MakeUppercase{\chaptertitlename}\ \thechapter}{0.5em}{}
\titlespacing*{\chapter}{0pt}{-10pt}{0pt}
\titleformat{\section}{\normalfont\large\bfseries}{\thesection}{0.5em}{}
\usepackage{slashed}

\usepackage[all]{nowidow}

\usepackage[nottoc, notlof, notlot]{tocbibind}
\usepackage[nonumberlist, acronym]{glossaries}
\usepackage{setspace}
\begin{document}


\doublespacing
\pagenumbering{roman}

\begin{titlepage}
  \begin{center}
    FLORIDA INTERNATIONAL UNIVERSITY \\
    Miami, Florida \\
    ~\\
    ~\\
    ~\\
    ~\\
    ~\\
    ~\\
    ~\\
    ~\\
    ~\\
    MULTINUCLEON  SHORT-RANGE CORRELATION MODEL FOR  NUCLEAR SPECTRAL  FUNCTIONS \\
    ~\\
    A dissertation submitted in partial fulfillment of the \\
    requirements of the degree of \\
    DOCTOR OF PHILOSOPHY \\
    in \\
    PHYSICS \\
    by \\
    Oswaldo Artiles \\
    ~\\
    2017
  \end{center}
\end{titlepage}

\setcounter{page}{2}

\newpage
\begin{singlespace}
\setlength{\parindent}{0.0cm}
To: Dean Michael R. Heithaus\\
\hangindent= 0.6cm{College of Arts, Sciences and Education}\\

This dissertation, written by
Oswaldo Artiles,
and entitled
Multinucleon Short-range Correlation Model for Nuclear Spectral Functions,
having been approved in respect to style and intellectual content, is referred to you for judgment.\\

We have read this dissertation and recommend that it be approved.

\flushright{
\vspace{0.3cm}
\rule{7.8cm}{0.4pt}\\
Oren W. Maxwell\\
\vspace{0.8cm}
\rule{7.8cm}{0.4pt}\\
Werner Boeglin\\
\vspace{0.8cm}
\rule{7.8cm}{0.4pt}\\
Gueo Grantcharov\\
\vspace{0.8cm}
\rule{7.8cm}{0.4pt}\\
Misak M. Sargsian, Major Professor\\
}
\flushleft{
Date of Defense: May 12, 2017\\
\vspace{0.3cm}
The dissertation of Oswaldo Artiles is approved.\\
}
\flushright{
  \rule{7.8cm}{0.4pt}\\
  Dean Michael R. Heithaus\\
  College of Arts, Sciences and Education\\
  \vspace{1cm}
  \rule{7.8cm}{0.4pt}\\
  Andr\'{e}s G.\ Gil\\
  Vice President for Research and Economic Development\\
  and Dean of the University Graduate School\\}
\vspace{2cm}
\begin{center}
  Florida International University, 2017
\end{center}
\end{singlespace}


\newpage
\vspace*{8cm}
\begin{center}
  DEDICATION
\end{center}

\begin{CJK}{UTF8}{ipxga}

This dissertation is dedicated to Rita Elisa and Ana Rita, my grandmother and my mother, who gave me love and  cared for me everyday of my childhood, and  who taught me the values of honesty, respect for life, and love for hard work, values which have guided me during my entire life. 

\end{CJK}

\newpage
\begin{center}
	ACKNOWLEDGMENTS
\end{center}

I would like to acknowledge all the  teachers and professors  that help me and taught me during all years of my life as a student.  

I am indebted to the members of my dissertation committee: 
Oren W. Maxwell, Werner Boeglin,  and Gueo Grantcharov, who helped me during all my research. My special acknowledgment to my adviser, Misak Sargsian, who taught me a lot of physics and was always patient with my ignorance. 

I wish to acknowledge the graduate program directors, Brian Raue, Rajamani Narayanan, and Jorge Rodriguez, 
as well the administrative staff of the  the FIU physics department,
specially
Elizabeth Bergano-Smith, Omar Tolbert, Maria Martinez, John Omara, Ofelia Adan-Fernandez, and Robert Brown
who support me  with  all their  administrative and technical expertise.

I also thank Frank Vera, Dhiraj Maheswari, and Christopher Leon, my fellow PhD students of the group of nuclear theoretical physics, for the very helpful 
physics and mathematics discussions that I have had with them.

Finally, my greatest debt to my wife Ivia, for giving me love and for taking care of me with total dedication, and to my daughters Livia and Claudia and all my grandchildren: Mauricio, Ana, Isabella e Ignacio,  for all their love  and for giving me the most beautiful reasons  to live. 

This research would not have been possible without financial support.
I would like to acknowledge the Department of Energy for providing the grant that supported my research assistantship.

\newpage
\begin{center}
  ABSTRACT OF THE DISSERTATION \\
  MULTINUCLEON  SHORT-RANGE CORRELATION MODEL FOR NUCLEAR SPECTRAL  FUNCTIONS\\
  by\\
  Oswaldo Artiles\\
  Florida International University, 2017\\
  Miami, Florida\\
  Professor Misak M. Sargsian, Major Professor
\end{center}
\vspace{-4mm}

The main goal of the  research presented in my dissertation was  to develop a theoretical model for relativistic nuclear  spectral functions at high missing momenta and removal energies  based on the 
multi-nucleon short-range correlation~(SRC) model. The nuclear spectral functions are necessary for the description of high energy nuclear processes currently being studied at different labs such as JLAB, LHC and FNAL. 

The model  followed the effective Feynman diagrammatic approach  in order  to account for the relativistic effects  important in the SRC domain. In addition to the two-nucleon (2N) SRC with center of mass motion contribution, the contribution of  the three-nucleon SRCs to the  spectral functions was also  derived. The latter was modeled based on the assumption that the 3N SRCs are a product of  two sequential short range nucleon-nucleon (NN) interactions. 

The nuclear  spectral functions models were derived from  two  theoretical frameworks for  evaluating covariant Feynman diagrams: In the first,  referred to as the virtual nucleon approximation,   the   Feynman diagrams were reduced to the time ordered  non-covariant diagrams by evaluating the  nucleon spectators in the SRC at their positive energy poles, neglecting  explicitly the contribution from vacuum diagrams. In the second approach, referred to  as the light-front  approximation, the boost invariant nuclear spectral  function  was  formulated in the  light-front reference frame in which case the vacuum diagrams are kinematically  suppressed and the bound nucleon is described by its light-front  variables such as  momentum  fraction,  transverse momentum and   invariant mass.

On the basis of  the derived  nuclear spectral functions, the corresponding computational models were  developed from which the numerical estimates   of  the SRC spectral functions, the SRC momentum distributions, and the SRC density matrices  were obtained.

\newpage

\renewcommand\cftchapfont{\mdseries}
\renewcommand\cftchappagefont{\mdseries}
\renewcommand{\cftpartleader}{\cftdotfill{\cftdotsep}} 
\renewcommand{\cftchapleader}{\cftdotfill{\cftdotsep}} 

\doublespacing
\renewcommand\contentsname{\normalfont\MakeUppercase{Table of contents}\hfill}
\setcounter{tocdepth}{1}
\tableofcontents 
\addtocontents{toc}{\normalfont\MakeUppercase{Chapter}\hfill\normalfont\MakeUppercase{Page}\par}

\singlespacing
\newpage
\listoftables 
\addtocontents{lot}{\normalfont\MakeUppercase{Table}\hfill\normalfont\MakeUppercase{Page}\par}

\newpage
\doublespacing
\newpage
\listoffigures 
\addtocontents{lof}{\normalfont\MakeUppercase{Figure}\hfill\normalfont\MakeUppercase{Page}\par}
\doublespacing


\chapter{Introduction}

\pagenumbering{arabic}




\pagenumbering{arabic}


Scattering experiments  have been the most important experimental tools to reveal the structure of visible matter, specially  the structure of atoms and  nucleus. In 1911, Rutherford, through scattering experiments of alpha particles off atomic targets,  discovered the atomic nucleus \cite{ Rut11}.  Experiments of disintegration of nitrogen nuclei in 1919, led Rutherford to conclude that  the nucleus was a composite system of particles  held together  by a strong force   \cite{ Rut19}. Rutherford also   gave the name of proton to the hydrogen nucleus,  and predicted the neutron in 1920  \cite{ Rut20}.  The prediction of Rutherford was confirmed by  Chadwick  \cite {Chad32}  and Curie-Joliot \cite{ CuJ32} who, independently,  showed  the existence of the neutron in 1932. The atomic nucleus has since been   considered a system composed of strong interacting nucleons \footnote{a common name for protons and neutrons} \cite {He32},  held together by strong short-range forces, of a different nature than those of  the electromagnetic and gravitational forces.  Since the atomic nucleus is the basic component of all the visible matter in the universe, a continuous and  very important experimental and theoretical efforts have been ongoing since 1932 to understand and explain  the nuclear and the nucleon structures at the fundamental level.  

In 1935, Yukawa described the  interaction of nucleons in the nucleus by means of a field of force associated with a  particle or a quantum which was a carrier of  this  interaction \cite {Yu35}.  Two particles, the pion+ and the pion-, were discovered in 1947  \cite {LMOP47}, and  the neutral pion was isolated  in 1950, completing a trilogy of particles that were identified as the carrier particles  predicted by  Yukawa.

The availability of high energy particle accelerators in the middle fifties  and early sixties, allowed  the discovery of a multitude of strong interacting particles collectively known as hadrons, that formed groups of particles with similar properties  \cite {Sa97}. The first formal  group classification  of the hadrons was given by  Gell Man \cite {GE61} and independently by Yuval Ne'Man \cite {NE61}  in the so-called "Eightfold Way", that classified  hadrons into subgroups identified with  octet representations of the SU(3) group. The classification of hadrons proposed by Gell Man was confirmed by the 1964 discovery of the omega-minus particle, predicted in 1962. After this discovery    Gell-Mann \cite {GE64} and George Zweig  \cite {ZW64a, ZW64b}  independently predicted  that hadrons were composed  of  elementary fermions named as quarks \cite {BRDH91}. 

 The ever increasing energy of modern particle accelerators and the complete knowledge of the electromagnetic interaction between electrons and  nucleons or nuclei were  essential to  prove  the  existence of the quarks as fundamental components of hadrons in general, and nucleons in particular.  In 1968, the first results from deep inelastic scattering experiments  demonstrated that protons and neutrons are composite structures  made up of particles  with fractional electric charge as well as neutral particles \cite  {Fr97, Bj97}. The charged particles were later identified with  quarks predicted by Gell-Mann \cite {GE64} and Zweig \cite {ZW64a, ZW64b}, and the neutral particles with the massless gluons, the carriers of the strong interaction between quarks in hadrons.  
 
 The above mentioned  experimental and theoretical efforts resulted  in the formalization and acceptance of  the quantum field theory of strong interactions: quantum chromodynamics (QCD), within QCD  the vast majority of physical hadrons are many-body, highly relativistic systems composed of light quarks and massless gluons  \cite {BRDH91, isosrc}.  Quantum Chromodynamics (QCD) is a non-abelian gauge field theory (Yang and Mills theory \cite {YM54}  ) that describes the strong interactions of colored quarks and gluons fields. Quantum Chromodynamics  is the SU(3)$_c$ component of the SU(3)$_c$xSU(2)$_L$xU(1)$_Y$ of the Standard Model of Particle Physics.
 
 Quarks are strongly interacting fermions with spin 1/2 and, by convention positive parity. The   electric charges of the quarks are  -1/3 and 2/3. There are six different types of flavors of quarks: up (u), down (d), strange (s), charm (c),  top (t) and bottom (b). In order to have antisymmetric wave functions  (Pauli exclusion principle), the quarks carry color charges, which for convention are called: red (R), green (G) and blue(B). The color of quarks is a quantum number that was first proposed by Fritzsch and  Gell-Man  \cite {FG73} who identified the extra symmetry of QCD with the color symmetry. As a result of  the color charges, quarks are said to be in the fundamental representation of the SU(3)$_c$ color  Lie group.   

Gluons are massless gauge bosons, with spin 1  and  two polarization states, which mediate color interactions among quarks.  They, represented through non-Abelian gauge fields in QCD, are responsible for binding the quarks together. Gluons carry color charges and hence interact with each other even in the absence of quarks \cite {HM84}. Gluons transform under the adjoint representation of the Lie group  SU(3)$_c$  \cite {PDG14b}. 

Even though hadrons are composite systems of quarks and gluons, the understanding of the nuclear structure is mainly based on the nucleonic degrees of freedom. Hence, a nucleus of mass number $A$ and atomic number $Z$,  is commonly represented as a system of  A nucleons, $Z$ of which are protons and $(A-Z)$ neutrons.  The main picture of nuclei as a composite system of nucleons is taken from the mean field approximation, on which each nucleon is  modeled as an independent particle moving in the mean field potential of the remaining nucleons \cite {RS04}.  However, the  picture  breaks down when the momentum  of the bound nucleon exceeds $k_F \simeq$ 250 MeV/$\it c$, where $k_F$ is the characteristic Fermi momentum of the nucleus described as a degenerate Fermi gas \cite {MSW71}. 

One of the major issues of modern nuclear physics is to understand the nuclear structure above the $k_F $  domain. The issue is directly related to  understanding  the  nucleon-nucleon interaction at short distances inside the nucleus. The   interaction  prevents two nucleons from coming very close together demanding the existence of high momentum components in the nuclear ground state wave function. The close nucleon-nucleon interactions cannot be described in the context of nuclear mean field models and are commonly called multi-nucleon short-range correlations (SRC) \cite {AHR12}.  

Deep inelastic scattering (DIS) of leptons off  the nucleus is one of the methods for probing nuclear structures at small distance scales. Before 1983, it was predicted that the quarks inside the  nucleus followed the rules of the mean field  model. According to which the quark-gluon dynamics in each nucleon  was not influenced by the nuclear mean field.   The fact that the quark momentum distributions for individual nucleons are modified in the nuclear medium was first observed by the European Muon Collaboration (EMC) experiment and is usually known as the EMC effect  \cite {Au83}. The discovery of the EMC effect is considered the starting event of a new era in nuclear QCD. 

The importance of the high momentum properties  of bound nucleon for  nuclear EMC effects  follows from the  recent observations of apparent correlation between the medium modification of the  quark momentum  distributions  and  the strength  of the two-nucleon short range correlations~(SRCs) in nuclei\cite{LW1,LW2}.  In order to improve our knowledge of the role of the QCD interactions in the nuclear dynamics, it is crucial to understand the role of the SRCs in the EMC effects \cite {MMSA03}. 

As a result of  the QCD evolution of nuclear parton distributions functions(PDFs)( see Eq. (\ref {strucFunc})), it 
is  expected  that at very large $Q^2$ the knowledge  of the  high momentum component of 
the bound nucleons becomes important because of  the contribution of   quarks  with 
momentum fractions ($x$) larger than the ones  provided by an isolated nucleon (i.e. partons with $x > 1$)\cite{FSS15,FS15}.

The same is true for the reliable interpretation of neutrino-nuclei scattering experiments  in which case both the medium modification of 
PDFs as well as the realistic treatment of SRCs are essential. One of the main results of the experiments is 
 the discrepancy found between measurements of $\sin^2 \theta_w$ ($ \theta_w$ is the weak  mixing angle)  involving free particles and those involving bound nucleons in the nuclear medium, the so-called  NuTev anomaly \cite {Nutev, CERN02, GPZ02, PDG12b}. A possible  explanation to the corresponding results is derived from  the presence of more energetics protons in neutron rich, large A,  asymmetric nuclei which implies that u-quarks are more modified than d-quarks, resulting in  a negative correction for the experimental value of  $\sin^2 \theta_W $ for bound nucleons \cite {mNutev}. 
 
All the above described areas of research require models to predict the high  momenta and binding  energies of bound nucleons in the nuclei. 
Such models are described  through the  nuclear spectral functions that  define the joint probability   of finding a nucleon in the nucleus  with momentum $\mathrm{p}$  and removal (binding) energy $E_m$. With the advent of the Large Hadron Collider and expected construction of electron-ion colliders as well 
as several ongoing neutrino-nuclei  experiments,  the knowledge of such  spectral functions will be an important part of the 
theoretical interpretation of the data involving nuclear targets.

The present dissertation is divided in four  major chapters. In chapter 2, the theoretical framework  for the multinucleon short-range correlation model for nuclear spectral functions is developed. The diagrammatic method of calculation of the spectral function is described, as well as the mathematical models for the spectral function in two approaches: Virtual nucleon and light front approximations. From the mathematical models of chapter 2, computational models to obtain numerical estimates for the spectral functions, density matrices, and nucleon momentum distributions  are developed in chapter 3. The numerical estimates  of the spectral functions, density matrices, and nucleon momentum distributions   are presented in chapter 4. Finally, the main conclusions and an outline of future projects  are included in chapter 5. 

In the rest of chapter 1, the relationship between lepton-nucleus  scattering cross section and the nuclear spectral function is presented. Then, a brief description of the EMC effect and its relation to 2N SRC is included. The assumptions for modeling  two and  three nucleon  short-range correlations are also described. Finally, some definitions and the range of validity of the nuclear spectral functions, including the treatment of the relativistic effects and the model for 2N SRC center of mass  motion are presented.

 \section{Lepton-nucleus  inclusive scattering cross section and nuclear spectral function}
\label { sec:inclusive}
\begin{figure}[tbh]
\centering\includegraphics[scale=0.90]{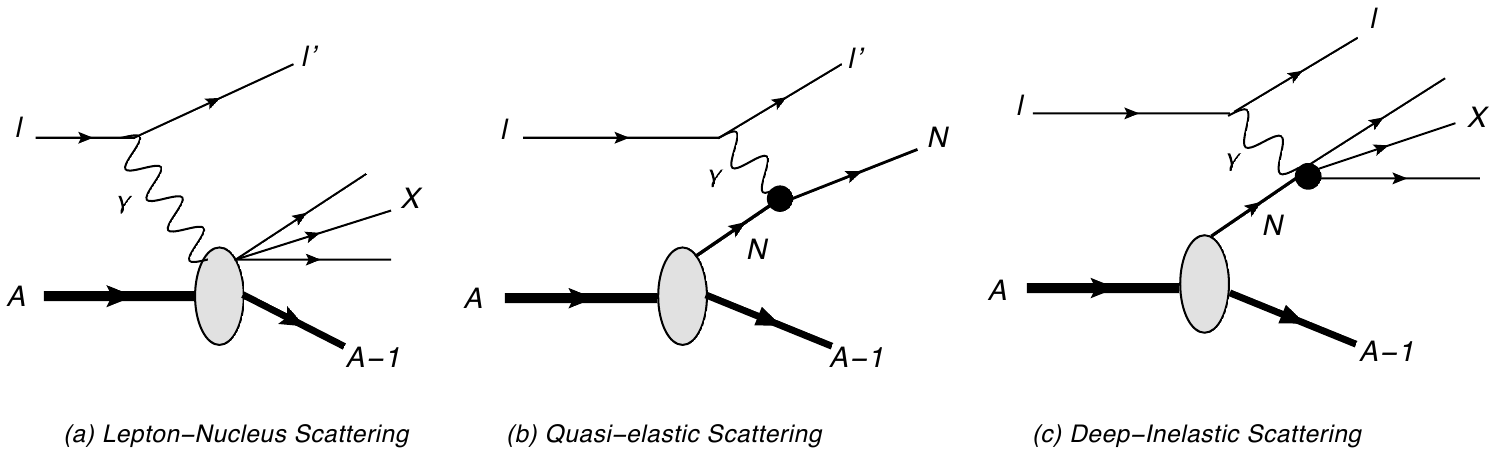}
\caption{Lepton-nucleus  inclusive scattering.} 
\label{LNIS}
\end{figure}

In the  lepton-nucleus inclusive scattering reaction ($l+A\longrightarrow l' + X + (A-1)^*$),  shown in  Fig.\ref {LNIS}(a),  X is the hadron final state   and  (A-1) the recoil system of  bound nucleons, and only the final lepton is detected. 

In the one-boson exchange approximation, if l and l' are the same leptons ( electrons or muons) the reaction is mediated by either the weak neutral gauge boson, $Z^0$,  or the photon $\gamma$.   If l and l' differs by one unit of charge, for instance (muon, neutrino) or (anti-muon, anti-neutrino),  the reaction is mediated by the weak charged  gauge bosons,  $W^-,$ or $W^+$. 

The kinematic variables for the lepton-nucleus inclusive  scattering in the lab frame (nucleus rest frame)  are:  $ p_A = (M_A, 0) $, for the four-momentum vector of the nucleus of mass $M_A$;   $ k^\mu_l = (E_l, \mathbf k_l) $ and  $ k^\mu_{l'} = (E_{l'}, \mathbf k_{l'}) $, for the four-momentum vectors of the initial l and final ${l'}$ leptons, respectively.   $ q^\mu  \equiv (q^0, \mathbf q) =(E_l-E_{l'},  \mathbf k_l-\mathbf k_{l'})$, is the four-momentum vector of the gauge boson,  with $Q^2 \equiv -(q^\mu)^2$ defined as the virtuality of the gauge boson.   If $q^0 \ll M_Z, M_W$, where $M_Z (M_W)$ is the mass of the Z ($W^+$) gauge boson, and l=l' (muon or electron) the reaction is defined by  one-photon exchange. 

Theoretical studies demonstrated that  If $Q^2 \geqslant M^2_N$, where $M_N$ is the mass of a nucleon.   the size of the probe ( $\sim 1/Q^2$) is smaller than the size of the nucleon, hence it  can be considered that the lepton scatters off a bound nucleon inside the nucleus (see Fig.\ref{LNIS}(b) and (c)). In  Fig \ref{LNIS}, it is assumed that the final state interactions of the hadron final state ($X$)  can be  neglected. Such an assumption is referred to   as the plane wave impulse approximation (PWIA) \cite {eheppn2,DF83}. The PWIA is justified for  inclusive processes with large $Q^2$($Q^2 \ge 2 \; \mathrm {GeV}^2$) such as deep inelastic scattering \cite {CMS14}. 

The total energy of the lepton-nucleon scattering (Fig. \ref{LNIS}(b)) is defined as $ W^2_N \equiv  (p^\mu_N + q^\mu)^2 $. If $W^2_N \sim M^2_N$ then the  hadron final state X is a nucleon and the corresponding  lepton-nucleus scattering is defined as quasi-elastic.  If $W^2_N > (M_N+M_\pi)^2$,  the  hadron final state X may consist of a nucleon and a pion ($\pi$) or baryonic excited resonances.  In the region of deep inelastic scattering (DIS), $W_N > $ 2.5 GeV and $Q^2 > $ 2.0 GeV$^2$,   the target nucleon breaks down in a  final state X consisting of a collection of hadrons. 

Deep inelastic lepton-nucleon scattering (DIS)   has been fundamental in unveiling the  structure of  nucleons. In the parton model with very large Q$^2$ \cite {BP69, RF72}, the virtual photon  interacts with one of the quarks of the bound nucleon (Fig \ref{LNIS}(c)). Such a picture is relevant in the infinite momentum frame (IMF) of the nucleon. The lepton-nucleon center of mass frame is a good approximation  to such  IMF, on  which the quarks (partons) within the nucleons are slowed down by Lorentz time dilation effects. In the IMF, therefore,   the struck quark can be considered free, and characterized by the momentum fraction of the fast nucleon defined as the invariant Bjorken scaling variable  $x = Q^2/(2pq)$,  (0 $\le x \le $ 1),   \cite {BP69}.

The   one-photon exchange differential cross section for  unpolarized  lepton-nucleon inclusive inelastic scattering can be expressed through two  independent nucleon structure functions $F^N_{1}$ and $F^N_{2}$ as \cite {RO50, DW64, MOTT, PDG12a}
\begin{align}
\frac{d^2\sigma}{dE_{l'}d\Omega_{l'}}\Big \vert_{lab}&=\Big ({\frac{d \sigma}{d \Omega}}\Big)_{Mott} 
\frac {1}{\nu}\left[ F^N_{2}(x, Q^2)+  \frac {2\nu}{M_N}\tan^2 (\theta_l/2)F^N_{1}(x, Q^2)\right ], 
\label{NucleonIS}
\end{align}
where $\nu = q^0 = Q^2/(2M_N x)$, and  $ \Big ({\frac{d \sigma}{d \Omega}}\Big)_{Mott} = \frac {\alpha^2 \cos^2(\theta_l/2)}{4 E^2_l \sin^4(\theta_l/2)}$ is the Mott cross section for electron-point charge scattering \cite {Moot29} , and $\theta_l$ the scattering angle.  In the  elastic scattering kinematics, $F^N_{1}$   and $F^N_{2}$  can be expressed as functions of the so called electric ($G_E$)  and magnetic ($G_M$) form factors of the nucleon\cite {HO58, HA63, BO79, HM84}. 

In the  inelastic scattering kinematics, $F^N_{1}$   and $F^N_{2}$ are functions of the two independent variables $x$ and $Q^2$. However within the partonic model of the nucleon, it was predicted that in deeps inelastic scattering (large $Q^2$): 
\begin{align}
F^N_{1}(x, Q^2) \to F^N_1(x), \nonumber \\
F^N_{2}(x, Q^2)\to F^N_2(x), 
\label{BjorkenScaling}
\end{align}
where  the dimensionless nucleon structure functions, $F^N_{1,2}$ are independent of $Q^2$, that is independent of any mass scale (scale invariant), signaling the presence of free point-like quarks in the nucleons by satisfying the  Bjorken scaling  property\cite {BP69}.  \\
The nucleon structure functions contain  the information about the parton's (quark's) momentum  distribution  in the nucleon, namely 
\begin{align}
 F^N_{2}(x) & = \sum_i e^2_i x f_i(x), \nonumber \\
F^N_{1}(x) & = \frac {1}{2x} F^N_{2}(x),
\label{strucFunc}
\end{align}
where $e_i$ is the quark electric charge, and  $f_i(x)$ is the parton longitudinal   momentum distribution function in the infinite momentum frame, that is the probability that the struck quark $i$ carries a fraction $x$ of the nucleon momentum $p$. Hence, the nucleon structure function $F^N_{2}$ gives the weighted, by the square of the parton electric charge,  probability of finding a parton in the nucleon that carries a fraction $x $ of the total nucleon momentum.\\
 The  parton  momentum distribution functions $f_i(x)$  are normalized as:
\begin{align}
 \sum_{i'} \int   x f_{i'}(x)  dx = 1,
\label{strucFuncSum}
\end{align}
where $i'$ sums over all the partons, not just the charged ones that interact with the photon \cite {CL79, HM84, HHM13}.
\begin{figure}[h]
\centering\includegraphics[scale=0.8]{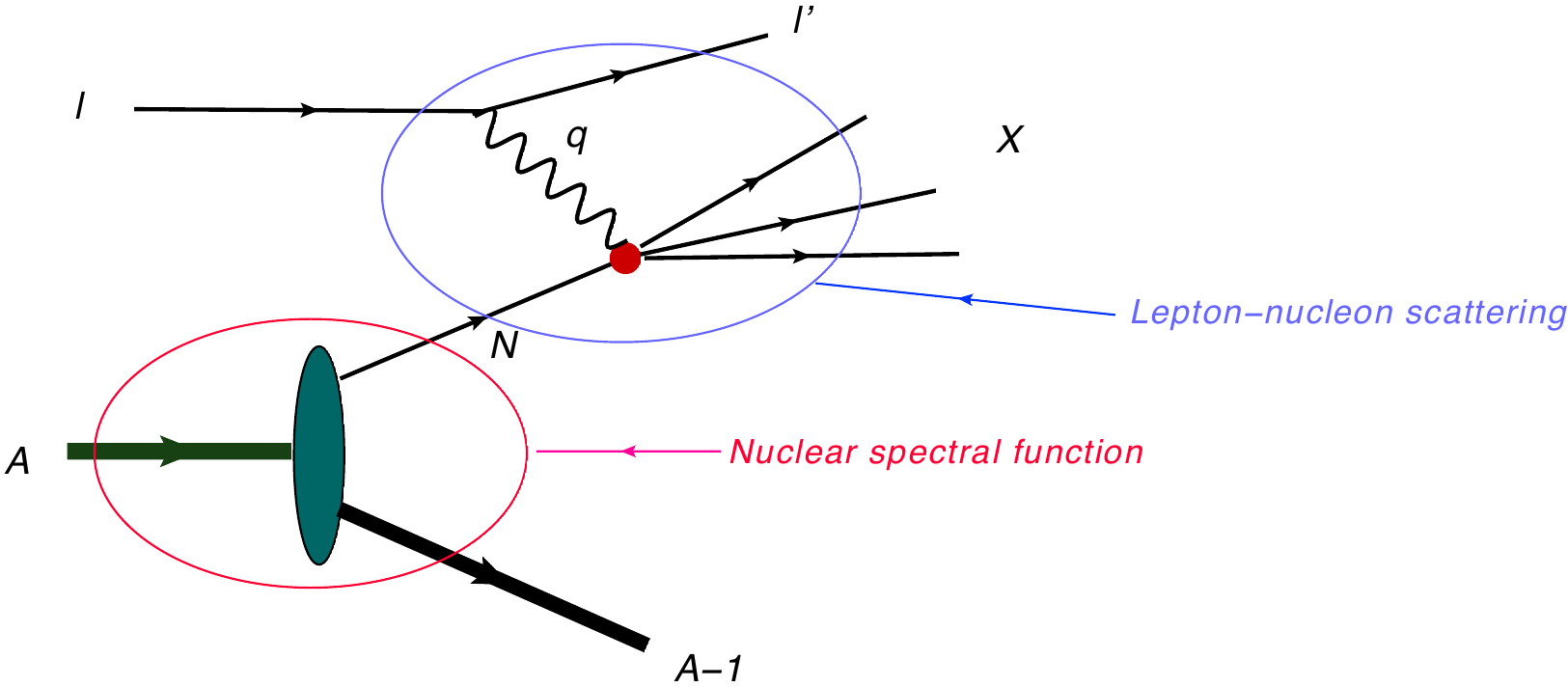}
\caption{Lepton-nucleus deep inelastic inclusive scattering regions}
\label {DISRegions}
\end{figure}

Similar to the free nucleon case, the   inclusive  cross section for   lepton-nucleus scattering  can be expressed as follows \begin{align}
\frac{d^2\sigma}{dE_{l'}d\Omega_{l'}}\Bigg\vert_{lab}&=\Big ({\frac{d \sigma}{d \Omega}}\Big)_{Mott} \frac {1}{\nu}
\left[ F^A_2(x_A, Q^2)+ \frac {2 \nu}{M_A} \tan^2 (\theta_l/2)F^A_1( x_A,Q^2)\right ], 
\label{NucleusDIS}
\end{align}
where     $F^A_1$ and $F^A_2$  are the nuclear structure functions, that, within the above discussed plane wave impulse approximation (PWIA),  can be expressed as a function  of the nucleon structure functions, $F^N_{1}$   and $F^N_{2}$ and of the nuclear spectral function (as depicted in Fig. \ref {DISRegions} for lepton-nucleus deep inelastic inclusive scattering),   by the following convolution integrals \cite {SSS02}  
\begin{align}
F^A_{1} (x,Q^2)&=\frac{AM_N}{M_A}\sum_N \int  S^N_A(\mathrm {p}, E_m)\Big [ F^N_1( \tilde x,Q^2) + \frac {p^2_\bot } {2M_N \nu} F^N_2( \tilde x,Q^2) \Big] d^4p,
\label{fa1dis}
\end{align}
\begin{align}
F^A_2(x,Q^2)&=\frac{AM_N}{M_A}\sum_N \int  S^N_A(\mathrm {p}, E_m)\frac {\nu}{\tilde \nu} \Big \{F^N_2 (\tilde {x},Q^2) \Big [\frac {(1 +  \cos\delta)^2}{M^2_N}{\Big (p^++q^+\frac{pq}{q^2}\Big )}^2\nonumber\\
& + \frac {{\mathbf p}^2_{\bot}}{2M^2_N}{\sin^2\delta}\Big] \Big\} d^4p,
\label{fa2dis}
\end{align}
where $S^N_A(\mathrm {p}, E_m)$  is  the nuclear  spectral function   which defines the joint probability   of finding a nucleon in the nucleus  with momentum $\mathrm{p}$  and removal  energy $E_m$, $M_A$ is the mass of the nucleus,  $x_A = Q^2/ (2q^0 M_A/A )$,  $\tilde \nu = \nu + (p^2 - M^2_N)/(2M_N)$, $\tilde x = Q^2/(2pq)$, and $\cos \delta = \nu /\sqrt{\nu^2-q^2}$. The four vector momentum of the bound nucleon  in the light-front coordinate frame is defined as $p^\mu = (p^-,p^+,  \mathbf p_\perp)$,  with  $ p^+= \mathrm p^0+\mathrm p_z,   {p^-}=\mathrm p^0-\mathrm p_z$, and $\mathbf p_\bot=(\mathrm p_x,\mathrm p_y)$.

 \section {The Nuclear European Muon Collaboration (EMC) Effect}
\label{sec:EMC}
\begin{figure}[h]
\centering\includegraphics[scale=0.42]{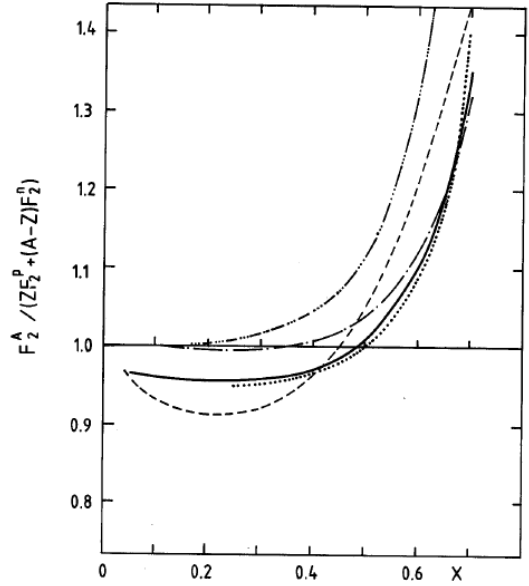}
\caption{ Theoretical predictions for the  Fermi motion corrections of  the bound nucleon structure functions $F^N_{2}$  for iron. Figure from \cite {Au83}.  }
\label {EMCPred}
\end{figure}

Since the binding energy of the nucleus is very small compared to the energy scales in deep inelastic scattering, it was  assumed that,  except for nucleon Fermi motion, the nucleus acted  as a collection of slowly moving weakly bound nucleons,  with their internal properties unchanged compared to the free nucleon case. Therefore  the following ratio  for  the  per nucleon inelastic structure functions $F^A_{2}$  to the  inelastic structure functions $F^p_{2}$ and $F^n_{2}$ of a free proton and neutron  
\begin{align}
R &= A F^A_{2}/[ZF^p_{2} + (A-Z)  F^n_{2}],
\label{R_ratio}
\end{align}
was expected to rise for x $\gtrsim $ 0.2, and be about  1.2 -1.3 for x = 0.65, as it is shown in Fig. \ref {EMCPred},  in which  Fermi motion corrections were included. Hence, it was expected that the nuclear deep inelastic inclusive scattering cross section   will be completely defined by partonic distributions of free nucleons, with the nucleon Fermi motion as the only nuclear effect.  It was also predicted that Eqs. (\ref {fa1dis}) and  (\ref {fa2dis}),  would give the same results for all nuclei. In summary, it was predicted that  the nuclear cross section would be the sum of the cross sections of the number of nucleons inside the target\cite {Au83, PN2003, HHM13}. 

\begin{figure}[h]
\centering\includegraphics[scale=0.42]{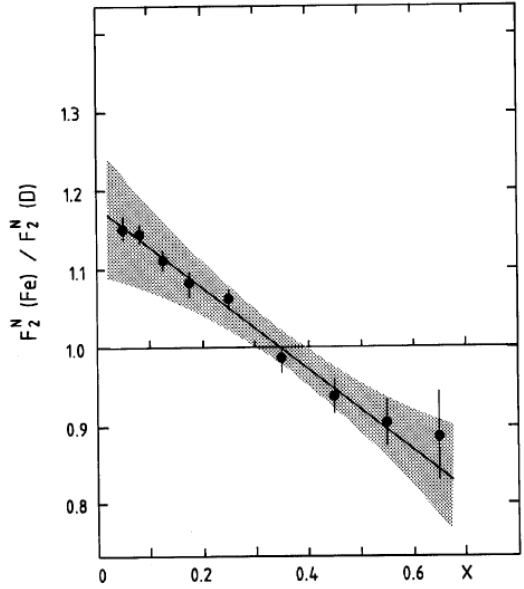}
\caption { EMC experimental results for the  ratio of the structure functions of iron nucleus and deuterium (F$^{N}_2$(Fe)/F$^N_2$(D)). Figure from \cite {Au83}. }
\label {EMCExpe}
\end{figure}

The above assumption  was demonstrated  to be wrong when the European Muon Collaboration (EMC) experiment discovered that in the muon-nucleus DIS, the ratio of the scattering cross section from nuclei to the deuteron, which is close to the ratio R in Eq. (\ref  {R_ratio}),  was in strong disagreement with the predictions of Fig. \ref {EMCPred}.  In the initial experiments, the  ratio of the structure functions of iron nucleus and deuteron (F$^{N}_2$(Fe)/F$^N_2$(D))  was experimentally found to be decreasing and substantially different  than unity  in the region $0.05 \le x \le 0.65$ (see Fig. \ref {EMCExpe}). The interpretation of the experimental results was that the  inelastic structure functions of nucleons measured in nuclei are different from those  of quasi-free nucleons in the deuteron \cite {Au83, As88}. It was also found that this difference was growing with the mass number A of the nucleus, which indicated that the nucleon structure modification is proportional to the nuclear density. The modification   of the quark  momentum distribution  of bound nucleons in the nucleus, as compared to that of free nucleons, became known  as  the nuclear European Muon Collaboration ( EMC) effect. 

\begin{figure}[h]
\centering
\includegraphics[scale=0.42]{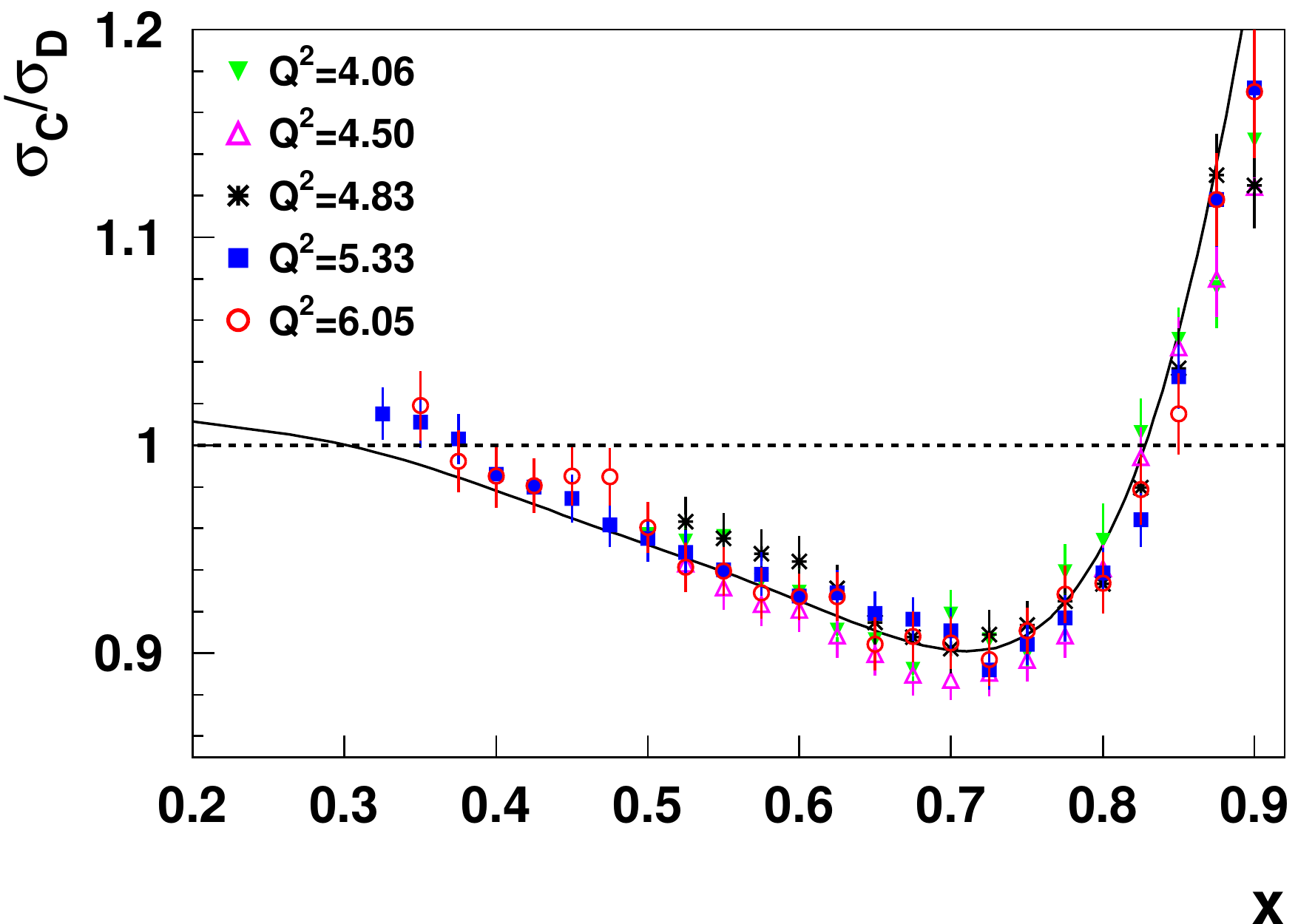}
\caption{  Carbon EMC ratios for the highest $Q^2$ settings ($Q^2$ quoted at x = 0.75). Uncertainties are the combined statistical and point-to-point systematic. The solid curve is  the SLAC fit to the Carbon EMC ratio. Figure from  \cite {Se09}. }
\label {CarbonEMC}
\end{figure}

Recent measurements  at the Jefferson lab (shown in Fig. \ref  {CarbonEMC} for Carbon) verified the nuclear EMC effect with unprecedented accuracy for a wide range of nuclei, confirming that the EMC  ratio R$_{EMC}$= 2 $(\sigma_{eA})_{is}$/ (A$ \sigma_{eD}$) ( where $(\sigma_{eA})_{is}$ is the average cross section for isoscalar nucleon and $\sigma_{eD}$ is the  cross section for deuteron)   is below unit for all the targets studied  \cite {Se09}. 

 \section {Multinucleon short-range correlations in nuclei}
 \label {sec:introSRC}
 The main theoretical picture that describes the bulk properties of nuclei, is that the nucleons are independent particles moving in an average or mean   field  generated by  the remaining $\;(A-1)\;$ nucleons in the nucleus. As a result each nucleon is independent of the exact instantaneous position of all other nucleons. The simple mean field model has been successful in correctly  predicting all nuclear magic numbers, as well as in describing  a large amount of nuclear data    \cite {MJ55,ST63, IT93}.
 \begin{figure}[tbh]
\centering\includegraphics[scale=0.42]{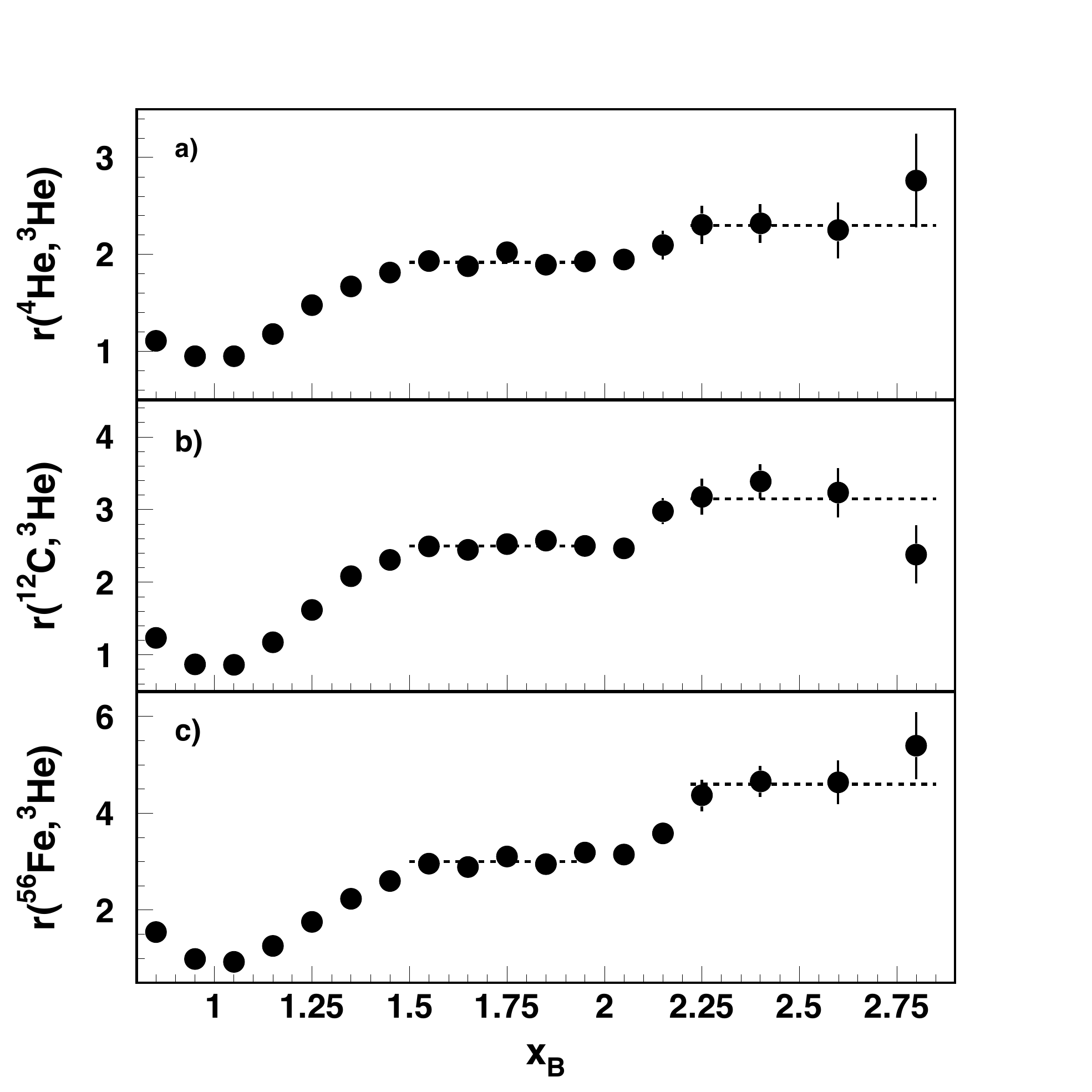}
\caption { {2N and 3N scaling regions for  nucleus A. Figure from Ref.\cite {Kim2}}}
\label{fig:fch4_ratio}
\end{figure} 

The nuclear shell model, based on the above described mean field picture of the nucleus,   is valid for  long range ($\ge$ 2 fm) mutual separation of nucleons, which in the momentum space corresponds to the nucleon momentum being less that $k_F \sim$ 250 MeV/$c$ (\cite{ FZC97}. The nuclear shell model however was found to  break down for inter nucleon distances smaller  
than $2r_N$, where $r_N\approx 0.85$ fm is the radius of the nucleon, on which  two nucleons start to overlap and the notion of the mean field become invalid. Their dynamics  are mainly defined by the NN interaction at short distances  which is dominated by the tensor interaction~($\sim 0.8-1.2$ fm)  and repulsive core ($\le 0.5-0.7$ fm) \cite {GR74, FZC97, RSRB07}. Such  configurations  are generally referred to as 2N Short Range Correlations (SRCs) in the nucleus \cite {FGross,FS92, FSS08A, OrDouglas, AHR12}. 
 
Theoretical analysis show that the nucleons belonging to such 2N SRCs  have large (greater than $k_F$) relative momentum and low (smaller than $k_F$) center of mass  momentum. There are  also
 lower  probability configurations such as three (3N) or multi-nucleon (MN) SRCs, a  very important high density structures  present in the ground state wave function of the nucleus, responsible for  high momentum  nucleons much  above the Fermi momentum $k_F$. Any experiment designed to access such MN SRCs must probe   the bound nucleon in the nucleus   at very large momenta.   Lepton-nucleus scattering,  at large values of the Bjorken parameter, $x_B$,  is  the most appropriate experiment to prove such  MN SRC nuclear structures. 

The kinematic region for lepton-free nucleon scattering is 0$ < x_B <$ 1, whereas for bound nucleon in a nucleus A is 
0$ < x_B <$ A. It is  expected  that scattering  from j-nucleon SRC will dominate at j-1$ < x_B <$ j\cite {FSDS93}. If a lepton scatters off the nucleon from j-nucleon SRCs, then it is expected that the   cross section ratio
\begin{align}
R (A_1, A_2)  =\frac{ \sigma (A_1, Q^2, x_B) /A_1}{ \sigma (A_2, Q^2, x_B) /A_2}, 
\label{crossect_ratio}
\end{align}
 where $\sigma (A_1, Q^2, x_B)$ and  $\sigma (A_2, Q^2, x_B)$ are the inclusive lepton scattering cross sections  of nucleus $A_1$ and $A_2$ respectively, will scale, that is to be constant.  The scaling results from the  dominance of MN SRCs in the high momentum component of the nuclear wave function. Hence,  plateaus are expected  in the ratio of the inclusive cross sections of heavy nucleus to light nuclei such as $^3$He, showing  that the momentum distributions at high momenta have the same shape for all nuclei differing  only by a scale factor \cite{FS88,Kim1,FSDS93}.

Recent experimental studies of  high energy $eA$ and $pA$ processes \cite{Kim1,Kim2,Fomin2011,Aclander1999,Yaron2002,isosrc,Eip3,Eip4} 
resulted in a  significant progress in understanding the dynamics  of 2N SRCs in nuclei.  The series of  electron-nucleus inclusive scattering experiments\cite{Kim1,Kim2,Fomin2011} have confirmed the 
prediction of the  scaling for the ratios of inclusive cross sections of a nucleus 
to the deuteron ($^3$He)  in the kinematic region $x_B >1$ (dominated by 
the scattering from the bound nucleons with momenta $p > k_{F}\sim250$~MeV/$c$)  Within the 2N SRC model, 
these ratios allowed to extract the parameter $a_2(A,Z)$ which characterizes the probability of finding 2N SRC in 
the nucleus relative to the deuteron.  Results of  the above mentioned  experiments with  $Q^2 \ge$ 1.5 GeV$^2$,  are  shown in Fig. \ref{fig:fch4_ratio}.The cross section ratios in Fig. \ref{fig:fch4_ratio} scales initially in  the region  1.5$ <$ x$ <$ 2.0, which indicates dominance of  2N SRC in this region, and scales a second time for x $>$ 2.25, indicating dominance of 3N SRC. 


 \subsection {Strong Correlation Between Nucleon-Nucleon  Short Range Correlations  and the EMC Effect}
\label {sec:SRC_EMC}
\begin{figure}[htb]
\begin{center}
\includegraphics[scale=0.50]{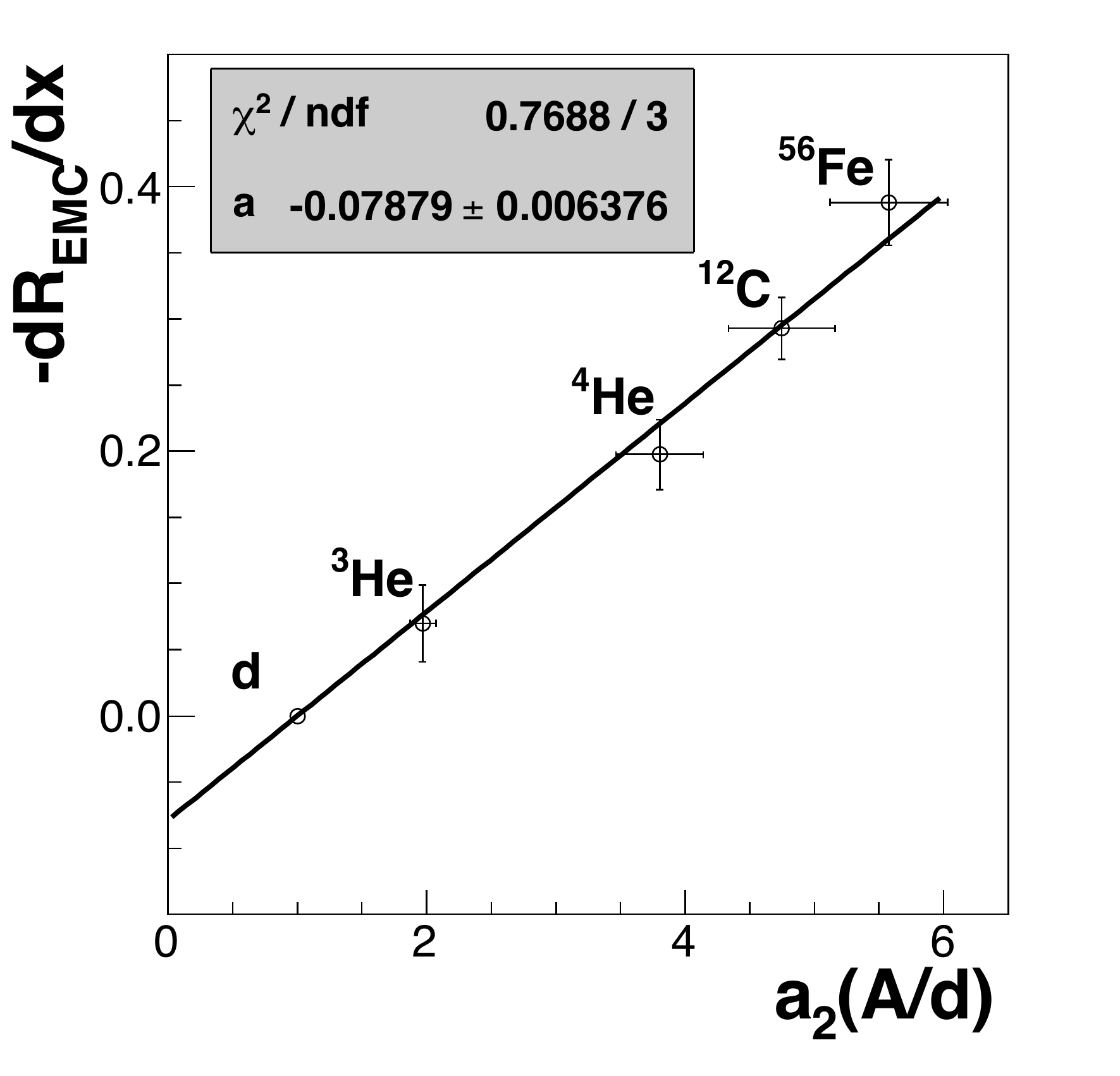}
\caption{ \label{fig:SRCEMC} The EMC slopes versus the SRC scale
  factors.   Figure from Ref. \cite {LW1}}
\end{center}
\end{figure}

Since both the EMC effect and the 2N SRCs  depend on  the mass number A and on the nuclear density, it was predicted that they were strongly correlated \cite {HGP10}. The prediction  was probed by experiments on which the correlation between the strength of the nuclear  EMC effect and the strength of 2N SRCs was observed as it is shown in Fig. \ref{fig:SRCEMC} \cite {LW1}.  
The strength of the EMC effect for a nucleus A in  Fig. \ref{fig:SRCEMC} is represented by  the slope of the EMC ratio (R$_{\textrm {EMC}}$) of the per-nucleon deep inelastic cross section of nucleus A relative to the deuteron, dR$_{\textrm {EMC}}$/dx, in the region 0.35 $\le$ x $\le$ 0.7 \cite {HGP10}; and the strength of the 2N SRCs is represented by  the nuclear scale factor of the nucleus A relative to the deuteron a$_2$ (A/d), which represents  the probability of having 2N SRCs in the nucleus A. 

The EMC-SRC correlation is  a very important experimental result that provides new insight into the origin of the EMC effect. Since the SRC structure in  the nucleus implies high momentum bound nucleons, it indicates that the EMC effect is only the result of  the high momentum component of the nuclear wave function, so that the possible modification of the parton distributions in nucleons in the nucleus occurs only in nucleons belonging to SRCs \cite {MS0912, HHM13}. 

Hence, a further understanding  of   the dynamical origin of the  observed EMC-SRC correlation requires   a theoretical model for  the nuclear spectral function,  $S^N_A(\mathrm {p}, E_m)$ , that  describes the 2N and 3N SRCs in a  consistent way.  The model will allow the calculation  of the cross sections for lepton -nucleus scattering  from Eqs. (\ref {NucleusDIS}), (\ref {fa1dis}), and  (\ref {fa2dis}). Thus,   the  main motivation of  the research presented in this dissertation  was to develop a self consistent theoretical model for calculation of the nuclear spectral functions
in the domain  of 2N and 3N short range correlations. One of the important requirements in developing such models was that the calculated spectral functions should  include all recent findings   that have been made in experimental and theoretical studies of SRCs in nuclei as those described in the following sections (\ref {sec:SRCModel} and \ref {Sec.Phenom}) of the present chapter.

  \subsection {Model for  2N and 3N  short-range correlations in nuclei}
 \label{sec:SRCModel}
 \begin{figure}[h]
\centering
\includegraphics[scale=0.90]{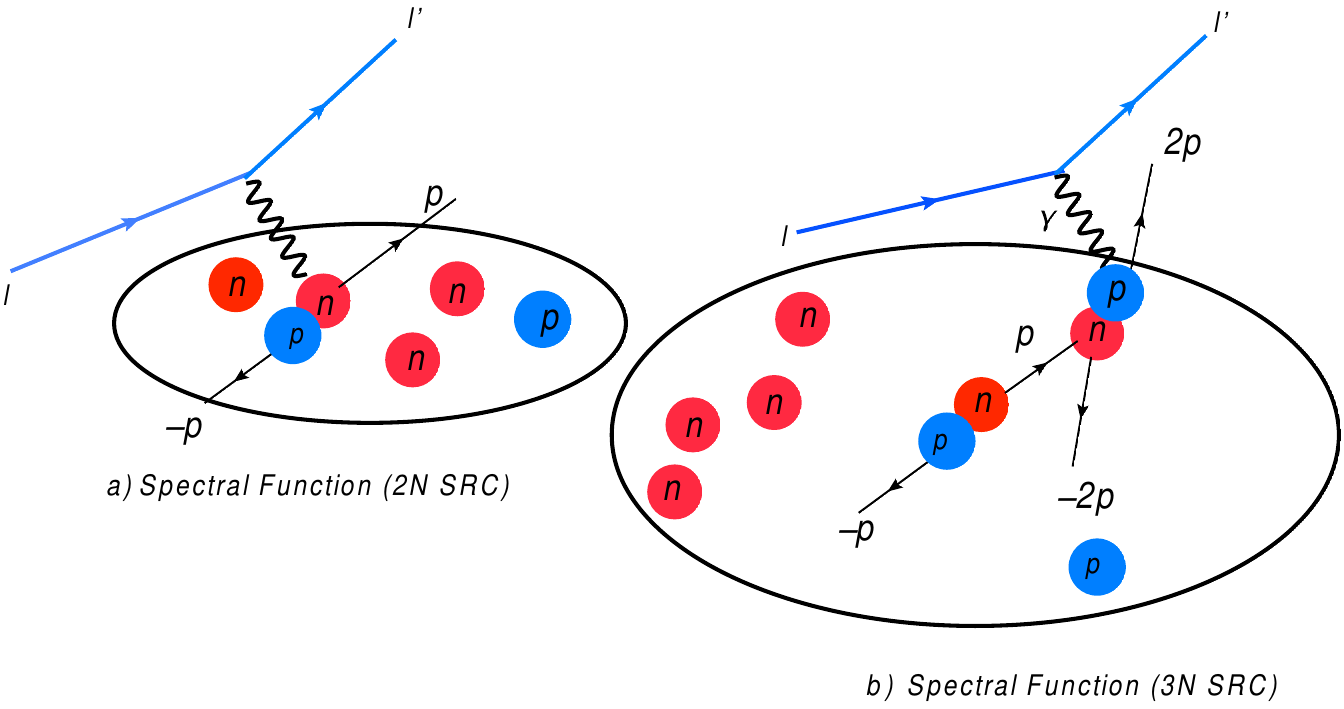}
\caption{2N and 3N SRC models for nuclear spectral functions} 
\label {2N3NSRCModel}
\end{figure}
Despite  impressive recent progress in \textit {ab initio} calculations of nuclear structure for the mean field (shell) model  (see e.g. Ref.\cite{Jansen:2014qxa}), their
relevance to the  development of the   spectral functions at large momenta and removal energies, where the dominance of SRCs is expected,  is rather limited. 
Not only the absence of  relativistic effects but also the impossibility of identifying  the  relevant nucleon-nucleon (NN) interaction potentials 
makes such a program unrealistic.  One way for  progress  is to  develop   theoretical models that use
the short-range NN correlation approach  in the description of the high momentum part of the nuclear wave function 
(see, e.g., \cite{FS81,Ciofi1991,CiofiSimula,Jan2012,Alvioli2012,Alvioli2013,Rios2013,Jan2014,Ciofi2015,Neff2015}).  
In such  an approach, it is possible  to consider the empirical knowledge  of SRCs acquired from  different high energy 
scattering experiments thus  reducing in some degree  the theoretical uncertainty related to the description of 
 high momentum nucleon  in the nucleus. 

The main goal of chapter 2 of  the present dissertation is to develop a model for  high momentum nuclear spectral function  using  the  several   phenomenological observations  
obtained in  recent years in studies of  the  properties of two-nucleon  SRCs\cite{Kim1,Kim2,isosrc,Eip3,srcrev,Fomin2011,srcprog,Cosyn2015,Eip4,contact}. 
A model is  first developed describing the nuclear spectral function at large momenta and missing energies 
dominated by 2N SRCs with their center of mass motion generated by the mean field of the $A-2$ residual nuclear system.

Even though some experiments have shown evidence for 3N SRCs \cite {Kim2}, there are other experiments \cite {Fomin2011} that did not see such evidence.  Considering  the  experimental ambiguity, a theoretical framework for calculating the contribution of 3N SRCs
to the nuclear spectral function is developed,  using a   model in which such correlations are generated by two sequential 2N short 
range correlations  as it is shown in Fig. \ref  {2N3NSRCModel}. Hence,  the  phenomenological knowledge of the properties of 2N SRCs is sufficient to 
calculate both the 2N and 3N  SRC contributions to the model of nuclear spectral function.

The  model is assumed  to be valid for nucleon momenta $k_{src} \le p \le$ 1000 MeV/c, where $k_{src}$ is a momentum  characteristic to the 
2N SRC. It is sufficiently large that 2N  SRCs can be factorized from residual mean field interaction. As a result the model  will have 
 limited validity in the transitional region of $k_{F} \le p < k_{src}$ where 
the role of the long-range correlations are more relevant.

 \subsection{Phenomenology of two nucleon short-range correlations in nuclei}
 \label{Sec.Phenom}

High energy  semi-inclusive experiments\cite{isosrc,Eip3}  probed   for the first time  the isospin composition of 
 2N SRCs, observing strong  (by factor of 20) dominance  of the  $pn$ SRCs in nuclei,   
as compared to  the $pp$ and $nn$  correlations, for internal momentum range  of $\sim250-650$~MeV/$c$.
The experimental results are understood by considering 
 the dominance of the tensor forces in the NN interaction at the 
momentum range corresponding to the average nucleon separations of $\sim1.1$~fm \cite{isosrc,eheppn2}.   
The tensor interaction projects the NN SRC part of the 
wave function to  the  isosinglet - relative angular momentum, $L=2$,  state,  almost identical to the high momentum part of the  $D$-wave component of 
the deuteron wave function.  As a result $pp$ and $nn$ components of the NN SRC are strongly suppressed since they 
are dominated by the  central NN potential with relative angular momentum $L=0$ \cite {PB75, FZC97, RSRB07} .

On the basis of  the above observation of the strong dominance of $pn$ SRCs,  it  was  predicted 
that single proton or neutron  momentum distributions in the 2N SRC domain are inversely proportional to their 
relative fractions in  nuclei  \cite{newprops, proa2}.  The prediction is in agreement with the results of  variational Monte-Carlo calculation
of momentum distributions of  light nuclei\cite{Wir2014} as well 
as for medium to heavy nuclei following   the  SRC model calculations of Ref.\cite{Jan2014}.  
The recent finding  of the $pn$ dominance in heavy nuclei (up to $^{208}$Pb)\cite{Eip4} 
validates the universality of the above prediction for  the whole spectrum of atomic nuclei.
The inverse proportionality of the high momentum component  to the relative fraction of 
the proton or neutron is  important for asymmetric nuclei and they need to be included in the 
modeling of nuclear spectral functions in the 2N  SRC region.

The $pn$ dominance in the SRC region  and its relation to the high momentum part of the deuteron wave function makes 
the studies of the deuteron structure at large internal momenta a very important part for the SRC studies in nuclei.
In this respect, the recent experiments \cite{Boeglin11,BS15} and planned new measurements \cite{Boeglinproposal}  of high energy exclusive electro-disintegration of the deuteron opens up  new  possibilities in the extraction of the deuteron momentum distribution 
at very large momenta. The measured  distributions  can then be  utilized in 
the calculation of the nuclear spectral functions in the multi-nucleon SRC region.

Finally, another progress relevant to the SRC studies was  the extraction of the center of mass momentum distribution of 2N SRCs 
from the data on  triple coincidence scattering in $A(p,ppn)X$ \cite{Tang}  and  $A(e,e^\prime,pn)X$ \cite{Shneor,Korover} reactions.
The Gaussian form and the width of the extracted distributions were in a good agreement with the predictions made in Ref. \cite{CiofiSimula},
which were based on the estimate of the mean kinetic energy of the NN pair in shell-model description of nuclei.
Similar results have been also  obtained within  the   correlated wave function method of Ref. \cite{Colle:2013nna}.


\section{Nuclear spectral function model}
\label{sec:SpectModelInt}

The above discussed phenomenology will provide  the necessary empirical input for modeling 
nuclear spectral functions in the SRC region. The model for high momentum nuclear spectral function, developed in chapter 2 of this dissertation, have  two regions determined by the range of momentum considered. For momenta below the Fermi momentum, $k_F$, a mean field spectral function is constructed  by using a nonrelativistic approach  to estimate the ground state wave functions   \cite {RS04}. For momenta above 400 MeV, a relativistic  multi-nucleon short range correlation model of the  spectral function is  obtained, which describes   the high momentum and high missing energy of   two and three   nucleons in short range correlations (2N and 3N SRC), for  symmetric and asymmetric nuclei. 

Since the domain of multi-nucleon SRCs is characterized by the relativistic momenta of the probed  nucleon, 
special care should be given to the treatment of  relativistic effects. 
To identify the relativistic effects,  in Sec.\ref{sec:Ch2Intro},   the nuclear spectral function is defined as 
a quantity which is  extracted in the semi-exclusive  high energy process 
whose scattering amplitude can be described through the covariant effective Feynman diagrams.  
The covariance here is important  to  consistently trace the relativistic effects related to the propagation of 
the bound nucleon. Then, the part of the covariant diagram  which reproduces  the nuclear spectral function is precisely identified.   Two approaches are adopted  for modeling the nuclear spectral function: 
virtual nucleon  and light-front approximations, general features of which are described in Sec.\ref{sec:Vacuum}. 
Section \ref{sec:Diagram} outlines the calculation of nuclear spectral functions  using the effective Feynman diagrammatic  method, identifying the diagrams 
corresponding to  the mean field, 2N SRC with center of mass motion and  3N SRC contributions.In  Secs.\ref{sec:SpecVNA} and \ref{sec:SpecLFA},  the detailed derivation of the nuclear spectral functions within the virtual nucleon and the light-front 
approximations are presented.  

Chapters 3 and 4 are dedicated to the development of computational models to evaluate nuclear spectral functions, density matrices and momentum distributions for  a wide range of light and heavy nuclei. The results of the  spectral function models  will be compared with \textit {ab initio}, nonrelativistic   quantum Monte Carlo calculations (QMC) (for A $\leq $ 11) \cite {AV1895, VMC01, PVW02, Wir2014}.   The values of $a_2$ and $a_3$ are also predicted,  which represent  the probability of having 2N and 3N  SRCs in the nuclear ground state wave function,  respectively.\\


 


\chapter{Multinucleon short-range correlation model for  nuclear spectral functions:  Theoretical framework}


\label{sec:Ch2Intro}
The definition of nuclear spectral functions used in the present dissertation  is derived by  identifying  a nuclear ``observable" which can be 
 extracted from the cross section of  the large momentum ($\gg$nucleon mass)  transfer  semi-inclusive 
 $h+A \rightarrow h^\prime + N + (A-1)^*$ reaction in which the $N$ can be unambiguously identified as a struck nucleon carrying almost all the energy and momentum transferred to the nucleus by the probe $h$. The reaction is specifically chosen to be semi-inclusive  so that it allows,  in the approximation in which no final state interactions are considered,  to relate  the missing momentum  and energy  of the reaction  to the properties of bound nucleon in the nucleus.  When those conditions are satisfied the extracted ``observable", referred to as a nuclear spectral function, represents a joint probability of finding a bound nucleon in the nucleus with given missing momentum $p$ and removal energy $E_m$. 

The  models  of nuclear spectral functions developed in the present   dissertation must be relativistic, since they will be used to describe  bound nucleons with  high momenta and high removal energies. The relativistic effects are accounted for  by using effective Feynman 
diagrammatic approaches similar to those developed in Refs. \cite{gea, geaproceed, ms01, eheppn1}.  One problem associated with the relativistic domain is the existence of vacuum fluctuations that implies the existence of negative energy components 
which are not related to the   probability amplitude of finding a nucleon with a given momentum
in the nucleus, and therefore, are not components of the nuclear spectral function.

Chapter 2 is organized as follows.  The two approaches to deal with vacuum fluctuations: virtual nucleon (VN) and light-front (LF) approximations are described in section \ref {sec:Vacuum}. Section \ref{sec:Diagram} outlines the modeling of nuclear spectral functions  using  the effective Feynman diagrammatic  method, identifying the diagrams and the corresponding amplitudes
which represent partial contributions to the total amplitude by nucleons in  the nuclear mean field, in two nucleons  short-range correlation with center of mass motion,  and  in  three nucleons  short-range correlation. The steps for the calculation of the models of the nuclear spectral function  are also defined in section \ref{sec:Diagram}.
Sections \ref{sec:SpecVNA} and \ref{sec:SpecLFA} include the detailed derivations of the models of nuclear spectral functions within the virtual nucleon and  the light-front 
approximations respectively.  
Section \ref{sec:Ch2Summ} summarizes the results of the chapter.

\section{Approaches to deal  with vacuum fluctuations}
 \label{sec:Vacuum}
The vacuum fluctuations are a purely relativistic phenomena associated with  the existence of particles and antiparticles that can pop out from the vacuum and then disappear into it.  The notion of the antiparticle was first proposed by Dirac in 1932 to explain the solutions of the Klein-Gordon equation with negative energies. The positron was predicted as the antiparticle of the electron with a positive charge and a negative energy. 

Stuckelberg in 1941 and Feynman in 1948 proposed that a negative energy solution describes a particle which propagates backward in time, or equivalently a positive energy antiparticle propagating forward in time. This concept  was  incorporated  in the Feynman diagrams,  a very powerful method of calculation in quantum field theory. With the help of Feynman diagrams it is possible to show that, for certain time ordering of process, a pair  particle- antiparticle may appear  spontaneously from the vacuum, and  unless there is some external energy carried by a probe, the pair will disappear back into it \cite {HM84}. 

\begin{figure}[h]
\centering\includegraphics[scale=0.78]{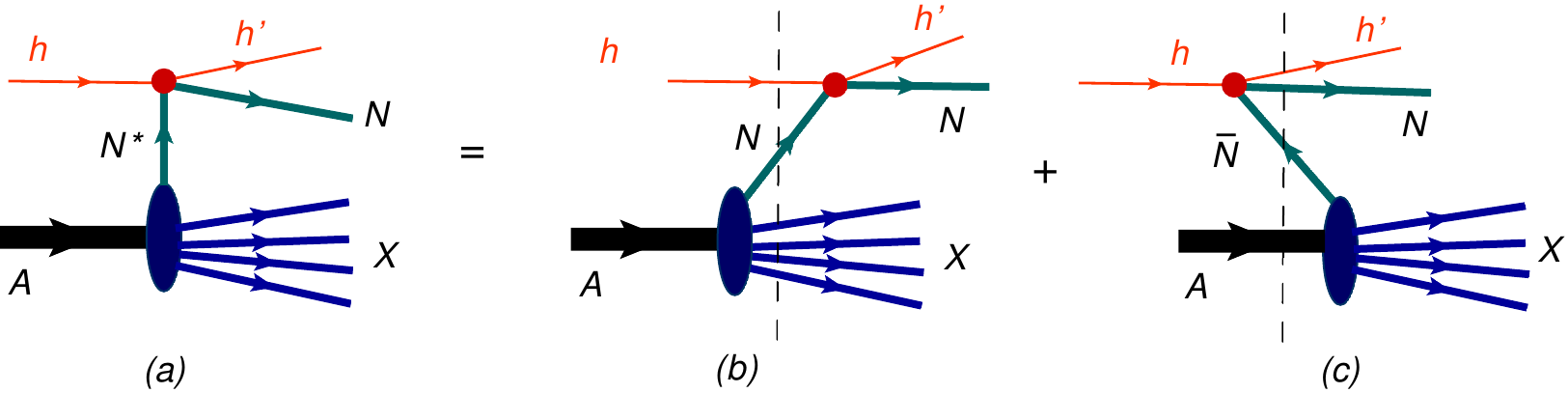}
\caption{Representation of the covariant Feynman amplitude through the sum of the 
time ordered amplitudes. Panel (b) corresponds to the scenario in which first, the bound 
nucleon is resolved in the nucleus which interacts  with the incoming probe $h$. In panel (c),  initially, the  incoming probe produces 
a $\bar N N$, anti-nucleon and  nucleon pair, with subsequent absorption of the $\bar N$ anti-nucleon in the nucleus.}
\label{Fig:R_diagram}
\end{figure}

The models, developed in the present   dissertation, for  relativistic  nuclear spectral functions  are derived from  an effective Feynman 
diagrammatic approach for calculation of the  $h+A \rightarrow h^\prime + N + (A-1)^*$ reactions (Fig.\ref  {Fig:R_diagram}) derived   in Refs. \cite{gea,geaproceed,ms01}. In the approach the 
covariant Feynman scattering amplitude is expressed through the effective nuclear vertices, vertices which are  related to the scattering 
of the probe $h$  with the bound nucleon, as well as vertices related to the final state of the reaction. 

The nuclear vertices related with the bound nucleon can not be associated  \textit {a priori} with the single nucleon  wave function of the nucleus, since they  contain negative energy components 
which are related to the  vacuum fluctuations rather than the  probability amplitude of finding nucleon with given momentum 
in the nucleus.
The problem of vacuum fluctuations  is illustrated in the diagrammatic representation of the reaction shown in Fig. \ref{Fig:R_diagram},  in which
the covariant diagram~(a) is  a sum of two non-covariant time ordered scattering diagrams (b) and (c).
Here, for the  calculation of the Lorentz invariant 
amplitude  of Fig. \ref{Fig:R_diagram}(a),  the  Feynman diagrammatic rules (given in Ref. \cite{ms01}) can be applied. 
However the nuclear  spectral function can only be formulated for the diagram of Fig. \ref{Fig:R_diagram}(b), where 
the time ordering is such that it first exposes the nucleus as being composed of a  bound nucleon and residual nucleus, 
followed by an  interaction of the incoming probe $h$ off the bound nucleon.  The 
other time ordering  [Fig. \ref{Fig:R_diagram}({c})] presents a  very different scenario of the scattering 
in which the probe produces a $\bar N N$, anti-nucleon and  nucleon   pair with subsequent absorption of the $\bar N$ anti-nucleon in the nucleus. The 
later is usually referred to as a $Z$-graph and is not related to the nuclear spectral function. 
It is worth noting that the $Z$-graph contribution is a purely relativistic effect and does not appear in the 
non-relativistic formulation of the nuclear spectral function. The $Z$-graph contribution however increases 
with an increase of the momentum of the bound nucleon [see e.g. Ref. \cite{FS81}]. 
 
 The above discussion  indicates that while defining the nuclear spectral function is straightforward in the non-relativistic domain 
(no $Z$-graph contribution), its definition becomes increasingly ambiguous  with an  increase of the momentum of the 
bound nucleon.   The ambiguity is reflected in the lack of uniqueness in defining the nuclear spectral function in the domain where it is  expected to  probe SRCs inside the nucleus.

In the present  dissertation two approaches are considered to deal with the vacuum fluctuations, so that a unique definition of  the nuclear spectral function from   the covariant scattering amplitude is obtained. In the first approach, referred to   as the virtual nucleon~(VN) approximation,   the $Z$-graph contribution is neglected \cite{noredepn,edepnx}.  In the  second  approach, referred to   as the  light front (LF) approximation,   the $Z$-graph contribution is kinematically suppressed \cite{Weinberg,FS81,FS92,GM00}.

\subsection{Virtual nucleon (VN) approximation approach}
 \label{Sec.VacuumVNA}

 In the  virtual nucleon~(VN) approximation approach, the $Z$-graph contribution is neglected by considering only the positive energy pole for the bound nucleon propagator in the nucleus.  The energy and momentum conservation  in the VN approach requires the interacting 
nucleon to be virtual which renders certain ambiguity in treating the propagator of the bound nucleon.  The ambiguity is solved   by  recovering the energy and momentum of the interacting nucleon from kinematic parameters of on-shell spectators 
[see Ref. \cite{Gross:1982} for general discussion of the spectator model of relativistic bound states].

The advantage of the VN approximation is that the spectral function is expressed through the 
nuclear  wave function defined in the rest frame of the nucleus which in principle can be calculated using 
conventional NN potentials.  

One  shortcoming of the VN approximation is that while it satisfies the baryonic number conservation law,  the momentum sum rule is not satisfied reflecting the virtual nature of the probed  nucleon in the nucleus.

 \subsection{Light-front (LF) approximation approach}
 \label{Sec.VacuumLFA}
 The light-front representation  of the space-time was first proposed by Dirac \cite {Dirac49} as a form of relativistic Hamiltonian dynamics. The three forms of Hamiltonian dynamics described  by Dirac were: the instant form, the front form, and the point form. These forms  differ in the hypersphere on which the fields are analyzed \cite {Brodsky:1997de}. In  the front form the hypersphere is a plane tangent to the light-cone, that is the three-dimensional surface in space-time formed by a plane wave frame advancing with the velocity of light, such a surface was called a \textit {front} by Dirac. 
 
 
 The space-time coordinates in the light-front are defined as \cite {Brodsky:1997de,AH98 }: \\
  \textbf{Lorentz Vectors:} The contravariant four-vectors of position $x^\mu$ are written as
 \begin{align}
 x^\mu &= (x^+, x^-, x^1, x^2) = (x^+,x^-, \mathbf {x_\bot}).
 \label{LBcvectors1}
\end{align}
 Its time-like and space-like components are related to the instant form by:
 \begin{align}
 x^+ &= x^0 + x^3,\nonumber\\
 x^-&= x^0 - x^3, 
 \label{LBcvectors2}
\end{align}
respectively, and referred to as the light-front  time and light-front longitudinal  position. The null plane is defined by x$^+$ = 0, that is, this condition defines the  hyperplane that is tangent to the light-cone. The initial boundary conditions for the dynamics in the light-front are defined on this hyperplane. The axis x$^+$ is perpendicular to the plane x$^+ $= 0. Therefore a displacement of such hyperplane for $x^+ > 0$ is analogous to the displacement of a plane in t = 0 to t $> 0$ of the four-dimensional space-time. With this analogy,  x$^+$ is recognized as the time in the light-front, or equivalently, as the light-front time $\tau = x^0 + x^3 $.
The contravariant four-vectors of momentum  $p^\mu$ are written as
 \begin{align}
 p^\mu &= (p^+, p^-, p^1, p^2) = (p^+, p^-,  \mathbf {p_\perp}).
 \label{LBmomentum}
\end{align}
 Its time-like and space-like components are related to the instant form by:
 \begin{align}
 p^+ &= p^0 + p^3,\nonumber\\
 p^-&= p^0 - p^3. 
 \label{LBcmomcomp}
\end{align}
The scalar product between two 4-vectors is defined by:
\begin{align}
 x·p=x^\mu p_\mu =x^+p_+ +x^-p_{-} +x^1p_1 +x^2p_2 = \frac {1}{2} (x^+p^{-} +x^{-}p^+) - \mathbf {x}_\bot \mathbf {p}_\bot.
 \label{LBc2}
\end{align}
if $ x = p $, then the following very useful relation is obtained:
\begin{align}
 p^+ =  \frac {M^2 + \mathrm p^2_\perp}{p^-}, 
 \label{DisperRelat}
\end{align}
where $M$ is the mass of the particle with momentum $p$.\\
The four-dimensional  phase space  differential element is  defined as 
\begin{align}
 d^4x &= dx^0 d^2\mathbf {x}_\bot dx^3= \frac {1}{2} dx^+ dx^-d^2\mathbf {x}_\bot.
 \label{LBVDE}
 \end{align}

In the light-front (LF) approximation  the nuclear spectral function is defined on the light front which corresponds to a
reference frame in which the nucleus has infinite momentum.   
 Weinberg \cite  {Weinberg} showed  that in the infinite momentum frame all diagrams with negative energy, like the
the $Z$-graph,  are  kinematically  suppressed. Then, as a result,  the invariant sum of 
the two light-cone time ordered amplitudes in Fig. \ref{Fig:R_diagram} is equal to   only the  contribution from 
the graph of Fig. \ref{Fig:R_diagram} (b).  Therefore,   the boost invariant LF nuclear spectral function defines   the joint probability of finding a   nucleon in  the nucleus with given light-front momentum fraction, transverse momentum, and invariant mass \cite {MI97 }.  

It is worth noting that the LF approximation satisfies  the baryonic conservation law, as well as   the momentum sum rules, thus providing  a  better framework for studies of  the effects associated with the nuclear medium modification of interacting particles.

The LF approach developed in the present dissertation is field-theoretical, that is the   Feynman diagrams are constructed with effective interaction vertices and the spectral functions are extracted from 
the imaginary part of the covariant forward scattering nuclear amplitude.    Another approach, in LF approximation,  is the 
construction of the  nuclear spectral function based on the relativistic Hamiltonian dynamics representing the interaction 
of fixed number on-mass shell constituents \cite{Salme}.

\section{Diagrammatic approach for modeling   nuclear spectral functions}
\label{sec:Diagram}

\begin{figure}[ht]
\centering\includegraphics[scale=0.99]{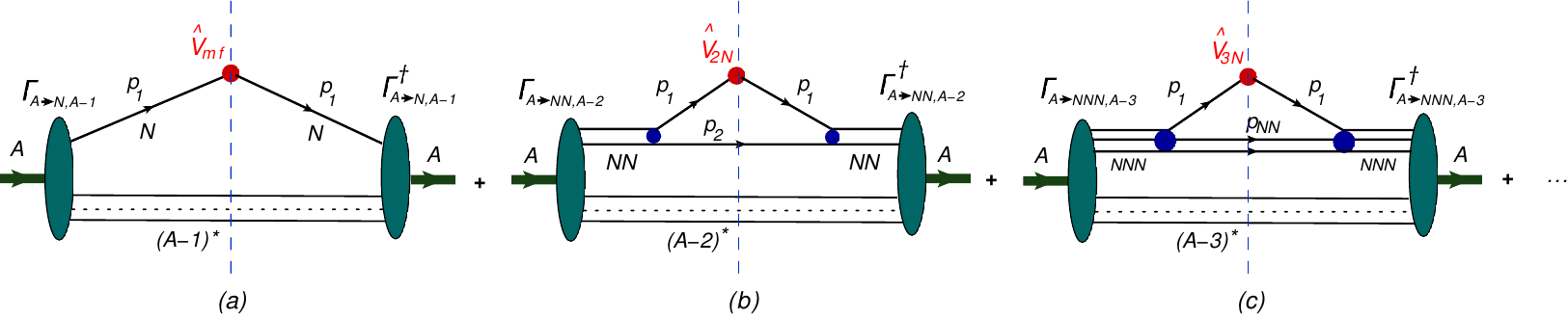}
\caption{Expansion of the nuclear spectral function into the contributions of the mean field (a),
the 2N (b) and the 3N ({c}) SRCs. For each case the initial nuclear transition vertices are different, corresponding to 
transition of $A\to N,A-1$; $A\to NN,A-1$ and $A\to NNN,A-3$ for the mean field, the 2N and the 3N SRCs respectively.
The NN (b)  and NNN (c)  labels identify the 2N and the 3N SRCs with effective vertices  elaborated in the text.}
\label{Fig:spectral_diagram}
\end{figure}

The application of the Feynman diagrammatic rules of Ref. \cite{ms01}   to 
obtain  the mathematical model of nuclear spectral functions starts by identifying   the effective interaction vertices  $\hat V$ shown in Fig. \ref{Fig:spectral_diagram},  such that the imaginary part of the covariant forward scattering  nuclear amplitude will reduce to the nuclear spectral function either in  VN or LF approximations.  The specific form of the vertices can be established by considering the amplitude of Fig. \ref{Fig:R_diagram}(b),  
taking into account the kinematics of the mean field, the 2N, and the 3N SRC scattering within VN and LF approximations, and
with subsequent factorization of the scattering factors related to the external probe $h$.  As a result the $\hat V$ vertices will be different for mean field, 2N, and 3N SRCs. They will also depend on the VN or the LF approximations used to calculate the scattering amplitude.

In applying  the diagrammatic approach,   the forward  nuclear scattering amplitude  $A$ can be expressed  as a sum of  
the   mean field and the multinucleon SRC contributions  as presented in Fig. \ref{Fig:spectral_diagram}, 
with (a), (b), and (c) corresponding to the contributions from nucleons in the nuclear mean-field, and from   the 2N and the 3N short-range correlations respectively:
\begin{equation}
A = A^{MF} + A^{2N} + A^{3N}, 
\end{equation}
where $A^{MF}$, $A^{2N}$, and $A^{3N}$  correspond to the contributions  from the 
diagrams of  Fig. \ref{Fig:spectral_diagram} (a)-\ref{Fig:spectral_diagram}(c) respectively.

Since the mean field contribution is dominated by the momenta of interacting nucleon below 
the characteristic Fermi momentum, $k_{F}$, it is valid to  approximate  the corresponding nuclear spectral function to the result following 
from nonrelativistic calculation.  Hence,  both the VN and the LF approximations are expected to give 
very close results.

For  the 2N and 3N SRCs, the  momenta of probed nucleon  is in  the range $k_{F} <p\le 600-1000$~MeV/$c$ 
and the nonrelativistic approximation is  increasingly  invalid.


\subsection{ Covariant amplitude for a nucleon in the nuclear mean field}
\label{sec:MfAmp}

In the mean field approximation,  the bound nucleon interacts with  the nuclear mean field induced by the  $A-1$ nuclear residual system.
In such approximation,  the  nuclear spectral function corresponds to a configuration in which 
the residual nuclear system is identified as a coherent $A-1$ state with  excitation energy in the order of 
tens of MeV.

Applying the effective Feynman rules \cite{ms01}  to the diagram of  Fig. \ref{Fig:spectral_diagram}(a) corresponding 
to the mean field contribution of nuclear spectral function, the following covariant amplitude is  obtained:
\begin{eqnarray}
ImA^{MF}   = & &    Im\int \chi^{s_A,\dagger}_A \Gamma_{A\rightarrow N,A-1}^\dagger 
{{\sh p}_1 + M_N\over p_1^2 - M_N^2 }\hat V_{MF} {\sh p_1 + M_N\over p_1^2 -  M^2_N}
\left[ { G_{A-1}(p_{A-1},\alpha)\over p_{A-1}^2 - M_{A-1}^2 + i\varepsilon}\right]^{on} \nonumber \\
& & \ \ \ \ \ \ \ \  \times \ \ \Gamma_{A\rightarrow N,A-1} \chi^{s_A}_A {dp^0_{A-1}\over i (2\pi)} 
{d^3 \mathbf p_{A-1}\over  (2\pi)^3},
 \label{eq:Smf}
\end{eqnarray}
where  $M_N$ and $M_{A-1}$ are the masses of the nucleon and  of the $A-1$ nuclear residual system respectively,  $\chi_A$ is the nuclear spin wave function, 
$\Gamma_{A\rightarrow  N,A-1}$ represents the covariant vertex of the 
$A\rightarrow N + (A-1)$ transition,  $G_{A-1}$ describes the propagation of the $A-1$ nuclear residual system in the 
intermediate state having an excitation $\alpha$. Following the effective Feynman rules \cite{ms01},   the  propagator of the form $(p_{A-1}^2 - M_{A-1}^2 + i\varepsilon)^{-1}$ and a factor $(2\pi)^{-4}$ have been assigned to  the spectator  $A-1$. The  label $[\cdots]^{on}$ indicates that   the cut diagram is  estimated  so that  the residual nuclear system is on mass shell. The abbreviated notation $\sh p_1 = \gamma^\mu p_\mu$, where  $\gamma^\mu$ are the Dirac $\gamma $ matrices, has been used \cite {HM84}

\subsection{Covariant amplitude  for two nucleons  in short-range correlation  } 
\label{sec:2NsrcAmp}
For   two-nucleons  in short-range correlation, it is  assumed that the intermediate nuclear state  consists of two correlated  fast ($> k_{F}$)  nucleons and a slow  ($< k_{F}$) 
coherent $A-2$ nuclear residual system.   

Applying the effective Feynman rules \cite{ms01}  to the diagram of  Fig. \ref{Fig:spectral_diagram}(b) corresponding 
to the 2N SRC contribution of nuclear spectral function, the following covariant amplitude is  obtained:
\begin{eqnarray}
& & ImA^{2N}  = \nonumber \\
& &   \ \ \ \  Im\int \chi^{s_A,\dagger}_A \Gamma_{A\rightarrow NN,A-2}^\dagger {G(p_{NN}, s_{NN})\over p_{NN}^2 - M_{NN}^2 }
 \Gamma^\dagger_{NN\rightarrow N,N} {\sh p_1 + M_N\over p_1^2 - M_N^2} \hat V_{2N}
 {\sh p_1 + M_N\over p_1^2 - M_N^2 } 
 \left[ {\sh p_2 + M_N\over p_2^2 - M_N^2 + i\varepsilon}\right]^{on}    \nonumber \\
& & \ \ \ \ \ \  \times \ \  \Gamma_{NN\rightarrow N,N}
 {G(p_{NN}, s_{NN})\over p_{NN}^2 - M_{NN}^2 } 
\left[ {G_{A-2}(p_{A-2}, s_{A-2})\over p_{A-2}^2 - M_{A-2}^2 + i\varepsilon}\right]^{on} \Gamma_{A\rightarrow NN,A-2} \chi^{s_A}_A  \nonumber \\
& & \ \ \ \ \ \  \times \ \  
 {dp^0_{2}\over i (2\pi)}{d^3 \mathbf p_{2}\over (2\pi)^3} {dp^0_{A-2}\over i (2\pi)} {d^3 \mathbf p_{A-2}\over  (2\pi)^3}, 
  \label{sp:2N_cov}
 \end{eqnarray}
 where  $M_{NN}$ is the mass of the 2N SRC system, $\Gamma_{A\rightarrow NN,A-2}$ now describes the  transition of the 
 nucleus A to  the  $NN$ SRC and  coherent $A-2$  nuclear residual  state, while 
 the  $\Gamma_{NN\rightarrow N,N}$  vertex describes the short range $NN$ interaction that generates  two-nucleon correlation in the nuclear spectral function.

\subsection{Covariant amplitude  for three nucleons  in short-range correlation } 
\label{sec:3NsrcAmp}
The nuclear spectral function that results from  3N short-range correlations is described in Fig. \ref{Fig:spectral_diagram}({c}) in which the 
intermediate state consists of three fast ($> k_F$) nucleons and a slow ($<k_{F}$) coherent $A-3$ nuclear residual  system.

The dynamics of the 3N SRCs allow more complex interactions than that of the 2N SRCs.  One of the complexities is 
the irreducible three-nucleon forces that can not be described by the  NN interaction only.  Such interactions may contain 
inelastic transitions such as the $NN\rightarrow N\Delta$ interaction.    Some studies demonstrated\cite{eheppn2} that 
irreducible three-nucleon forces predominantly contribute at very large magnitudes of missing energy characteristic to 
the $\Delta$ excitations $\sim 300$~MeV/$c$.  Thus for nuclear spectral functions for which the missing energy does not exceed the 
$\Delta$ resonance threshold $\sim M_{\Delta}- M_{N}$,   only  the   contributions of the $NN\rightarrow NN$  interactions need to be   considered.
\begin{figure}[h]
\centering\includegraphics[scale=0.8]{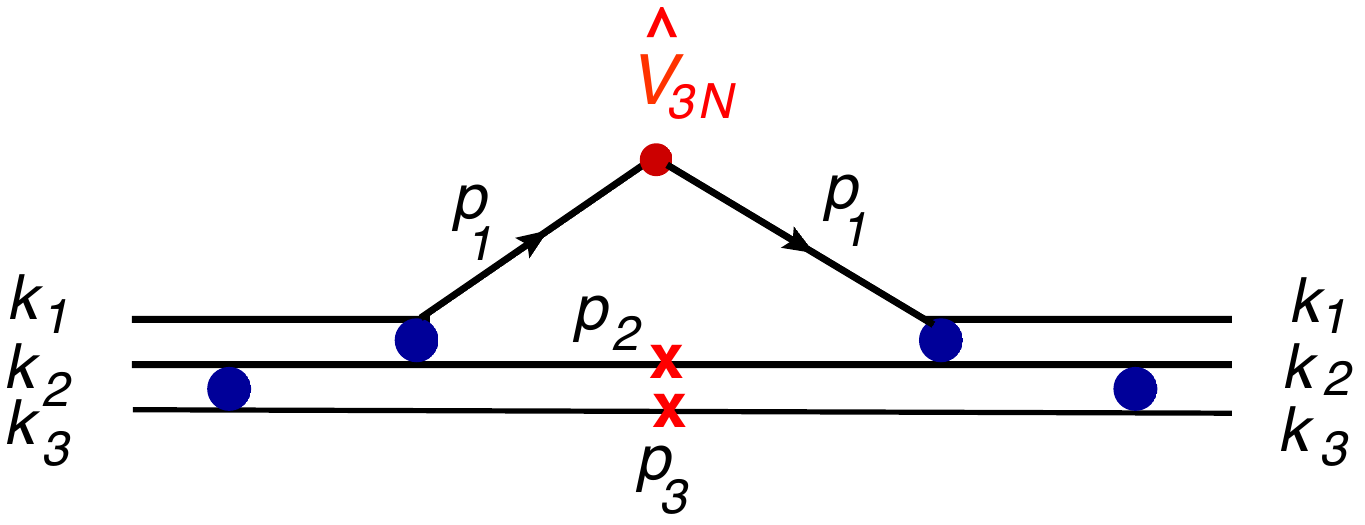}
\caption{Diagram corresponding to the 3N SRC  contribution to the nuclear spectral functions.}
\label{Fig:3nSpectral}
\end{figure}

 In the    two  sequential NN short-range interaction scenario for the   generation  of  3N SRCs,    
     the  spectral function can 
be represented through the diagram of Fig. \ref{Fig:3nSpectral}, where  it is assumed that     the contribution of the 
low momentum  $A-3$ nuclear residual system is  negligible. The assumption results from the fact that  a much larger momenta are  involved in the 3N SRCs as compared
to the momenta  in  the 2N SRCs discussed in the previous section. As a result the effects due to the center of mass motion of  the $A-3$ system can be safely  neglected. 

In  the collinear approximation,  the initial three collinear nucleons undergo two short-range NN interactions generating one nucleon with much larger momenta  than the other two.  It is  assumed  that  the momentum fraction of the 3N SRCs carried by each initial nucleon is unity  and that their total transverse momenta is  neglected [see Eq. (\ref {coll_initial})].
Within the VN approximation,  the collinear approximation assumes that the initial total  momentum of  the three nucleons 
is much smaller than  $k_{src}$ -momenta  characteristic to NN SRC- and therefore can be neglected ($\mathbf k_{1} + \mathbf k_{2} + \mathbf k_{3} = 0 $).
The collinear approximation is commonly used in the  calculation of 
the quark structure function of the nucleon in the valence quark region. Hence the  calculations in the present dissertation 
for the  LF approximation are analytically similar to the QCD calculation of the nucleon structure function.

Applying the effective Feynman rules \cite{ms01}  to the diagram of  Fig. \ref{Fig:3nSpectral}  corresponding 
to the 3N SRC contribution of the nuclear spectral function, the following covariant amplitude is  obtained:
\begin{eqnarray}
ImA^{3N} \vspace{-0.9cm}    &  = &  Im\int \bar u(k_1,\lambda_1) \bar u(k_2,\lambda_2) \bar u (k_3,\lambda_3) 
  \Gamma^\dagger_{NN\rightarrow N,N} {\sh p_{2^\prime} +M_N  \over p_{2^\prime}^2 - M_N^2}
 \Gamma^\dagger_{NN\rightarrow N,N} 
 {\sh p_1 + M_N\over p_1^2 - M_N^2} \hat V_{3N}
 {\sh p_1 + M_N\over p_1^2 - M_N^2}  \nonumber  \\
& &  \times \  \left[  {\sh p_2 + M_N\over p_2^2 - M_N^2 + i\varepsilon} \right]^{on}\Gamma_{NN\rightarrow N,N}
{\sh p_{2^\prime} + M_N\over p_{2^\prime}^2 - M_N^2} 
\left[ {\sh p_{3} + M_N\over p_{3}^2 - M_N^2 + i\varepsilon}\right]^{on}  \Gamma_{NN\rightarrow N,N}  \nonumber \\
& & \times  \ u(k_1,\lambda_1)  u(k_2,\lambda_2)  u(k_3,\lambda_3) 
  {d p^0_{2}\over i (2\pi)} {d^3 \mathbf  p_{2}\over (2\pi)^3} {dp^0_{3}\over i (2\pi)} {d^3 \mathbf  p_{3}\over (2\pi)^4}, 
 \label{sp:3N_cov}
 \end{eqnarray}
where ``$2^\prime$" labels the intermediate state of the nucleon 2 after the first short-range NN interaction, 
$\lambda_i$ is the spin of the $i$th nucleon 
and the $\Gamma_{NN\rightarrow N,N}$ is the  same short range NN interaction vertex  included in  Eq. (\ref{sp:2N_cov}).

Note that there are several other 3N SRC diagrams which differ from that of Fig. \ref{Fig:3nSpectral} by the ordering of 
the two sequential NN short range interactions.  In collinear approximation these diagrams result in the same analytic form both 
in VN and LF approximations (see, e.g., Ref. \cite{FSS15}), thus their contribution can be absorbed in the definition of the parameter 
$n^N_{3N}$ [see Eq. (\ref{norm_VNA_3NSRC})],   which defines the contribution of the norm of the 3N SRCs to the total normalization of the nuclear spectral function.

\subsection{Models calculation  of nuclear spectral functions}
\label{sec:ModCal}
The first step to calculate the nuclear spectral functions from the forward scattering amplitudes given  by 
Eqs. (\ref{eq:Smf})-(\ref{sp:3N_cov}) is to  define
the effective vertices  $\hat V$ which identify the bound nucleon in the mean field, in the 2N or in the  3N SRCs, as well as to define the poles
at which  the cut propagators of the intermediate states are  estimated. 
Both will depend on  the approach used to deal with vacuum fluctuations, that is virtual nucleon or light-front approximation.

After the vertex definition, the effective Feynman rules \cite{ms01} are applied to the covariant  forward scattering 
amplitudes  corresponding to the
mean field, the two-, or the three-nucleon SRC contributions (Fig. \ref{Fig:spectral_diagram}) separately.  Within the VN or the LF approximation,    the loop-integrals are calculated through the on-mass shell conditions of intermediate states, by integration through the positive poles of the   cut propagators of the corresponding intermediate states. Finally, the numerators of the propagators are estimated by using sum rules representing the completeness relation for the wave functions of the intermediate states. 

The following step is the definition of transition wave functions for the nucleus to nucleons and to nuclear residuals systems.  Such  definitions are formulated by  the identification of the interaction diagrams for the bound states with 
the corresponding equations for the bound state wave function.  For example in the  non-relativistic limit,  the 
interaction diagrams for the bound state, calculated based on the effective Feynman diagrammatic rules,   are  identified with 
the Lippmann-Schwinger equation \cite{Gribov,Bertocchi} in the non-relativistic limit.  In the relativistic case,  similar 
identifications are made with the Bethe-Salpeter type \cite{SB51, Gross:1982} (for VN approximation) or the Weinberg type \cite{Weinberg} (for LF approximation) equations  for the relativistic bound state wave function.

Finally, the mathematical models  for the SRC nuclear spectral functions   are simplified by introducing effective momentum distributions for 2N SRC which are function of the deuteron momentum distribution, as well as approximate Gaussian distribution models for the center of mass motion of the 2N SRC. For the mean field nuclear spectral functions, a nonrelativistic approximation obtained from the conventional mean field calculations for a  single nucleon  is  applied.

\section{Nuclear spectral function in virtual nucleon approximation} 
\label{sec:SpecVNA}
The main assumptions of the model for nuclear spectral function in virtual nucleon approximation are that the nucleus is in the laboratory frame,  the  interacting bound nucleon is described as a virtual particle, and  the spectators are put  on their mass-shells. 

The nuclear  spectral function, $S^{N}_A(\mathrm p,E_m)$, in virtual nucleon approximation is  defined as  the joint probability of finding  a nucleon in the nucleus  with momentum $\mathrm p$ and removal energy $E_m$. The conventional definition of the removal energy  is
 \begin{equation}
 E_m = E_{A-1} + M_N - M_{A} - {\mathrm p^2\over 2M_{A-1}},
\label{EM_def}
\end{equation}
where $E_{A-1}$  and $M_{A-1}$ are the energy and the mass of the  $A-1$  residual nuclear system respectively, and  the nonrelativistic expression for the kinetic energy of $A-1$ system is subtracted.  However, in practice  the kinetic energy of the $A-1$ system depends on the mean field, the 2N-SRC or the 3N- SRC picture of the nuclear wave function, hence the corresponding  kinetic energy will be accordingly defined for  each particular case.
 
 The following normalization condition for the nuclear spectral function in  virtual nucleon approximation is defined  with base on  the conservation of baryonic number of the nucleus in hadron-nucleus scattering\cite{FS87}:
\begin{equation}
\sum\limits_{N=1}^A\int  S^{N}_A(\mathrm p,E_m)  \alpha d^3\mathbf p  dE_m    = A,
\label{bnorm}
\end{equation}
where $\alpha$ is  the ratio of the flux factors of the (external probe)-(bound nucleon) and  (external probe)-(nucleus) systems, 
which in the high momentum limit of  the probe (hadron or virtual photon) yields 
\begin{equation}
\alpha = {E_N + \mathrm p_{z} \over M_A/A} = A{p^{+}\over p^+_{A}}.
\label{alpha}
\end{equation}
Here, $p^{+}$ and $p^+_{A}$ are the light-front longitudinal momenta of the bound nucleon and of the nucleus respectively,  
$E_N$ is the energy of the bound nucleon and the $z$ direction is defined opposite to the direction of the
incoming probe.

Following the decomposition of Fig. \ref{Fig:spectral_diagram},  the  mean field, the 2N, and the 3N SRC contributions 
to the nuclear spectral function are separately considered. 
In the VN approximation,  the cut diagrams of Figs. \ref{Fig:spectral_diagram} and  \ref{Fig:3nSpectral} will be evaluated at the positive energy poles of the spectator residual system. 
For the mean field contribution,  it  corresponds to the positive energy pole of the coherent $A-1$ nuclear residual  system.  For the  2N SRC contribution,  they correspond  to the positive energy 
poles of the correlated nucleon and the $A-2$ nuclear residual system, whereas for the 3N SRC contribution,  they correspond to the positive energy poles of the two correlated nucleons.

 \subsection{Nuclear spectral function in virtual nucleon approximation for a nucleon in the nuclear mean field} 
\label{sec:MfVNA}

In the mean-field approximation [Fig. \ref{Fig:spectral_diagram}(a)] the  missing momentum ${\bf p_m} \equiv - {\bf p_1}$ and the missing energy $E_m$ characterizes the total momentum and  the excitation energy of the  $A-1$  residual nuclear system.  In the nuclear shell model, $E_m$ also defines  the energy needed to remove the nucleon from a particular nuclear shell. 
For such a situation,  the energy and the momentum of the interacting nucleon can be recovered from the kinematic parameters of the on-shell spectator. 

The calculation of  the nuclear spectral function in virtual nucleon approximation for a nucleon in the  mean field of the $A-1$  residual nuclear system, starts with the following definition of  the effective vertex $\hat V_{MF}$ in the mean field covariant amplitude Eq.(\ref{eq:Smf})  
\begin{equation}
\hat V_{MF}  = i \bar a(p_1,s_1) \delta({ \mathbf p_1} + {\mathbf p_{A-1}})\delta(E_m - E_\alpha) a(p_1,s_1),
\label {MF_VNA_vertex}
\end{equation}
where $E_\alpha$ is the characteristic energy of the given nuclear shell, and the delta functions represent the momentum and the energy conservation in the vertex. The creation, $\bar a$ ($p_1,s_1)$,   and the annihilation, $a(p_1,s_1)$,  operators in Dirac space
are defined in such a way that they obey the following relation:
\begin{equation}
a(p_1,s_1)(\sh p_1 + M_N) = \bar u(p_1,s_1) \ \ \ \mbox{and}  \ \ \  (\sh p_1 + M_N) \bar a(p_1,s_1) =  u(p_1,s_1),
\label{aops}
\end{equation}
where $u(p_1,s_1)$ is the spinor of the virtual nucleon.  

Since in the VN approximation the vacuum fluctuations are neglected, the integral   in Eq. (\ref{eq:Smf}), needs  only  to be calculated through  the positive energy pole of the propagator of the on-shell $A-1$  residual nuclear system. Since this pole is displaced slightly  below the real axis by the infinitesimal factor $i\varepsilon$, the integral by   $dp^0_{A-1}$ is obtained by applying the Cauchy's residue theorem to the lowest-half complex plane semicircle enclosing the displaced pole in the positive sense (clockwise direction), namely
\begin{equation}
\oint {dp^0_{A-1}\over p^2_{A-1} - M^2_{A-1} + i\varepsilon}   =  \oint {dp^0_{A-1}\over E^2_{A-1}- (M^2_{A-1} +\mathbf p^2_{A-1}- i\varepsilon ) } = - {2\pi i\over 2 E_{A-1}}\left |_{E_{A-1} = \sqrt{M^2_{A-1} + \mathbf p^2_{A-1}}}\right.,
\label{pole_VNA_MF}
\end{equation}
where $p^0_{A-1} = E_{A-1}$.

Let  $\chi_{A-1}$ be the  the spin wave function of the  on-shell  $A-1$ residual nuclear system, then   the numerator of the propagator in Eq. (\ref{eq:Smf}), can be expressed by  the following sum rule representing the completeness relation for the wave function: 
\begin{equation}
G_{A-1}(p_{A-1},\alpha) = \sum\limits_{s_{A-1}} \chi_{A-1}(p_{A-1},s_{A-1},E_\alpha) \chi^\dagger_{A-1}(p_{A-1},s_{A-1},E_\alpha)
\label {Gsp_VNA_MF}.
\end{equation}
It is important to note that in the relativistic treatment,  the spin wave functions are momentum dependent as it is     indicated in the argument of $\chi_{A-1}$. Such a momentum dependence is also  accounted for the spin wave function of other particles discussed in the present dissertation.

The definition of the vertex (\ref  {MF_VNA_vertex}) follows the 
convention (see, e.g., Ref. \cite{Brodsky:1997de}) for which the  annihilation operator projects the   nucleon propagator  to 
the positive energy state,  and the nuclear transition vertex produces the Fock component of the nuclear wave function.
Note that the above  definition is different from the conventional definition (see e.g. Ref. \cite{Ciofi2015})  in which the annihilation operator 
acts over  the nuclear wave function to produce nucleon-hole states. However the final results in both approaches are similar  in
the   nonrelativistic limit. 

Hence,  using the above convention and from Eqs. (\ref {aops})- (\ref {Gsp_VNA_MF}),  the single nucleon wave function, $\psi_{N/A}$ ,  for the given nuclear shell $E_\alpha$ is defined as 
\begin{equation}
\psi^{s_A}_{N/A}(p_1,s_1,p_{A-1},s_{A-1},E_\alpha) = { \bar u(p_1,s_1)\chi^\dagger_{A-1}(p_{A-1},s_{A-1},E_\alpha)\Gamma_{A \to N,A-1} 
\chi^{s_A}_A\over (M_{N}^2 - p_{1}^2) \sqrt{(2\pi)^3 2E_{A-1}}}.
\label {wf_VNA_MF}
\end{equation}
Inserting the above wave function and the mean field vertex (\ref {MF_VNA_vertex})
  into Eq.(\ref{eq:Smf}),  and summing over all possible nuclear shells, $\alpha$, and spin projection, $s_1$ and $s_{A-1}$,  gives the following expression for the nuclear spectral function in virtual nucleon approximation for a nucleon in the mean field
\begin{equation}
S^{N}_{A,MF} (\mathrm p_1,E_m) = \sum\limits_{\alpha}\sum\limits_{s_1,s_{A-1}} \int \mid 
\psi^{s_A}_{N/A}(p_1,s_1,p_{A-1},s_{A-1},E_\alpha) \mid^2 \delta(E_m - E_\alpha) \delta^3({ \mathbf p_1} 
+ {\mathbf p_{A-1}})d^3 \mathbf p_{A-1},
\label {MF_spe_VNA_1}
\end{equation}
which defines the joint probability of finding  a nucleon in the mean field of the nucleus  with momentum $\mathrm p_1$ and removal energy $E_m$. 

Integration of Eq. (\ref {MF_spe_VNA_1}) over  $d^3 \mathbf p_{A-1}$ and  through the delta function, $\delta^3(\mathbf p_1+\mathbf p_{A-1})$ ,
yields
\begin{equation}
S^{N}_{A,MF} (\mathrm  p_1,E_m) = \sum\limits_{\alpha}\sum\limits_{s_1,s_{A-1}} \mid 
\psi^{s_A}_{N/A}(p_1,s_1,p_{A-1},s_{A-1},E_\alpha) \mid^2 \delta(E_m - E_\alpha)
\label {MF_spe_VNA_2}.
\end{equation}

In order to obtain  numerical estimates of the  nuclear spectral function (\ref {MF_spe_VNA_2}), it is important to consider    that  in the mean field approximation, the substantial strength of the  wave function $\psi^{s_A}_{N/A}$  comes from the momentum range  $\mathrm p_1\le \mathrm k_{F}$. Hence,  the nonrelativistic approximation is valid,  so that  the mean field  wave function (\ref {wf_VNA_MF})  can be approximated by the nonrelativistic wave function obtained  from the conventional mean field calculations of the single nucleon wave functions. 

Additionally, in the nonrelativistic limit
$\alpha \approx 1 + {\mathrm p_{1,z}\over M_A/A}$, and, in  
Eq.(\ref{bnorm}), the  ${\mathrm p_{1,z}\over M_A/A}$ part does not contribute to the integral, resulting 
in  the condition for the nonrelativistic normalization:
\begin{equation}
\int  S^{N}_{A,MF}(\mathrm p,E_m) dE_m d^3\mathbf p     = n^N_{MF},
\label{nrnorm}
\end{equation}
where $n^N_{MF}$ is  the  mean field contribution    to  the total normalization of the nuclear spectral  function.

\subsection{Nuclear spectral function in virtual nucleon approximation for two nucleons in  short-range correlation  } 
\label{sec:2NsrcVNA}
The calculation of  the nuclear spectral function in virtual nucleon approximation for two nucleons in  short-range correlation,  starts with the definition of  the removal energy $E_m^{2N}$  for the 2N SRC in which the correlated NN pair has a total momentum $\mathbf p_{NN} = \mathbf p_{1} +  \mathbf p_{2}  = -{\mathbf p_{A-2}}$,  in the mean field of the $A-2$ residual nuclear system.  The magnitude of   $E_m^{2N}$ is therefore defined as 
\begin{equation}
E_m^{2N} = E^{(2)}_{thr} + T_{A-2} + T_{2} - T_{A-1}=   E^{(2)}_{thr} + {\mathrm p_{A-2}^2\over 2 M_{A-2}} + T_2  - {\mathrm p_{1}^2\over 2 M_{A-1}},
\label {2NSRC_removal}
\end{equation}
where $E^{(2)}_{thr}$ is the threshold energy needed to remove two nucleons from the nucleus, with an approximated value of 
  $E^{(2)}_{thr} \approx 2M_N + M_{A-2} - M_{A}$. Furthermore, $T_{A-1}$ and  $T_{A-2}$  are the nonrelativistic  kinetic energies of the   $A-1$ and the $A-2$    residual  nuclear systems  respectively, and $T_{2}$ is the relativistic kinetic energy of the correlated nucleon 2. The expression for $T_{A-1}$ follows from the fact that in the rest frame of the nucleus $ \mathbf p_1 = -\mathbf p_{A-1}$, as well as  from the definition of the removal energy in Eq. (\ref {EM_def}).

 The effective vertex $\hat V_{2N}$  in the 2N SRC covariant amplitude 
 (\ref{sp:2N_cov}) is defined as
\begin{equation}
\hat V_{2N}  = i  \bar a(p_1,s_1) \delta^3( {\mathbf p_1} + {\mathbf p_2} + { \mathbf p_{A-2}})\delta(E_m - E_m^{2N}) a (p_1,s_1),
\label {2NSRC_VNA_vertex}
\end{equation}
where   the creation, $\bar a(p_1,s_1)$,  and the annihilation, $a(p_1,s_1)$, operators of the nucleon with 
four-momentum $p_1$ and spin $s_1$ satisfy  the relations of Eq. (\ref{aops}). The delta functions represent the momentum and energy conservation in the vertex. 

Similarly to the propagator  integration  described in section \ref {sec:MfVNA}, the  integrations by $dp^0_2$ and $dp^0_{A-2}$  in   Eq. (\ref{sp:2N_cov})  through the
positive energy poles of the  propagators of the on-shell particle $2$,  and  $A-2$ nuclear residual  system respectively, yields
\begin{align}
\oint  {dp^0_2\over p_2^2 - M_N^2 + i\varepsilon}  & =  \oint  {dp^0_2\over E_2^2 - (M_N^2 + \mathbf p^2_2- i\varepsilon)} =  - {2\pi i\over 2E_2}\left |_{E_2 = \sqrt{M^2_N  +  \mathbf  p^2_2}}\right., \nonumber \\
\oint  {dp^0_{A-2}\over p_{A-2}^2 - M^2_{A-2} + i\varepsilon}  & =   \oint  {dp^0_{A-2}\over E^2_{A-2} - (M^2_{A-2}+ \mathbf p^2_{A-2}- i\varepsilon)} =  - {2\pi i\over 2E_{A-2}}\left |_{E_{A-2} = \sqrt{M^2_{A-2}  +  \mathbf  p^2_{A-2}}}\right..
\label{pole_VNA_2NSRC}
\end{align}

Let  $\chi_{A-2}$ be the  the spin wave function of the  on-shell $A-2$  residual nuclear system, then   the numerator of its  propagator  in Eq. (\ref{sp:2N_cov}), can be expressed by  the following sum rule representing the completeness relation for the wave function:
\begin{eqnarray}
 G(p_{A-2}, s_{A-2})  & = &  \sum\limits_{s_{A-2}}\chi_{A-2}(p_{A-2},s_{A-2})\chi^\dagger_{A-2}(p_{A-2},s_{A-2}),
\label{GA-2_VNA_2NSRC}
\end{eqnarray}
where   
 $s_{A-2}$ is the spin projection of the   $A-2$ residual  nuclear  system. Similarly,  
the numerator of the propagator for the on-shell particle $2$ is given by the completeness relation 
\begin{eqnarray}
\sh p_2 + M_N & = &  \sum\limits_{s2}u(p_2,s_2)\bar u(p_2,s_2),
\label{2_VNA_2NSRC}
\end{eqnarray}
where  $u(p_2,s_2)$ is the spinor of particle $2$  with momentum $p_2$ and spin projection $s_2$ \cite {HM84}.   

It  is assumed that  the center of mass momentum of the 2N SRC is small compared to the relative motions of  the nucleons in this type of correlation, then the numerator of its propagator in    Eq. (\ref{sp:2N_cov}), can be expressed by  the following sum rule representing the completeness relation for the wave function:
\begin{equation}
G(p_{NN},s_{NN}) = \sum\limits_{s_{NN}} \chi_{NN}(p_{NN}, s_{NN})\chi^\dagger_{NN}(p_{NN},s_{NN}),
\label{GNN_VNA_2NSRC}
\end{equation}
where $\chi_{NN}$ is  the  spin wave function, and  $s_{NN}$ is the projection of the total spin of the NN correlation with the three-momentum,  ${\bf p_{NN}} =  - {\bf p_{A-2}}$ in the rest frame of the nucleus.

Inserting  Eqs. (\ref{aops}) and  (\ref{pole_VNA_2NSRC}) -(\ref{GNN_VNA_2NSRC})  in  the covariant 2N SRC amplitude [Eq. (\ref{sp:2N_cov})], and summing over all possible spin projections  $s_2,s_{A-2}$ and $s_{NN}$,   reduces the latter to the 
2N SRC part of the nuclear spectral function in VN approximation, namely:
\begin{eqnarray}
S_{A,2N}^{N}(\mathrm p_1,E_m) & = & \sum\limits_{s-1,s_2,s_{A-2},s_{NN},s^\prime_{NN}}
\int \chi^{s_A,\dagger}_A \Gamma_{A\to NN,A-2}^\dagger {\chi_{NN}(p_{NN}, s_{NN})
\chi_{NN}^\dagger (p_{NN},s_{NN}) \over p_{NN}^2 - M_{NN}^2 }  \Gamma^\dagger_{NN\rightarrow N,N} \nonumber \\
&& \times {u(p_1,s_1)\over p_1^2 - M_N^2 }   \delta^3(\mathbf  p_1 + \mathbf  p_{2} + \mathbf  p_{A-2}) \delta(E_m - E_m^{2N}){\bar u(p_1,s_1)
\over p_1^2 - M_N^2 }   \nonumber \\
&& \times{u(p_2,s_2)\bar u(p_2,s_2) \over 2E_2} \Gamma_{NN\rightarrow N,N}
 {\chi_{NN}(p_{NN}, s_{NN})\chi^\dagger_{NN}(p_{NN},s_{NN}) \over p_{NN}^2 - M_{NN}^2 }  \nonumber \\
& & \times {\chi_{A-2}(p_{A-2},s_{A-2}) \chi^\dagger_{A-2}(p_{A-2},s_{A-2})\over 2E_{A-2}} \Gamma_{A\to NN,A-2} \chi^{s_A}_A 
{d^3\mathbf p_{2}\over  (2\pi)^3} {d^3 \mathbf p_{A-2}\over  (2\pi)^3},
\label {spe_VNA_2NSRC_1}
 \end{eqnarray}
which defines the joint probability of finding  a nucleon in NN SRC  with momentum $\mathrm p_1$ and removal energy $E_m$. 

 The introduction  of the transition  wave function of the center of mass of the 2N SRC  Eq. (\ref {GNN_VNA_2NSRC}), implies the  decoupling of the slow moving center of mass  of the 2N SRC from the fast moving   relative motion of the nucleons in the correlation.  Then,  the 2N SRC center of mass  wave function for  the  transition  $A\rightarrow  (NN)+(A-2)$ in the rest frame of the nucleus is defined as 
 \begin{eqnarray}
  \psi^{s_A}_{CM}(p_{NN},s_{NN},p_{A-2},s_{A-2}) & =  & {\chi^\dagger_{NN}(p_{NN},s_{NN})\chi^{\dagger}_{A-2}(p_{A-2},s_{A-2}) 
 \Gamma_{A\to NN,A-2} \chi^{s_A}_A  
\over (M_{NN}^2 - p_{NN}^2)\sqrt{2E_{A-2}(2\pi)^3}},
\label{wfsCM_VNA_2NSRC}
 \end{eqnarray}
 and the wave function for the transition   $(NN)\rightarrow N+N$,  that represents the relative motion of nucleons in 2N SRC in the rest frame of  the 2N SRC  (see e.g. Ref. \cite{ms01}),  is   defined as 
 \begin{eqnarray}
 \psi_{NN}^{s_{NN}}(p_1,s_1,p_2,s_2) & =  & {\bar u(p_1,s_1)\bar u(p_2,s_2) \Gamma_{NN\rightarrow N,N} \chi_{NN}(p_{NN},s_{NN})\over
 (M_N^2 - p_1^2)\sqrt{2E_2(2\pi)^3}}.  
\label{wfsRel_VNA_2NSRC}
 \end{eqnarray} 
inserting the above defined  wave functions in the  2N SRC nuclear spectral function (\ref {spe_VNA_2NSRC_1})  yields\begin{eqnarray}
 S_{A,2N}^{N}(\mathrm p_1,E_m)  & = &  \sum\limits_{s_1,s_2,s_{A-2},s_{NN},s^\prime_{NN}}
\int \psi^{s_A,\dagger}_{CM}(p_{NN},s^\prime_{NN},s_{A-2})\psi^{s^\prime_{NN}, \dagger}_{NN}(p_1,s_1,p_2,s_2)\nonumber \\
& & \times \psi^{s_A}_{CM}(p_{NN},s_{NN},s_{A-2})\psi^{s_{NN}}_{NN}(p_1,s_1,p_2,s_2) \nonumber \\
& & \times \delta^3(\mathbf  p_1 + \mathbf  p_2 - \mathbf  p_{NN})\delta(E_m - E_m^{2N})
d^3 \mathbf p_2  d^3 \mathbf  p_{NN},
\label {spe_VNA_2NSRC_2}
\end{eqnarray}
where the relation ${\bf p_{NN}} = - {\bf p_{A-2}}$, in the nucleus rest frame, has been used. 

The nuclear spectral function (\ref {spe_VNA_2NSRC_2}) can be integrated  by $d^3 \mathbf  p_2$ through the $\delta^3(\mathbf  p_1 + \mathbf  p_2 - \mathbf  p_{NN})$ delta function. Furthermore 
using  the  2N SRC model in which the  wave function of the relative motion is dominated by the $pn$ component with spin equal to 1,   and 
with the  low momentum CM wave function being  in the  S state,  the summation by  
$s_{A-2}$ results in $\delta_{s_{NN},s^\prime_{NN}}$. Then, the 2N SRC nuclear spectral function (\ref {spe_VNA_2NSRC_2}) can be expressed as 
\begin{eqnarray}
 S^N_{A,2N}(\mathrm p_1,E_m) =   \sum\limits_{s_1,s_2,s_{A-2},s_{NN}}\int
\mid \psi^{s_A}_{CM}(p_{NN},s_{NN},s_{A-2})\mid^2 \mid \psi^{s_{NN}}_{NN}(p_1,s_1,p_2,s_2)\mid^2 \nonumber \\
\times\delta(E_m - E_m^{2N}) d^3 \mathbf p_{NN}, 
\label{spe_VNA_2NSRC_3}
\end{eqnarray}
where $ {\bf p_{NN}} = \bf {p_2} + {\bf p_1}$. 

The above expression can be represented in a more simple form by noticing that the 2N SRC center of mass wave function (\ref {wfsCM_VNA_2NSRC}) depends on the center of mass momentum, and  that the wave function for  the relative motion of nucleons in 2N SRC depends on their relative momentum. Thus, by  introducing effective momentum distribution of the 2N SRC center  of mass, $n_{CM}$, as well for the relative momentum of  the nucleons in 2N SRC, $n_{NN}$, and by summing over the final  and  averaging all possible initial polarization configurations in (\ref {spe_VNA_2NSRC_3}), the following expression  is obtained for the 2N SRC  nuclear spectral function
\begin{equation}
S^N_{A,2N}(\mathrm p_1,E_m) = \int n_{CM}(\mathrm p_{NN}) n_{NN}(\mathrm  p_{rel})\delta(E_m - E_m^{2N}) d^3 \mathbf p_{NN},
\label{spe_VNA_2NSRC_4}
\end{equation}
where ${\bf p_{rel}} ={{\bf p_1} - {\bf p_2}\over 2}$. 

The normalization of the 2N SRC nuclear  spectral function (\ref  {spe_VNA_2NSRC_4} ) should be related to the total probability of finding a nucleon in such a correlation, and  can be defined from the normalization condition  of Eq. (\ref{bnorm}),  namely:
\begin{equation}
\int S^N_{A,2N}(\mathrm p_1,E_m) \alpha_1 dE_m  d^3\mathbf p_1 = n^N_{2N},
\label {norm_VNA_2NSRC}
\end{equation}
where for the 2N SRC model 
$\alpha\equiv \alpha_1 = {M_N - E_m - T_{A-1} + \mathrm p_{1,z}\over  M_A/A}$,  and $n^N_{2N}$ is the  contribution 
to  the total normalization of the nuclear spectral function.

If  the relative and the center of mass momentum distributions of the NN correlations are given,   the 2N SRC part of the nuclear spectral function  (\ref{spe_VNA_2NSRC_4}) can be numerically calculated.   Then, since it is assumed that the center of mass momenta of the NN SRCs are small,   $n_{CM}$ is approximated by the distribution obtained in Ref. \cite{CiofiSimula} through the overlap  of two   Fermi momentum distributions which results in the simple Gaussian distribution:
\begin{equation}
n_{CM}(\mathrm p_{NN}) = N_0(A) e^{-\beta(A) \mathrm p_{NN}^2},
\label{2NSRC_CMdist}
\end{equation}
normalized to unity: $\int n_{CM}(p_{NN})d^3\mathbf p_{NN}  = 1$.  The parameter  $\beta(A)$ is estimated from the nuclear 
mean field distribution, while $N_{0}(A)$ is found from the normalization condition.

The relative momentum distribution of the NN SRC, $n_{NN}(\mathrm p_{rel})$,  can be   modeled according to 
Ref.\cite{newprops,proa2}, where the high momentum strength of the nucleon momentum distribution is predicted to be 
inverse proportional to the relative fraction of the nucleon  in the nucleus.  Such a distribution is in agreement with 
the recently observed dominance of $pn$ SRCs\cite{isosrc,Eip3,Eip4} and can be expressed in the form:
\begin{equation}
n^N_{NN}(\mathrm p_{rel}) = {a_2(A,Z)\over (2x_N)^\gamma} {n_d(\mathrm p_{rel})\Theta(\mathrm p_{rel}-\mathrm k_{src})\over {M_N- E_m - T_{A-1}\over M_A/A} },
\label{2NSRC_relmd}
\end{equation}
where $\Theta$ is the step function;  $x_{N} = N/A$ is the nucleon relative fraction \cite {newprops}, with $N$ being the number of   protons ($Z$) and neutrons ($A-Z$) in the nucleus $A$;  
the parameter $a_2(A,Z)$ is related to the probability of finding 2N SRC in the nucleus, $A$,  relative to the 
deuteron; and 
$\gamma$ is a free parameter $\gamma \lesssim 1$.  $n_d(\mathrm p_{rel})$ is the  high momentum distribution in the deuteron,   
and $k_{src}\gtrsim k_{F}$ is the momentum threshold at which a NN system with such relative momentum can be considered in  short-range correlation. The factor ${M_N- E_m - T_{A-1}\over M_A/A}$ is the  generalization of the normalization scheme of  \cite{FS87} which enforces  the normalization condition of Eq. (\ref{bnorm}).  The normalization of the above defined distribution,  $\int n^N_{NN}(p)d^3\mathbf p = n^N_{2N}$, defines the contribution of the 2N SRCs to the total norm of the momentum distribution for  the nucleon $N$.

The  2N SRC  nuclear spectral function in VN approximation [Eq.(\ref {spe_VNA_2NSRC_4})], together with the momentum distribution for the NN SRC center of mass [Eq.(\ref {2NSRC_CMdist})], the relative momentum distribution of the NN SRC [Eq.(\ref {2NSRC_relmd})], and the normalization condition [Eq.(\ref {norm_VNA_2NSRC})], constitute the mathematical model from which  the corresponding computational models and numerical estimates are obtained  in the chapters 3 and 4  of the present dissertation.  

It is worth mentioning that in the non-relativistic  limit, and  assuming an  equal 2N SRC contributions from proton and neutron:
$n^N_{NN}(\mathrm p_{rel}) = a_2(A) n_d(\mathrm p_{rel})$,  the 2N SRC nuclear spectral function 
[Eq.(\ref{spe_VNA_2NSRC_4})] reduces to the "NN SRC-CM motion"  model of Ciofi-Simula \cite{CDL92,CiofiSimula}.


\subsection{Nuclear spectral function in virtual nucleon approximation for three nucleons in  short-range correlation } 
\label{sec:3NsrcVNA}
The calculation of the nuclear spectral function in virtual nucleon approximation for three nucleons in  short-range correlation,  starts with the definition of  the magnitude of the  removal energy $E_m^{3N}$  which is calculated considering the   3N SRC model in which the recoil nuclear system consists of two fast nucleons and a slow $A-3$  residual nuclear system  whose excitation energy is neglected. The magnitude of   $E_m^{3N}$ is therefore defined as 
\begin{equation}
E^{3N}_m = E^{(2)}_{thr} + T_{2} + T_{3} - T_{A-1}=  E^{(3)}_{thr} + T_{3} + T_{2} - {\mathrm p_{1}^2\over 2 M_{A-1}},
 \label{3NSRC_removal}
 \end{equation}
where $E^{(3)}_{thr}$ is the threshold  energy needed to remove three nucleons from the nucleus, with an approximated value of  $E^{(3)}_{thr} \approx 3M_N + M_{A-3} - M_{A}$ where $M_{A-3}$ is the mass of the $A-3$  residual nuclear system.
Within the considered 3N SRC model,  the kinetic energy of the $A-1$  residual nuclear system is the result of  the  
kinetic energies of the correlated spectator nucleons, $T_2$ and $T_3$ which are  treated relativistically.
Here, as in the case of 2N SRC,  the expression for the kinetic energy $T_{A-1}$ follows from the fact that in the rest frame of the nucleus $ \mathbf p_1 = -\mathbf p_{A-1}$, as well as  from the definition of the missing energy in Eq. (\ref {EM_def}).

For the  3N SRC model in collinear approximation,  the effective vertex $\hat V_{3N}$ in the 3N SRC covariant amplitude 
Eq. (\ref{sp:3N_cov})   is defined as
\begin{equation}
\hat V_{3N}  = i \bar a (p_1,s_1) \delta^3(\mathbf  p_1 + \mathbf  p_2 + \mathbf  p_{3})\delta(E_m - E_m^{3N}) a (p_1,s_1),
\end{equation}
where   the creation, $\bar a(p_1,s_1)$,  and the annihilation, $a(p_1,s_1)$, operators of a nucleon with 
four-momentum $p_1$ and spin $s_1$ satisfy  the relations of Eq. (\ref{aops}). The delta functions represent the momentum and energy conservation in the vertex. 
 
Similar to the propagator  integration  described in section \ref {sec:2NsrcVNA}, the  integrations by $dp^0_2$ and $dp^0_{3}$  in   Eq. (\ref{sp:3N_cov}), are done through the
positive energy poles of the  propagators of the on-shell particles  $2$ and $3$ yielding:
\begin{align}
\oint  {dp^0_2\over p_2^2 - M_N^2 + i\varepsilon}  & =  \oint  {dp^0_2\over E^2_2 - (M^2_N + \mathbf p^2_2- i\varepsilon)} =  - {2\pi i\over 2E_2}\left |_{E_2 = \sqrt{M^2_N  +  \mathbf  p^2_2}}\right. \nonumber \\
\oint  {dp^0_3\over p^2_3 - M^2_N + i\varepsilon}  & =  \oint  {dp^0_3\over E^2_3 - (M^2_N+ \mathbf p^2_3- i\varepsilon)} =  - {2\pi i\over 2E_3}\left |_{E_3 = \sqrt{M^2_N  +  \mathbf  p^2_3}}\right..
\label{pole_VNA_3NSRC}
\end{align}   
The numerator of the propagator for the on-shell particle $i$ is defined, as in the 2N SRC case,   by
\begin{eqnarray}
\sh p_i + M_N & = &  \sum\limits_{si}u(p_i,s_i)\bar u(p_i,s_i),
\label{i_VNA_3NSRC}
\end{eqnarray}
where  $u(p_i,s_i)$ is the spinor of particle $i$  with momentum $p_i$ and spin projection $s_i$, for $i =2,2',3,$ \cite {HM84}.   

Inserting  Eqs. (\ref{aops}),  (\ref{pole_VNA_3NSRC}), and (\ref{i_VNA_3NSRC})  in the covariant 3N SRC amplitude  [Eq. (\ref{sp:3N_cov})], and summing over all possible spin projections  $s_2,s_{2'}$ and $s_{3}$,   reduces the latter to the 
3N SRC part of the nuclear spectral function, namely:
\begin{eqnarray}
& & S^{N}_{A,3N}( \mathrm p_1,E_m)  =  \sum\limits_{s_{2^\prime},\tilde s_{2^\prime},s_{2},s_{3}} \int \bar u(k_1,\lambda_1) \bar u(k_2,\lambda_2) 
\bar u (k_3,\lambda_3)  \Gamma^\dagger_{NN\rightarrow N,N} 
  {u(p_{2^\prime},s_{2^\prime})\bar u(p_{2^\prime},s_{2^\prime})  \over p_{2^\prime}^2 - M_N^2 } \nonumber \\
& & \ \ \ \ \ \ \ \ \times \ \Gamma^\dagger_{NN\rightarrow N,N} {u(p_1,s_1)\over p_1^2 - M_N^2} 
\delta^3( \mathbf  p_1 +  \mathbf  p_2 +  \mathbf  p_3)\delta(E_m - E_m^{3N}) {\bar u(p_1,s_1)\over p_1^2 - M_N^2} 
 \nonumber \\ 
&& \ \ \ \ \ \ \ \ \times \  {u(p_2,s_2)\bar u(p_2,s_2)\over 2E_2} \Gamma_{NN\rightarrow N,N} 
 {u(p_{2^\prime},\tilde s_{2^\prime})\bar u(p_{2^\prime},\tilde s_{2^\prime})\over  p_{2^\prime}^2 - M_N^2 } 
 {u(p_3,s_3)\bar u(p_3,s_3)\over 2E_3}  \nonumber \\
& &  \ \ \ \ \ \ \ \ \times \   \Gamma_{NN\rightarrow N,N} u(k_1,\lambda_1) u(k_2,\lambda_2) u (k_3,\lambda_3) 
 {d^3 \mathbf  p_{3}\over (2\pi)^3} {d^3 \mathbf  p_{2}\over (2\pi)^3},
 \label{spe_VNA_3NSRC_1}
 \end{eqnarray}
which defines the joint probability of finding  a nucleon in 3N SRC  with momentum $\mathrm p_1$ and removal energy $E_m$. 
 
In analogy with Eq. (\ref{wfsRel_VNA_2NSRC}), the following $2N$ SRC wave functions can be introduced  for  the nuclear spectral function (\ref {spe_VNA_3NSRC_1}), as follows:
  \begin{align}
  \psi_{NN}(p_{2'},s_{2'};p_3,s_3;k_2,\lambda_2,k_3,\lambda_3) & =  
  \frac {\bar u(p_{2'},s_{2'})\bar u(p_3,s_3) \Gamma_{NN\to N,N} u(k_2,\lambda_{2})u(k_{3},\lambda_{3})}
  {(M^2_N - p^2_{2'})\sqrt{2E_3(2\pi)^3}}, 
  \label {wfs_VNA_3NSRC_1}
 \end{align}
 \begin{align}
 \psi_{NN}(p_1,s_1;p_2,s_2;k_1,\lambda_1,p_{2^\prime},s_{2^\prime}) & =  
  \frac {\bar u(p_1,s_1)\bar u(p_2,s_2) \Gamma_{NN\to N,N} u(k_1,\lambda_{1})u(p_{2'},s_{2'})}
  {(M^2_N - p^2_1)\sqrt{2E_2(2\pi)^3}}, 
  \label {wfs_VNA_3NSRC_2}
  \end {align}
 where the wave function (\ref {wfs_VNA_3NSRC_1}) represents the first 2N SRC, shown in Fig. \ref {Fig:3nSpectral}, between particle $2$ with momentum $k_2$,  and particle $3$ with momentum $k_3$, which results in particle $2'$ with momentum $p_{2'}$,  and particle $3$ with momentum $p_3$. Whereas  the wave function (\ref {wfs_VNA_3NSRC_2}) represents the second 2N SRC between particle $1$ with momentum $k_1$,  and particle $2'$ with momentum $p_2'$, which results in particle $1$ with momentum $p_1$,  and particle $2$ with momentum $p_2$.
 
Inserting Eqs. (\ref {wfs_VNA_3NSRC_1}) and (\ref {wfs_VNA_3NSRC_2}) in Eq.(\ref{spe_VNA_3NSRC_1}), the 3N SRC nuclear spectral function   can be  expressed as follows:
\begin{eqnarray}
S^{N}_{A,3N}(\mathrm p_1,E_m) & = & \sum\limits_{s_{2^\prime}, \tilde s_{2^\prime},s_2,s_3}
\int \psi^\dagger_{NN}(p_1,s_1,p_2,s_2;k_1,\lambda_1,p_{2^\prime},s_{2^\prime})
\psi^\dagger_{NN}(p_{2^\prime},s_{2^\prime},p_3,s_3;k_2,\lambda_2,k_3,\lambda_3) \nonumber \\
&&\times  \psi_{NN}(p_1,s_1;p_2,s_2;k_1,\lambda_1,p_{2^\prime},s_{2^\prime})
      \psi_{NN}(p_{2^\prime},\tilde s_{2^\prime};p_3,s_3;k_2,\lambda_2,k_3,\lambda_3)\nonumber \\
 &&\times  \delta^3(\mathbf p_1 + \mathbf  p_2 + \mathbf  p_3)\delta(E_m - E_m^{3N}) d^3 \mathbf p_{3} d^3 \mathbf p_{2}.
 \label{spe_VNA_3NSRC_2}
 \end{eqnarray}

The nuclear spectral function (\ref {spe_VNA_3NSRC_2}) can be integrated  by $d^3 \mathbf  p_2$ through the $\delta^3(\mathbf  p_1 + \mathbf  p_2 + \mathbf  p_{3})$ delta function. Furthermore, using the  2N SRC model in which the  wave function of the relative motion is dominated by the $pn$ component with spin equal to 1, the summation over  the polarizations of 2 and 3 particles,  $s_{2}$ and $s_{3}$, results in in $s_2= s_{2^\prime} = \tilde s_{2^\prime}$. Then, the 3N SRC nuclear spectral function ( \ref {spe_VNA_3NSRC_2}) can be expressed as 
 \begin{eqnarray}
S^{N}_{A,3N}(\mathrm p_1,E_m)  & =  & \sum\limits_{s_{1},s_2,s_3}
\int \mid \psi_{NN}(p_{2^\prime},s_{2},p_3,s_3;k_2,\lambda_2,k_3,\lambda_3)\mid^2\nonumber \\
& &\times \mid \psi_{NN}(p_1,s_1;p_2,s_2;k_1,\lambda_1,p_{2^\prime},s_{2})\mid^2 
 \delta(E_m - E_m^{3N}) d^3 \mathbf p_{3}.
 \label {spe_VNA_3NSRC_3}
\end{eqnarray}

The above expression can be represented in a more simple form by noticing that the  wave function   (\ref {wfs_VNA_3NSRC_1}), for  the first 2N SRC,  depends on the relative  momentum of particles $2'$ and $3$, and  the  wave function   (\ref {wfs_VNA_3NSRC_2}), for  the second 2N SRC,  depends on the relative  momentum of particles $1$ and $2$. Thus, by  introducing effective relative  momentum distribution for   the nucleons in the first and second 2N SRC, $n_{NN}$, and by summing over the final  and averaging all possible initial polarization configurations in (\ref {spe_VNA_3NSRC_3}), the following simplified expression is obtained for the 3N SRC  nuclear spectral function
\begin{equation}
S^{N}_{A,3N}(\mathrm p_1,E_m)  = \int n_{NN}(\mathrm p_{2^\prime,3})  n_{NN}(\mathrm p_{12})\delta(E_m - E_m^{3N}) 
d^3\mathbf p_{3},
\label{spe_VNA_3NSRC_4}
\end{equation}
where in the collinear approximation:  ${\bf p_{12}} = {{\bf p_1}- {\bf p_2}\over 2} = {\bf p_1} + {{\bf p_3}\over 2}$ and 
${\bf p_{2^\prime,3}} = {{\bf p_{2^\prime}} - {\bf p_3} \over 2} \approx   {{-\bf p_{3}} - {\bf p_{3}} \over 2} \approx - {\bf p_3}$.  

The normalization of the 3N SRC nuclear  spectral function (\ref  {spe_VNA_3NSRC_4} ) should be related to the total probability of finding a nucleon in such a correlation, and can be defined from the normalization condition  of Eq. (\ref{bnorm})  which yields:
\begin{equation}
\int S^{N}_{A,3N}(\mathrm p_1,E_m) \alpha_1 d^3\mathbf p_1 dE_m = n^N_{3N},
\label {norm_VNA_3NSRC}
\end{equation}
where  
$\alpha$ was defined in Eq. (\ref  {norm_VNA_2NSRC}), and $n^N_{3N}$ is the  contribution    to  the total normalization of the nuclear spectral function.
 
Within the model of  $pn$ dominance of  two-nucleon SRCs,    the 
3N SRCs are predicted to be generated predominantly through two sequential short-range $pn$ interactions. 
Then,  the overall probability of finding such correlations is predicted to be proportional to the factor
${a^2_2(A,Z)}$, where $a_2(A,Z)$ is defined in Eq. (\ref {2NSRC_relmd}). Then, by using  relations similar to  those defined in Eq. (\ref {2NSRC_relmd}),  the following approximation is obtained for the product of the relative  momentum distributions in Eq. (\ref {spe_VNA_3NSRC_4}) 
\begin{equation}
n_{NN}(\mathrm p_{2^\prime,3}) n_{NN}(\mathrm p_{12}) = a^2_2(A,Z) C^N(A,Z) 
{n_d(\mathrm p_{2^\prime,3}) n_d(\mathrm p_{12})\over  {M_N - E_m - T_{A-1}\over M_A/A}}\Theta(\mathrm p_{2^\prime,3}-\mathrm k_{src})\Theta(\mathrm p_{12}-\mathrm k_{src}),
\label{3NSRC_relmd}
\end{equation}
where $\mathrm k_{src}> \mathrm k_{F}$ is the relative momentum threshold at which the $NN$ system can be considered in short-range correlation.
Here $C^N(A,Z)$ is  a suppression factor  which accounts  for the suppression of the 3N configurations with 
 two identical spectators like $pp$ and $nn$ pairs, that is,  effects associated with the isospin structure of two-nucleon recoil
system. Namely, in the collinear approximation  two recoil nucleons emerge with small relative momenta (or invariant mass).
In Ref. \cite{eheppn2} it was demonstrated that the $NN$ system with small  relative momenta is strongly dominated in the 
isosinglet $pn$ channel.  The  dominance  introduces an additional restriction on the isospin composition of the 3N SRCs, in which the recoil $NN$ system predominately consists of a $pn$ pair.    For example, one direct consequence of such dynamics is that high momentum neutrons in $^3$He nucleus  can not be generated in 3N SRCs while protons can.

The  3N SRC  nuclear spectral function in VN approximation [Eq.(\ref {spe_VNA_3NSRC_4})], together with the product of  momentum distribution of the first and second NN SRC [Eq.(\ref {3NSRC_relmd})], and the normalization condition 
[Eq.(\ref {norm_VNA_3NSRC})], constitute the mathematical model from which the corresponding computational models and numerical estimates are obtained in the  chapters 3 and 4  of the present dissertation.  

 \section{Nuclear spectral function in light-front approximation} 
\label{sec:SpecLFA}

The nuclear spectral function on the light-front (LF)  was for the first time  defined in Ref. \cite{FS88},  however  its 
calculation from  first principles is impossible due to the lack of   knowledge of  the LF nuclear wave functions. 
In the present dissertation,  two assumptions are formulated   to calculate  the  LF nuclear spectral functions. The first assumption is that  the nuclear  mean field contribution to  the LF  nuclear spectral function corresponds to the 
nonrelativistic limit of  the  momentum  and  missing energy of a bound nucleon.  As a result the mean field part of the LF nuclear spectral function 
can be related to the mean field contribution of the conventional nuclear spectral functions as  discussed in  section \ref {sec:MfVNA}. The second assumption is that the dynamics of the LF nuclear spectral function in the high momentum  domain is mainly defined by the $pn$ interaction at short distances. Thus, a model for the  LF deuteron wave function at short distances is sufficient to obtain the models for the 2N and the 3N SRC LF nuclear  spectral functions.
 
The  first step to obtain the LF nuclear spectral functions is the  definition of  the kinematic parameters that characterize 
the bound nucleon in the light front,  as well as the sum rules  that the  LF nuclear spectral functions should satisfy.

In defining the  LF nuclear spectral function the primary requirement is that it should be  a  Lorentz boost  invariant function  in the 
direction of the large CM momentum of the nucleus $p_A$.  To satisfy this condition,  the bound nucleon $N$ must be 
 described by a light-front  "+" momentum fraction $\alpha_N = A{p^+_{N}\over p^+_{A}}$,  transverse (to ${\bf p_A}$) momentum $\bf p_{N, \perp}$ and an invariant mass $\tilde M_N^2 = p^-_{N}p^+_{N} - \mathbf p_{N,\perp}^2$.  

For  future  derivations,  it is useful to present the invariant phase space of bound nucleon, $d^4 {p_N}$, as a function of the light-front variables $(\alpha_N,\mathbf p_{N,\perp}, \tilde M_N^2)$. Thus, to calculate the   Jacobian for  the change of variables $(\mathrm {p^+_{N}, p^-_{N}}, \mathbf p_{N,\perp})$ to $(\alpha_N, \tilde M_N^2,\mathbf p_{N,\perp})$ the following derivatives are used
\begin{align}
\frac {\partial \alpha_N}{\partial \mathrm {p^+_{N}}}  & = \frac {A}{\mathrm {p^+_{A}}},  \quad
\frac {\partial \alpha_N}{\partial \mathrm {p^-_{N}}} = 0,  \quad
\frac {\partial \mathrm {\tilde   M^2_{N} }}{\partial \mathrm {p^+_{N}}}   = \mathrm {p^-_{N}},  \quad
\frac {\partial \tilde M_N^2}{\partial \mathrm {p^-_{N}}}  =  \mathrm {p^+_{N}},
\label{alpha_M_N_derivatives}
\end{align} 
 hence, the Jacobian is given by
\begin{align}
 J & =\Big( \frac {\partial \alpha_N}{\partial \mathrm  {p^+_{N}}}  \frac {\partial \mathrm {\tilde   M^2_{N} }}{\partial \mathrm  {p^-_{N}}} - \frac {\partial \alpha_N}{\partial  \mathrm {p^+_{N}}} \frac {\partial \mathrm {\tilde   M^2_{N} }}{\partial   \mathrm {p^-_{N}}}\Big)^{-1} = \Big ( \frac {A \mathrm  {p^+_{N}}}{\mathrm  {p^+_{A}}} \Big )^{-1}= \frac {1}{ \alpha_N}.
\label{alpha_M_N_Jacobian}
\end{align}
 Then,  the  invariant phase space $d^4 { {p_N}}$ in the light-front can be expressed 
 as follows:
\begin{equation}
d^4 {p_{N}} = {1\over 2} d \mathrm  {p}^-_{N}d\mathrm  {p}^+_{N}d\mathbf {p}_{N\perp} =  {d\alpha_N\over 2 \alpha_N}d\mathbf {p}_{N\perp}d {\tilde   M^2_{N} }.
\label{LF_phspace}
\end{equation} 

After identifying the  kinematic variables  describing  the bound nucleon, the LF nuclear spectral function, $P_A(\alpha_N,p_{N,\perp},\tilde M_N^2)$, is defined as the joint probability of finding a bound nucleon in the nucleus with light-front  momentum fraction $\alpha_N$, transverse momentum $\bf p_{N,\perp}$   and invariant mass $\tilde M_N^2$.  The normalization 
condition for such nuclear spectral function is defined from the requirements of the baryonic number  and the total light-front momentum conservation laws \cite{FS88,FS87}:
\begin{eqnarray}
& & \sum\limits_{N=1}^A\int P^N_A(\alpha_N,\mathrm p_{N,\perp},\tilde M_N^2)  {d\alpha_N\over 2 \alpha_N}d^2\mathbf p_{N,\perp}d \tilde M_N^2 =  A,
\nonumber \\
& &   \sum\limits_{N=1}^A\int \alpha_N P^N_A(\alpha_N,\mathrm p_{N,\perp},\tilde M_N^2)  {d\alpha_N\over 2 \alpha_N}d^2\mathbf  p_{N,\perp}d \tilde M_N^2 = A,
\label{LF_sumrules}
\end{eqnarray}
where the second relation is  exact if it is  assumed  that   all the momentum in the nucleus is carried by the constituent nucleons.

The LF density matrix $\rho^N_{A}(\alpha_N,\mathrm  p_{N,\perp})$ is defined as the joint probability of finding a bound nucleon in the nucleus with light-front  momentum fraction $\alpha_N$ and  transverse momentum $\bf p_{N,\perp}$.  Hence,  the following  relation between the LF nuclear  spectral function and  the LF density matrix is obtained
\begin{equation}
\rho^N_{A}(\alpha_N,\mathrm  p_{N,\perp}) = \int P^N_A(\alpha_N,\mathrm p_{N,\perp},\tilde M_N^2) {1\over 2} d \tilde M_N^2.
\label{LF-density}
\end{equation}
Then, from Eqs. (\ref {LF_sumrules}) and (\ref {LF-density}) it follows that the normalization of the LF density matrix is given by:
\begin{eqnarray}
& & \sum\limits_{N=1}^A\int \rho^N_{A}(\alpha_N,\mathrm  p_{N,\perp})  {d\alpha_N\over  \alpha_N}d^2\mathbf p_{N,\perp}=  A,
\nonumber \\
& &   \sum\limits_{N=1}^A\int \alpha_N \rho^N_{A}(\alpha_N,\mathrm  p_{N,\perp})  {d\alpha_N\over  \alpha_N}d^2\mathbf  p_{N,\perp}= A.
\label{LF_density_sumrules}
\end{eqnarray}
 
To obtain  the models  of the LF nuclear spectral functions,  similar to the VN approximation case, the  mean field, the 2N, and the 3N SRC contributions are separately considered by following  the decomposition of Fig. \ref{Fig:spectral_diagram}. 
In the LF  approximation, the cut diagrams of Fig. \ref{Fig:spectral_diagram} and \ref{Fig:3nSpectral} will be evaluated at the positive  light-front (\textquotedblleft-" component) energy poles of the spectator residual system. 
For the mean field contribution,  it  corresponds to the positive light-front energy pole of the coherent $A-1$ residual  nuclear  system.  For the  2N SRC contribution,  they correspond  to the positive light-front energy 
poles of the correlated nucleon and the $A-2$ residual nuclear  system, whereas for the 3N SRC contribution,  they correspond to the positive light-front energy poles of the two correlated nucleons.

\subsection{Nuclear spectral function in light-front approximation for a nucleon in the  nuclear mean field } 
\label{sec:MfLFA}

To calculate the light-front nuclear spectral function for a nucleon in the  nuclear mean field a procedure similar to the VN approximation case may be applied. Thus, after defining the effective vertex $\hat V_{MF}$ in the mean field covariant amplitude Eq.(\ref{eq:Smf}),    the integral of the  denominator of the propagator of the on-shell ($A-1$)  residual nuclear system will be  calculated.  With the result of the  integration plus the definition of the numerator of the propagator of  the on-shell ($A-1$) system, similar to Eq. (\ref {Gsp_VNA_MF}),  the mean field LF nuclear spectral function will be expressed through the  unknown light-front mean field  wave function of the nucleus. 

 A different approach  is used in the present  dissertation that is developed because of  the fact that  the mean field  nuclear spectral function  is  dominating  
at small momenta and  small removal energies ($< k_F$) of the bound nucleon,
for which the application of the nonrelativistic limit of the light-front approximation is well justified.  
Then, what is only needed is a  relation between  the mean field LF spectral function  $P^N_{A,MF}(\alpha_1,\mathrm p_{1,\perp},\tilde M_N^2)$  and the 
 VN approximation mean field spectral function, $S^N_{A,MF}(\mathrm p, E_m)$ [Eq.\ref {MF_spe_VNA_2}],  in the nonrelativistic limit\footnote{Note that hereafter 
$\alpha_N$ and $p_{N,\perp}$ will be identified with $\alpha_1$ and $p_{1,\perp}$ respectively, giving the 
subscript "1" to the bound nucleon.}.
The relation between  $P^N_{A,MF}$ and $S^N_{A,MF}$ can be found by using the normalization condition: 
\begin{equation}
\int P^{N}_{A,MF}(\alpha_1,\mathrm p_{1,\perp},\tilde M_N^2) {d\alpha_1\over 2 \alpha_1}d^2\mathbf p_{1,\perp}d \tilde M_N^2 = \int S^{N}_{A,MF}(\mathrm p_1, E_m)dE_m d^3\mathbf  p_1 = n^{N}_{MF},
\label{mfnorms}
\end{equation}
where $n^N_{MF}$ is  the  contribution   to  the total normalization of the nuclear spectral  function. The Jacobian for  the change of variables from  the LF phase space [Eq.\ref {LF_phspace}] to  the VN phase space,  $dE_m d^3\mathbf p_1$, is calculated by considering that  the total energy of the on-shell $A-1$ residual nuclear system  is given by
\begin{equation}
E_{A-1} = \sqrt{M_{A-1}^2 + p_{A-1}^2} = M_A - M_N + E_m + {\mathrm p_1^2\over 2 M^0_{A-1}},
\label{EA-1}
\end{equation}
where the last part of the equation follows from  the definition of the removal energy $E_m$ [Eq.(\ref{EM_def})] which is inherently  nonrelativistic. In the above expression $M_{A-1}$ is the mass of the $A-1$  system, 
which can be in the excited state, while $M^{0}_{A-1}$ represents the ground state mass of the residual nucleus. The above equation results in the following expressions for $\alpha_1$
\begin{align}
\alpha_1 & = \frac {A p^+_1}{p^+_A} = \frac {A (p^+_A- p^+_{A-1})}{p^+_A}  = A - \frac {E_{A-1} + \mathrm  p_{A-1,z}}{p^+_A} = A - {E_{A-1} - \mathrm  p_{1,z}\over p^+_A},
\label{alpha_mf}
\end{align}
and for $\tilde M_N^2$:
\begin{align}
\tilde M_N^2   & = p^-_{1}p^+_{1} - \mathbf p_{1,\perp}^2 = p^+_{1}\Big (p^-_{A}-p^-_{A-1}\Big) - \mathbf p_{1,\perp}^2 = p^+_{1}\left (\frac {M^2_A}{p^+_{A}}-\frac {M^2_{A-1} + \mathbf p_{1,\perp}^2}{p^-_{A-1}}\right) - \mathbf  p_{1,\perp}^2 \nonumber \\
&  =  \frac {p^+_{1}} {p^+_{A}}\left (M^2_A -\frac {M^2_{A-1} + \mathbf p_{1,\perp}^2}{p^-_{A-1}/ p^+_{A}}\right) - \mathbf p_{1,\perp}^2  = {\alpha_1\over A}\left (M_A^2 - {M_{A-1}^2 + \mathbf p_{1,\perp}^2\over (A-\alpha_1)/A}\right) - \mathbf p_{1,\perp}^2,
\label{mn_mf}
\end{align}
where  $\mathbf p_{A,\perp} = 0 $, and   the relation $\alpha_{A-1} = A-\alpha_1 $ has been applied. 

Using  Eqs. (\ref {alpha_mf}) and (\ref {mn_mf}) yields
\begin{align}
\frac {\partial \alpha_1}{\partial \mathrm  { p_{1,z}} } & =  \frac {A}{p^+_{A}}  \left ( 1- \frac {\partial   {E_{A-1}}}{\partial \mathrm  {p_{1,z}}} \right ) = \frac {A}{\mathrm  { p^+_{A}}}\left (\frac { {M^0_{A-1}- p_{1,z}} }{ {M^0_{A-1}}}\right), \nonumber\\
\frac {\partial \alpha_1}{\partial   { E_{m}}} & = \frac {\partial \alpha_1}{\partial   {E_{A-1}}}  \frac {\partial   {E_{A-1}}}{\partial   {E_m}} = - \frac {A}{\mathrm  { p^+_{A}}}, \nonumber\\
\frac {\partial  {\tilde   M^2_{N} } }{\partial \mathrm  { p_{1,z}}}  & = \frac {\partial  {\tilde   M^2_{N} }}{\partial \alpha_1 }  \frac {\partial \alpha_1}{\partial\mathrm  { p_{1,z}}}+ \frac {\partial  {\tilde   M^2_{N} }}{\partial   {M^2_{A-1}}} \frac {\partial  {  M^2_{A-1} }}{\partial  \mathrm  { p_{1,z}} }\nonumber\\
& =  \left (\frac {  {M^0_{A-1}}- \mathrm  { p_{1,z }}}{M^0_{A-1}}\right) \left [\frac {  { M^2_A}}{\mathrm  { p^+_{A}}} -\frac {A^2}{\mathrm  { p^+_{A}}} \frac {   {M^2_{A-1}}+\mathrm  { p^2_{1, \perp}}}{(A- \alpha_1)^2} \right] 
 - \left ( \frac {2 \alpha_1} {A - \alpha_1} \right ) \ \left (\frac {  {E_{A-1}}-  { M^0_{A-1}}}{M^0_{A-1}}\right) \mathrm  { p_{1,z}}, \nonumber\\
\frac {\partial  {\tilde   M^2_{N} } }{\partial { E_{m}}} & = \frac {\partial{\tilde   M^2_{N} }}{\partial \alpha_1}  \frac {\partial \alpha_1}{\partial   {E_m} }+ \frac {\partial  {\tilde   M^2_{N} }}{\partial   {M^2_{A-1}}} \frac {\partial  {   M^2_{A-1} }}{\partial   {E_m} }=-\frac {  { M^2_A}}{\mathrm  { p^+_{A}}} +\frac {A^2}{\mathrm  { p^+_{A}}} \frac {  {M^2_{A-1}}+\mathrm  { p^2_{1, \perp}}}{(A- \alpha_1)^2} - \left ( \frac {2 \alpha_1} {A - \alpha_1} \right )    {E_{A-1}},
\label{alpha_M_N_derivatives2}
\end{align}
thus the Jacobian for  the change of variables $(\alpha_1, \tilde M_N^2)$  to $( { E_{m}}, \mathrm  { p_{1,z}})$ is given by
\begin{align}
 J & = \frac {\partial \alpha_1}{\partial \mathrm  {p_{1,z}}}  \frac {\partial  {\tilde   M^2_{N} }}{\partial   {E_m}} - \frac {\partial \alpha_1}{\partial   {E_m}} \frac {\partial {\tilde   M^2_{N} }}{\partial   \mathrm {p_{1,z}}} =  -\frac {2 \alpha_1  }{\alpha_{A-1}} \left (\frac {A \mathrm { p^+_{A-1}}}{ \mathrm  { p^+_{A}}} \right )=  -2 \alpha_1.
\label{alpha_M_N_Jacobian2}
\end{align}
Then
\begin{equation}
d\tilde M_N^2 d\alpha_1 = |J | d E_m d p_{1,z} =  2\alpha_1 d E_m d p_{1,z}.
\label {diff}
\end{equation}
The above equation,  together with Eq. (\ref{LF_phspace}), results in  
\begin{align}
d^4p_1 = {d\alpha_1\over 2 \alpha_1}d^2\mathbf p_{1,\perp}d \tilde M_N^2 = d E_m d^3 \mathbf p_1,
\label {phase-space-equi}
\end {align}
which inserted   in  Eq. (\ref{mfnorms}) results in the following relation between  the mean field  nuclear spectral functions  $P^N_{A,MF}(\alpha_1,\mathrm p_{1,\perp},\tilde M_N^2)$  and  $S^N_{A,MF}(\mathrm p_1, E_m)$
\begin{equation} 
 P^{N}_{A,MF}(\alpha_1,\mathrm p_{1,\perp},\tilde M_N^2) =  S^{N}_{A,MF}(\mathrm p_1, E_m),
 \end{equation}
 where $\alpha_1$ and $\tilde M_N^2$ are expressed through $E_{m}$ and $\mathrm p_{1}$ according to Eqs. (\ref{EA-1})-(\ref{mn_mf}).
Note that the above equation is valid for up to the overall normalization factor, since VN and LF approximations  result 
in different normalizations  for the mean field contribution to the nuclear spectral function. 

\subsection{Nuclear spectral function in light-front  approximation for two nucleons in  short-range correlation} 
\label{sec:2NsrcLFA}
The calculation of  the nuclear spectral function in light-front approximation for two nucleons in  short-range correlation starts with the definition of the effective vertex $\hat V_{2N}$  in the 2N SRC covariant amplitude 
 (\ref{sp:2N_cov}) \cite{FSS15}:
\begin{equation}
\hat V_{2N} = i \bar a(p_1,s_1) 2\alpha_1^2\delta(\alpha_1 + \alpha_2 + \alpha_{A-2}-A)\delta^2(\mathbf p_{1,\perp}+\mathbf p_{2,\perp} + \mathbf p_{A-2,\perp})
\delta(\tilde M_N^2 - \tilde M^{(2N),2}_N)a(p_1,s_1),
\label {2NSRC_LF_vertex}
\end{equation}
where $(\alpha_2,\mathbf p_{2,\perp})$, ($\alpha_{A-2},\mathbf p_{A-2,\perp}$) are the light-front momentum fractions and the transverse momenta 
of the correlated second nucleon and  of the $A-2$  residual nuclear system respectively, and $\mathbf p_{A,\perp} = 0$. The creation, $\bar a(p_1,s_1)$,  and the annihilation, $a(p_1,s_1)$ , operators of a  nucleon with 
four-momentum $p_1$ and spin $s_1$ satisfy  the relations of Eq. (\ref{aops}). The delta functions represent the momentum and the invariant mass conservation in the vertex.

Since  in the light-front NN correlation model, the particle 2   and  the  $A-2$  residual nuclear system  are considered on light-front energy shells,   the magnitude of the invariant mass $  \tilde M^{(2N),2}_N$ is calculated as follows
\begin{align}
 \tilde M^{(2N),2}_N& =  {p^-_{1}p^+_{1} }- \mathrm  {p}^2_{1,\perp} =   {p^+_{1} } ( {p^-_{A} -p^-_{2} -  p^-_{A-2} } ) - \mathrm  {p}^2_{1,\perp}\nonumber \\
& =   {p^+_{1} }  {\left (\frac  {M^2_A} { {p}^+_{A}} - \frac  {M^2_N + \mathrm  {p}^2_{2,\perp} } { {p}^+_{2}}  -  \frac  {M^2_{A-2} + \mathrm  {p}^2_{A-2,\perp}} { {p}^+_{A-2}} \right) }-\mathrm  {p}^2_{1,\perp}\nonumber \\
& =  { \frac { {p}^+_{1}} { {p}^+_{A } }}{\left (  {M^2_A} - \frac  {M^2_N + \mathrm {p}^2_{2\perp} } {  {p}^+_{2}/p^+_{A }} -  \frac  {M^2_{A-2} + \mathrm {p}^2_{A-2,\perp}} { {p}^+_{A-2}/{p}^+_{A }} \right) }- \mathrm  {p}^2_{1\perp}\nonumber \\
& =   { \frac {\alpha_{1}} {A }}{\left (  {M^2_A} -A \frac  { M^2_N + \mathrm {p}^2_{2,\perp} } { \alpha_2} -  A\frac  {M^2_{A-2} + \mathrm  {p}^2_{A-2,\perp}} {\alpha_{A-2}} \right) }- \mathrm {p}^2_{1,\perp}\nonumber \\
& =   { \frac {\alpha_{1}} {A }} {\left [ {M^2_A} - A \frac  {M^2_N + \Big(\mathbf  {p}_{A-2,\perp} - \mathbf {p}_{1,\perp}  \Big )^2 } { A - \alpha_1 -\alpha_{A-2} } -  A \frac  {M^2_{A-2} + \mathrm {p}^2_{A-2,\perp}} {\alpha_{A-2}} \right ] }- \mathrm {p}^2_{1,\perp}
\label{Inv-mass-2N},
\end{align}
where  $\mathbf  {p}_{A-2,\perp} =  \mathbf {p}_{1,\perp} + \mathbf {p}_{2,\perp}$, and the $z$ axis  in the reference frame is taken in the $\mathbf p_{A}$ direction.

Similar to the propagator  integration  described in section \ref {sec:MfVNA}, the  integrations by $dp^-_2$ and $dp^-_{A-2}$  in   Eq. (\ref{sp:2N_cov})  are performed through the
$"-"$ components of the positive  poles of the  propagators of the on-shell particle $2$  and the  $A-2$ residual nuclear  system respectively,   provided that their $"+"$ components 
are large and positive, so that the $Z$ diagram in this scheme will be suppressed by the inverse power of large $"+"$ component of 
the nucleon's four-momentum  \cite{Weinberg}. The integration  by $dp^-_2$  of   the propagator of particle 2 yields 
\begin{align}
\oint  \frac {d^4  p_{2}} {p^2_2 -  {M_N} + i \varepsilon } & = \oint  \frac {  {d p^-_2 d p^+_2  d^2} \mathrm {p}_{2,\perp} }{2  { \Big (p^+_2p^-_2 } - \mathrm {p}^2_{2,\perp} -  {M_N} \Big) + i \varepsilon} \nonumber \\
& = \oint  \frac {  {d p^-_2 d p^+_2  d^2} \mathrm {p}_{2,\perp} }{2  { p^+_2 \Big [ p^-_2 }- \Big (\mathrm {p}^2_{2,\perp} +  {M_N \Big )/ p^+_2  \Big ]   + i \varepsilon}} 
 =   - {2\pi i  {d p^+_2  d^2} \mathbf {p}_{2,\perp}\over 2  {p^+_2}} \left |_{ {p^-_2} = \frac { \mathrm {p}^2_{2\perp} + {M_N }}{ p^+_2 }}  \right.  \nonumber \\
& =   - \pi i  \frac {d \alpha_2 }{\alpha_2} d^2\mathbf {p}_{2,\perp}  \left |_{ {p^-_2} = \frac { \mathrm {p}^2_{2,\perp} +  {M_N }}{ p^+_2 }}  \right..
\label{lc_poles-p2}
\end{align}
Similarly, integration by $d p^-_{A-2}$ in Eq.(\ref{sp:2N_cov}) over the positive  pole of the propagator of the  $A-2 $ residual nuclear system  yields
\begin{align}
\oint  \frac {d^4  p_{A-2}} {p^2_{A-2} -  {M_{A-2}} + i \varepsilon } & =\oint   \frac { \mathrm {d^2 p^-_{A-2} d p^+_{A-2}  d} \mathbf {p}_{{A-2},\perp} }{2 \mathrm { \Big (p^+_{A-2}p^-_{A-2} } - \mathrm {p}^2_{{A-2},\perp} -  {M_{A-2}} \Big) + i \varepsilon} \nonumber \\
 & =  - \pi i \frac { d \alpha_{A-2}}{\alpha_{A-2}} d^2\mathbf {p}_{{A-2},\perp}    \left |_{\mathrm {p^-_{A-2}} = \frac { \mathrm {p}^2_{{A-2},\perp} +  {M_{A-2}}}{ p^+_{A-2} }}  \right.
\label{lc_poles-pA-2}
\end{align}
The  above integrations project the intermediate state to the positive light-front energy state thus excluding the contribution from the Z  graph of Fig. \ref{Fig:R_diagram}(c).  With the diminished contribution from the $Z$ graph, the 2N SRC covariant amplitude  [Eq. (\ref{sp:2N_cov}) ] will result in the 2N SRC  light-front spectral function, $P^N_{A,2N}(\alpha_1,p_{1,\perp},\tilde M_N^2)$.

As in the VN approximation, if  $\chi_{A-2}$ is  the  LF  spin wave function of the  on-shell  $A-2$  residual nuclear  system, then   the numerator of its propagator  in Eq. (\ref{sp:2N_cov}), can be expressed by  the following sum rule representing the completeness relation for the wave function:
\begin{eqnarray}
 G(p_{A-2}, s_{A-2})  & = &  \sum\limits_{s_{A-2}}\chi_{A-2}(p_{A-2},s_{A-2})\chi^\dagger_{A-2}(p_{A-2},s_{A-2}),
\label{GA-2_LF_2NSRC}
\end{eqnarray}
where   $s_{A-2}$ is the spin projection of the   $A-2$  residual nuclear  system. Similarly,  the numerator of the propagator for the on-shell particle $2$ is given by
\begin{eqnarray}
\sh p_2 + M_N & = &  \sum\limits_{s2}u(p_2,s_2)\bar u(p_2,s_2),
\label{2_LF_2NSRC}
\end{eqnarray}
where  $u(p_2,s_2)$ is the spinor of particle $2$  with momentum $p_2$ and spin projection $s_2$ \cite {HM84}.
    
In  the  nonrelativistic limit for the center of mass motion of the 2N SRC, $k_{CM}\ll M_{NN}$ [for $k_{CM}$ see Eq. (\ref{k-CM}) below],  the propagator $G(p_{NN},s_{NN})$ can be approximated as 
\begin{equation}
G(p_{NN},s_{NN}) \approx  \sum\limits_{s_{NN}} \chi_{NN}(p_{NN},s_{NN}) \chi_{NN}^\dagger(p_{NN},s_{NN}),
\label{GNN_LF_2NSRC}
\end{equation}
where $\chi_{NN}$ is the spin wave function of the center of mass motion  of the 2N SRC.

Inserting  Eqs. (\ref{aops}) and  (\ref{lc_poles-p2}) -(\ref{GNN_LF_2NSRC})  in  the covariant 2N SRC amplitude [Eq. (\ref{sp:2N_cov})], and summing over all possible spin projections  $s_2,s_{A-2}$ and $s_{NN}$,   reduces the latter to the 
2N SRC part of the nuclear spectral function in LF approximation, namely:
\begin{align}
P_{A,2N}^{N}(\alpha_1, \mathrm p_{1,\perp},\tilde M_N^2)  & =  
  \sum\limits_{s_2,s_{NN},s_{A-2}}\int \chi^{s_A,\dagger}_A  
\Gamma_{A\to NN,A-2}^\dagger \chi_{A-2}(p_{A-2},s_{A-2}) { \chi_{NN}(p_{NN},s_{NN})  \over p_{NN}^2 - M_{NN}^2}\nonumber \\
&  \times  
 \chi_{NN}^\dagger(p_{NN},s_{NN}) \Gamma^\dagger_{NN\rightarrow NN} {u(p_1,s_1)u(p_2,s_2)\over p_1^2 - M_N^2}
  \nonumber \\
&    \times 
 2\alpha_1^2\delta(\alpha_1+ \alpha_2 + \alpha_{A-2}-A)\delta^2(\mathbf p_{1,\perp}+\mathbf p_{2,\perp} + \mathbf p_{A-2,\perp})
\delta(\tilde M_N^2 - \tilde M_{N}^{(2N),2})
\nonumber \\ 
& \times {\bar u(p_1,s_1)\bar u(p_2,s_2)\over p_1^2 - M_N^2 } \Gamma_{NN\rightarrow N,N}\chi_{NN}(p_{NN},s_{NN})
\nonumber \\
&  \times {  \chi_{NN}^\dagger(p_{NN},s_{NN})  \over p_{NN}^2 - M_{NN}^2} \chi^\dagger_{A-2}(p_{A-2},s_{A-2})  \Gamma_{A\to NN,A-2}\chi^{s_A}_A \nonumber \\
&  \times {d\alpha_2\over \alpha_2}{d^2\mathbf p_{2,\perp}\over 2(2\pi)^3}
{d\alpha_{A-2}\over \alpha_{A-2}}{d^2\mathbf p_{A-2,\perp}\over 2(2\pi)^3}.
\label{spe_LF_2NSRC_1}
\end{align} 
\subsubsection {Light-front  wave function of the NN SRC} 
The   calculation of the light-front  wave function of the NN SRC  is based on the consideration of  the following  expression, as well as its complex conjugated form,   in Eq. (\ref{spe_LF_2NSRC_1}):
\begin{align}
{\bar u(p_1,s_1)\bar u(p_2,s_2)\over p_1^2 - M_N^2 } \Gamma_{NN\rightarrow NN}  \chi_{NN}(p_{NN},s_{NN}).
\label{2NSRCpart}
\end{align}
On the basis of   the light-front momentum and energy   conservation 
at the $\Gamma_{NN\rightarrow N,N}$  vertex, and on the on-shell condition of particle 2, the propagator in Eq. (\ref{2NSRCpart}) can be expressed as 
\begin{align}
p_1^2 - M_N^2 & = (p_{NN} - p_2)^2 - M_N^2 = (p^+_{NN}-p^+_{2})(p^-_{NN}-p^-_{2}) - (\mathbf p_{NN,\perp} - \mathbf p_{2,\perp})^2 - M_N^2 \nonumber \\
&= (p^+_{NN}-p^+_{2})\left(p^-_{NN}-p^-_{2}- {M_N^2 + \mathrm p_{1\perp}^2\over p^+_{NN}-p^+_{2}}\right) 
\nonumber \\
& = p^+_1  \left( {M_{NN}^2  + \mathrm p_{NN,\perp}^2\over p^+_{NN}} - 
{M_N^2 + \mathrm p_{2,\perp}^2\over p^+_{2}} - {M_N^2 + \mathrm p_{1,\perp}^2\over p^+_{1}}\right)  \nonumber \\
& = \frac {\alpha_1}{ \alpha_{NN}} \left\{ M_{NN}^2  + \mathrm p_{NN,\perp}^2  - \frac {\alpha_{NN}}{\alpha_{1}\alpha_{2}}\left[
(M_N^2 + \mathrm p_{2,\perp}^2)\alpha_{1} - (M_N^2 + \mathrm p_{1,\perp}^2) \alpha_{2}\right] \right\},
\label{LCprop1_1}
\end{align}
where it has been assumed that   $k_{CM}\ll M_{NN}$, which  justifies 
the approximation, $p^-_{NN} \approx {M^2_{NN}+ \mathrm p_{NN,\perp}^2\over p^+_{NN}}$.
Eq. (\ref{LCprop1_1})  can be further simplified by using the relations 
$\alpha_1 + \alpha_2 = \alpha_{NN}$ and $\mathbf p_{1,\perp} + \mathbf p_{2,\perp} = \mathbf p_{NN,\perp}$,   namely
\begin{align}
p_1^2 - M_N^2 & =  \frac {\alpha_1}{ \alpha_{NN}} \left [M_{NN}^2   - \frac {\alpha^2_{NN}}{\alpha_{1}\alpha_{2}}\left(
M_N^2 + \mathrm p^2_{1,\perp} + \frac {\alpha^2_{1}}{\alpha^2_{NN}}\mathrm p^2_{NN,\perp}  - \frac {2\alpha_1}{\alpha_{NN} }\mathbf p_{1,\perp}\mathbf p_{NN,\perp} \right) \right]
 \nonumber \\
& = \frac {\alpha_1}{ \alpha_{NN}}\left \{M^2_{NN} -\frac  {\alpha^2_{NN}}{ \alpha_1\alpha_2}\left [M^2_N + \left (\mathbf p_{1\perp} - \frac {\alpha_1}{ \alpha_{NN}}\mathbf p_{NN,\perp}\right)^2\right] \right \}.\nonumber \\
& = \frac {\alpha_1}{ \alpha_{NN}}\Big (M^2_{NN} - s_{NN} \Big ),
\label{LCprop1_2}
\end{align}
where the invariant squared energy of the $NN$  system, $s_{NN}$, is given by 
\begin{align}
s_{NN} & =  p^2_{NN} = (p_1 + p_2) ^ 2 = (p^+_1 + p^+_2)(p^-_1 + p^-_2) - (\mathbf p_{1,\perp} + \mathbf p^+_{2,\perp})^2 \nonumber \\
& = \frac  {\alpha^2_{NN}}{ \alpha_1\alpha_2}\left [M^2_N + \left (\mathbf p_{1\perp} - \frac {\alpha_1}{ \alpha_{NN}}\mathbf p_{NN,\perp}\right)^2\right].
\label{sNN_1}
\end{align}

The  propagator [Eq.(\ref {LCprop1_2})]  can be completely expressed though the internal momenta of the $NN$ system by introducing the  momentum fraction of 
the 2N SRC carried by the nucleon 1, $\beta_1$, and the relative transverse momentum, $\mathbf k_{1,\perp}$, as follows:
\begin{align}
\beta_1 = 2- \beta_2 = {2\alpha_1\over \alpha_{NN}},\\
\mathbf k_{1,\perp}  = \mathbf p_{1,\perp} - {\beta_1\over 2} \mathbf p_{NN,\perp}.
\label {rel_trans_mom}
\end{align}
The propagator [Eq.(\ref {LCprop1_2})] can therefore be expressed as 
\begin{align}
p_1^2 - M_N^2 & =  \frac {\beta_1}{ 2} \left [M_{NN}^2  -\frac {4}{ \beta_{1} (2-\beta_{1})}\Big(M^2_N + \mathrm k^2_{1, \perp} \Big) \right ],
\label{LCprop1_3}
\end{align}
and the invariant squared energy $s_{NN}$ as 
\begin{align}
s_{NN} & =  \frac {4}{ \beta_1(2-\beta_1)}(M_N^2 + \mathrm k_{1,\perp}^2) = 4(M_N^2 + \mathrm k_1^2),
\label{sNN_2}
\end{align} 
where the relative momentum,  $\mathbf k_1$,  in the $NN$ center of mass frame, is invariant with respect to the Lorentz  boost in the $\bf p_{NN}$ direction.  The relative momentum $\mathbf k_1$ will be used to set a  momentum scale for  the  2N SRCs, requiring 
$\mathrm k_1\ge \mathrm k_{src}$ similar to Eq. (\ref{2NSRC_relmd}).

Inserting Eqs. (\ref {LCprop1_3}) and  (\ref {sNN_2}) in Eq. (\ref{2NSRCpart}) yields
\begin{align}
\frac {\bar u(p_1,s_1)\bar u(p_2,s_2)}{ \frac {\beta_1}{ 2} \left [M_{NN}^2  -\frac {4}{ \beta_{1} (2-\beta_{1})}\Big(M^2_N + \mathrm k^2_1 \Big) \right ]} \Gamma_{NN\rightarrow NN}  \chi_{NN}(p_{NN},s_{NN})  \nonumber \\
 = \frac {\bar u(p_1,s_1)\bar u(p_2,s_2)}{ \frac {\beta_1}{ 2} \left [M_{NN}^2  -4(M_N^2 + \mathrm k_1^2) \right ]} \Gamma_{NN\rightarrow NN}  \chi_{NN}(p_{NN},s_{NN}).
 \label{2NSRCpart2}
\end{align}

The expression in  Eq. (\ref{2NSRCpart2}) can be presented in a more compact form by  introducing the light-front wave function of the $NN$ SRC \cite{FS81,FSS15}
in the form:
\begin{equation}
\psi^{s_{NN}}_{NN}(\beta_{1},k_{1,\perp},s_1,s_2) = -{1\over  \sqrt{2(2\pi)^3}} {\bar u(p_1,s_1)\bar u(p_2,s_2)
\Gamma_{NN\rightarrow NN} \chi_{NN}(p_{NN},s_{NN})\over {1\over 2}[ M_{NN}^2 - 4(M_N^2 + \mathrm k_1^2)]}.
\label{srclcwf}
\end{equation}
The above defined wave function results in the following expression for Eq. (\ref{2NSRCpart}) 
\begin{equation}
{\bar u(p_1,s_1)\bar u(p_2,s_2)\over p_1^2 - M_N^2 } \Gamma_{NN\rightarrow NN} \chi_{NN}(p_{NN},s_{NN}) = 
-  {\sqrt{2(2\pi)^3}\over \beta_1} \psi^{s_{NN}}_{NN}(\beta_1,\mathrm k_{1,\perp},s_1,s_2). 
\label{SRCprop}
\end{equation}
\subsubsection{ Light-front wave function of the NN - (A-2)  System}
The light-front wave function of the $NN - (A-2)$ system describes the motion of the 
center of mass of the NN correlation. The calculation of this wave function  depends  on the consideration of  the following  expression, as well as its complex conjugated form,  in Eq. (\ref{spe_LF_2NSRC_1}):
\begin{equation}
{ \chi_{NN}^\dagger(p_{NN},s_{NN}) \chi^\dagger_{A-2}(p_{A-2},s_{A-2}) \over p_{NN}^2 - M_{NN}^2}
 \Gamma_{A\rightarrow NN,A-2}\chi_A.
\label{NNA-2}
\end{equation}
On the basis of  the  light-front momentum and energy   conservation 
at the $\Gamma_{A\rightarrow NN,A-2}$  vertex, and on the on-shell conditions  of the initial nucleus $A$ and the $A-2$ residual nuclear system, the propagator in Eq. (\ref{NNA-2}) can be expressed as
\begin{align}
p_{NN}^2 - M_{NN}^2 & = (p_{A}-p_{A-2})^2 - M_{NN}^2 = (p^+_{A}-p^+_{A-2})(p^-_{A}-p^-_{A-2}) - (\mathbf p_{A, \perp}-\mathbf p_{A-2, \perp})^2- M_{NN}^2 \nonumber \\
& = (p^+_{A}-p^+_{A-2})\left (p^-_{A}-p^-_{A-2} -\frac {(\mathbf p_{A, \perp}-\mathbf p_{A-2, \perp})^2}{  p^+_{A}-p^+_{A-2}}  - \frac {M_{NN}^2}{  p^+_{A}-p^+_{A-2}}    \right )            \nonumber \\
& = p^+_{NN}\left (\frac {M^2_A +\mathrm p^2_{A,\perp} }{p^+_A} - \frac {M^2_{A -2}+\mathrm p^2_{A-2,\perp} }{p^+_{A-2}} 
 -\frac {\mathrm p^2_{A, \perp}-2 \mathbf p_{A, \perp} \mathbf p_{A-2, \perp} + \mathrm p^2_{A-2, \perp}}{  p^+_{NN}}
- \frac {M_{NN}^2}{  p^+_{NN}}\right) \nonumber \\
& = \frac {\alpha_{NN}}{A}\left \{ M^2_A - \frac {A^2}{ \alpha_{NN}(A-\alpha_{NN})}\left[ M_{NN}^2 +\frac {\alpha_{NN}}{ A}\Big (M_{A-2}^2 - M_{NN}^2 \Big) +  \Big (p_{NN,\perp} - \frac {\alpha_{NN}}{ A}p_{A,\perp} \Big)^2\right]
 \right \} \nonumber \\
 & ={\alpha_{NN}\over A}(M_A^2 - s_{NN,A-2}),
\label{propNN}
\end{align}
where the invariant squared energy  $s_{NN,A-2}$, of the $NN - (A-2)$ system is given by 
\begin{align}
s_{NN,A-2} & = (p_{NN} + p_{A-2})^2 = (p^+_{NN} + p^+_{A-2})(p^-_{NN} + p^-_{A-2}) - (\mathbf p_{NN, \perp} + \mathbf p_{A-2, \perp})^2 \nonumber \\
& =  \frac {A^2}{ \alpha_{NN}(A-\alpha_{NN})}\left[ M_{NN}^2 +\frac {\alpha_{NN}}{ A}\Big (M_{A-2}^2 - M_{NN}^2 \Big) +  \Big (\mathbf p_{NN,\perp} - \frac {\alpha_{NN}}{ A} \mathbf p_{A,\perp} \Big)^2\right].
\label {sNN_A-2}
 \end{align}

The relative three-momentum, $\mathbf k_{CM} $,  in the center of mass of the $NN-(A-2)$ system is defined such that $\mathbf k_{CM}= \bf p_{NN} = -  \bf p_{A-2} $, then 
\begin {align}
E_{NN} & = \sqrt{M^2_{NN}  + \mathrm k^2_{CM}},\nonumber \\
 E_{A-2} & = \sqrt{M^2_{A-2} + \mathrm k^2_{CM}},
\end {align}
that implies that the invariant squared energy,  $s_{NN,A-2}$ [Eq. (\ref {sNN_A-2})], is given by 
\begin {align}
s_{NN,A-2} & = (p_{NN} + p_{A-2})^2 = (E_{NN} + E_{A-2}) ^2  = M^2_{NN}  + 2 \mathrm k^2_{CM} + 2 \sqrt{(M^2_{NN}  + \mathrm k^2_{CM})(M^2_{A-2} + \mathrm k^2_{CM})},
\label {sNN_A-2_CM}
\end {align}
 the magnitude of $\mathbf k_{CM} $ is therefore given by: 
\begin{align}
\mathrm k_{CM} & = \frac {\sqrt{[s_{NN,A-2} - (M_{NN} + M_{A-2})^2][s_{NN,A-2} - (M_{NN} - M_{A-2})^2]}}{2\sqrt{s_{NN,A-2}}},
\label{k-CM}
\end{align}
and the  
light-front momentum fraction of the $NN$ pair, in the center of mass of the $NN-(A-2)$ system,  by 
\begin{align}
\alpha_{NN} & = A \frac{p^+_{NN}}{p^+_{A} } = \frac {A(E_{NN} + \mathrm k_{CM,z})}{ E_{NN} + E_{A-2}}.
\end{align}

With the above definitions, similar to Eq. (\ref{srclcwf}),    the light-front wave function of the $NN-(A-2)$ system is defined as:
\begin{equation}
\psi^{s_A}_{CM}(\alpha_{NN},\mathrm k_{NN,\perp},s_{NN},s_{A-2})= - {1\over \sqrt{{A-2\over 2}}} {1\over \sqrt{2 (2\pi)^3}} 
{  \chi_{NN}^\dagger(p_{NN},s_{NN})  \chi^\dagger_{A-2}(p_{A-2},s_{A-2}) \Gamma_{A\rightarrow NN,A-2}\chi^{s_A}_A
\over {2\over A}\left[ M_{A}^2 - s_{NN,A-2} \right]},
\label{cmlcwf}
\end{equation}
where $\mathbf k_{NN,\perp} = \mathbf p_{NN,\perp} - \frac {\alpha_{NN}}{ A} \mathbf p_{A,\perp}$.
The coefficients in this wave function  are chosen such that in the nonrelativistic limit, $k_{CM}\ll M_{NN}$:
\begin{equation}
\psi^{s_A}_{CM}(\alpha_{NN},\mathrm p_{NN,\perp},s_{NN},s_{A-2})\approx 2M_N \psi^{NR}_{CM}(\mathrm  k_{CM}),  
\end{equation}
where $\psi^{NR}_{CM}(\mathrm  k_{CM}) $ is the nonrelativistic wave function of the center of mass of the  $NN-(A-2)$ system.

Substituting Eqs.(\ref{cmlcwf}) and (\ref{propNN}) in  Eq.(\ref{NNA-2}) yields
\begin{equation}
{ \chi_{NN}^\dagger(p_{NN},s_{NN}) \chi^\dagger_{A-2}(p_{A-2},s_{A-2}) \over p_{NN}^2 - M_{NN}^2}
\Gamma_{A\to NN,A-2}\chi_A  =  -{\sqrt{{A-2\over 2}}\sqrt{2 (2\pi)^3}\over \alpha_{NN}/2}\psi^{s_A}_{CM}(\alpha_{NN},\mathrm  k_{NN,\perp},s_{NN},s_{A-2}).
\label{CMprop}
\end{equation}
\subsubsection {2N SRC light-front  nuclear spectral function}
Inserting  Eqs. (\ref{SRCprop}) and (\ref{CMprop}) in Eq.(\ref{spe_LF_2NSRC_1}),  the following expression for the 2N SRC  light-front nuclear spectral function is obtained 
\begin{align}
 P_{A,2N}^{N}(\alpha_1, \mathrm p_{1,\perp},\tilde M_N^2) & =  {A-2\over 2} \sum\limits_{s_2,s_{NN},s_{A-2}}\int 
\psi^{s_A,\dagger}_{CM}(\alpha_{NN},\mathrm k_{NN,\perp},s_{NN},s_{A-2})  \psi^{s_{NN},\dagger}_{NN}(\beta_{1},\mathrm k_{1,\perp},s_{1},s_2) \nonumber \\
&   \times 2\delta(\alpha_1+ \alpha_2 + \alpha_{A-2}-A)\delta^2(\mathbf p_{1,\perp}+\mathbf  p_{2,\perp} -\mathbf   p_{NN,\perp})
\delta(\tilde M_N^2 - \tilde M_{N}^{(2N),2}) \nonumber \\
& \times \psi^{s_A}_{CM}(\alpha_{NN},\mathrm k_{NN,\perp},s_{NN},s_{A-2})  \psi^{s_{NN}}_{NN}(\beta_{1},\mathrm k_{1,\perp},s_{1},s_2) \nonumber \\
& \times \frac {d\beta_2}{2 - \beta_1}d^2\mathbf p_{2,\perp}
\frac {d\alpha_{NN}}{ A-\alpha_{NN}}d^2\mathbf p_{NN,\perp},
\label {spe_LF_2NSRC_2}
\end{align}
where the relations $\beta_2 = 2- \beta_1$, $d\beta_2/\beta_2 = d\alpha_2/\alpha_2$, $\bf p_{NN,\perp} = -  \bf p_{A-2,\perp}$,  and $\alpha_{A-2} = A-\alpha_{NN}$ have been applied. 

The  nuclear spectral function [Eq.(\ref {spe_LF_2NSRC_2})] can be further simplified by introducing the spin averaged density matrices for the relative motion  in the 2N SRC  \cite{FS81,FSS15}:
\begin{equation}
\rho^{N}_{NN}(\beta_1,\mathrm k_{1,\perp}) = {1\over 2}{1\over 2 s_{NN}+1}\sum\limits_{s_{NN},s_1,s_2}
{\mid \psi^{s_{NN}}_{NN}(\beta_1,\mathrm k_{1,\perp},s_1,s_2)\mid^2\over 2-\beta_1},
\label{rel_denmatrix_2NSRC}
\end{equation}
and for the center of mass motion of the 2N SRC:
\begin{equation}
\rho_{CM}(\alpha_{NN},\mathrm k_{NN,\perp}) = {1\over 2}{A-2\over 2s_{A} + 1}\sum\limits_{s_{NN},s_{A-2}} 
{\mid\psi^{s_A}_{CM}(\alpha_{NN},\mathrm k_{NN,\perp},s_{NN},s_{A-2})\mid^2\over A - \alpha_{NN}}.
\label{cm_denmatrix_2NSRC}
\end{equation}
With the above definitions, the  2N SRC light-front  nuclear spectral function is given by
\begin{eqnarray}
P_{A,2N}^{N}(\alpha_1,\mathrm p_{1,\perp},\tilde M_N^2) = {1\over 2}\int \rho^{N}_{NN}(\beta_1, \mathrm k_{1,\perp})\rho_{CM}(\alpha_{NN},\mathrm k_{NN,\perp}) 
2\delta(\alpha_1+ \alpha_2 - \alpha_{NN}) \nonumber \\ 
\delta^2(\mathbf p_{1,\perp}+\mathbf p_{2,\perp} - \mathbf p_{NN,\perp}) 
\delta(\tilde M_N^2 - \tilde M_{N}^{(2N),2}) d\beta_2 d^2\mathbf p_{2,\perp}
d\alpha_{NN}d^2\mathbf p_{NN,\perp}.
\label{spe_LF_2NSRC_3}
\end{eqnarray} 

The normalization conditions for the above introduced density matrices are 
defined from the sum-rule conditions of Eqs. (\ref{LF_sumrules}) and (\ref {LF_density_sumrules}).
For the density matrices of  NN SRC, $\rho^{N}_{NN}$,  the  normalization conditions to satisfy the baryonic  and momentum sum rules \cite{FS81}  yields
\begin{equation}
\int \rho^{N}_{NN}(\beta,\mathrm k_\perp){d\beta\over \beta}d^2\mathbf k_\perp = \int \rho^N_{NN}(\beta,\mathrm k_\perp)\beta {d\beta\over \beta}d^2\mathbf k_\perp = n^N_{2N},
\label {NN_LF_2NSRC_norms}
\end{equation}
where $n^N_{2N}$ is the contribution   to the total norm of the spectral function.
Similar to Eq.(\ref{2NSRC_relmd}), the light-front density matrix, $ \rho^{N}_{NN}$,  can be  modeled  through the high momentum 
part of the  light-front density matrix of the deuteron $\rho_d(\beta_1,k_{1,\perp})$ in the form
\begin{equation}
\rho^N_{NN}(\beta_1,\mathrm k_{1,\perp}) = {a_2(A,Z)\over (2x_N)^\gamma} \rho_d(\beta_1,\mathrm k_{1,\perp})\Theta(\mathrm k_1-\mathrm k_{src}),
\label{2NSRC_NNdens}
\end{equation}
where $\mathrm k_1$ is defined in Eq. (\ref{sNN_2}), and  $a_2(A,Z) $ in Eq. (\ref {2NSRC_relmd}). A model for the deuteron density  matrix $\rho_d$ is given in Eq. (\ref {rho_d}). \\
For the light-front density matrix,  $\rho_{CM}$,  the conditions of Eq. (\ref{LF_sumrules}) require the following 
normalization relations:
\begin{align}
\int \rho_{CM}(\alpha_{NN},\mathrm k_{NN,\perp}){d\alpha_{NN}\over \alpha_{NN}} d^2\mathbf k_{NN,\perp} & = 1, 
\label {CM_LF_2NSRC_norm1}
\end{align}
\begin{align}
 \int \rho_{CM}(\alpha_{NN},k_{NN,\perp})\alpha_{NN}{d\alpha_{NN}\over \alpha_{NN}} d^2\mathbf k_{NN,\perp} & = 2.
\label{CM_LF_2NSRC_norm2}
\end{align}
Since in the considered 2N SRC model the CM motion is nonrelativistic ($\mathbf k_{NN}\ll 2M_{N}$),  the momentum 
distribution    $\rho_{CM}$ can be approximated as 
\begin{equation}
\rho_{CM}(\alpha_{NN},\mathrm k_{NN,\perp})  = {E_{NN} E_{A-2}\over (E_{NN} + E_{A-2})/A} {n_{CM}(\mathrm k_{CM})\over A-\alpha_{NN}}\approx 2M_{N}n_{CM}(\mathrm k_{CM}),
\label {2NSRC_CMdens}
\end{equation}
where $n_{CM}$ is defined in Eq. (\ref{2NSRC_CMdist}).
Then,  for the \textquotedblleft middle" form of  (\ref {2NSRC_CMdens}),  the  normalization condition of Eq. (\ref{CM_LF_2NSRC_norm1}) is exact,  while  Eq. (\ref{CM_LF_2NSRC_norm2}) is exact 
 only in the non-relativistic limit.    

The  2N SRC  nuclear spectral function in LF approximation [Eq.(\ref {spe_LF_2NSRC_3})], together with the density  matrix  for the 2N SRC center of mass [Eq.(\ref{2NSRC_CMdens})], the density matrix  for the relative motion  in the 2N SRC 
[Eq.(\ref {2NSRC_NNdens})], and the normalization conditions [Eq.(\ref {NN_LF_2NSRC_norms})], 
[Eq.(\ref {CM_LF_2NSRC_norm1})] and [Eq.(\ref {CM_LF_2NSRC_norm2})], constitute the mathematical model from which  the corresponding computational models and numerical estimates are obtained in the chapters 3 and 4  of the present  dissertation.  

Finally, it is worth mentioning that in the non-relativistic limit of the density matrix of 2N SRC, $\rho(\beta_1,k_{1,\perp}) \approx n_{NN}(p_1) M_N$, the 2N SRC LF nuclear spectral function 
   [Eq.(\ref{spe_LF_2NSRC_3})], similar to the VN approximation,  reduces to the SRC  model of spectral function of Ref. \cite{CiofiSimula}.

\subsection{Nuclear spectral function in light-front  approximation for three nucleons in  short-range correlation} 
\label{sec:3NsrcLFA}

The calculation of  the nuclear spectral function in light-front approximation for three nucleons in  short-range correlation starts with the definition of the effective vertex $\hat V_{3N}$  in the 3N SRC covariant amplitude 
 (\ref{sp:3N_cov})
\begin{equation}
\hat V_{3N} = i \bar a(p_1,s_1) 2\alpha_1^2\delta(\alpha_1 + \alpha_2 + \alpha_{3}-3)\delta^2(\mathbf p_{1,\perp}+\mathbf p_{2,\perp} + \mathbf p_{3,\perp}) \delta(\tilde M_N^2 - \tilde M^{(3N),2}_N)a(p_1,s_1),
\end{equation}
where $(\alpha_2,\mathbf p_{2,\perp})$, ($\alpha_{3},\mathbf p_{3,\perp}$) are light-front momentum fractions and transverse momenta 
of correlated second and third nucleon respectively. The creation, $\bar a(p_1,s_1)$,  and the annihilation, $a(p_1,s_1)$, operators of nucleon with four-momentum $p_1$ and spin $s_1$ satisfy  the relations of Eq. (\ref{aops}). The delta functions represent the momentum and the invariant mass conservation in the vertex.

Since  in the light-front 3N correlation model, the particles 2   and  3  are considered on light-front energy shells,   the magnitude of the invariant mass $  \tilde M^{(3N),2}_N$ is calculated as follows
\begin{align}
\tilde M^{(3N),2}_N & =  {p^-_{1}p^+_{1} }- \mathrm  {p}^2_{1,\perp}  = p_{1+}(p_{A-} - p_{2-} - p_{3-} -
p_{A-3,-}) - \mathrm p_{1,\perp}^2 \nonumber \\
& =  p^+_{1}   \left (\frac  {M^2_A} { {p}^+_{A}} - \frac  {M^2_N + \mathrm  {p}^2_{2,\perp} } {{p}^+_{2}}  -  \frac  {M^2_{N} + \mathrm  {p}^2_{3,\perp}} {{p}^+_{3}}-\frac  {M^2_{A-3}} { {p}^+_{A-3}} \right) -\mathrm  {p}^2_{1,\perp}\nonumber \\
& =  \frac { p^+_1}{3p^+_A} \left ( \frac{M^2_A }{3}- \frac  {M^2_N + \mathrm  p^2_{2,\perp} } { p^+_2 /{3p^+_A} }  -  \frac  {M^2_N + \mathrm  p^2_{3,\perp} } { p^+_3 /{3p^+_A}} -    \frac  {M^2_{A-3} }  { p^+_{A-3} /{3p^+_A} }\right)    -\mathrm  p^2_{1,\perp} \nonumber \\
& = {\alpha_1\over 3}\left( M_{3N}^2 - {M_N^2 + \mathrm p_{2,\perp}^2\over \alpha_{2}/3} - 
{M_N^2 + \mathrm p_{3,\perp}^2\over \alpha_{3}/3}\right) - \mathrm p_{1,\perp}^2,
\label{Inv-mass-3N}
\end{align}
with the mass of the 3N SRC  defined as:
\begin{equation}
M^2_{3N} = {3\over A} M_A^2 - 3 {M_{A-3}^2\over \alpha_{A-3}},
\label {M_3N_sq}
\end{equation}
where  $ \alpha_{A-3} = A-3$.

Similarly to the propagator  integration  described in section \ref {sec:2NsrcLFA}, the  integrations by $dp^-_2$ and $dp^-_3$  in   Eq. (\ref{sp:3N_cov})  are performed through the
$"-"$ components of the positive  poles of the  propagators of the on-shell particles $2$  and $3$  respectively,   provided that their $"+"$ components 
are large and positive, so that the $Z$ diagram in this scheme will be suppressed by the inverse power of large $"+"$ component of 
the nucleon's four-momentum  \cite{Weinberg}. The integration  by $dp^-_2$  of   the propagator of the on-shell particle 2 yields 
\begin{align}
\oint  \frac {d^4  p_{2}} {p^2_2 -  {M_N} + i \varepsilon } & = \oint   \frac {  {d p^-_2 d p^+_2  d^2} \mathbf {p}_{2,\perp} }{2  { \Big (p^+_2p^-_2 } - \mathrm {p}^2_{2,\perp} -  {M_N} \Big) + i \varepsilon} \nonumber \\
& = \oint  \frac {  {d p^-_2 d p^+_2  d^2} \mathbf {p}_{2,\perp} }{2  { p^+_2 \Big [ p^-_2 }- \Big (\mathrm {p}^2_{2,\perp} +  {M_N \Big )/ p^+_2  \Big ]   + i \varepsilon}} 
 =   - {2\pi i  {d p^+_2  d^2} \mathbf {p}_{2,\perp}\over 2  {p^+_2}} \left |_{ {p^-_2} = \frac { \mathrm {p}^2_{2,\perp} + {M_N }}{ p^+_2 }}  \right.  \nonumber \\
& =   - \pi i  \frac {d \alpha_2 }{\alpha_2} d^2\mathbf {p}_{2,\perp}  \left |_{ {p^-_2} = \frac { \mathrm {p}^2_{2,\perp} +  {M_N }}{ p^+_2 }}  \right.,
\label{3NSRC_poles-p2}
\end{align}
and integration  by $dp^-_3$   of the propagator of the on-shell particle $3$ yields 
\begin{align}
\oint  \frac {d^4  p_{3}} {p^2_3 -  {M_N} + i \varepsilon } & = \oint  \frac {  {d p^-_3 d p^+_3  d^2} \mathbf {p}_{3,\perp} }{2  { \Big (p^+_2p^-_3 } - \mathrm {p}^2_{3,\perp} -  {M_N} \Big) + i \varepsilon} \nonumber \\
& =\oint   \frac {  {d p^-_3 d p^+_3  d^2} \mathbf {p}_{3,\perp} }{2  { p^+_3 \Big [ p^-_2 }- \Big (\mathrm {p}^2_{3,\perp} +  {M_N \Big )/ p^+_3  \Big ]   + i \varepsilon}} 
 =   - {2\pi i  {d p^+_3  d^2} \mathbf {p}_{3,\perp}\over 2  {p^+_3}} \left |_{ {p^-_3} = \frac { \mathrm {p}^2_{3,\perp} + {M_N }}{ p^+_3 }}  \right.  \nonumber \\
& =   - \pi i  \frac {d \alpha_3 }{\alpha_3} d^2\mathbf {p}_{3,\perp}  \left |_{ {p^-_3} = \frac { \mathrm {p}^2_{3,\perp} +  {M_N }}{ p^+_2 }}  \right..
\label{3NSRC_poles-p3}
\end{align}
The  above integrations project the intermediate state to the positive light-front energy state thus excluding the contribution from the Z  graph of Fig. \ref{Fig:R_diagram}(c).  With the diminished contribution from the $Z$ graph,  the covariant 3N SRC amplitude  [Eq. (\ref{sp:3N_cov})], will result in the  light-front spectral function, $P^N_{A,3N}(\alpha_1,p_{1,\perp},\tilde M_N^2)$  of the 3N SRC.

As in the LF 2N SRC approximation, the numerator of the propagator for the on-shell particles $2$ and $3$  in Eq. (\ref{sp:3N_cov}), can be expressed by  the following on-shell sum rule 
\begin{align}
\sh p_2 + M_N & =   \sum\limits_{s_2}u(p_2,s_2)\bar u(p_2,s_2), 
\label{2_LF_3NSRC}
\end{align}
\begin{align}
\sh p_3 + M_N & =   \sum\limits_{s_3}u(p_3,s_3)\bar u(p_3,s_3), 
\label{3_LF_3NSRC}
\end{align}
where  $u(p_i,s_i)$ is the spinor of the particle $i$  with momentum $p_i$ and 
spin projection $s_i$.
Using a similar approximate relation for the 2$^\prime$  propagator which represents the nucleon between consecutive $NN$ interaction vertices yields 
\begin{align}
\sh p_{2'} + M_N & =   \sum\limits_{s_{2'}}u(p_{2'},s_{2'})\bar u(p_{2'},s_{2'}). 
\label{2'_LF_3NSRC}
\end{align} 

Inserting  Eqs. (\ref{aops}) and  (\ref {3NSRC_poles-p2}) -(\ref {2'_LF_3NSRC})  in   the covariant 3N SRC amplitude [Eq. (\ref{sp:3N_cov})], and summing over all possible spin projections  $s_2,s_{3}$ and $s_{2'}$,   reduces the latter to the 
3N SRC part of the nuclear spectral function in LF approximation, namely
\begin{align}
P^N_{A,3N}(\alpha_1,\mathrm p_{1,\perp},\tilde M^2_N) & = \sum\limits_{s_2,s_3,s_{2^\prime},\tilde s_{2^\prime}} \int \bar u(k_1, \lambda_1) \bar u(k_2, \lambda_2)\bar u(k_3, \lambda_3) \Gamma^\dagger_{NN\rightarrow N,N} 
 {u(p_{2^\prime},\tilde s_{2^\prime}) \bar u(p_{2^\prime},\tilde s_{2^\prime})  \over p_{2^\prime}^2 - M_N^2} 
 \Gamma^\dagger_{NN\rightarrow N,N} \nonumber \\
  &\times     2\alpha_1^2 \delta(\alpha_1 + \alpha_2  + \alpha_3 - 3) \delta^2(\mathbf p_{1,\perp } + \mathbf p_{2,\perp} + \mathbf p_{3,\perp}) \delta(\tilde M_N^2 - 
  \tilde M_N^{(3N),2}) \nonumber \\
& \times  { u(p_1,s_1) \over p_1^2 - M_N^2}  
  u(p_2,s_2) \bar u(p_2,s_2) {\bar u(p_1,s_1) \over p_1^2 - M_N^2} \Gamma_{NN\rightarrow N,N} {u(p_{2^\prime}, s_{2^\prime}) \bar u(p_{2^\prime},s_{2^\prime})  \over p_{2^\prime}^2 - M_N^2} 
 \nonumber \\
&   \times  u(p_3,s_3) \bar u(p_3,s_3) \Gamma^\dagger_{NN\rightarrow NN} u(k_1, \lambda_1) u(k_2, \lambda_2) u(k_3, \lambda_3) \nonumber \\
&   \times  {d\alpha_2\over \alpha_2} {d^2 \mathbf p_{2,\perp}\over 2(2\pi)^3}{d\alpha_3\over \alpha_3}{d^2 \mathbf p_{3,\perp}\over 2(2\pi)^3}.
 \label{spe_LF_3NSRC_1}
\end{align}.

\subsubsection {Light-front  wave function of the first 2N SRC}
The   calculation of the light-front  wave function of the first 2N SRC, shown in Fig. \ref {Fig:3nSpectral},  between particle $2$ with momentum $k_2$,  and particle $3$ with momentum $k_{3}$, which results in particle $2'$ with momentum $p_{2'}$,  and particle $3$ with momentum $p_3$,  depends on  the consideration of  the following  expression, as well as its complex conjugated form,   in Eq. (\ref{spe_LF_3NSRC_1}):
\begin{equation}
 {\bar u(p_{2^\prime},s_{2^\prime}) \bar u(p_3,s_3) \Gamma_{NN\rightarrow N,N} u(k_2,\lambda_2) u(k_3, \lambda_3)  \over p_{2^\prime}^2 - M_N^2}.
\label{first_2NSRC_1}
\end{equation}

Before calculating the LF wave functions, it is convenient to define important kinematic variables and relations  that result from the collinear approximation. The assumptions of  collinear approximation in the diagram of Fig. \ref {Fig:3nSpectral} implies  that the 
momentum fractions of the 3N SRCs carried by each initial nucleon is unity  and their total transverse momenta is  neglected, then
\begin{align}
  k^+ & = k^+_{1} +   k^+_{2}+   k^+_{3},  \nonumber \\
 \alpha_k & = \frac {3 k^+_{1} } { k^+ } = \frac {3 k^+_{2} } { k^+ } = \frac {3 k^+_{3} } { k^+ } = 1 \nonumber \\
 \mathbf k_{1,\perp} +   \mathbf k_{2,\perp} +   \mathbf k_{3,\perp} & = 0,
\label{coll_initial}
\end{align}
also from the energy and momentum conservation it follows that: 
\begin{align}
k_{1} +   k_{2}+   k_{3} & =   p_{1} +   p_{2}+   p_{3}, \nonumber \\
 k^+ &  =   p^+_{1} +   p^+_{2}+   p^+_{3}, \nonumber \\
 \mathbf p_{1,\perp} +   \mathbf p_{2,\perp} +   \mathbf p_{3,\perp} = 0, \nonumber \\
  k_{2}+   k_{3} & =     p_{2'}+   p_{3}, \nonumber \\
  k_{1}+   p_{2'} & =     p_{2}+   p_{3},  
\label{coll_conser}
\end{align}
and the momentum fractions of the initial as well as  those of the correlated nucleons are
\begin{align}
\alpha_i & = \frac {3 p^+_{i} } { k^+ } \; \;\; i = 1,2,3  \nonumber \\ 
\alpha_1 + \alpha_2 + \alpha_3 & = 3,   \nonumber \\
\alpha_{2'} & = \frac {3 p^+_{2'} } { k^+ }  =  \frac { 3 (k_{2}+   k_{3} - p_{3})}  { k^+ } = 2 - \alpha_3, \nonumber \\
\beta_1 & = \frac {2 \alpha_1} {\alpha_{12} }  = \frac {2 \alpha_1} {\alpha_{1} + \alpha_{2}} = \frac {2 \alpha_1} {3- \alpha_3}, \nonumber \\
\beta_2 & = \frac {2 \alpha_2} {\alpha_{12} }  = \frac {2 \alpha_2} {\alpha_{1} + \alpha_{2}} = \frac {2 \alpha_2} {3- \alpha_3},  \nonumber \\
\beta_3 & = \frac {2 \alpha_3} {\alpha_{2'3} }  = \frac {2 \alpha_3} {\alpha_{2'} + \alpha_{3}} =  \alpha_3,  \nonumber \\
\beta_{2'} & = \frac {2 \alpha_{2'}} {\alpha_{2'3} }  = \frac {2(2- \alpha_3)} {2} = 2- \alpha_3.
\label{coll_momFrac}
\end{align}

By applying  the above defined relations, as well as the  light-front momentum and energy   conservation 
at the $\Gamma_{NN\rightarrow N,N}$  vertex, and  the on-shell condition of initial particles as well of the correlated particles 2 and 3, the propagator in Eq. (\ref{first_2NSRC_1}) can be expressed as 
\begin{align}
p_{2^\prime}^2 - M_N^2  & = (k_{2}+   k_{3} -   p_{3} )^2- M_N^2 \nonumber \\
& = (  k^+_{2}+   k^+_{3} -    p^+_{3} )(k^-_{2}+   k^-_{3} - p^-_{3} ) -   \mathrm p^2_{3,\perp} )-M^2_N \nonumber \\
& =  p^+_{2'}  \left(    \frac {M^2_N}{k^+_{2}} + \frac {M^2_N}{k^+_{3}} - \frac {M^2_N + \mathrm p^2_{3,\perp}}{p^+_{3}} - \frac {M^2_N + \mathrm p^2_{3,\perp}}{p^+_{2'}}     \right) \nonumber \\
& =  (2-\alpha_3)  \left(    2 M^2_N - \frac {M^2_N + \mathrm p^2_{3,\perp}}{\alpha_3} - \frac {M^2_N + \mathrm p^2_{3,\perp}}{2-\alpha_3}     \right) \nonumber \\
& = \frac {2-\beta_3}{ 2}\left[M_{23}^2 - 4 \frac {M_N^2 + \mathrm p_{3,\perp}^2}{ \beta_3(2-\beta_3)}\right],
\label{3LCprop1_1}
\end{align}
where the  invariant mass $M^2_{23}$ squared is 
\begin{align}
 M_{23}^2  & = (k_2 + k_3 )^2   = (k^+_{2} +   k^+_{3} )( k^-_{2}+   k^-_{3} ) 
\nonumber \\
& =  M^2_N \left ( 2 + \frac { k^+_{3} }{k^+_{2} }+ \frac { k^+_{2} }{k^+_{3}}\right ) \approx 4M_N^2.
 \label{M23}
\end{align}

The  propagator (\ref {3LCprop1_1})  can be completely expressed though the internal momenta of the $NN$ system by introducing the relative momentum,  $\mathbf {\tilde k}_3$,  in the $NN$ center of mass frame, as follows:
\begin{align}
\mathrm {\tilde k^2}_3 = \frac {M^2_N + \mathrm p^2_{3,\perp}}{ \beta_3(2-\beta_3)} - M^2_N.
\label{tilde_k3_LF}
\end{align} 
The relative momentum $\mathbf {\tilde k}_3$  is invariant with respect to the Lorentz  boost in the $\bf p_{NN}$ direction, and   
will be used to set a  momentum scale for  the  2N SRCs, requiring  $\mathrm {\tilde k}_3\ge \mathrm k_{src}$ similar to Eq. (\ref{2NSRC_relmd}).\\
The propagator (\ref {3LCprop1_1}) can therefore be expressed as
\begin{align}
p_{2^\prime}^2 - M_N^2   
& = \frac {2-\beta_3}{ 2}\left[M_{23}^2 - 4 (M_N^2 +\mathrm {\tilde k^2}_3) \right].
\label{3LCprop1_2}
\end{align}

Inserting   Eq. (\ref{3LCprop1_2}) in Eq. (\ref {first_2NSRC_1}),  the light-front wave function of the  first  $NN$ SRC is defined  as 
\begin{align}
\psi_{NN}(\beta_{2'},\mathrm p_{3,\perp},s_{2'},s_3) & =  \frac {1}{ \sqrt{2(2\pi)^3}} \frac { \bar u(p_{2'},s_{2'})\bar u(p_3,s_3)
\Gamma_{NN\rightarrow N,N}  u(k_2,\lambda_2) u(k_3,\lambda_3)}{\frac {1}{2}\Big [ M_{23}^2 - 4 (M_N^2 +\mathrm {\tilde k^2}_3)\Big]}.
\label{NNlcwfb}
\end{align}
The above defined wave function results in the following simplified expression for Eq.(\ref{first_2NSRC_1})
\begin{align}
{\bar u(p_{2^\prime},s_{2^\prime}) \bar u(p_3,s_3) \Gamma_{NN\rightarrow NN} u(k_2) u(k_3)  \over p_{2^\prime}^2 - M_N^2}  & = 
 \sqrt{2(2\pi)^3} \frac {\psi_{NN}(\beta_{2'},\mathrm p_{3,\perp},s_{2'},s_3)}{ 2 - \beta_3}.
\label{first_2NSRC_2}
\end{align}

\subsubsection {Light-front  wave function of the second 2N SRC}
The   calculation of the light-front  wave function of the second 2N SRC, shown in Fig. \ref {Fig:3nSpectral},  between particle $1$ with momentum $k_1$,  and particle $2'$ with momentum $p_{2'}$, which results in particle $1$ with momentum $p_{1}$,  and particle $2$ with momentum $p_2$,  depends on the consideration of  the following  expression, as well as its complex conjugated form,   in Eq. (\ref{spe_LF_3NSRC_1}):
\begin{align}
 {\bar u(p_1,s_1) \bar u(p_2,s_2) \Gamma_{NN\rightarrow N,N} u(p_{2^\prime}, s_{2^\prime}) u(k_1,\lambda_1)  \over p_{1}^2 - M_N^2}. 
\label{second_2NSRC_1}
\end{align}

By applying  the relations in Eqs. (\ref {coll_initial}) to (\ref {coll_momFrac} ), as well as   the   light-front momentum and energy   conservation at the $\Gamma_{NN\rightarrow N,N}$  vertex, and  the on-shell condition of the initial particles as well of the  correlated particles 2 and 3, the propagator in Eq. (\ref{second_2NSRC_1}) can be expressed as 
\begin{align}
p_1^2 - M_N^2 & = (k_{1} +   k_{2}+   k_{3} - p_{2}-   p_{3} )^2- M^2_N \nonumber \\
& = (k^+_{1} +   k^+_{2}+   k^+_{3} - p^+_{2}-   p^+_{3} )(k^-_{1} +   k^-_{2}+   k^-_{3} - p^-_{2}-   p^-_{3} ) - (\mathbf p_{2,\perp} +   \mathbf p_{3,\perp} )^2-M^2_N \nonumber \\
& =  p^+_{1}  \left( \frac {M^2_N}{k^+_{1}} +   \frac {M^2_N}{k^+_{2}} + \frac {M^2_N}{k^+_{3}} - \frac {M^2_N + \mathrm p^2_{2,\perp}}{p^+_{2}} - \frac {M^2_N + \mathrm p^2_{3,\perp}}{p^+_{3}}     -\frac{ M_N^2 +\mathrm p^2_{1,\perp} }{p^+_{1}}\right) \nonumber \\
& =  \alpha_{1}  \left( 3M^2_N - \frac {M^2_N + \mathrm p^2_{2,\perp}}{\alpha_{2}} - \frac {M^2_N + \mathrm p^2_{3,\perp}}{\alpha_{3}}     -\frac{ M_N^2 +\mathrm p^2_{1,\perp} }{\alpha_{1}}\right) \nonumber \\ 
& = \frac {\beta_1}{ 2}\left [ M^2_{12} - \frac { 4[M^2_N + (\mathbf p_{1,\perp} - \frac {\beta_1}{ 2} \mathbf  p_{12,\perp})^2]}{ (2-\beta_1)\beta_1}\right ], 
\label{3LCprop2_1}
\end{align}
where $\bf  p_{12,\perp} = \bf p_{1,\perp} + \bf p_{2,\perp}$, and the invariant mass  $M^2_{12} $ squared  is
\begin{align}
M_{12}^2 & =   (k_1 + k_2 + k_3 - p_3)^2   \nonumber \\ 
& = (k^+_{1} +   k^+_{2}+   k^+_{3} -   p^+_{3} )(k^-_{1} +   k^-_{2}+   k^-_{3} -    p^-_{3} ) -    \mathrm p^2_{3,\perp} 
\nonumber \\
& = \left ( 1- \frac {\alpha_3}{3} \right ) \left (9 M^2_N - \frac {M^2_N + \mathrm p^2_{3,\perp}}{\alpha_{3}/3}\right ) - \mathrm p^2_{3,\perp}\nonumber \\ 
& \approx M_{3N}^2\left (1 -  {\alpha_3\over 3}\right ) - 3{M_N^2 +  \mathrm p_{3,\perp}^2\over \alpha_3} + M_N^2.
\end{align}

The  propagator (\ref {3LCprop2_1})  can be completely expressed though the internal momenta of the $NN$ system by introducing the relative momentum,  $\mathbf {\tilde k}_1$,  in the $NN$ center of mass frame, as follows:
\begin{align}
\mathrm {\tilde k}_1^2 & = {M^2_N + \mathrm {\tilde k^2}_{1,\perp} \over \beta_1(2-\beta_1)} - M^2_N,
\label{tilde_k1_LF}
\end{align}
where the relative transverse momentum is
\begin{align}
\mathbf {\tilde k}_{1,\perp} & = \bf p_{1,\perp} - \frac {\beta_1}{ 2}\bf p_{12,\perp}. 
\label{tilde_k1_perp_LF}
\end{align}
The relative momentum $\mathbf {\tilde k}_1$  is invariant with respect to the Lorentz  boost in the $\bf p_{NN}$ direction, and   
will be used to set a  momentum scale for  the  2N SRCs, requiring  $\mathrm {\tilde k}_1\ge \mathrm k_{src}$ similar to Eq. (\ref{2NSRC_relmd}).\\
The propagator (\ref {3LCprop2_1}) can therefore be expressed as
\begin{align}
p_1^2 - M_N^2 & = \frac {\beta_1}{ 2}\left [ M^2_{12} - \frac { 4(M^2_N + \mathrm k^2_{1,\perp})}{ (2-\beta_1)\beta_1}\right ] \nonumber \\
& = \frac {\beta_1}{ 2}\left [ M^2_{12} - 4(M^2_N + \mathrm {\tilde k}^2_{1}) \right ].
\label{3LCprop2_2}
\end{align}

Inserting   Eq. (\ref{3LCprop2_2}) in Eq. (\ref {second_2NSRC_1}),  the light-front wave function of the  second  $NN$ SRC is defined  as 
\begin{align}
\psi_{NN}(\beta_1, \mathrm  {\tilde k}_{1,\perp},s_1,s_2) & =  \frac {1}{ \sqrt{2(2\pi)^3}} \frac { \bar u(p_1,s_1)\bar u(p_2,s_2)
\Gamma_{NN\rightarrow N,N}  u(p_{2^\prime},s_{2^\prime})u(k_1,\lambda_1)}{\frac {1}{2}\Big [ M^2_{12} - 4(M^2_N + \mathrm {\tilde k}^2_{1})\Big]}.
\label{NNlcwfa}
\end{align}
The above defined wave function  results in the following simplified expression for  Eq.(\ref{second_2NSRC_1})
\begin{align}
{\bar u(p_1,s_1) \bar u(p_2,s_2) \Gamma_{NN\rightarrow N,N} u(p_{2^\prime}, s_{2^\prime}) u(k_1)  \over p_{1}^2 - M_N^2} & = \sqrt{2(2\pi)^3} {\psi_{NN}(\beta_1,\mathrm {\tilde k}_{1,\perp},s_1,s_2) \over \beta_1}.
\label{second_2NSRC_2}
\end{align}

\subsubsection {3N SRC light-front  nuclear spectral function}
Inserting  Eqs. (\ref{first_2NSRC_2}) and (\ref{second_2NSRC_2}) in Eq.(\ref{spe_LF_3NSRC_1}),  the following expression for the 3N SRC  light-front nuclear spectral function is obtained 
\begin{align}
P^N_{A,3N}(\alpha_1,\mathrm p_{1,\perp},\tilde M^2_N) & =\sum\limits_{s_1,s_2,s_3,s_{2^\prime}} \int 
{\psi^\dagger_{NN}(\beta_{2'},\mathrm p_{3,\perp}, s_{2^\prime},s_3)\over 2-\beta_3}{\psi^\dagger_{NN}(\beta_1, \mathrm {\tilde k}_{1,\perp}, s_{1},s_2) 
\over \beta_1} \nonumber \\
& \times 2\alpha_1^2 \delta(\alpha_1 + \alpha_2  + \alpha_3 - 3) \delta^2(\mathbf p_{1,\perp} + \mathbf p_{2,\perp} + \mathbf p_{3,\perp}) 
\delta(\tilde M_N^2 - \tilde M_N^{(3N),2}) 
\nonumber \\
&  \times {\psi_{NN}(\beta_1, \mathrm {\tilde k}_{1,\perp}, s_{1},s_2) \over \beta_1} {\psi_{NN}(\beta_{2'}, \mathrm p_{3,\perp}, s_{2'},s_3)\over 2-\beta_3} 
{d\alpha_2\over \alpha_2} d^2 \mathbf p_{2,\perp}{d\alpha_3\over \alpha_3}d^2 \mathbf p_{3,\perp}.
 \label{spe_LF_3NSRC_2}
\end{align}

The  nuclear spectral function (\ref  {spe_LF_3NSRC_2}) can be further simplified  by summing over the final  and average of 
all possible initial polarization configurations,  similar to the one in VN approximation [Eq. (\ref{spe_VNA_2NSRC_3})], 
 as well as by introducing  the light-front spin averaged density matrices for the relative motion  in the first and second  2N SRC,   similar to those defined in Eq.(\ref{rel_denmatrix_2NSRC}). Then,  the 3N SRC light-front nuclear spectral function is obtained  as the  convolution of two NN light-front density matrices as follows:
\begin{align}
P^N_{A,3N}(\alpha_1,\mathrm p_{1,\perp},\tilde M^2_N)  &  =   \int  {3-\alpha_3\over 2(2-\alpha_3)^2} \rho_{NN}(\beta_{2'}, \mathrm p_{3,\perp}) 
\rho_{NN}(\beta_1, \mathrm {\tilde k}_{1,\perp}) 
2 \delta(\alpha_1 + \alpha_2  + \alpha_3 - 3)\nonumber \\
& \times  \delta^2(\mathbf p_{1,\perp} + \mathbf p_{2,\perp} + \mathbf p_{3,\perp}) 
 \delta(\tilde M_N^2 - \tilde M_N^{(3N),2})  d\alpha_2 d^2  \mathbf p_{2,\perp}d\alpha_3d^2  \mathbf p_{3,\perp}.
 \label{spe_LF_3NSRC_3}
\end{align}

The normalization conditions for the above introduced density matrices are 
defined from the sum-rule conditions of Eqs. (\ref{LF_sumrules}) and (\ref {LF_density_sumrules}).
For the density matrices of  NN SRC, $\rho^{N}_{NN}$,  the  normalization conditions to satisfy the baryonic  and momentum sum rules \cite{FS81}  yields
\begin{equation}
\int \rho^{N}_{NN}(\beta,\mathrm k_\perp){d\beta\over \beta}d^2\mathbf k_\perp = \int \rho^N_{NN}(\beta,\mathrm k_\perp)\beta {d\beta\over \beta}d^2\mathbf k_\perp = n^N_{3N},
\label {NN_LF_3NSRC_norms}
\end{equation}
where $n^N_{3N}$ is the contribution   to the total norm of the nuclear spectral function.

 In the nuclear spectral function [Eq.(\ref {spe_LF_3NSRC_3})],  similar to Eq. (\ref{3NSRC_relmd})  for VN approximation, the product of the two density matrices is expressed through the product of high momentum parts of the  deuteron density matrices in the form:
\begin{equation}
\rho_{NN}(\beta_{2'},\mathrm p_{3,\perp}) \rho_{NN}(\beta_1, \mathrm  {\tilde k}_{1,\perp})  =  a^2_2(A,Z) C^N(A,Z) \rho_{d}(\beta_{2'},\mathrm  p_{3,\perp})\Theta( \mathrm {\tilde k}_1-k_{src}) \rho_{d}(\beta_1, \mathrm  {\tilde k}_{1,\perp}) \Theta(\tilde k_3-k_{src}),
\label{3N_LFanzats}
\end{equation}
where $\tilde k_1$ and $\tilde k_3$ are defined in Eqs. (\ref {tilde_k1_LF}) and (\ref {tilde_k3_LF}) respectively.  The factors $a_2(A,Z) $ and $C^N(A,Z) $ are the same as those defined in Eq. (\ref{3NSRC_relmd}) for the case of VN approximation.  A model for the deuteron density  matrix $\rho_d$ is given in Eq. (\ref {rho_d}).

The  3N SRC  nuclear spectral function in LF approximation [Eq.(\ref {spe_LF_3NSRC_3})], together with the product of  the density  matrices   for the relative momentum distribution of the NN SRC [Eq.(\ref {3N_LFanzats})], as well as  the normalization conditions [Eq.(\ref {NN_LF_3NSRC_norms})], constitute the mathematical model from which  the corresponding computational models and numerical estimates are  obtained in the chapters 3 and 4  of the present  dissertation.

\section{Summary.} 
\label{sec:Ch2Summ}

On the basis of  the NN short-range correlation picture of the high-momentum component of nuclear wave function,  a model has been  developed
for the nuclear spectral functions in  the domain of high momentum and  high removal energy of bound nucleon in the nucleus.
In the high momentum  domain,  the models of nuclear spectral functions   should  be relativistic. 

A main focus therefore  in  developing the  models, in the present chapter, is given to the   treatment of  the relativistic effects which are important for the bound nucleon momenta   exceeding the characteristic 
Fermi momentum, $k_{F}$, in the nucleus.  The relativistic effects in the present  dissertation  are treated by using the effective Feynman 
diagrammatic approach, in which  the  Lorentz covariant amplitudes are reduced  to the nuclear spectral functions, in a process of calculation designed  to trace the relativistic effects entering in such functions.

One of the main ambiguities related to the treatment of  relativistic effects is the account for the vacuum fluctuations  ($Z$ graphs) which ultimately alter the  definition of the spectral function as a probability of finding a bound nucleon in the nucleus with the given momentum and positive removal energy. Two approaches are employed in the present  dissertation to deal with the vacuum fluctuations:  virtual nucleon and light-front approximations, both   suitable for an adequate  treatment of  the corresponding  relativistic effects.  

The results obtained in the present chapter within VN [Eq. (\ref {spe_VNA_2NSRC_4})], and LF [Eq. (\ref {spe_LF_2NSRC_3})], represent an attempt to account for the relativistic effects in the domain of 2N SRCs with center of mass motion  of the NN pair. The results  agree with the 2N SRC (with center of mass motion) model of Ref. \cite{CiofiSimula} in the non-relativistic limit. 

 The above defined approaches to deal with the vacuum fluctuations are  extended to calculate  the contributions from three-nucleon short-range correlations. The 3N SRCs are modeled as two  sequential NN short-range correlations. The corresponding models of nuclear spectral functions are based on the collinear approach in which a   negligible center of mass momentum for the residual or uncorrelated $A-3$ nuclear system is assumed. The derived spectral functions within  VN [Eq. (\ref{spe_VNA_3NSRC_4})]and LF [Eq. (\ref {spe_LF_3NSRC_3})] approximations represent  the results for 3N SRC  contribution to the nuclear spectral functions.

The models for nuclear spectral function developed in the present chapter  also account for   the experimental results of the strong  (by factor of 20) dominance  of the  $pn$ SRCs in nuclei,   
as compared to  the $pp$ and $nn$  correlations, for internal momentum range  of $\sim250-650$~MeV/$c$. The dominance of the pn correlations implies that to quantitatively describe the obtained models of nuclear spectral functions, in a high momentum domain, only what is needed is the knowledge  of  the high momentum deuteron wave function, either in the laboratory frame (for VN approximation) or on the light front (for LF approximation).


\chapter{Multinucleon short-range correlation model for nuclear spectral functions: Models for numerical computation}


\label{sec:Ch3Intro}

In order to obtain numerical estimates of the nuclear spectral functions,  the  computational models  are  developed in the present  chapter by integration of the delta functions in the mathematical models obtained in chapter 2. 
The integrals are  calculated by using the following formulas \cite {AW05} 
\begin{align}
\int f(x) \delta(g(x)) dx & =  \sum_a  \int f(x)  \frac { \delta(x-x^a)} {{| g'(x^a)|}_{x = x^a}} dx  = \sum_a \frac {f(x^a) } {{| g'(x^a)|}_{x = x^a}}.
\label{delta_integral}
\end{align}
where  $x^a$ are the roots of  $ g(x) = 0$. 

The prediction that single proton or neutron  momentum distributions in the 2N SRC domain are inversely proportional to their 
relative fractions in  nuclei Ref.\cite{newprops,proa2},  is represented in the computational  models of the nuclear spectral functions by the proton and neutron relative fraction factor $x_{p/n}$, introduced  in Eq. (\ref {2NSRC_relmd}) and    given by: 
 \begin{align}
x_p  &  = \frac {Z}{A}, \nonumber\\ 
x_n &  = \frac {A-Z}{A}.
 \label{2N-fraction}
 \end{align}

In the case of 3N SRC for asymmetric nuclei, the 3N SRC suppression factor $C^N(A,Z)$, defined in Eq.  (\ref {3NSRC_relmd}),  which accounts for the effects associated with the isospin structure of two-nucleon recoil system, are given by
 \begin{align}
C^{p}(A,Z) &  = \frac {2(Z-1)}{A}, \nonumber\\ 
C^{n}(A,Z) &  = \frac {2(A-Z-1)}{A}.
 \label{3N-C}
 \end{align} 
 
The computational models developed in the present chapter for 2N SRC and 3N SRC in asymmetric nuclei are specifically associated to proton or to neutron through the   factors given by  Eqs. (\ref {2N-fraction}) and (\ref {3N-C}) respectively. 

The outline of the chapter is as follows. Section  \ref {sec:ComSpecVNA}  presents  the computational  models   for 2N and 3N SRC  nuclear spectral functions  in VN  approximation.  The computational  models   for 2N and 3N SRC  nuclear spectral functions  in LF approximation are developed in section  \ref{sec:ComSpecLFA}.  Section  \ref{sec:ComSpecNRA}   includes the  computational models   for 2N and 3N SRC  nuclear spectral functions  in  the non-relativistic limit of the VN approximation. Section \ref {sec:Ch3Summ} summarizes the results of the chapter.

 \section{Computational\; models\; for\;SRC\; nuclear\; spectral\; function\; in\; virtual\; nucleon\; approximation} 
 \label{sec:ComSpecVNA}
 The computational models for high momentum  nuclear  spectral function in virtual nucleon (VN) approximation, with two  and three nucleons in short range correlation,  are developed in the present section.

 \subsection { Computational\;model\; for\; two\; nucleon\; SRC \;nuclear\;spectral\;function \;in virtual\;nucleon\;approximation }
 \label{subsec:Com2NsrcVNA}

The following expression for  the  2N SRC  nuclear spectral function in VN approximation  is obtained by inserting      the relative momentum distribution of the NN SRC [Eq.(\ref {2NSRC_relmd})] in Eq. (\ref {spe_VNA_2NSRC_4}), namely
\begin{align}
 S^{p/n} _{A,2N}( \mathrm p_1,  E_m) & =  \frac {a_2(A,Z)}{ (2x_{p/n})^{\gamma}}  \int \frac {n_{d}( \mathrm {p_{rel} })\Theta ( \mathrm  {p_{rel} - k_{src}}) }{\frac {M_N-E_m-T_{A-1}}{{M_A} /{A}}} n_{CM} (\mathrm {p_{NN}})\delta [ E_m - E^{2N}_m  ] d \mathbf p_{\mathrm {NN}},
\label{Snn_vn_1}
\end{align}
where  $p/n$ indicates proton and neutron respectively,  the parameter $a_2(A,Z)$   is defined in Eq. (\ref {2NSRC_relmd}), $x_{p/n}$ are given by  Eq. (\ref {2N-fraction}), and  $n_d$ is the deuteron momentum distribution calculated from a nonrelativistic wave function. 

 The relativistic  kinetic energy of the on-shell correlated nucleon 2 [Fig. \ref {Fig:spectral_diagram} (b)], is given by 
\begin{align}
  T_{2}& = E_2 - M_N =  \sqrt {\mathrm p^2_2 + M^2_N} - M_N\nonumber\\
  & = \sqrt {(\mathbf p_{NN} - \mathbf p_1)^2 + M^2_N} - M_N\nonumber\\
  & = \sqrt {\mathrm p^2_1 - 2 \mathrm p_1 \mathrm p_\mathrm {NN}\cos {\theta_\mathrm {1NN}} + \mathrm p^2_\mathrm {NN} + M^2_N} - M_N,
\label {T2}
\end{align}
where $ \theta_\mathrm {1NN}$ is the angle between $\mathbf p_1$ and $\mathbf p_\mathrm {NN} $. Hence,  the removal energy $E_m^{2N}$  for the 2N SRC [Eq. (\ref {2NSRC_removal})] can be expressed as
\begin{align}
  E^{2N}_m& = E^{(2)}_{thr}+ \sqrt {\mathrm p^2_1 - 2 \mathrm p_1 p_{NN}\cos {\theta_{1NN}} + \mathrm p^2_{NN} + M^2_N} - M_N+  \frac {\mathrm {p^2_{NN}}}{2 M_{A-2}} - \frac {\mathrm p_{1}^2}{2M_{A-1}},
\label {Em-rel2}
\end{align}
and the argument of the delta function in the nuclear spectral function [Eq.(\ref {Snn_vn_1})] is therefore   
\begin{align}
 E_m - E^{2N}_{m}   & =  E_m  - E^{(2)}_{thr} +M_N- \sqrt {\mathrm p^2_1 - 2 \mathrm p_1 \mathrm {p_{NN}}\cos {\theta_ \mathrm  {1NN}} + \mathrm {p_{NN}} + M^2_N} -  \frac {\mathrm {p^2_{NN}}}{2 M_{A-2}} + \frac {\mathrm p_{1}^2}{2M_{A-1}},
 \label {arg-2N-deltaf2-rel}
\end{align}
where   $E^{(2)}_{thr} \approx 2M_N + M_{A-2} - M_{A}$. \\
If  $\cos {\tilde \theta_ \mathrm  {1NN}}$ is the root of  $E_m - E^{2N}_{m} = 0$,  it follows that 
\begin{align}
 E_{20}   & =  \sqrt {\mathrm p^2_1 - 2 \mathrm p_1 \mathrm {p_{NN}}\cos {\tilde \theta_ \mathrm  {1NN}} + \mathrm {p_{NN}} + M^2_N} = E_m  - E^{(2)}_{thr} +M_N-   \frac {\mathrm {p^2_{NN}}}{2 M_{A-2}} + \frac {\mathrm p_{1}^2}{2M_{A-1}},
 \label {E_20}
\end{align}
where $E_{20}$ is the relativistic  energy of the nucleon 2 for $\cos {\tilde \theta_ \mathrm  {1NN}}$.\\
$\cos {\tilde \theta_ \mathrm  {1NN}}$ is then given by 
\begin{align}
 \cos \tilde  \theta_ \mathrm  {1NN} & =   \frac {\mathrm p^2_1 +\mathrm p^2_ \mathrm  {NN} + M^2_N -E^2_{20}  }{2\mathrm {p_1}\mathrm {p_{NN}}}, 
 \label {cos-theta-1NN}
 \end{align} 
 with the restrictions 
 \begin{align}
 E^2_{20} \ge 0, \nonumber \\
 \cos^2 \tilde \theta_ \mathrm  {1NN} \le 1.  
 \label {cossqtile20_restr}
 \end{align}
 
The derivative of Eq. (\ref {arg-2N-deltaf2-rel}) yields
\begin{align}
\frac{d( E_m - E^{2N}_{m})}{d \cos {\theta_ \mathrm  {1NN}} } |_{  \theta_ \mathrm  {1NN} = \tilde \theta_ \mathrm  {1NN}} & =  \frac { \mathrm p_1 \mathrm {p_{NN}}}{  \sqrt {\mathrm p^2_1 - 2 \mathrm p_1 \mathrm {p_{NN}}\cos {\tilde \theta_ \mathrm  {1NN}} + \mathrm {p_{NN}} + M^2_N} } = \frac { \mathrm p_1 \mathrm {p_{NN}}}{ E_{20}}.
 \label {der_arg-2N-delta}
\end{align}
Then, from Eqs.(\ref {delta_integral}), and (\ref {E_20}) to (\ref {der_arg-2N-delta} ),  the integration  of the delta function in the nuclear spectral functions [Eq.(\ref {Snn_vn_1})]  over $\cos \theta_ \mathrm  {1NN}$ yields
 \begin{align}
 \int  \delta^0 [ E_m - E^{2N}_m ] d\mathbf  {p_{NN}} & =  2 \pi  \int  \delta^0 [ E_m - E^{2N}_m ] \mathrm {p^2_{NN}}d\mathrm {p_{NN}} d \cos {\theta_ \mathrm  {1NN}} = 2 \pi  \int \frac {E_{20}} {p_1}  \mathrm  {p_  {NN}} d \mathrm {p_{NN}}. 
\label{int-2N-deltaf2-rel}
\end {align} 

 The 2N SRC nuclear spectral functions [Eq.(\ref {Snn_vn_1})] are therefore given by: 
\begin{align}
  S^{p/n} _{A,2N} ( \mathrm p_1,  E_m) & =  2\pi \frac {a_2(A,Z)}{ (2x_{p/n})^{\gamma}} {\int_0^1}\frac {n_{d} ( \mathrm {p_{rel}})}{\omega_{rel}} n_{CM}(\mathrm  {p_{NN}})\frac {E_{20} } {p_1} \Theta ( \mathrm  {p_{rel} - k_{src}}) \mathrm  { p_{NN}} d \mathrm {p_{NN}},
 \label{Snn_vn_2}
\end{align}
where  the center of mass momentum distribution is given by Eq.(\ref {2NSRC_CMdist}),  and the relative momentum of the correlated  nucleons  by 
 \begin{align}
\mathrm { p_{rel}}&  =   \sqrt {\frac {(\mathbf{p_1 -p_2})^2}{4}} = \sqrt {\frac {(\mathbf{2p_1 -p_{NN}})^2}{4}}\nonumber\\
&=\sqrt {\mathrm {p^2_1+ \frac {p^2_{NN}}{4} - p_1p_{NN}} \cos \tilde \theta_{1NN} }.  
 \label{2N-p-rel}
 \end{align}
 Also
 \begin{align}
\mathrm { \omega_{rel}}&  = \frac {M_N-E_m-T_{A-1}}{\frac {M_A} {A}} \nonumber\\
& =   \frac {A}{M_A} \Big(M_N-E_m- \frac {\mathrm p_{1}^2}{2M_{A-1}} \Big),
 \label{omega-rel}
 \end{align}
 and the cosine of the angle $\theta_\mathrm{1rel}$, between $\mathbf p_1$ and $\mathbf {p_{rel}}$, is given by 
 \begin{align}
 \cos  \theta_\mathrm{1rel} & =   \frac {2\mathrm p_1 -\mathrm {p_{NN}} \cos \tilde  \theta_\mathrm{1NN} }{\mathrm {2p_{rel}} },
 \label {cos-theta-1rel}
 \end{align}
 with the restriction 
 \begin{align}
 \cos^2  \theta_\mathrm {1rel} \le 1.
 \label {cossq1relvna_restr}
 \end{align}
 The momentum of the correlated nucleon 2 is
  \begin{align}
\mathrm { p_{2}}&  =  \sqrt {(\mathbf{p_{1} -2p_{rel}})^2} = \sqrt {\mathrm {p^2_1+ 4p^2_{rel}- 4p_1p_{rel}} \cos \theta_{1rel} },  
 \label{2N-p-2}
  \end{align}
  with the restriction 
 \begin{align}
 \mathrm {p_2 >   k_F \sim 0.250 \; GeV/c}.
  \label {p2vna_restr}
 \end{align}
 
 The  2N SRC  nuclear spectral functions  [Eq.(\ref {Snn_vn_2})], together with Eqs. (\ref {2NSRC_CMdist}),  (\ref {E_20}) to (\ref {cossqtile20_restr}), and (\ref {2N-p-rel}) to  (\ref {p2vna_restr} ), and the normalization condition  [Eq.(\ref {norm_VNA_2NSRC})],  constitute the computational  model from which   numerical estimates are obtained in the chapter  4  of the present dissertation. 

 \subsection{Computational\;model\; for\; three\; nucleon\; SRC \;nuclear\;spectral\;function \;in virtual\;nucleon\;approximation}
\label {subsec:sec:Com3NsrscVNA}
The following expression for  the  3N SRC  nuclear spectral function in VN approximation is obtained by inserting    the  product of the  relative momentum distributions of the NN SRC [Eq.(\ref {3NSRC_relmd})] in Eq. (\ref {spe_VNA_3NSRC_4}), namely
\begin{align}
  S_{A,3N}^{p/n}(\mathrm  p _1,E_m)  = a^2_2(A,Z)C^{p/n} (A,N) \int {n_d(\mathrm {p}_{3}) n_d(\mathrm {p}_{12})\over  {M_N - E_m - T_{A-1}\over M_A/A}}\Theta(\mathrm {p} _{3}-{k} _{src})\Theta(\mathrm {p}_{12}-{k} _{src})\delta(E_m - E^{3N}_m) 
d^3 \mathbf {p} _{3},
\label{S3n_vn_1}
\end{align}
where   $p/n$ indicates proton and neutron respectively, and   the $C^{p/n} (A,N)$  factors are given by Eq. (\ref {3N-C}).
  
The relativistic  kinetic energies of the on-shell correlated nucleons 2  and 3 [Fig. \ref {Fig:3nSpectral}], aregiven by 
\begin{align}
  T_{2}& =  \sqrt {\mathrm p^2_2 + M^2_N} - M_N =  \sqrt {(\mathbf p_1 + \mathbf p_3)^2+ M^2_N} - M_N \nonumber\\ 
  &= \sqrt {\mathrm p^2_1 +2 \mathrm p_1 \mathrm p_{3}\cos {\theta_{13}} + \mathrm p^2_{3} + M^2_N}  - M_N
\label {T2_3},
\end{align}
\begin{align}
  T_{3}& =  \sqrt {\mathrm p^2_3 + M^2_N} - M_N,
  \label {T3_3}
\end{align}
where $ \theta_\mathrm {13}$ is the angle between $\mathbf p_1$ and $\mathbf p_{3} $. Then, the removal energy of the bound nucleon $E^{3N}_m$ [Eq. (\ref {3NSRC_removal})] can be expressed as 
\begin{align}
  E^{3N}_m & = E^{(3)}_{thr}  - 2M_N  - \frac {\mathrm p_{1}^2}{2M_{A-1}}+  \sqrt {\mathrm p^2_3 + M^2_N}  + \sqrt {\mathrm p^2_1 +2 \mathrm p_1 \mathrm p_{3}\cos {\theta_{13}} + \mathrm p^2_{3} + M^2_N}.
\label {e3n_1}
\end{align}
The argument of the delta function in the nuclear spectral function [Eq.(\ref {S3n_vn_1})] is therefore 
\begin{align}
 E_m - E^{3N}_m  & =  E_m  - E^{(3)}_{thr}  +2M_N + \frac {\mathrm p_{1}^2}{2M_{A-1}}- \sqrt {\mathrm p^2_3 + M^2_N}  -\sqrt {\mathrm p^2_1 +2 \mathrm p_1 \mathrm p_{3}\cos {\theta_{13}} +\mathrm p^2_{3} +M^2_N},
 \label {arg-3N-deltaf2-rel}
\end{align}
where  $E^{(3)}_{thr} \approx 3M_N + M_{A-3} - M_{A}$.  \\
Similar to the case of the 2N SRC nuclear spectral function in section (\ref {subsec:Com2NsrcVNA}), the integration  of the delta function in the spectral functions [Eq.(\ref {S3n_vn_1})] over $\cos \theta_{13}$ yields
 \begin{align}
 \int  \delta^0 [ E_m - E^{3N}_m ] d\mathbf  {p_{3}} 
  & =  2 \pi    \int  \delta^0 [ E_m - E^{3N}_m ] \mathrm  {p^2_{3}} d\mathrm  {p_{3}}\cos \theta_{13}  = 2 \pi  \int \frac {E_{30}} {\mathrm p_1}  \mathrm p_{3} d \mathrm {p_{3}} ,
\label{int-3N-deltaf2-rel}
\end {align}
where   $ E_{30}$ is  the relativistic energy of the nucleon 3 for $\cos {\tilde \theta_ \mathrm  {13}}$, given by 
\begin{align}
E_{30}    & =   \sqrt {\mathrm p^2_1 +2 \mathrm p_1 \mathrm p_{3}\cos {\tilde  \theta_{13}} +\mathrm p^2_{3} +M^2_N} = E_m  - E^{(3)}_{thr}  +2M_N + \frac {\mathrm p_{1}^2}{2M_{A-1}}- \sqrt {\mathrm p^2_3 + M^2_N},
 \label {E_30}
\end{align}
and $\cos \tilde  \theta_{13}$, the root of $E_m - E^{3N}_m = 0$,  is 
\begin{align}
 \cos \tilde  \theta_{13} & =   \frac { E^2_{30} -\mathrm p^2_1 -\mathrm p^2_{3} - M^2_N }{2\mathrm {p_1}\mathrm {p_{3}}}, 
 \label {cos-theta-13-3N}
 \end{align} 
 with the restrictions 
 \begin{align}
 E^2_{30} \ge 0, \nonumber \\
 \cos^2  \tilde \theta_\mathrm {13} \le 1.
 \label {cossqtile30_restr}
 \end{align}
 
 The 3N SRC nuclear spectral functions  [Eq.(\ref {S3n_vn_1})]  are  therefore given by: 
\begin{align}
  S^{p/n} _{A,3N} ( \mathrm p_1,  E_m) & =  2\pi a^2_2(A,Z) C^{p/n}(A,N) {\int_0^1}\frac {n_{d} ( \mathrm {p_{3}})}{\omega_{3N}} n_{d}(\mathrm p_{12})\frac {E_{30}} {\mathrm p_1}  \Theta(\mathrm {p} _{3}-{k} _{src})  \Theta(\mathrm {p} _{12}-{k} _{src}) \mathrm p_3 d \mathrm {p_3},
 \label{S3n_vn_2}
\end{align}
where, by using the collinear assumption  within the VN approximation  [$\mathbf {p_1 + p_2 +p_3 }= 0$], 
 \begin{align}
\mathrm { p_{12}}&  = \sqrt {\frac {(\mathbf{p_1 -p_2})^2}{4}} = \sqrt {\frac {(\mathbf{2p_1 +p_{3}})^2}{4}}\nonumber\\
&= \sqrt { \mathrm {p^2_1+ \frac {p^2_{3}}{4} + p_1p_{3}} \cos \tilde \theta_{13} },  
 \label{3N-p-12}
 \end{align} 
and 
  \begin{align}
\mathrm { \omega_{3N}}&  = \frac {M_N-E_m-T_{A-1}}{\frac {M_A} {A}} \nonumber\\
& =   \frac {A}{M_A} \Big(M_N-E_m- \frac {\mathrm p_{1}^2}{2M_{A-1}} \Big). 
 \label{omega_3N}
 \end{align}
 The cosine of the angle  $\theta_\mathrm{112}$,   between $\mathbf p_1$ and $\mathbf {p_{12}}$,  is given by 
 \begin{align}
 \cos  \theta_\mathrm{112} & =   \frac {2\mathrm p_1 +\mathrm {p_{3}} \cos \tilde  \theta_\mathrm{13} }{\mathrm {2p_{12}} }, 
 \label {cos-theta-112}
 \end{align} 
 with the restriction 
 \begin{align}
 \cos^2  \theta_\mathrm {112} \le 1.
 \label {cossq112_restr}
 \end{align}
 The momentum of the correlated nucleon $2$  is 
 \begin{align}
\mathrm { p_{2}}&  =  \sqrt {(\mathbf{p_{1} -2p_{12}})^2} = \sqrt {\mathrm {p^2_1+ 4p^2_{12}- 4p_1p_{12}} \cos \theta_{112} }, 
 \label{3N-p-2}
  \end{align}
with the restriction 
 \begin{align}
 \mathrm  {p_2 > k_F \sim 0.250 \;GeV/c}
 \label {p2vna3_restr}
 \end{align}
  
  The  3N SRC  nuclear spectral functions  [Eq.(\ref {S3n_vn_2})], together with Eqs. (\ref {E_30}) to (\ref {cossqtile30_restr}), and (\ref {3N-p-12}) to  (\ref {p2vna3_restr} ), and the normalization condition  [Eq.(\ref {norm_VNA_3NSRC})], constitute the computational  model from which   numerical estimates are obtained in the chapter  4  of the present dissertation. 
 
\section{Computational\;models\;for\;SRC\;nuclear\;spectral\;function\;in\;light-front\\ approximation} 
\label {sec:ComSpecLFA}
The computational models for high momentum  nuclear  spectral function in light-front (LF) approximation, with two   and three  nucleons in short range correlation, are developed in the present section.

\subsection { Computational \;model\; for\; two\;nucleon\; SRC\; nuclear\; spectral\; function\; in\;light-front\; approximation }
\label {sec:Com2NsrcLFA}

The following expression for  the  2N SRC  nuclear spectral function in LF approximation is obtained by inserting     the light-front density matrix for the relative motion in the 2N SRC  [Eq.(\ref {2NSRC_NNdens})] in Eq. (\ref {spe_LF_2NSRC_3}), namely
\begin{align}
P^{p/n}_{A,2N}(\alpha_1,  \mathrm p_{1,\perp},\tilde M_N^2) & = {a_2(A,Z)\over (2x_{p/n})^\gamma} M_N \int \rho_d(\beta_1, \mathrm k_{1,\perp}) \Theta(  \mathrm k_1 -  \mathrm k_{src}) n_{CM}(\mathrm k_{CM}) 2\delta(\alpha_1+ \alpha_2 - \alpha_{NN}) 
\nonumber \\
&\delta^2(\mathbf p_{1,\perp}+\mathbf p_{2,\perp} - \mathbf p_{NN,\perp}) 
\delta(\tilde M_N^2 - \tilde M_{N}^{(2N),2}) d\beta_2 d^2  \mathbf p_{2,\perp}
d\alpha_{NN}d^2  \mathbf p_{NN,\perp},
\label{spe_LF_2NSRC_4}
\end{align} 
where  $p/n$ indicates proton and neutron respectively, $\beta_1 $ is defined in Eq. (\ref {rel_trans_mom}),  $x_{p/n}$ are given by  Eq. (\ref {2N-fraction}), and  the light-front density matrix for the deuteron is defined as 
\begin{align}
 \rho_d(\beta_1, \mathrm k_{1,\perp}) & = \frac {n_d (\mathrm k_1)}{2 - \beta_1} \sqrt {M^2_N + \mathrm k_1} = \frac {n_d (\mathrm k_1)}{2 - \beta_1} \sqrt {\frac {s_{NN}}{4}} \Theta (\mathrm {k_1 - k_{src}}),
\label{rho_d}
\end{align}
where the invariant energy $s_{NN}$ and the relative momentum $\mathbf k_1$ are defined  in Eq. (\ref {sNN_2}) and the parameter $a_2(A,Z)$   in Eq. (\ref {2NSRC_relmd}). $n_d$ is the deuteron momentum distribution calculated from a nonrelativistic wave function. The normalization of this density matrix is given by
\begin{align}
\int  \rho_d(\beta_1, \mathrm k_{1,\perp}) \frac {d \beta_1}{\beta_1} d^2 \mathbf k_1 & = 1.
\label{rho_d_norm}
\end{align}

Integration in the nuclear spectral function [Eq. (\ref {spe_LF_2NSRC_4})]  by  $d^2  \mathbf p_{2,\perp}$ through $\delta^2(\mathbf p_{1,\perp}+\mathbf p_{2,\perp} - \mathbf p_{NN,\perp})$, and by $d\beta_2 $ through $ \delta(\alpha_1+ \alpha_2 - \alpha_{NN})= 2 \delta(\beta_1+ \beta_2 - 2)/\alpha_{NN} $, and considering that $d^2  \mathbf p_{NN,\perp} = \mathrm {p}_{NN,\perp} d \mathrm {p}_{NN,\perp}d\phi_{NN}$   yields
\begin{align}
P_{A,2N}^{p/n}(\alpha_1,  \mathrm p_{1,\perp},\tilde M_N^2) & = {a_2(A,Z)\over (2x_{p/n})^\gamma} M_N \int \rho_d(\beta_1, \mathrm k_{1,\perp}) \Theta(  \mathrm k_1 -  \mathrm k_{src}) n_{CM}(\mathrm k_{CM}) 
\nonumber \\
&\delta(\tilde M_N^2 - \tilde M_{N}^{(2N),2}) \frac {4d\alpha_{NN} }{\alpha_{NN}}\mathrm {p}_{NN,\perp} d \mathrm {p}_{NN,\perp}d\phi_{NN},
\label{spe_LF_2NSRC_5}
\end{align} 
where $ \phi_{NN}$ is the  angle between $\mathbf  p_{1,\perp}$ and $\mathbf  p_{NN,\perp}$, and 
\begin{align}
\beta_1 + \beta_2  & =  2, 
\label{beta_1}
\end{align}
\begin{align}
\mathbf p_{1,\perp}+\mathbf p_{2,\perp}  & =  \mathbf p_{NN,\perp}.
\label {p-perp}
\end{align}

From Eqs.  (\ref  {beta_1}) and (\ref {p-perp}) ,   the invariant mass $  \tilde M^{(2N),2}_{N}$ [Eq.(\ref {Inv-mass-2N})] can be expressed as 
\begin{align}
 \tilde M^{(2N),2}_{N} & = {p^-_{1}p^+_{1} }- \mathrm  {p}^2_{1,\perp}   = p^+_{1}  (p^-_{NN}- p^-_{2} ) -  \mathrm  p^2_{1,\perp} \nonumber \\
& =  \alpha_{1} \left  (\frac  {M^2_{NN} + \mathrm  {p}^2_{NN,\perp} } { \alpha_{NN}} -  \frac  {M^2_{N} + \mathrm  {p}^2_{2,\perp} } { \alpha_{2}} \right ) -  \mathrm  p^2_{1,\perp} \nonumber \\
& =  \frac {2 - \beta_2}{2} \left  \{M^2_{NN} + \mathrm  p^2_{NN,\perp} - \frac {2 }{ \beta_2}  \left [ M^2_{N} + 
\Big (\mathbf  p_{NN,\perp} - \mathbf  {p}^2_{1,\perp} \Big )^2 \right ]   \right \}-  \mathrm  p^2_{1,\perp} \nonumber \\
& = \frac {\beta^2_1}{2(2 - \beta_1)}\left ( \frac {2- \beta_1}{ \beta_1}M^2_{NN} -  \frac {2}{ \beta_1}M^2_{N} -  
\mathrm  p^2_{NN,\perp} + \frac {4} {\beta_1} \mathrm  p_{NN,\perp} \mathrm  p_{1,\perp}\cos \phi_{NN} -  
\frac {4}{ \beta^2_1}\mathrm  p^2_{1,\perp}  \right ),
\label{Inv-mass-2N_1}
\end{align}
where $M_{NN}$ is taken as the deuteron mass.

Inserting Eq.  (\ref {Inv-mass-2N_1}) in the argument of the delta function in the nuclear spectral function [Eq.(\ref {spe_LF_2NSRC_5})] yields
\begin{align}
\tilde M_{N}^{2}  - \tilde M^{(2N),2}_{N} & = 
\frac  {\beta^2_1}{ 2(2-\beta_1)}  \left ( \mathrm p^2_{NN,\perp} - \frac  {4}{\beta_1} \mathrm {p}_{1,\perp} \mathrm {p}_{NN,\perp}  \cos \phi_{NN} +  \mathrm p^2_{02,\perp}   \right),
\label{Arg_delta}
\end{align}
where 
\begin{align}
\mathrm p^2_{02,\perp} & = 
\frac  {2}{ \beta_1}\left( \frac  {2}{ \beta_1} \mathrm  p^2_{1,\perp} + \frac  {2 - \beta_1 }{ \beta_1}  \tilde M_{N}^{2} - 
\frac  {2-\beta_1}{ 2} M_{NN}^{2} + M^2_N \right),
\label{p_02}
\end{align}
with the restriction 
 \begin{align}
\mathrm p^2_{02,\perp} \ge 0.
 \label {p2sq112lfa2_restr}
 \end{align}
The roots of 
$\tilde M_{N}^{2}  - \tilde M^{(2N),2}_{N} = 0 $  are
\begin{align}
\mathrm {\tilde {p}}^a_{NN,\perp} & = \frac {2}{\beta_1} \left ( \mathrm {p}_{1,\perp} \cos \phi_{NN} +\frac{1}{4} \sqrt {16 \mathrm {p}^2_{1,\perp} \cos^2 \phi_{NN} - \beta^2_1 \mathrm p^2_{02,\perp}} \right ), \nonumber \\
\mathrm {\tilde {p}}^b_{NN,\perp} & = \frac {2}{\beta_1} \left ( \mathrm {p}_{1,\perp} \cos \phi_{NN} -\frac{1}{4} \sqrt {16 \mathrm {p}^2_{1,\perp} \cos^2 \phi_{NN} - \beta^2_1 \mathrm p^2_{02,\perp}} \right ),
\label{roots_pNN}
\end{align}
 with the restriction 
 \begin{align}
16 \mathrm {p}^2_{1,\perp} \cos^2 \phi_{NN} - \beta^2_1 \mathrm p^2_{02,\perp} > 0.
 \label {p2sqperplf2_restr}
 \end{align}
 
Then, from Eqs. (\ref {delta_integral}) and (\ref {roots_pNN}), integration of the delta function in  the nuclear spectral function 
[Eq.(\ref {spe_LF_2NSRC_5})]  over $d \mathrm {p}_{NN,\perp}$, yields
\begin{align}
P^{p/n}_{A,2N}(\alpha_1,  \mathrm p_{1,\perp},\tilde M_N^2) & = {a_2(A,Z)\over (2x_{p/n})^\gamma} M_N \int^A_{\alpha_1} \int^{2\pi}_0 \Big[ \rho_d(\beta_1, \mathrm k^a_{1,\perp}) n_{CM}(\mathrm k^a_{CM}) \tilde {\mathrm   p}^a_{NN,\perp}\Theta(  \mathrm k^a_1 -  \mathrm k_{src}) \nonumber\\
& +   \rho_d(\beta_1, \mathrm k^b_{1,\perp})  n_{CM}(\mathrm k^b_{CM})  \tilde {\mathrm   p}^b_{NN,\perp} \Theta(  \mathrm k^b_1 -  \mathrm k_{src})  \Big]\nonumber \\
& \left (\frac  {2-\beta_1}{ \beta_1}\right)\frac {2 }{\sqrt {16 \mathrm p^2_{1,\perp} \cos^2 \phi_{NN}-\beta^2_1\mathrm p_{02,\perp}^2}} \frac {4d\alpha_{NN} }{\alpha_{NN}}d\phi_{NN},
\label{spe_LF_2NSRC_6}
\end{align} 
where from Eqs. (\ref {rel_trans_mom}) and (\ref {sNN_2}) it follows that 
\begin{align}
 \mathrm k^{a(b)}_{1,\perp} & = \sqrt {\mathrm p^2_{1,\perp} + \left( \frac {\beta_1}{2}\tilde {\mathrm   p}^{a(b)}_{NN,\perp}\right)^2-  \beta_1 \mathrm p_{1,\perp}\tilde {\mathrm   p}^{a(b)}_{NN,\perp} \cos\phi_{NN}  },
 \label{kab_1perp} 
\end{align}
\begin{align}
s^{a(b)}_{NN} & = {4\over \beta_1(2-\beta_1)}\Big [M_N^2 + (\mathrm k^{a(b)}_{1,\perp})^2 \Big],
\label{sab_nn}
\end{align}
\begin{align}
 \mathrm k^{a(b)}_{1} & = \frac {1}{2}\sqrt {s^{a(b)}_{NN} - 4 M^2_N  },
 \label{kab_1} 
\end{align}
and from Eqs. (\ref {sNN_A-2}) and (\ref {k-CM}) with $ \mathbf p_{A,\perp}  = 0$
\begin{align}
s^{a(b)}_{NN,A-2} & = A^2{\left[ M_{NN}^2 + {\alpha_{NN}\over A}(M_{A-2}^2 - M_{NN}^2) + ( \tilde {\mathrm   p}^{a(b)}_{NN,\perp} )^2\right]
\over \alpha_{NN}(A-\alpha_{NN})},
\label {ssn-ab-nn}
\end{align}
\begin{align}
k^{a(b)}_{CM} & = {\sqrt{\left[ s^{a(b)}_{NN,A-2} - (M_{NN} + M_{A-2})^2\right]\left[ s^{a(b)}_{NN,A-2} - (M_{NN} - M_{A-2})^2\right]}\over 2\sqrt{s^{a(b)}_{NN,A-2}}}.
\label {Kcm_2N_com}
\end{align}

Following a similar calculation as above, integration of the delta function in the nuclear spectral function 
[Eq.(\ref {spe_LF_2NSRC_5})] by $d  \phi_{NN}$ yields
\begin{align}
P^{p/n}_{A,2N}(\alpha_1,  \mathrm p_{1,\perp},\tilde M_N^2) & = {a_2(A,Z)\over (2x_{p/n})^\gamma} M_N \int^A_{\alpha_1} \int^{1}_0  \rho_d(\beta_1, \mathrm k_{1,\perp}) \Theta(  \mathrm k_1 -  \mathrm k_{src}) n_{CM}(\mathrm k_{CM}) 
\nonumber \\
&\left (\frac  {2-\beta_1}{ \beta_1}\right)\frac {2}{\sqrt {16\mathrm p^2_{1,\perp} \mathrm p^2_{NN,\perp} -\beta^2_1(\mathrm p^2_{NN,\perp}  +\mathrm p^2_{02,\perp}})^2} \frac {4d\alpha_{NN} }{\alpha_{NN}}\mathrm {p}_{NN,\perp} d \mathrm {p}_{NN,\perp},
\label{spe_LF_2NSRC_7}
\end{align}
with the restriction 
 \begin{align}
16\mathrm p^2_{1,\perp} \mathrm p^2_{NN,\perp} -\beta^2_1(\mathrm p^2_{NN,\perp}  +\mathrm p^2_{02,\perp})^2 > 0,
 \label {p1sqperplf2_restr}
 \end{align}
and 
\begin{align}
\cos \tilde \phi_{NN}& = 
 \left(\frac  {\beta_1}{ 4}\right)\frac {\mathrm  p^2_{NN,\perp} + \mathrm p_{02,\perp}^2}{\mathrm p_{1,\perp} \mathrm p_{NN,\perp} },
 \label{cos-phi_NN}
\end{align}
is the root of $\tilde M_{N}^{2}  - \tilde M^{(2N),2}_{N} = 0 $ for $\cos  \phi_{NN}$,   with the restriction 
 \begin{align}
\cos^2 \tilde \phi_{NN} \le 1.
 \label {cossqtilphinn_restr}
 \end{align}
From Eq. (\ref {rel_trans_mom})    it follows that 
\begin{align}
 \mathrm k_{1,\perp} & = \sqrt {\mathrm p^2_{1,\perp} + \left( \frac {\beta_1}{2} {\mathrm   p}_{NN,\perp}\right)^2-  \beta_1 \mathrm p_{1,\perp} {\mathrm   p}_{NN,\perp}  \cos \tilde \phi_{NN}  }.
 \label{kphi_1perp} 
\end{align}
 The momenta $k_1$ and $k_{CM}$ are calculated from Eqs. (\ref {sNN_2}),  and (\ref {sNN_A-2}) and (\ref {k-CM}) respectively, with $ \mathbf p_{A,\perp}  = 0$.
 
 The  2N SRC  nuclear spectral functions  [Eq.(\ref {spe_LF_2NSRC_7})], together with Eqs.(\ref {2NSRC_CMdist}),
 (\ref {sNN_2}), (\ref {sNN_A-2}), (\ref {k-CM}), (\ref {p_02}), and  (\ref {p1sqperplf2_restr}) to (\ref {kphi_1perp}), and the normalization conditions [Eq.(\ref {NN_LF_2NSRC_norms})], [Eq.(\ref {CM_LF_2NSRC_norm1})] and [Eq.(\ref {CM_LF_2NSRC_norm2})],    constitute the simplest computational  model from which   numerical estimates are obtained in the chapter  4  of the present dissertation. The Eqs. (\ref {2NSRC_CMdist}),  (\ref {p_02})  to (\ref {Kcm_2N_com}), (\ref {NN_LF_2NSRC_norms}), (\ref {CM_LF_2NSRC_norm1}) and (\ref {CM_LF_2NSRC_norm2}), also constitute a computational  model but more calculations are needed to obtain the same numerical estimates.

\subsection { Computational \;model\; for\; three\;nucleon\; SRC\; nuclear\; spectral\; function\; in\;light-front\; approximation }
\label {sec:Com3NsrcLFA}
The following  expression for   the  3N SRC  nuclear spectral function in LF approximation is obtained by inserting     the product of light-front density matrices  for the relative motion in the 2N SRC  [Eq.(\ref {3N_LFanzats})] in Eq. (\ref {spe_LF_3NSRC_3}), namely
\begin{align}
P^{p/n}_{A,3N}(\alpha_1, \mathrm  p_{1,\perp},\tilde M_N)  &  =  a^2_2(A,Z) C^{p/n}(A,Z)  \int {3-\alpha_3\over 2(2-\alpha_3)^2} \rho_{d}(\beta_{2'}, \mathrm p_{3,\perp})\Theta(  \mathrm {\tilde k}_1-\mathrm k_{src}) \rho_{d}(\beta_1, \mathrm   {\tilde k}_{1,\perp}) \Theta( \mathrm  {\tilde k}_3- \mathrm k_{src})
\nonumber \\
&  2 \delta(\alpha_1 + \alpha_2  + \alpha_3 - 3) \delta^2(\mathbf p_{1,\perp} + \mathbf p_{2,\perp} + \mathbf p_{3,\perp}) 
 \delta(\tilde M_N^2 - M_N^{(3N),2})  d\alpha_2 d^2   \mathbf  p_{2,\perp}d\alpha_3d^2   \mathbf p_{3,\perp},
 \label{spe_LF_3NSRC_4}
\end{align}
where  $p/n$ indicates proton and neutron respectively,  $\beta_{2'}$ and $\beta_1$ are given by Eq. (\ref {coll_momFrac} ), the light-front density matrix for the deuteron by Eq. (\ref {rho_d}), and  the factors $C^{p/n}(A,Z)$  by Eq. (\ref {3N-C}).

Integration in the nuclear spectral function [Eq. (\ref {spe_LF_3NSRC_4})] by  $d^2  \mathbf p_{2,\perp}$ through $\delta^2(\mathbf p_{1,\perp}+\mathbf p_{2,\perp} + \mathbf p_{3,\perp})$, and by $d\alpha_{2}$ through $ \delta(\alpha_1+ \alpha_2 + \alpha_{3} - 3)$,  and considering that $d^2  \mathbf p_{3,\perp} = \mathrm {p}_{3,\perp} d \mathrm {p}_{3,\perp}d\phi_{13}$   yields

\begin{align}
P^{p/n}_{A,3N}(\alpha_1, \mathrm  p_{1,\perp},\tilde M_N)  &  =  a^2_2(A,Z)C^{p/n}(A,Z)  \int {3-\alpha_3\over (2-\alpha_3)^2} \rho_{d}(\beta_{2'}, \mathrm p_{3,\perp})\Theta(  \mathrm {\tilde k}_1-\mathrm k_{src}) \rho_{d}(\beta_1, \mathrm   {\tilde k}_{1,\perp}) \Theta( \mathrm  {\tilde k}_3- \mathrm k_{src})
\nonumber \\
&    \delta(\tilde M_N^2 - M_N^{(3N),2})  d\alpha_3 \mathrm {p}_{3,\perp} d \mathrm {p}_{3,\perp}d\phi_{13},
 \label{spe_LF_3NSRC_5}
\end{align} 
where $\phi_{13}$ is the angle between $\mathbf {p}_{1,\perp}$ and  $\mathbf {p}_{3,\perp}$, and 
\begin{align}
 \alpha_2   & =  3 - \alpha_1 - \alpha_3, 
\label{alpha_123}
\end{align}
\begin{align}
\mathbf p_{2,\perp}  & =  -\mathbf p_{1,\perp} -\mathbf p_{3,\perp}. 
\label{p2_perp}
\end{align}

From Eqs. (\ref {alpha_123}) and (\ref {p2_perp}),  the invariant mass $  \tilde M^{(3N),2}_{N}$ [Eq.(\ref {Inv-mass-3N})] can be expressed as 
\begin{align}
 \tilde M^{(3N),2}_{N} & = \frac {\alpha_1}{ 3}\left( M^2_{3N} - \frac {M^2_N + \mathrm p^2_{2,\perp}}{ \alpha_{2}/3} - 
\frac {M^2_N + \mathrm p^2_{3,\perp}}{ \alpha_{3}/3}\right) - \mathrm p^2_{1,\perp} \nonumber \\
& =  \frac {\alpha_1}{ 3}\left[ M^2_{3N} - \frac {M^2_N +( \mathbf p_{1,\perp} +\mathbf p_{3,\perp})^2}{ (3 - \alpha_1 - \alpha_3)/3} - \frac {M^2_N + \mathrm p^2_{3,\perp}}{ \alpha_{3}/3}\right] - \mathrm p^2_{1,\perp} \nonumber \\
& =  \frac {\alpha_1 (3- \alpha_1)}{ \alpha_3(3 - \alpha_1 - \alpha_3)}\nonumber \\ 
& \times \left[ \frac {\alpha_3 (3- \alpha_1- \alpha_3)}{ 3(3 - \alpha_1)} M^2_{3N} - M^2_N - \frac { \alpha_3(3- \alpha_3) }{\alpha_1(3- \alpha_1)}\mathrm p^2_{1,\perp} 
-  \mathrm p^2_{3,\perp}  + \frac {\alpha_3}{  \alpha_1 -3 } \mathrm p_{1,\perp} \cos \phi_{13}  \mathrm p_{3,\perp} \right], 
\label{Inv-mass-3N_1}
\end{align}
where $M^2_{3N} $ is given by Eq. (\ref {M_3N_sq}).

Inserting Eq.  (\ref {Inv-mass-3N_1}) in the argument of the delta function in the nuclear spectral function [Eq.(\ref {spe_LF_3NSRC_5})] yields
\begin{align}
\tilde M_{N}^{2}  - \tilde M_{N}^{(3N),2} & = 
 \frac  {\alpha_1(3-\alpha_1 )} { \alpha_3(3-\alpha_1- \alpha_3)}\left (\mathrm p^2_{3,\perp}  - \frac  {\alpha_3}{ \alpha_1 -3 }\mathrm {p}_{1,\perp} \cos \phi_{13} \mathrm {p}_{3,\perp} - \mathrm p^2_{03,\perp}  \right),
\label{Arg-3}
\end{align}
where 
\begin{align}
\mathrm p^2_{03,\perp} & = 
 \frac  { \alpha_3(3- \alpha_1 - \alpha_3)}{\alpha_1( \alpha_1-3 )}\left[\tilde M_{N}^{2}   - \frac  {\alpha_1}{ 3} M^2_{3N} + \frac {\alpha_1(3-\alpha_1 )} {\alpha_3 (3- \alpha_1- \alpha_3)} M^2_N + \frac  {3- \alpha_3}{ 3-\alpha_1-\alpha_3}\mathrm  p^2_{1,\perp}\right],
\label{p_03}
\end{align}
with the restriction 
 \begin{align}
\mathrm p^2_{03,\perp} \ge 0.
 \label {p3sq112lfa3_restr}
 \end{align}
Similar to the case of 2N SRC in section (\ref {sec:Com2NsrcLFA}), integration of the delta function in the nuclear spectral function [Eq.(\ref {spe_LF_3NSRC_5})] by $d  \mathrm p_{3,\perp}$ yields
\begin{align}
P^{p/n}_{A,3N}(\alpha_1, \mathrm  p_{1,\perp},\tilde M_N)  &  =  a^2_2(A,Z)C^{p/n}(A,Z)  \int^{3-\alpha_1}_0 \int^{2 \pi}_0 {3-\alpha_3\over (2-\alpha_3)^2} \nonumber \\
&\Big[ \rho_{d}(\beta_{2'},  \tilde {\mathrm   p}^a_{3,\perp})\Theta(  \mathrm {\tilde k}^a_1-\mathrm k_{src}) \rho_{d}(\beta_1, \mathrm   {\tilde k}^a_{1,\perp}) \Theta( \mathrm  {\tilde k}^a_3- \mathrm k_{src}) \tilde {\mathrm   p}^a_{3,\perp}\nonumber \\
& + \rho_{d}(\beta_{2'},  \tilde {\mathrm   p}^b_{3,\perp})\Theta(  \mathrm {\tilde k}^b_1-\mathrm k_{src}) \rho_{d}(\beta_1, \mathrm   {\tilde k}^b_{1,\perp}) \Theta( \mathrm  {\tilde k}^b_3- \mathrm k_{src})\tilde {\mathrm   p}^b_{3,\perp} \Big]\nonumber \\
&  \frac { \alpha_3(3-\alpha_1- \alpha_3)|\alpha_1- 3|} {\alpha_1(3-\alpha_1)}  \frac {1 }{\sqrt {\alpha^2_3\mathrm p^2_{1,\perp} \cos^2 \phi_{13}+4(\alpha_1-3)^2\mathrm p_{03,\perp}^2} }  d\alpha_3d\phi_{13},
 \label{spe_LF_3NSRC_6}
\end{align}
with the restriction 
 \begin{align}
\alpha^2_3\mathrm p^2_{1,\perp} \cos^2 \phi_{13}+4(\alpha_1-3)^2\mathrm p_{03,\perp}^2 > 0,
 \label {alpha3sq1_restr}
 \end{align}
and 
\begin{align}
\tilde {\mathrm   p}^a_{3,\perp}& = 
 \frac  {\alpha_3}{ 2(\alpha_1 - 3)}\mathrm  p_{1,\perp} \cos \phi_{13}+ \frac  {1}{ 2|\alpha_1 -3|}\sqrt {\alpha^2_3\mathrm p^2_{1,\perp} \cos^2 \phi_{13}+4(\alpha_1-3)^2\mathrm p_{03,\perp}^2},
 \label{pa_3}
\end{align}
\begin{align}
\tilde {\mathrm   p}^b_{3,\perp}& = 
 \frac  {\alpha_3}{ 2(\alpha_1 - 3)}\mathrm  p_{1,\perp} \cos \phi_{13}- \frac  {1}{ 2|\alpha_1 -3|}\sqrt {\alpha^2_3\mathrm p^2_{1,\perp} \cos^2 \phi_{13}+4(\alpha_1-3)^2\mathrm p_{03,\perp}^2},
 \label{pb_3}
\end{align}
are the roots of $\tilde M_{N}^{2}  - \tilde M_{N}^{(3N),2} = 0$.  \\
From Eqs. (\ref {tilde_k1_LF}), (\ref {tilde_k1_perp_LF}), and (\ref {tilde_k3_LF}) it follows that
\begin{align}
\tilde{ \mathrm k}^{a(b)}_{1} & = \sqrt { \frac {M^2_N + (\tilde{ \mathrm k}^{a(b)}_{1,\perp})^2 }{\beta_1(2-\beta_1)}- M^2_N},
 \label{kab3_1perp} 
\end{align}
\begin{align}
\tilde{ \mathrm k}^{a(b)}_{1,\perp} & = \sqrt {\mathrm p^2_{1,\perp} + \left( \frac {\beta_1}{2}\tilde {\mathrm   p}^{a(b)}_{3,\perp}\right)^2 +  \beta_1 \mathrm p_{1,\perp}\tilde {\mathrm   p}^{a(b)}_{3,\perp} \cos\phi_{13}  },
\label{kab_3perp} 
\end{align}
\begin{align}
\tilde{ \mathrm k}^{a(b)}_{3} & =  \sqrt {\frac {M^2_N + (\tilde{ \mathrm p}^{a(b)}_{3,\perp})^2 }{\beta_3(2-\beta_3)}- M^2_N}.
 \label{kab3_3perp} 
\end{align}

Following a similar calculation as above, integration of the delta function in Eq.(\ref {spe_LF_3NSRC_5}) by $d  \phi_{13}$ yields
\begin{align}
P^{p/n}_{A,3N}(\alpha_1, \mathrm  p_{1,\perp},\tilde M_N)  &  =  a^2_2(A,Z)C^{p/n}(A,Z) \int^{3-\alpha_1}_{3} \int^1_0 {3-\alpha_3\over (2-\alpha_3)^2} \rho_{d}(\beta_{2'},  {\mathrm   p}_{3,\perp})\Theta(  \mathrm {\tilde k}_1-\mathrm k_{src}) \rho_{d}(\beta_1, \mathrm   {\tilde k}_{1,\perp})  \nonumber \\
&  \Theta( \mathrm  {\tilde k}_3- \mathrm k_{src}) \frac { \alpha_3(3-\alpha_1- \alpha_3)|\alpha_1- 3|} {\alpha_1(3-\alpha_1)} \nonumber \\
&\frac {1 }{\sqrt {\alpha^2_3\mathrm p^2_{1,\perp}\mathrm p^2_{3,\perp} - (\alpha_1-3)^2 (\mathrm p^2_{3,\perp}-\mathrm p^2_{03,\perp})^2 }} d\alpha_3\mathrm {p}_{3,\perp} d \mathrm {p}_{3,\perp},
 \label{spe_LF_3NSRC_7}
\end{align}
with the restriction 
 \begin{align}
\alpha^2_3\mathrm p^2_{1,\perp}\mathrm p^2_{3,\perp} - (\alpha_1-3)^2 (\mathrm p^2_{3,\perp}-\mathrm p^2_{03,\perp})^2 > 0,
 \label {alpha3sq2_restr}
 \end{align}
and 
\begin{align}
\cos \tilde \phi_{13}& = 
\left ( \frac  {\alpha_1- 3}{\alpha_3}\right)\frac {\mathrm  p^2_{3,\perp} - \mathrm p_{03,\perp}^2}{\mathrm p_{1,\perp} \mathrm p_{3,\perp}},
 \label{cos-phi_13}
\end{align}
is the root of $\tilde M_{N}^{2}  - \tilde M_{N}^{(3N),2} = 0$, with the restriction 
 \begin{align}
\cos^2 \tilde \phi_{13} \le 1.
 \label {costilsqtilphi13_restr}
 \end{align} 
From Eq (\ref {tilde_k1_perp_LF}) it follows that
\begin{align}
\tilde { \mathrm k}_{1,\perp} & = \sqrt {\mathrm p^2_{1,\perp} + \left( \frac {\beta_1}{2} {\mathrm   p}_{3,\perp}\right)^2 + \beta_1 \mathrm p_{1,\perp} {\mathrm   p}_{3,\perp}  \cos \tilde \phi_{13}  }.
 \label{ktilde_1perp} 
\end{align} 
The momenta $\mathrm {\tilde k}_1$ and $\mathrm {\tilde k}_3$ are calculated from Eqs. (\ref {tilde_k1_LF}), and (\ref {tilde_k3_LF}) respectively.

The  3N SRC  nuclear spectral functions  [Eq.(\ref {spe_LF_3NSRC_7})], together with Eqs. (\ref {tilde_k1_LF}), (\ref {tilde_k3_LF}),  (\ref {alpha3sq2_restr}) to (\ref {ktilde_1perp}), and  the normalization conditions (\ref {NN_LF_3NSRC_norms}),    constitute the simplest computational  model from which   numerical estimates will be calculated in the chapter  4  of the present dissertation. Eqs. (\ref {p_03}) to (\ref {kab3_3perp}),  and  (\ref {NN_LF_3NSRC_norms}), also constitute a computational  model but more calculations are needed  to obtain the same numerical estimates.

 \section{Computational models for SRC  nuclear spectral function  in nonrelativistic \\ approximation } 
\label {sec:ComSpecNRA}
 The computational models for high momentum  nuclear  spectral function in  the non-relativistic limit of the VN approximation, $\alpha \approx 1 $, with two  and three   nucleons in short range correlation, are developed in the present section.
 
 \subsection { Computational\;model\;for\;two\;nucleon\;SRC \;nuclear\;spectral\;function\; \\
in\;nonrelativistic\;approximation }
\label {sec:Com2NsrcNRA}
The non relativistic 2N SRC   nuclear  spectral functions for proton and neutron are obtained   from Eq.(\ref {Snn_vn_1}), in the non-relativistic limit of the VN approximation, namely
\begin{align}
 S^{p/n} _{A,2N}( \mathrm p_1,  E_m) & =  \frac {a_2(A,Z)}{ (2x_{p/n})^{\gamma}}  \int n_{d}( \mathrm {p_{rel} })\Theta ( \mathrm  {p_{rel} - k_{src}}) n_{CM} (\mathrm {p_{NN}})\delta [ E_m - E^{2N}_m  ] d \mathbf p_{\mathrm {NN}},
\label{Snn_nr}
\end{align}
where  $p/n$ indicates proton and neutron respectively,  the parameter $a_2(A,Z)$   is defined in Eq. (\ref {2NSRC_relmd}), $x_{p/n}$ are given by  Eq. (\ref {2N-fraction}), and  $n_d$ is the deuteron momentum distribution calculated from a nonrelativistic wave function.

The removal energy $E_m^{2N}$  for the 2N SRC [Eq. (\ref {2NSRC_removal})] in the nonrelativistic limit is given by 
\begin{align}
  E^{2N}_m& = E^{(2)}_{thr}+ \frac {\mathrm  p^2_{A-2}}{2(M_{A-1}-M_N)} +  \frac {\mathrm {p^2_{2}}}{2 M_{N}} - \frac {\mathrm p_{1}^2}{2M_{A-1}}\nonumber \\
  & = E^{(2)}_{thr}+ \frac {\mathrm  p^2_{NN}}{2(M_{A-1}-M_N)} +  \frac {(\mathbf {p}_{NN} - \mathbf {p}_{1})^2}{2 M_{N}} - \frac {\mathrm p_{1}^2}{2M_{A-1}}\nonumber \\
  & = E^{(2)}_{thr}+ \frac {M_{A-1}-M_N} {2M_N M_{A-1}}\left  ( \mathbf {p}_{1} -  \frac {M_{A-1}}{M_{A-1}-M_N} \mathbf {p}_{NN} \right )^2,
\label {Em-nrel1}
\end{align}
where $\mathbf {p}_{A-2} = -\mathbf {p}_{NN} $, $\mathbf {p}_{2} = \mathbf {p}_{NN} - \mathbf {p}_{1}$, and 
$M_{A-1} \approx M_{A-1}-M_N$, have been used. The argument of the delta function in the nuclear spectral function 
[Eq. (\ref {Snn_nr}] is therefore given by 
\begin{align}
 E_m - E^{2N}_m& = \frac {M_{A-1}-M_N} {2M_N M_{A-1}}\Big  (\mathrm p^2_{0NN} - \tilde {\mathrm p}^2_{NN} \Big ),
\label {delta-arg}
\end{align}
where  $\mathrm { p_{0NN}}$ is 
\begin{align}
 \mathrm {p_{0NN}} & = \sqrt{2 M_N  \frac {M_{A-1} } {M_{A-1}-M_N} \Big (E_m  - E^{(2)}_{thr}\Big)},   
 \label{p-0NN}
 \end{align}
 with the restriction 
 \begin{align}
 \mathrm {p^2_{0NN}} \ge 0,
 \label {p20nn_restr}
 \end{align}
 and $\tilde {\mathbf  {p}}_{NN} $ is 
 \begin{align}
\tilde {\mathbf  {p}}_{NN} & =   \mathbf {p_1} -   \frac {M_{A-1}}{M_{A-1}-M_N} \mathbf {p_{NN}}. 
\label {tilde_pNN}
 \end{align}
 
The   delta function in Eq. (\ref {Snn_nr})   is integrated  over $\mathrm {p_{NN}}$   by using the change of   variables  given by Eq. (\ref {tilde_pNN}), hence 
\begin{align*}
2 \pi \mathrm {p^2_{NN}} d \mathrm {p_{NN}} d\cos  \theta_{1NN}  & =  2 \pi \Big ( \frac {M_{A-1}-M_N}{M_{A-1}} \Big)^3 
 \tilde {\mathrm { p}}^2_{NN} d \tilde { \mathrm { p}}_{NN} d\cos \tilde {  \theta}_{1NN}, 
 \end{align*}
where $ \tilde { \theta}_{1NN} $ is the angle between $ \mathbf{p_1} $ and $  \tilde { \mathbf { p}}_{NN}$. \\
The integration of the delta function in the nuclear spectral function [Eq. (\ref {Snn_nr}] yields 
\begin{align}
  \int \delta ( E_m - E^{2N}_m  ) d \mathbf p_{\mathrm {NN}} & = 2 \pi  \Big ( \frac {M_{A-1}-M_N}{M_{A-1}} \Big)^3\int \delta (\mathrm {p^2_{0NN}} - \mathrm {  \tilde {p}^2_{NN}}  ) \tilde {\mathrm { p}}^2_{NN} d \tilde { \mathrm { p}}_{NN} d\cos \tilde {  \theta}_{1NN}\nonumber \\
  & = 2 \pi M_N \Big ( \frac {M_{A-1}-M_N}{M_{A-1}}  \Big)^2  \int \mathrm {p_{0NN}}    d\cos \tilde {  \theta}_{1NN},
   \label{Snn_delta}
\end{align}
and therefore
\begin{align}
  S^{p/n} _{A,2N}( \mathrm p_1,  E_m) & =  2 \pi M_N \Big ( \frac {M_{A-1}-M_N}{M_{A-1}} \Big)^2   \frac {a_2(A,Z)}{ (2x_{p/n})^{\gamma}} \int_{-1}^{1} n_{d}( \mathrm {p_{rel} })\Theta ( \mathrm  {p_{rel} - k_{src}}) n_{CM} (\mathrm {p_{NN}}) \mathrm { p_{0NN}} d\cos  \widetilde {  \theta}_{1NN}. 
 \label{Snn_nr_1}
\end{align}

 The momentum of the correlated nucleons, $\mathrm {p_{NN}} $,   is given by
 \begin{align} 
 \mathrm {p_{NN}} & =  \frac {M_{A-1}-M_N}{M_{A-1}} \sqrt {  (\mathbf {p_1}  - \widetilde {\mathbf  {p}}_{NN} )^2 }\nonumber\\
 & =  \frac {M_{A-1}-M_N}{M_{A-1}} \sqrt {  \mathrm  {p}^2_{0NN} +\mathrm {p^2_1}  -2 \mathrm  {p}_{0NN}\mathrm{p_1} \cos  \widetilde {  \theta}_{1NN}  }, 
 \label {pNN_nr}
 \end{align} 
 with the restriction 
 \begin{align}
 \mathrm  {p}^2_{0NN} +\mathrm {p^2_1}  -2 \mathrm  {p}_{0NN}\mathrm{p_1} \cos  \widetilde {  \theta}_{1NN} \ge 0.
 \label {psq0nn_restr}
 \end{align}
 The cosine of the angle $ \theta_{1NN}$, between $\mathbf p_1$ and $\mathbf {p_{NN}}$, is given  by
 \begin{align} 
 \cos \theta_{1NN} & =  \frac {M_{A-1} -M_N}{M_{A-1}\mathrm p_{NN}}  \big (  \mathrm  {p}_{0NN} \cos  \widetilde {  \theta}_{1NN}   -\mathrm {p_1}\big )  
  \nonumber \\
  & = \frac { \mathrm {p^2_{0NN}}  - \mathrm {p^2_{1}} -  \left (\frac {M_{A-1} } {M_{A-1}-M_N}\right)^2 \mathrm {p^2_{NN} } }{2.0 \left ( \frac {M_{A-1} } {M_{A-1}-M_N} \right) \mathrm {p_{1}}\mathrm {p_{NN}}},
   \label{2N-cosTheta1NN}
 \end{align}
 with the restriction 
 \begin{align}
  \cos^2 \theta_{1NN} \le 1.
 \label {cossqth1nn_restr}
 \end{align}
 The relative momentum  of the correlated  pair, $\mathrm { p_{rel}}$,  is
  \begin{align}
\mathrm { p_{rel}}&  =   \sqrt {\frac {(\mathbf{p_1 -p_2})^2}{4}} = \sqrt {\frac {(\mathbf{2p_1 -p_{NN}})^2}{4}}\nonumber\\
&=\sqrt {\mathrm {p^2_1+ \frac {p^2_{NN}}{4} - p_1p_{NN}} \cos \theta_{1NN} },   
 \label{2N-p-relNR1}
 \end{align} 
  with the restriction 
 \begin{align}
  \mathrm {p^2_1+ \frac {p^2_{NN}}{4} - p_1p_{NN}} \cos \theta_{1NN} \ge 0.
   \label {psqnn_restr}
 \end{align}
  The cosine of the angle $\theta_\mathrm{1rel} $, between $\mathbf p_1$ and $\mathbf {p_{rel}}$, is
 \begin{align}
 \cos  \theta_\mathrm{1rel} & =   \frac {2\mathrm p_1 -\mathrm {p_{NN}} \cos \theta_\mathrm{1NN} }{\mathrm {2p_{rel}} }, 
 \label {cos-theta-1relNR1}
 \end{align} 
 with the restriction 
 \begin{align}
  \cos^2  \theta_\mathrm {1rel} \le 1.
 \label {cossqth1rel_restr}
 \end{align}
The momentum   of the correlated nucleon $2$  is given by 
  \begin{align}
\mathrm { p_{2}}&  =    \sqrt {(\mathbf{p_1 -2p_{rel}})^2} \nonumber\\
&=\sqrt {\mathrm {p^2_1+ 4{p^2_{rel}}- 4p_1p_{rel}} \cos \theta_{1rel} },
 \label{2N-p-2NR1}
  \end{align}
  with the restrictions 
 \begin{align}
 \mathrm {p^2_1+ 4{p^2_{rel}}- 4p_1p_{rel}} \cos \theta_{1rel} > 0, \nonumber \\
 p_2 > \mathrm  {k_F \sim 0.250 \; GeV/c}.
 \label {p2sq_restr}
 \end{align}
  
  The  2N SRC  nuclear spectral functions [Eq.(\ref {Snn_nr_1})], together with Eqs. (\ref {2NSRC_CMdist}),  (\ref {p-0NN}), and (\ref {pNN_nr}) to  (\ref {p2sq_restr}), and the normalization condition [Eq.(\ref {norm_VNA_2NSRC}) ]with $\alpha_1 = 1$,  constitute the computational  model from which   numerical estimates are obtained in the chapter  4  of the present dissertation. 
 
 \subsection { Computational\;model\;for\;three\;nucleon\;SRC \;nuclear\;spectral\;function\; \\
in\;nonrelativistic\;approximation }
\label {sec:Com3NsrcNRA}
The non relativistic 3N SRC   nuclear  spectral functions for proton and neutron are  obtained   from Eq.(\ref {S3n_vn_1}), in the non-relativistic limit of the VN approximation, namely
  \begin{align}
  S_{A,3N}^{p/n}(\mathrm  p _1,E_m)  = a^2_2(A)C^{p/n} (A,N) \int n_d(\mathrm {p}_{3}) n_d(\mathrm {p}_{12})\Theta(\mathrm {p} _{3}-{k} _{src})\Theta(\mathrm {p}_{12}-{k} _{src})\delta(E_m - E^{3N}_m) 
d^3 \mathbf {p} _{3},
\label{S3n_nr}
\end{align}
where  $p/n$ indicates proton and neutron respectively,    the factors $C^{p/n}(A,Z)$  by Eq. (\ref {3N-C}), and and  $n_d$ is the deuteron momentum distribution calculated from a nonrelativistic wave function. 

The removal energy $E_m^{3N}$  for the 2N SRC [Eq. (\ref {3NSRC_removal})] in the nonrelativistic limit is given by 
\begin{align}
  E^{3N}_m& = E^{(3)}_{thr}+ \frac {\mathrm  p^2_{3}}{2 M_N} +  \frac {\mathrm {p^2_{2}}}{2 M_{N}} - \frac {\mathrm p_{1}^2}{2M_{A-1}}\nonumber \\
  & = E^{(3)}_{thr}+ \frac {\mathrm  p^2_{3}}{2M_N} +  \frac {(\mathbf {p}_{1} + \mathbf {p}_{3})^2}{2 M_{N}} - \frac {\mathrm p_{1}^2}{2M_{A-1}}\nonumber \\
  & = \frac {1} {M_N} \left [ {M_N} E^{(3)}_{thr}- \left  ( \frac {2 M_N - M_{A-1}}{M_{A-1}} \right ) \frac {\mathrm p^2_1}{4} + 
  \Big (\frac {\mathbf  p_{1}}{2} + \mathbf  p_{3} \Big )^2 \right ],
\label {Em-nrel3}
\end{align}
where   the VN collinear approximation $\mathbf {p}_{1} + \mathbf {p}_{2} + \mathbf {p}_{3} =0$, have been used. The argument of the delta function in the nuclear spectral function 
[Eq. (\ref {S3n_nr}] is therefore given by 
\begin{align}
 E_m - E^{3N}_m & = \frac {1} {M_N} \Big  (\mathrm p^2_{03N} - \mathrm p^2_{3N} \Big ),
\label {delta3-arg}
\end{align}
where $\mathrm {p_{03N}}$ is  
\begin{align}
p_{03N} = \sqrt {M_N \big (E_m  -E^{(3)}_{thr}\big) -\frac {M_{A-1}-2M_N }{M_{A-1}}\frac {\mathrm p^2_{1}}{4}},
\label {p-03N}
\end{align}
with the restriction 
 \begin{align}
 \mathrm {p^2_{03N}} \ge 0,
 \label {p03n_restr}
 \end{align}
and   $\mathbf { p_{3N}}$ is 
\begin{align}
 \mathbf  p_{3N} = \mathbf p_1/2 + \mathbf p_{3}.
 \label{p-3N}
 \end{align}

The   delta function in Eq. (\ref {S3n_nr})   is integrated  over $\mathrm {p_{3}}$   by using the  change of  variables  given by 
Eq. (\ref {p-3N}), hence 
\begin{align}
  \int \delta ( E_m - E^{3N}_m  ) d \mathbf p_{\mathrm {3}} & = 2 \pi \int \delta (\mathrm p^2_{03N} - \mathrm { p}^2_{3N}   ) \mathrm { p}^2_{3N} d \mathrm { p}_{3N} d\cos  \theta_{13N}\nonumber \\
 & =\pi M_N  \int \mathrm {p_{03N}}    d\cos   \theta_{13N},
 \label{S3n_delta}
\end{align}
where $\theta_{13N}$ is the angle between $\mathbf p_1$ and $\mathbf p_{3N}$, and  therefore
\begin{align}
S^{p/n} _{A,3N} ( \mathrm { p_1, E_m}) & = \pi M_N a^2_2(A,y)C^{p/n} (A,Z)    \int_{-1}^{1} n_d(\mathrm {p}_{3}) n_d(\mathrm {p}_{12})\Theta(\mathrm {p} _{3}-{k} _{src})\Theta(\mathrm {p}_{12}-{k} _{src})\mathrm {p_{03N}} d\cos   \theta_{13N}.
\label{S3n_nr_1}
\end{align}
The momenta $ \mathrm p_{3} $ and $ \mathrm p_{12}$ are given by 
\begin{align}
  \mathrm p_3 &= \sqrt {  \Big (\mathbf p_{3N} -\frac {\mathbf p_{1}}{2}\Big)^2}  \nonumber\\
&= \sqrt { \mathrm {p^2_{03N}} -\mathrm{p_{03N}} \mathrm p_1\cos   \theta_{13N}+ \frac{ \mathrm p^2_1}{4}},
\label {3N-SRC_p3NR}
\end {align}
\begin{align}
 \mathrm {p_{12}}& = \sqrt { \Big(\frac {\mathbf p_{1} - \mathbf p_{2}}{2}\Big)^2} = \frac{1}{2}  \sqrt { \Big(\mathbf p_{3N}+\frac {3\mathbf p_{1}}{2}\Big)^2} \nonumber\\ 
&=\frac{1}{2} \sqrt {\mathrm {p^2_{03N}} +3\mathrm {p_{03N}} \mathrm p_1\cos   \theta_{13N}+  \frac{9\mathrm p^2_1}{4}},
 \label{3N-SRC-p12NR}
 \end{align}
 with the restriction 
 \begin{align}
 \mathrm {p^2_{03N}} -\mathrm{p_{03N}} \mathrm p_1\cos   \theta_{13N}+ \frac{ \mathrm p^2_1}{4} \ge 0.
 \label {psq12nr_restr}
 \end{align}
 The cosine of the angle $ \theta_{13}$,  between $\bf p_1$ and $ \bf p_3$, is 
 \begin{align} 
 \cos \theta_{13} & =  \frac {2\mathrm {p_{03N}}\cos   \theta_{13N} -\mathrm p_1  }{2\mathrm p_3},  
 \label{3N-cosTheta13NR} 
 \end{align}
 with the restriction 
 \begin{align}
  \cos^2 \theta_{13} \le 1.
 \label {cossqth13_restr}
 \end{align}
 The cosine of the angle $\theta_\mathrm{112}$, between $\mathbf p_1$ and $\mathbf {p_{12}}$, is
 \begin{align}
 \cos  \theta_\mathrm{112} & =   \frac {3\mathrm p_1 +2\mathrm {p_{03N} } \cos \theta_\mathrm{13N} }{4\mathrm{ p_{12} }},
 \label {cos-theta-112NR}
 \end{align} 
 with the restriction 
 \begin{align}
  \cos^2 \theta_{112} \le 1.
 \label {cossqth1nn_restr}
 \end{align}
 The momentum of the correlated nucleon $2$ is
   \begin{align}
\mathrm p_{2}&  =    \sqrt {(\mathbf{p_1 -2p_{12}})^2} \nonumber\\
&=\sqrt {\mathrm {p^2_1+ 4p^2_{12} - 2p_1p_{12}} \cos \theta_{112 }},
 \label{3N-p-2NR1}
 \end{align}
  with the restrictions 
 \begin{align}
 \mathrm {p^2_1+ 4p^2_{12} - 2p_1p_{12}} \cos \theta_{112 } \ge 0, \nonumber \\
  \mathrm  {p_2 > k_F \sim 0.250 \; GeV/c}.
 \label {p2sqnr_restr}
 \end{align}
 
 The  3N SRC  nuclear spectral functions  (\ref {S3n_nr_1}), together with Eqs. (\ref {p-03N}),  and (\ref {3N-SRC_p3NR}) to  (\ref {p2sqnr_restr}), and the normalization condition (\ref {norm_VNA_3NSRC}) with $\alpha_1 = 1$, constitute the computational  model from which   numerical estimates will be calculated in chapter  4  of the present  dissertation. 
 
\section{Summary}
\label{sec:Ch3Summ}
   The computational models for 2N and 3N SRC  high momentum  nuclear  spectral functions in VN and LF approximation, as well as in  the non-relativistic limit of the VN approximation,  have been obtained in the present chapter. The computational models have been developed  by integration of the delta functions in the mathematical models of chapter 2.    

The computational models are  a set of computable integrals  that represent the high momentum nuclear spectral functions,  as well as equations for    momenta and  cosine of angles between momentum vectors,  and for auxiliary variables.  The restrictions in the values of momenta and  cosine of angles have been also defined to ensure that the results are physically valid. The computational models are used   to obtain the numerical estimates of the nuclear spectral functions, density matrices and momentum distributions, given in the next chapter of the present dissertation.

\chapter{Numerical\;estimates\;of\;nuclear\;spectral\;functions,\;density\;
matrices\;and\;momentum\\distributions }

\label{sec:Ch4Intro}
The computational models developed in chapter 3 of the present  dissertation are used to obtain numerical estimates of 2N and 3N SRC nuclear spectral functions for LF, VN and VN in the nonrelativistic limit approximations. The  numerical estimates for the LF density matrix for light and heavy nuclei are also included. The numerical estimates and the parametrization of the momentum distribution of the bound nucleon  are also presented. The momenta distribution  estimates are obtained by the integration of the corresponding nuclear spectral functions for VN and VN in the nonrelativistic limit approximations.
  
 The outline of the chapter is as follows. Section \ref {sec:Parameters} contains a  brief discussion about  the range of validity and set of parameters which are  used for  the numerical estimates of the the nuclear spectral functions. The deuteron momentum  distribution used in the nuclear spectral function calculations is described in section \ref {sec:DeuteronMd}, which also includes the numerical estimates for the deuteron LF density matrix. The numerical estimates for the nuclear spectral functions in  VN, VN in the nonrelativistic limit, and LF approximations are included in the sections \ref {sec:NumSpecVNA}  and \ref {sec:NumSpecLFA} respectively.  The section \ref{sec:NumDmLFA} includes the numerical estimates for the SRC density matrices  in LF approximation. The section \ref {sec:NumMdVNA}   presents the numerical estimates for the SRC momentum distribution in VN  approximation, and VN approximation in the nonrelativistic limit. The section \ref {sec:NumMdVNA} also contains the numerical estimates  for the SRC nonrelativistic momentum distributions for light nuclei (A $\le 12$) which  are compared with \textit{ab initio} quantum Monte Carlo calculations \cite {RW91,VMCpc,PPC97, WPC00, VMC01, PVW02}, and for some heavy nuclei (A $> 12$)  with the calculations for proton given in\cite {CiofiSimula}.  
 
 The parametrization of  the SRC and the mean field  momentum distributions in VN and VN in the nonrelativistic approximations are presented in section  \ref {sec:ParMd}. The section  \ref {sec:Ch4Summ} summarizes the results of the chapter.

\section{Parameters for numerical estimates of the nuclear spectral functions } 
\label{sec:Parameters}
Before presenting  the numerical estimates of the nuclear spectral functions, it is convenient to  discuss briefly the range of validity and set of parameters which will be used to obtain such numerical estimates. 
 
 The main assumption for the nuclear spectral function  models  is the dominance of NN SRC in the nuclear dynamics for internal momenta $\mathrm {p\gtrsim k_{src} > k_F}$, where $\mathrm { k_{src} }$ is the relative momentum threshold at which a NN system with such relative momentum can be considered in the short-range correlation. The  next major assumption is the dominance of the isosinglet $pn$ component in the NN SRC. The empirical evidence of the dominance of NN SRCs was accumulated during the  last several decades  (see e.g. \cite{FS88,FSDS93,Aclander1999,Yaron2002,Kim1,Kim2,Fomin2011}) in high energy electro- and hadroproduction reactions.  Recent triple-coincident experiments \cite{isosrc,Eip3,Eip4} indicated that the $pn$ dominance in the nucleon-nucleon SRC persists for up to the heavy nuclei such as $A$ = 208.  Hence, it is  expected that the models developed in the present  dissertation  should be valid for a wide  range of atomic nuclei.  

The parameters discussed  below  are independent of the use of the VN or LF approximations. Therefore,   further refinements are  achieved in their values for lightest nuclei ($A\le $ 12), by considering the nonrelativistic  limit of the VN approximation model  and comparing them with  $ab\;initio$ calculations that result from  the  variational Monte Carlo methods given in \cite{Wir2014}.  These refinements are presented in section \ref {sec:NumSpecVNA}. 
 
 With the parameters for 2N and 3N SRCs fixed,  the normalization factors $n^N_{2N}$ and $n^N_{3N}$ within VN and LF approximations can be calculated. Note that the normalization factors will be model dependent since the  2N and 3N momentum distributions predicted in VN and LF approximations are  different.  Once these normalizations are calculated,   the norm of the mean field distributions can be estimated from the relation $n^N_{MF}  = 1 - n^N_{2N}-n^N_{3N}$. Thus the estimates for the normalization of mean field distributions will be indirectly  VN or LF model dependent.
 
  \subsection {Parameter $a_2(A,Z)$}
 The most  important parameter that defines the strength of 2N SRC is $a_2(A,Z)$. Within  the short-range correlation framework, the  parameter can be extracted from the ratios of  the cross sections of high momentum transfer inclusive electronuclear scattering off nuclei $A$ and the deuteron \cite{FS88, FSDS93}. Recent measurements at Jefferson Lab \cite{Kim1,Kim2,Fomin2011}  provided the magnitudes  of $a_{2}(A,Z)$ for a  rather wide spectrum of atomic nuclei.   

 \subsection {Relative momentum threshold $k_{src}$ }
 The 2N SRC relative momentum threshold,  $k_{src}$, defines the momentum distribution of NN SRC
 in Eqs. (\ref {2NSRC_relmd}) and (\ref{2NSRC_NNdens}).  The value of  $k_{src}$ is  fixed from the condition that it  ishould be  sufficiently large 
 for mean-field contribution to be insignificant, as well as close to the threshold value  for which $pn$ dominance is 
 observed empirically \cite{isosrc,Eip3,Eip4}. Another condition in defining $k_{src}$ is the onset of the dominance of the $D$-wave contribution in the high momentum 
 part of the deuteron wave function as described in section  \ref {sec:DeuteronMd} below. The conditions imply  a value for  $k_{src}\sim$400~MeV/$c$ to be  applied for the calculation of numerical estimates of the SRC nuclear spectral function.  
 
  \subsection {Parameter $\gamma$}
 In the nuclear spectral function model developed in the present dissertation, the contributions of $pp$  and  $nn$ SRCs were neglected, however they are expected to increase with the mass number A \cite {newprops}. To account for the effects due to the $pp$  and  $nn$ SRCs  correlations,  the  parameter $\gamma$ is   introduced. On the basis  of  the experimental observation \cite{Eip4} that in the 2N SRC regions $pn$ dominates by almost a factor of 20 for a wide range of 
 nuclei (up to $A$ = 208), the following approximations are adopted: $\gamma\approx 0.8$ for asymmetric nuclei, and $\gamma\approx 1.0$ for symmetric nuclei.   

\subsection {Parameters for 2N SRC center of mass momentum distribution}
 The width of the 2N SRC center of mass distribution, $\beta(A)$, and  the parameter $N_0$  in Eq. (\ref{2NSRC_CMdist}),   are chosen from the  estimates that result from  the convolution of the mean-field distribution of  two independent nucleons presented in Ref. \cite{CiofiSimula}.  
 
\subsection {3N SRC suppression factor  $C^N(A,Z)$}
 For the case of 3N SRCs,  the only additional parameter needed to estimate  the nuclear  spectral function is the suppression factor $C^N(A,Z)$ defined in 
 Eq. (\ref{3NSRC_relmd}). The factor $C^N(A,Z)$ accounts for the suppression of the 3N configurations with 
 two identical spectators like $pp$ and $nn$ pairs, and affects only  the distribution of the minority component in the asymmetric nucleus. For example, according to the considered model, the neutron can not be generated from 3N SRC in the $^3$He nucleus, since it will produce two \textquotedblleft parallel" protons in the final state.  The factors $C^p(A,Z) $ and $C^n(A,Z) $ are  defined  in  Eqs. (\ref {3N-C}) for proton and neutron respectively.


\section{Momentum distribution  and light-front density matrix for deuteron} 
\label{sec:DeuteronMd}
 The numerical estimates of the momentum distribution  and the light-front density matrix for deuteron are included in the present  section,   

\subsection{Momentum distribution   for deuteron}  
\begin{figure}[H]
\centering\includegraphics[scale=0.42]{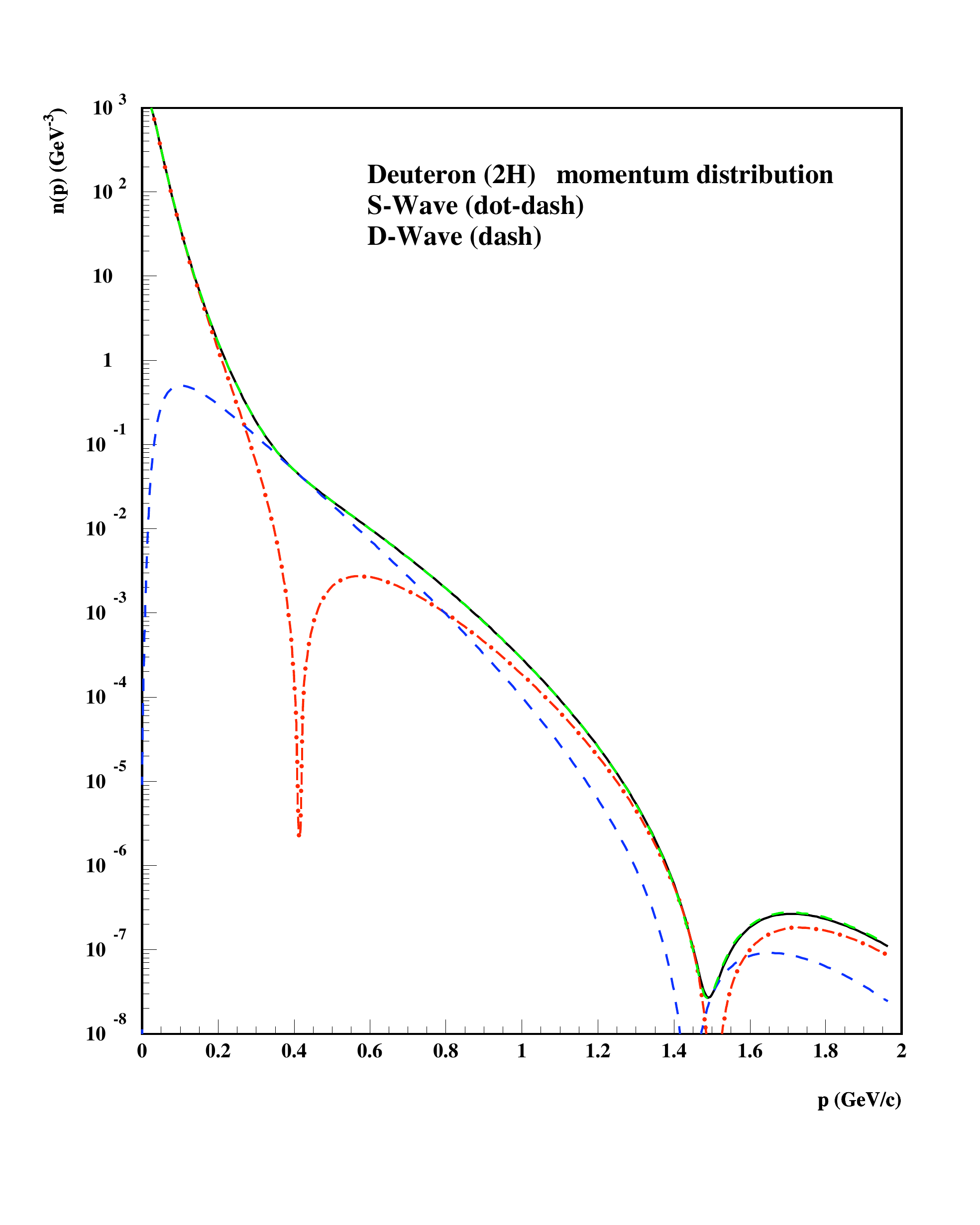}
\caption{The momentum distribution of deuteron (s-wave: dot-dash, d-wave:dash) .} 
\label{deuteron}
\end{figure}

 The  D-wave deuteron momentum distribution ($n_{d}$) used for the nuclear spectral function  calculations, results from a  wave function modeled with  the Argone V18 (AV18) two-nucleon potential, that is specially suitable for the spectral function model since fits both $pp$ and $np$ data \cite {AV1895}. The total momentum distribution, as well as the S-wave and the D-wave momentum distribution components, calculated with the AV18 potential  is shown in Fig. \ref {deuteron}  \cite {VMCpc}.

 The D-wave of the deuteron momentum distribution, shown in Fig. \ref {deuteron}, starts to dominate (by a factor of $\approx$ 8)  the S-wave momentum distribution at p $\approx$ 350 MeV/c, and  for p $\approx$ 400 MeV/c is more than two orders of magnitude greater than the S-wave momentum distribution.  Hence, for  a proper modeling of the  D-wave deuteron momentum distribution ($n_{d}$),  the value of   400 MeV/c for the 2N SRC relative momentum threshold,  $k_{src}$, was chosen for the  numerical estimates of the nuclear spectral function. 
  
 \subsection{Light-front density matrix for deuteron}  
 
\begin{figure}[H]
\centering\includegraphics[scale=0.42]{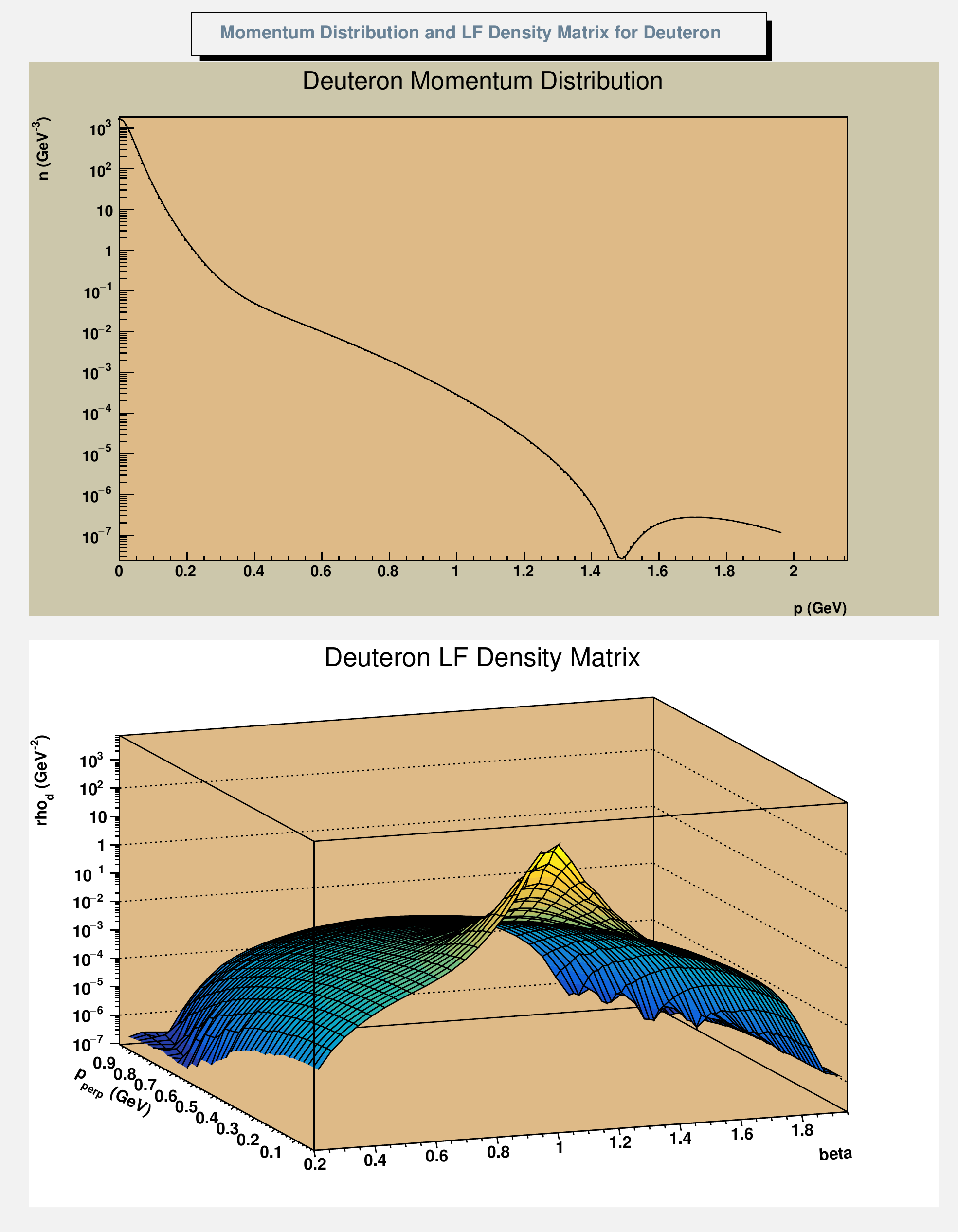}
\caption{Upper graph: Deuteron momentum distribution. Lower graph: Deuteron light-front density matrix} 
\label{deuteron-DM2}
\end{figure}

  The numerical estimates for   light-front density matrix of the  deuteron used for the LF nuclear spectral function  calculations, was obtained by using the model for this density matrix given by Eq. (\ref {rho_d}). The graph for the density matrix as well as the graph of the  momentum distribution used in its calculation is given in Fig. \ref {deuteron-DM2}. 


The graph shows that the deuteron density matrix  is the joint probability of the transverse momentum ($\mathrm {p_\perp}$)  and the  momentum fraction of the 2N SRC, $\beta$, carried by the bound nucleon,  with a maximum value for $\beta = 1$, as it is expected. The momentum distribution graph reproduces the total momentum distribution showed in the Fig. \ref {deuteron}  above, as it was also expected.

 
\section{Numerical\;  estimates\;  for\;  SRC\;  nuclear\; spectral\;  function\; in\; virtual\; nucleon\\ approximation } 
\label{sec:NumSpecVNA}

The 2N and 3N  SRC nuclear spectral functions for proton and neutron in VN approximation  are calculated with the computational models given by 
 Eqs.(\ref {E_20}) to (\ref {p2vna_restr}), and Eqs.(\ref {E_30}) to (\ref {p2vna3_restr}) respectively. The same calculations  for 2N and 3N SRC nuclear spectral function  in VN approximation  in the  nonrelativistic limit  are made with Eqs.(\ref {p-0NN}) to (\ref {p2sq_restr}), and with Eqs.(\ref {p-03N}) to (\ref {p2sqnr_restr}) respectively. The SRC parameters for the calculation of the nuclear spectral functions are given in the Tables \ref  {table2NSRC} and \ref {table3NSRC}  described in  the section \ref {sec:NumMdVNA} of the present chapter.
  
  \begin{figure}[H]
\centering\includegraphics[scale=0.45]{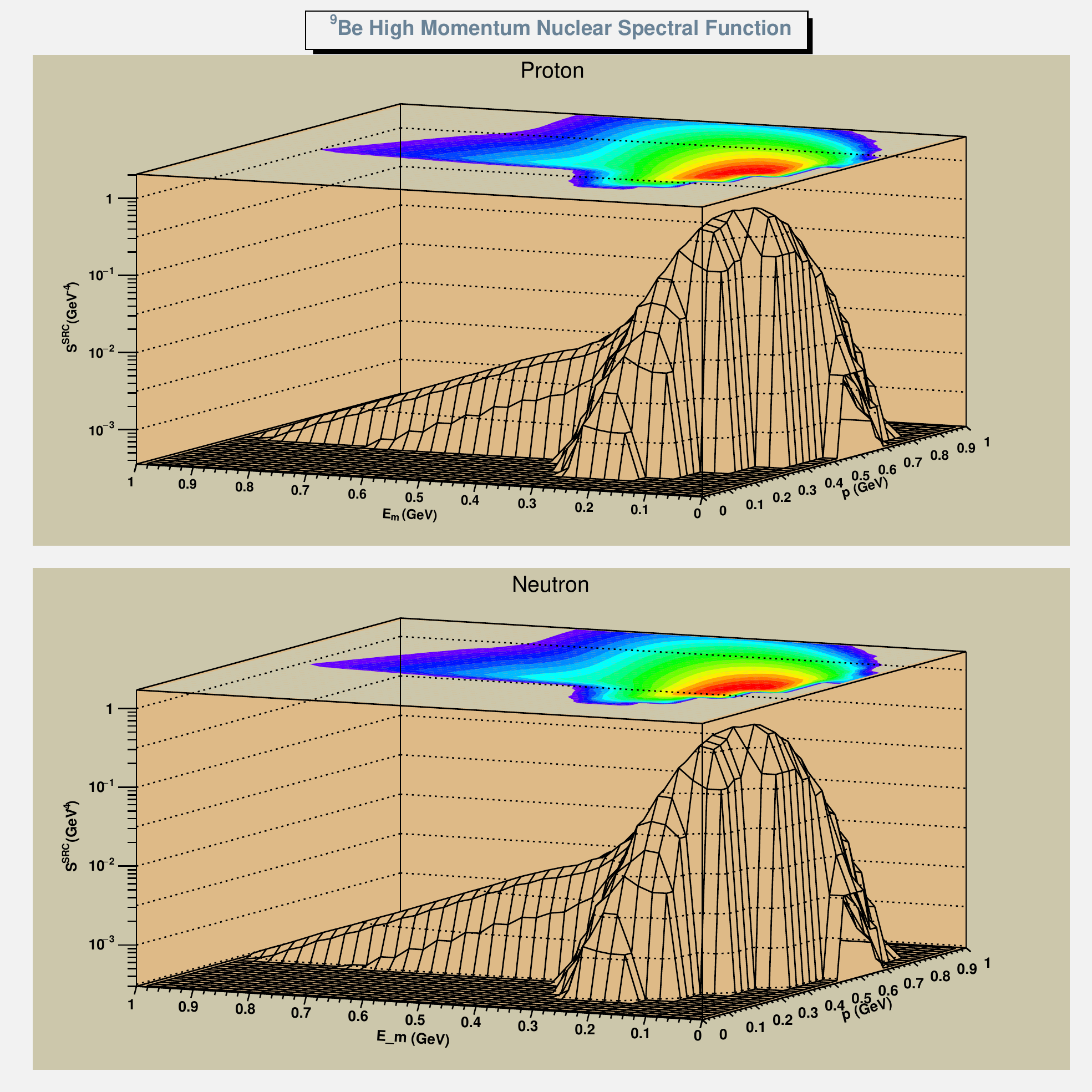}
\caption{Nuclear spectral function for $^{9}$Be in the nonrelativistic limit of the  VNA  }
\label{Fig_NRSpectFunc_Be9}
\end{figure}

\begin{figure}[H]
\centering\includegraphics[scale=0.45]{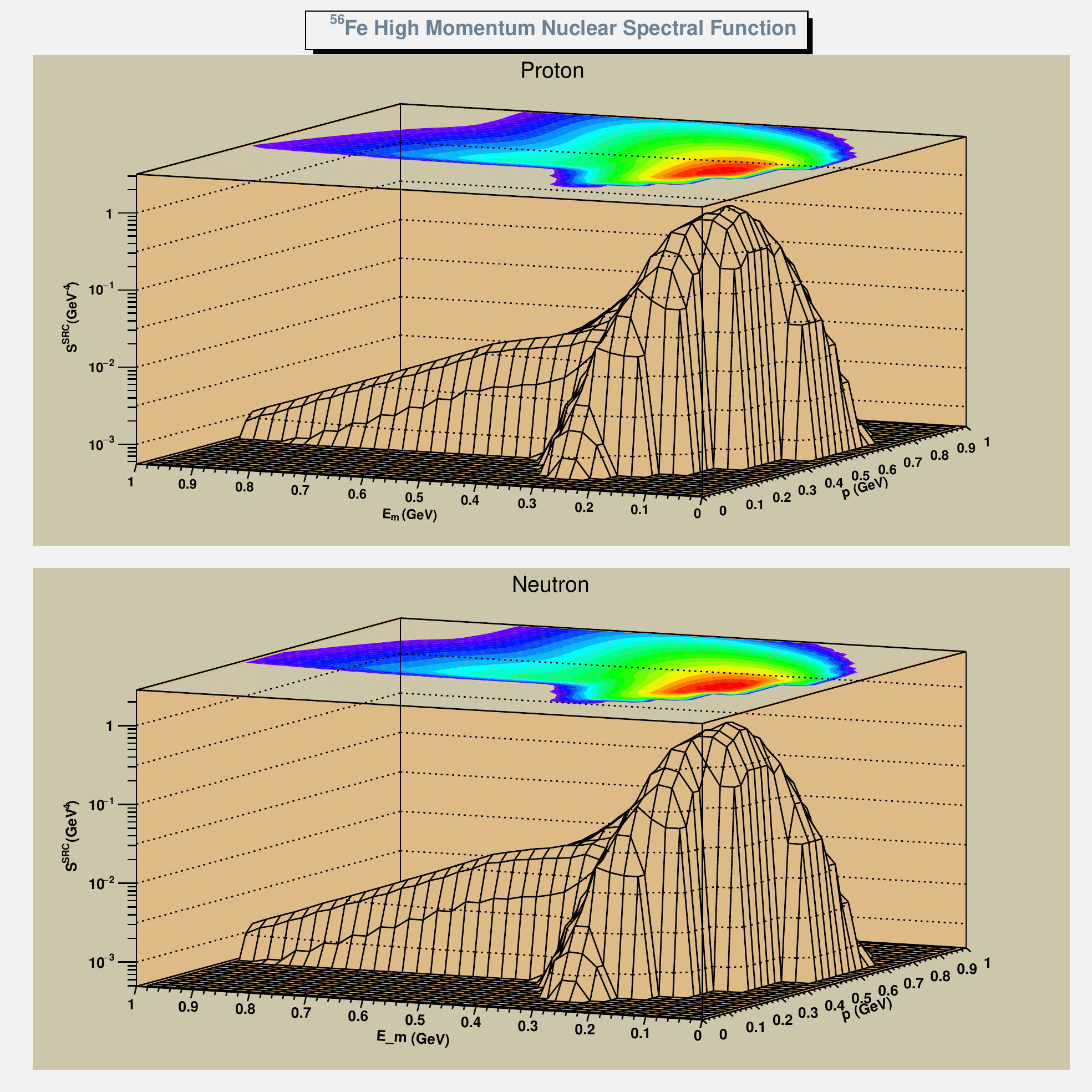}
\caption{Nuclear spectral function for $^{56}$Fe in the nonrelativistic limit of the  VNA }
\label{Fig_NRSpectFunc_Fe56}
\end{figure}

 The numerical results for nuclear spectral function for  $^{9}$Be  and $^{56}$Fe in the non relativistic limit of the VN approximation are represented by  the  three dimensional graphs depicted  in Figs.  \ref {Fig_NRSpectFunc_Be9} and \ref {Fig_NRSpectFunc_Fe56} respectively.  The graphs show that the nuclear spectral function as expected, is the joint probability of the momentum ($\mathrm {p}$)  and removal energy ($\mathrm {E_m}$)  of the bound nucleon.
 
The spectral function graphs also present a region of maximum values  for  0.3 $\le p \le$ 0.6 GeV/c, that, as expected, corresponds to the dominance of the  D-wave of the deuteron momentum distribution (see Fig. \ref {deuteron}),  over the S-wave momentum distribution, that implies maximum probability of having 2N SRC in the nucleus. 

\begin{figure}[H]
\centering\includegraphics[scale=0.45]{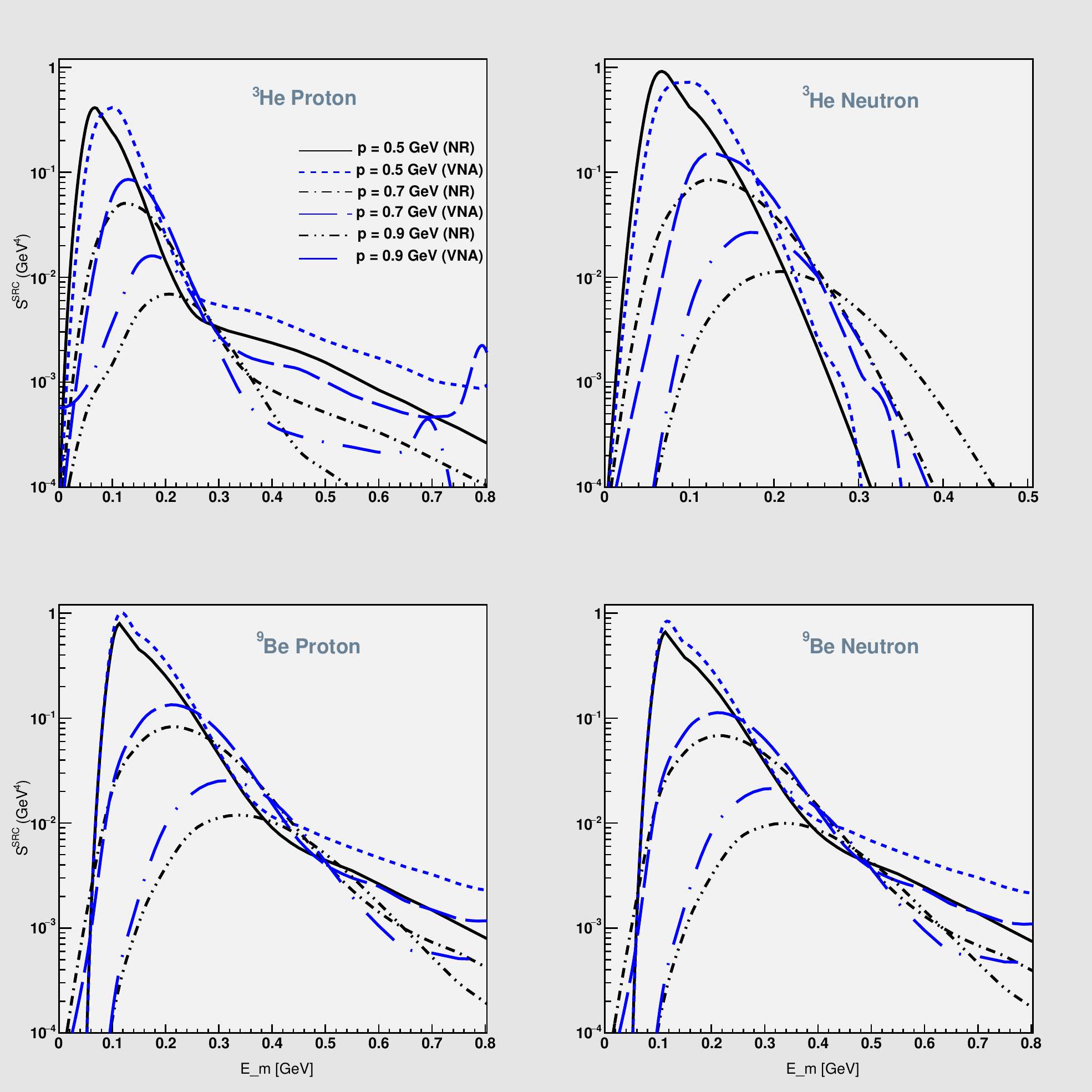}
\caption{  Nuclear spectral function,  with constant values
               of momentum for $^{3}$He and  $^{9}$Be, in VNA  and in  the nonrelativistic limit  of the  VNA }
\label{Fig_SpectFunc_He3Be9}
\end{figure}

The graphs for the numerical estimates of     nuclear spectral function  with constant values of momentum for  VNA and VNA in the nonrelativistic limit, are shown in Fig.  \ref {Fig_SpectFunc_He3Be9} for  $^{3}$He and  $^{9}$Be.  The  spectral function for proton   in $^{3}$He,  and for proton and neutron in  $^{9}$Be   shows a tail for p $>$ 0.45 GeV/c due to the presence of 3N SRC for  these nucleons. This tail is not present  for the  spectral function for the neutron in  the $^{3}$He nucleus, due to the fact that 3N SRC is not allowed for this nucleon due to the Pauli exclusion principle.  The maximum value of the spectral function has a monotonic decrease with the values of momentum,  also, in general,  the dispersion of the   removal energy E$_m$ increases when the momentum increases. 

\section{Numerical\; estimates\; for\; SRC\; nuclear \; spectral\;  function\; in\; light-front\\ approximation} 
\label{sec:NumSpecLFA}

The 2N and 3N  SRC nuclear spectral functions for proton and neutron in the LF approximation  are calculated with the computational models given by Eq.(\ref {spe_LF_2NSRC_7})  (to (\ref {kphi_1perp}), and Eqs.(\ref {spe_LF_3NSRC_7}) to (\ref{ktilde_1perp} ) respectively. The SRC parameters for the calculation of the spectral function are given in Tables \ref  {table2NSRC} and \ref {table3NSRC}  described in  the section \ref {sec:NumMdVNA} of the present chapter.
  
  \begin{figure}[H]
\centering\includegraphics[scale=0.55]{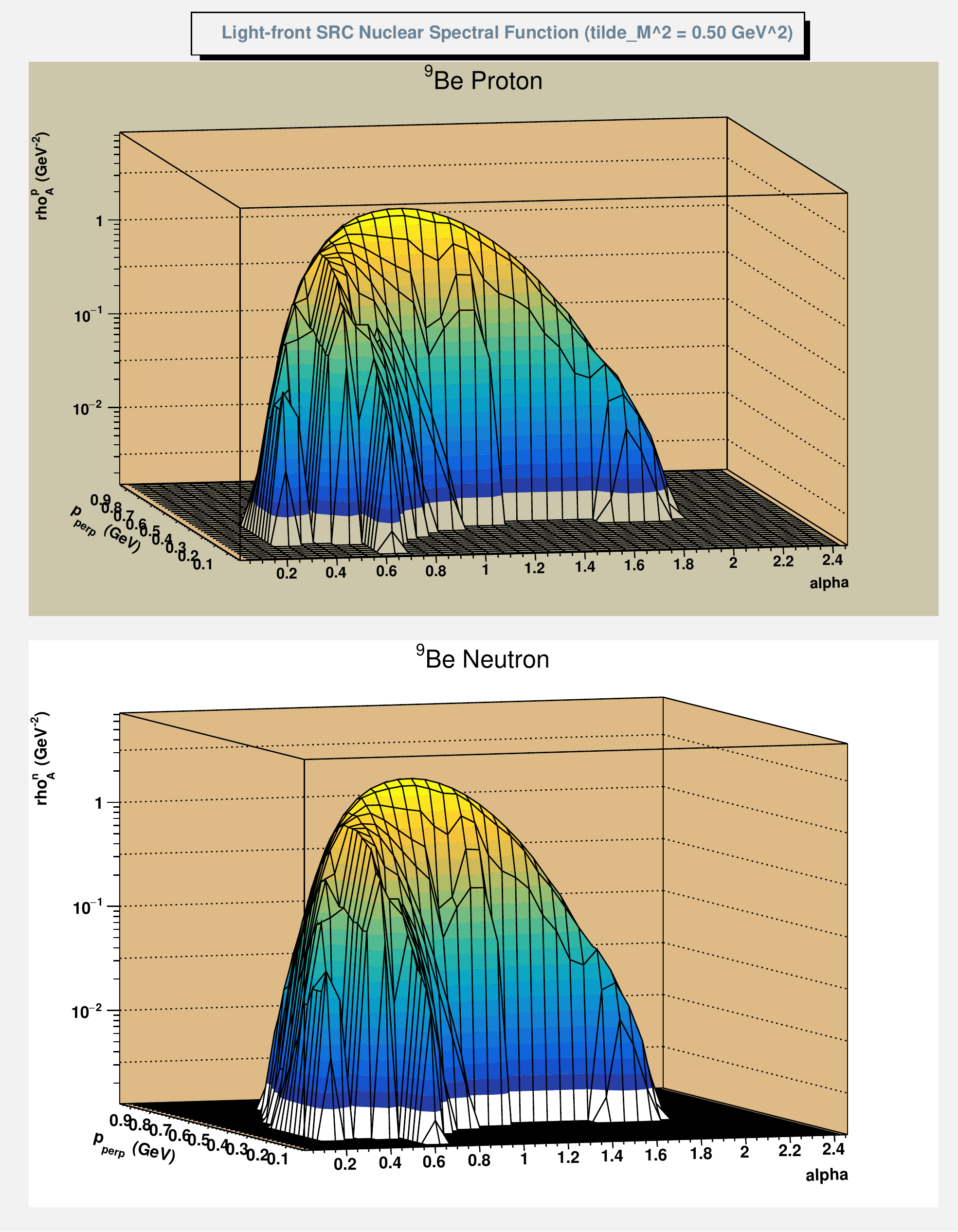}
\caption{Nuclear spectral function for $^{9}$Be in light-front approximation, for $\tilde M^2_N = 0.5 \;\mathrm {GeV}^2$}
\label{Fig_LFSpectFunc_Be9}
\end{figure}

\begin{figure}[H]
\centering\includegraphics[scale=0.55]{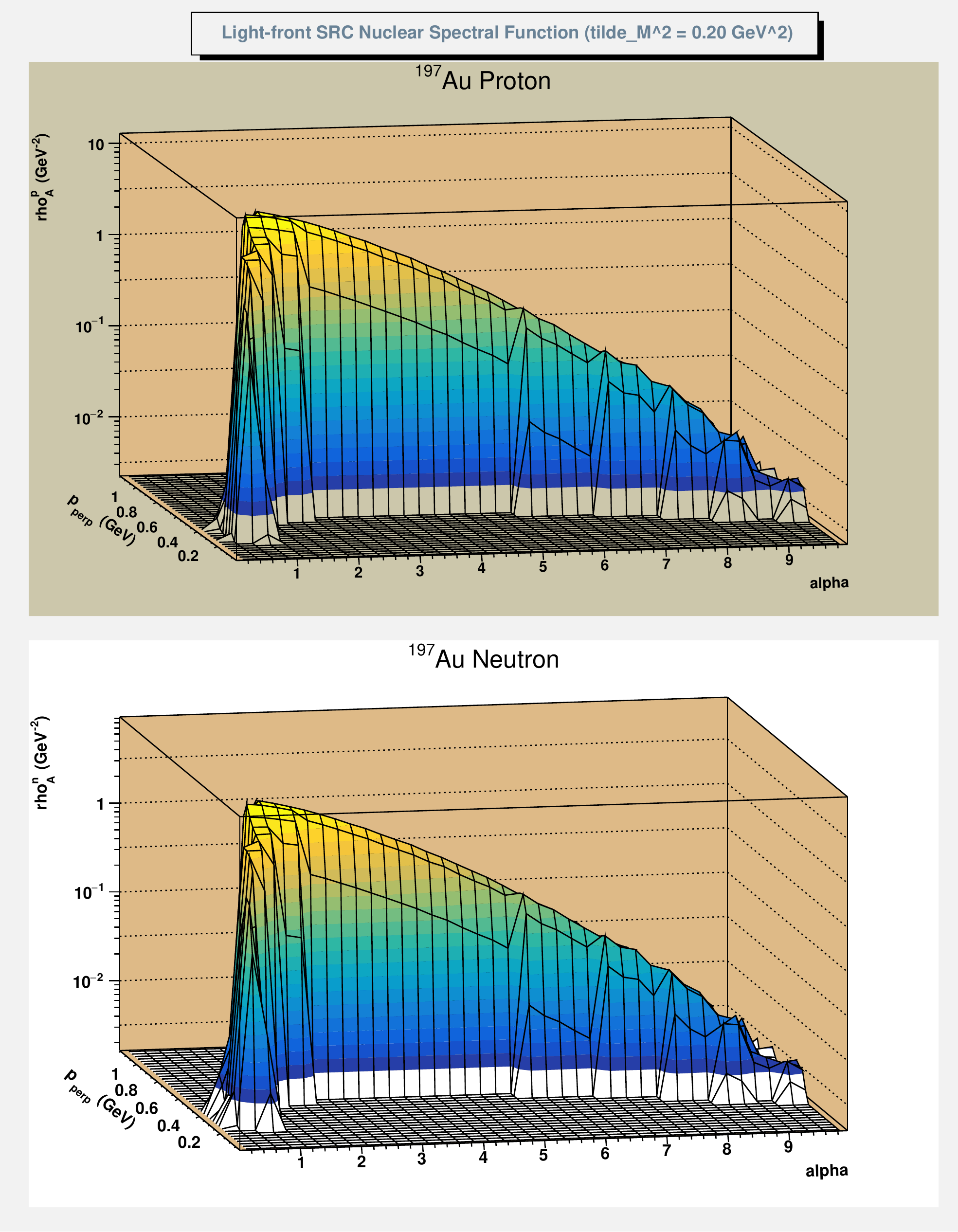}
\caption{Nuclear spectral function for $^{197}$Au in light-front approximation, for $\tilde M^2_N = 0.2 \;\mathrm {GeV}^2$}
\label{Fig_LFSpectFunc_Au197}
\end{figure}

 The numerical results for the SRC nuclear spectral function for  $^{9}$Be   in the LF approximation are represented by  the  three dimensional graphs depicted  in Fig.  \ref {Fig_LFSpectFunc_Be9}, for an invariant mass  
 $\tilde M^2_N = 0.5 \;\mathrm {GeV}^2$.  A similar graph is shown in  Fig.  \ref {Fig_LFSpectFunc_Au197}, for the SRC nuclear spectral function for  $^{197}$Au   in the LF approximation, for an invariant mass  $\tilde M^2_N = 0.2 \;\mathrm {GeV}^2$.
The graphs show that the LF nuclear spectral function for constant values of the invariant mass,  is the joint probability of  finding a bound nucleon in the nucleus with   light-front  "+" momentum fraction  ($\alpha$)  and  transverse momentum  ($\mathrm {p_\perp}$), as it is expected,.  

The SRC nuclear spectral function graphs show that the range of values of $\alpha$ for which the spectral function is greater than zero, increases with the value of the mass number $A$ of the nucleus. The graphs  also show that the maximum values of   the LF nuclear spectral function for $^{9}$Be are in the range  $ 0.8 < \alpha < 1.2$, whereas for $^{197}$Au  are in the range $ 1.0 < \alpha < 3.0$, so that the range also increases with the mass number $A $ of the nucleus. The results are as expected, since   the range of possible values of the light-front  "+" momentum fraction for any nucleus with mass number $A$ is $0 < \alpha < A$.

 
\section{Numerical\; estimates\; for\; SRC\; density\; matrices\; in\; light-front\; approximation} 
\label{sec:NumDmLFA}

The   SRC density matrices  for proton and neutron in the LF approximation  are calculated with 
Eq. (\ref {LF-density}),  on which the density matrix is obtained by the integration of the SRC nuclear spectral function in the LF
approximation over the invariant mass $\tilde M^2_N$. 
  
  \begin{figure}[H]
\centering\includegraphics[scale=0.55]{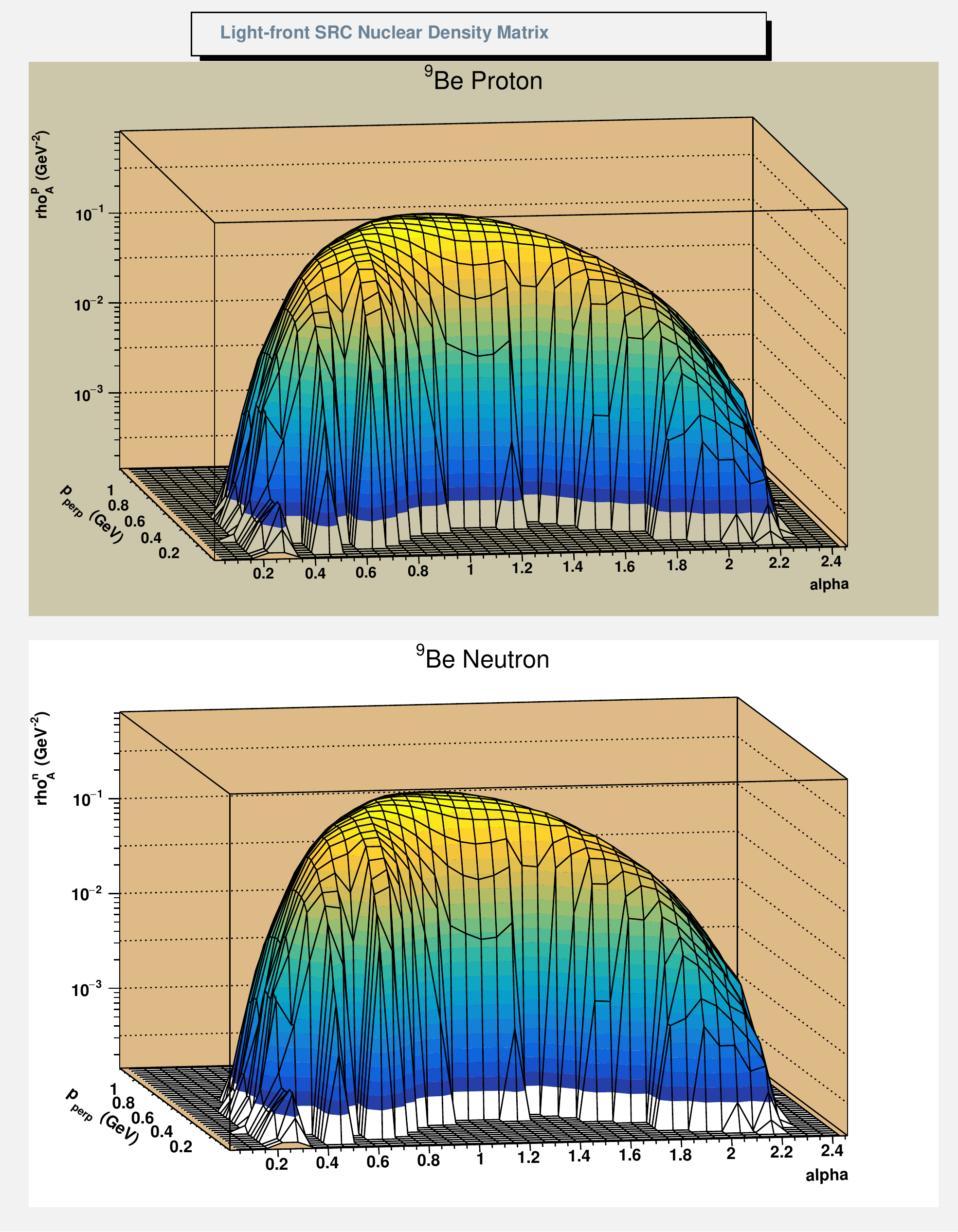}
\caption{SRC density matrix  for $^{9}$Be in light-front approximation}
\label{Fig_LFDensMat_Be9}
\end{figure}

\begin{figure}[H]
\centering\includegraphics[scale=0.55]{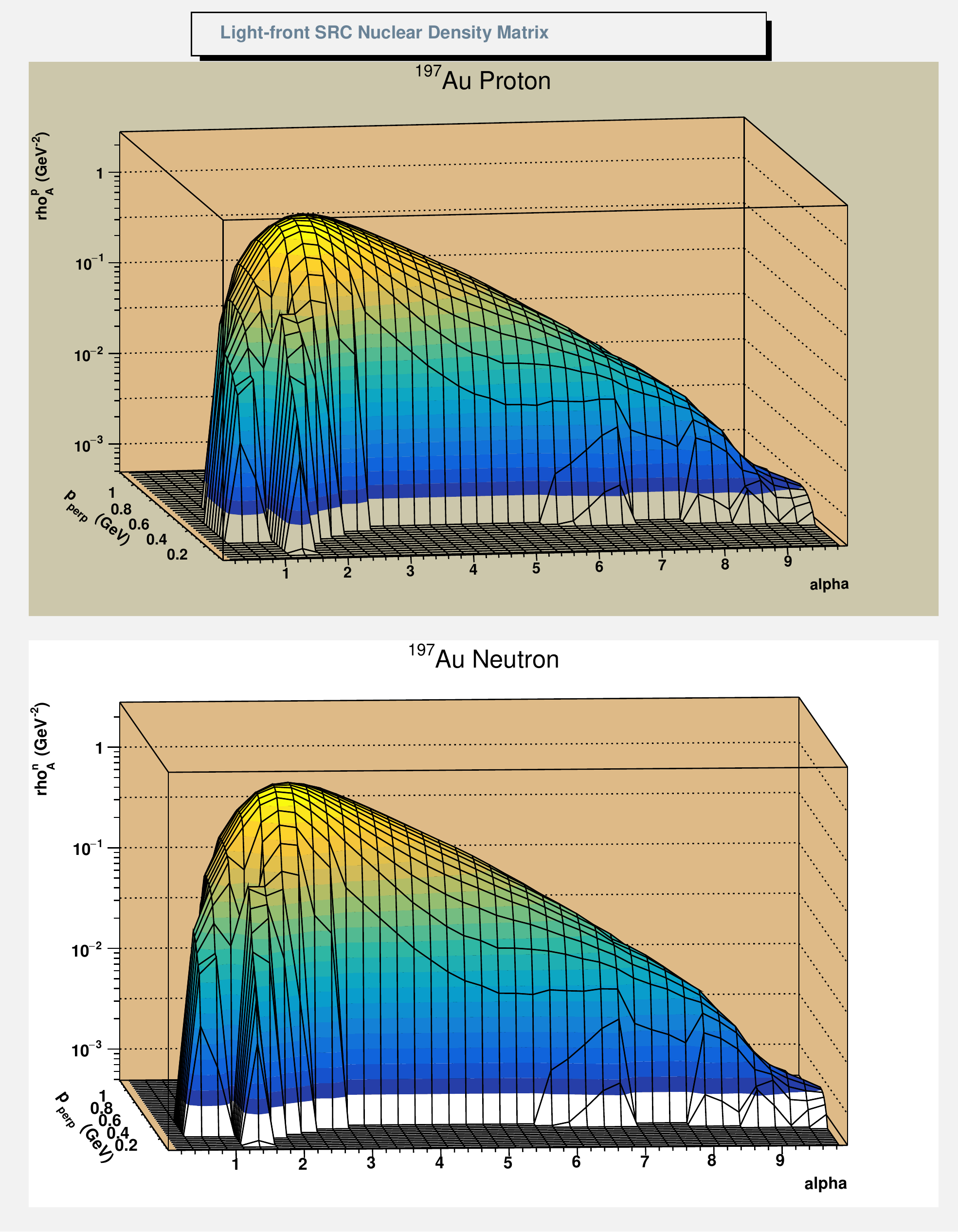}
\caption{SRC density matrix  for $^{197}$Au in light-front approximation}
\label{Fig_LFDensMat_Au197}
\end{figure}

 The numerical results for the SRC density matrices  for  $^{9}$Be  and $^{197}$Au in the LF approximation are represented by  the  three dimensional graphs depicted  in Figs.  \ref {Fig_LFDensMat_Be9} and  \ref {Fig_LFDensMat_Au197}. 
The graphs show that the LF SRC density matrix  is the joint probability of finding a bound nucleon in the nucleus with  the light-front  "+" momentum fraction  ($\alpha$)  and the  transverse momentum  ($\mathrm {p_\perp}$), as it is expected. 
Similar to the case of the  SRC nuclear spectral function, the  graphs for the SRC density matrices  show that the range of values of $\alpha$ for which the spectral function is greater than zero, increases with the value of the mass number $A$ of the nucleus. The graphs also show that the maximum values of   the LF nuclear spectral function for $^{9}$Be are in the range  $ 0.8 < \alpha < 1.2$, whereas for $^{197}$Au  are in the range $ 1.0 < \alpha < 3.0$, so that the range also increases with the mass number $A $ of the nucleus. The results are as expected, since   the range of possible values of the light-front  "+" momentum fraction for any nucleus with mass number $A$ is $0 < \alpha < A$.

 
\section{Numerical\;  estimates\;  for\;  SRC\;  momentum\;  distribution\; in\; virtual\; nucleon\\ approximation } 
\label{sec:NumMdVNA}

The numerical estimates for  2N and 3N SRC momentum distribution for proton and neutron  are calculated with the following equations: 
   \begin{align}
   n^{p/n}_{A,2N} (\mathrm  p_1) & =   \int^1_0 S^{p/n}_{A,2N} ( \mathrm p_1,  E_m)  dE_m \nonumber\\
  n^{p/n}_{A,3N} (\mathrm  p_1) & =   \int ^1_0S^{p/n}_{A,3N} ( \mathrm p_1,  E_m) dE_m
   \label {SRC-pnmd2}
  \end{align}
 where the nuclear spectral functions $S^{p/n}_{A,2N}  $ and $S^{p/n}_{A,3N}  $  are given by Eqs. (\ref {Snn_vn_2}) and  (\ref {S3n_vn_2})  for the VN approximation approach respectively, and by Eqs. (\ref {Snn_nr_1}) and  (\ref {S3n_nr_1}) for the VN approximation in the  nonrelativistic  limit respectively. 
 
 The SRC parameters for the calculation of the spectral function are given in Tables \ref  {table2NSRC} and \ref {table3NSRC},   where the auxiliary parameters $a_{\textrm {2p(n)-eff}} $,  $a_{\textrm {3p(n)}}$,  and $a_{\textrm {3p(n)-eff}}$ have been included to simplify the notations. The equations used to calculate the auxiliary parameters are 
\begin {align}
a_{\textrm {2p(n)-eff}} & = \frac { a_{2} (A,Z)}{{(2X_{p(n)})} ^{\gamma}} \nonumber\\
a_{\textrm {3p}} & = a^2_{2}(A,Z)f_p \nonumber\\
a_{\textrm {3n}}& = a_{\textrm {3p}}f_n\nonumber\\
a_{\textrm {3p-eff}} & = a_{3p}C^p(A,Z)\nonumber\\
a_{\textrm {3n-eff}} & = a_{3n}C^n(A,Z)
\label{a2-a3eff}
\end {align}
where p(n) stands for proton and neutron respectively,  $a_2(A,Z)$   is defined in Eq. (\ref {2NSRC_relmd}), $x_{p/n}$ are given by  Eq. (\ref {2N-fraction}), and the $C^{p/n} (A,N)$  factors  by Eq. (\ref {3N-C}). The proportionality factors  f$_{p/n}$ were chosen to verify the prediction  that the overall probability of finding 3N SRC in the nucleus   is  proportional to the factor
${a^2_2(A,Z)}$ (Eq. (\ref {3NSRC_relmd}).  The expected range of values for the proportionality  factors was defined as  $.85 \le f_{p/n}\le1.15$. 
 
  \begin{table}[H]
\caption{2N SRC parameters for  light and heavy Nuclei ($ \mathrm {k_{F}}$  in GeV)   } 
\centering
\renewcommand {\arraystretch}{1.00}
\begin{tabular}{|c || c|  c | c | c| c| c |c| }
\hline
Nucleus & $a_{2}$ & $a_{\mathrm { 2peff}}$ & $a_{\mathrm { 2neff}}$& $a_{\mathrm { 2exp}}$ \cite{Fomin2011}  & $\alpha_{cm}$ & $ \gamma$ & $ \mathrm {k_{F}}$ \\ \hline
$^{3}$He & 1.65 & 1.31 &2.28 &2.13 & 3.70&0.80&0.15 \\
$^{4}$He  & 3.10 &3.10 &3.10 &3.60& 2.40&1.00&0.16 \\
$^{6}$He & 2.35 & 3.25 &1.87& & 2.20&0.80&0.18 \\ 
$^{8}$He & 2.10 & 3.66 &1.52 & & 2.20&0.80&0.20 \\
$^{6}$Li & 2.75 & 2.75 &2.75&& 2.20&1.00&0.20 \\
$^{7}$Li & 2.72 & 3.08 &2.44 && 2.10&0.80&0.20\\
$^{8}$Li & 2.60 & 3.27&2.17 && 2.00&0.80&0.20 \\
$^{9}$Li& 2.65 & 3.67&2.11 & & 1.90&0.80&0.20 \\
$^{8}$Be& 3.40 & 3.40 &3.40&&1.50&1.00&0.20 \\
$^{9}$Be & 3.10 & 3.41 &2.85&3.91& 1.20&0.80&0.20 \\
$^{10}$Be & 2.90& 3.47&2.50 & & 1.10&0.80&0.22 \\
$^{10}$B & 3.00 & 3.00 &3.00 && 1.10&1.00&0.22 \\
$^{11}$B & 3.20 & 3.45 &2.98&& 1.10&0.80&0.22\\
 $^{12}$C&4.25 & 4.25 &4.25 &4.75& 1.00&1.00 &0.221\\
 $^{16}$O & 4.20 & 4.20 &4.20 & & 1.20&1.00&0.220 \\
 $^{27}$Al & 4.50 & 4.64 &4.37 && 1.00&0.80&0.235 \\
 $^{40}$Ca & 4.40 & 4.40 &4.40 && 1.00&1.00&0.251 \\
$^{56}$Fe & 4.95 &5.25 &4.68& & 1.10&0.80&0.256\\ 
$^{64}$Cu & 5.02 & 5.43 &4.67 &5.21& 1.00&0.80&0.260\\
$^{197}$Au & 4.56 & 5.44 &3.95 &5.16& 1.00&0.80&0.265\\
$^{208}$Pb & 4.80 & 5.80 &4.12 & & 1.00&0.80&0.265 \\ \hline
\end{tabular}
\label{table2NSRC}
\end{table}
The criteria used to estimate  the values of $a_2$ and $a_3$  for light nuclei  (A$\le$  12)  was  that, in  the momentum range 0.45 $<$ p $<$ 1.00 GeV/c,  the allowed maximum deviation of the  momentum distribution numerical estimates, obtained with Eq. \ref {SRC-pnmd2},   from  the quantum Monte Carlo  (QMC) results given in \cite {VMCpc}   was of $\pm$ 15 $\%$. For nuclei with experimental values, $a_{2ep}$  for $a_2$, the values $a_2$ were chosen   to be, up to 30 $\%$, equal to the experimental  values  given in references \cite {Fomin2011, Kim2}. For heavier nuclei with A$ >$12, the values of $a_2$ were chosen equal to the available experimental values in \cite {Fomin2011, Kim2}.
\begin{table}[H]
\caption{3N SRC parameters for  light and heavy nuclei  } 
\centering
\renewcommand {\arraystretch}{1.00}
\begin{tabular}{|c || c | c  |c | c |c| c |c |}
\hline
Nucleus & $f_p$  &$a_{\textrm {3p}}$  & $f_n$&  $a_{\textrm {3n}}$  &   $a_{\textrm {3p-eff}}$ & $a_{\textrm {3n-eff}}$  \\ \hline
 $^{3}$He &3.00&8.17&1.00& 8.17 & 5.45 &0.0 \\
 $^{4}$He & 2.00&19.22 &1.00&19.22 & 9.61 &9.61   \\
$^{6}$He &5.70&31.48&0.20 &6.30 & 10.49&6.30 \\ 
$^{8}$He &12.00 &52.92 &0.08&4.23 & 13.23 &5.29  \\
$^{6}$Li &1.50 &11.34 &1.00&11.34 & 7.56 &7.56\\
$^{7}$Li &2.25 &16.65 &1.00&16.65& 9.51 &14.27\\
$^{8}$Li &3.16 &21.36&0.75 &16.02 & 10.68 &16.02  \\
$^{9}$Li&4.30&30.20 &0.35&10.57 & 13.42 &11.74\\
$^{8}$Be&1.15&13.29&1.00&13.29& 9.97 &9.97 \\
$^{9}$Be &1.60 &15.38 &0.70&10.76 & 10.25&9.57\\
$^{10}$Be &2.20&18.50 &0.70&12.95&11.10 &12.95  \\
$^{10}$B &1.25&11.25&1.00 &11.25 & 9.00 &9.00\\
$^{11}$B &3.50&35.84 &0.85&30.46 & 26.07 &27.69 \\
$^{12}$C&1.20 &21.68&1.00&21.68 & 18.06 &18.06\\
 $^{16}$O&1.00&17.64&1.00& 17.64 & 15.43 &15.43 \\
$^{27}$Al &1.00&20.25&1.00 &20.25 & 18.00&19.50\\ 
$^{40}$Ca &1.00 &19.36&1.00&19.36& 18.39 &18.39  \\
$^{56}$Fe &1.00 &24.50 &1.00&24.50 & 21.88 &25.38\\
$^{64}$Cu&1.00&25.20&1.00 &25.20 & 22.05 &26.77 \\
$^{197}$Au&1.00&20.79 &1.00&20.79 & 16.47&24.70\\
$^{208}$Pb &1.00&23.04 &1.00&23.04 & 17.94&27.69\\ \hline
\end{tabular}
\label{table3NSRC}
\end{table}
The values of the proportionality factors  f$_{n}$ in the Table \ref {table3NSRC} are within the expected range $.85 \le f_{n}\le1.15$. However, the values of the proportionality factors  f$_{p}$  are outside the expected range for some light nuclei. 

\begin{figure}[H]
\centering\includegraphics[scale=0.50]{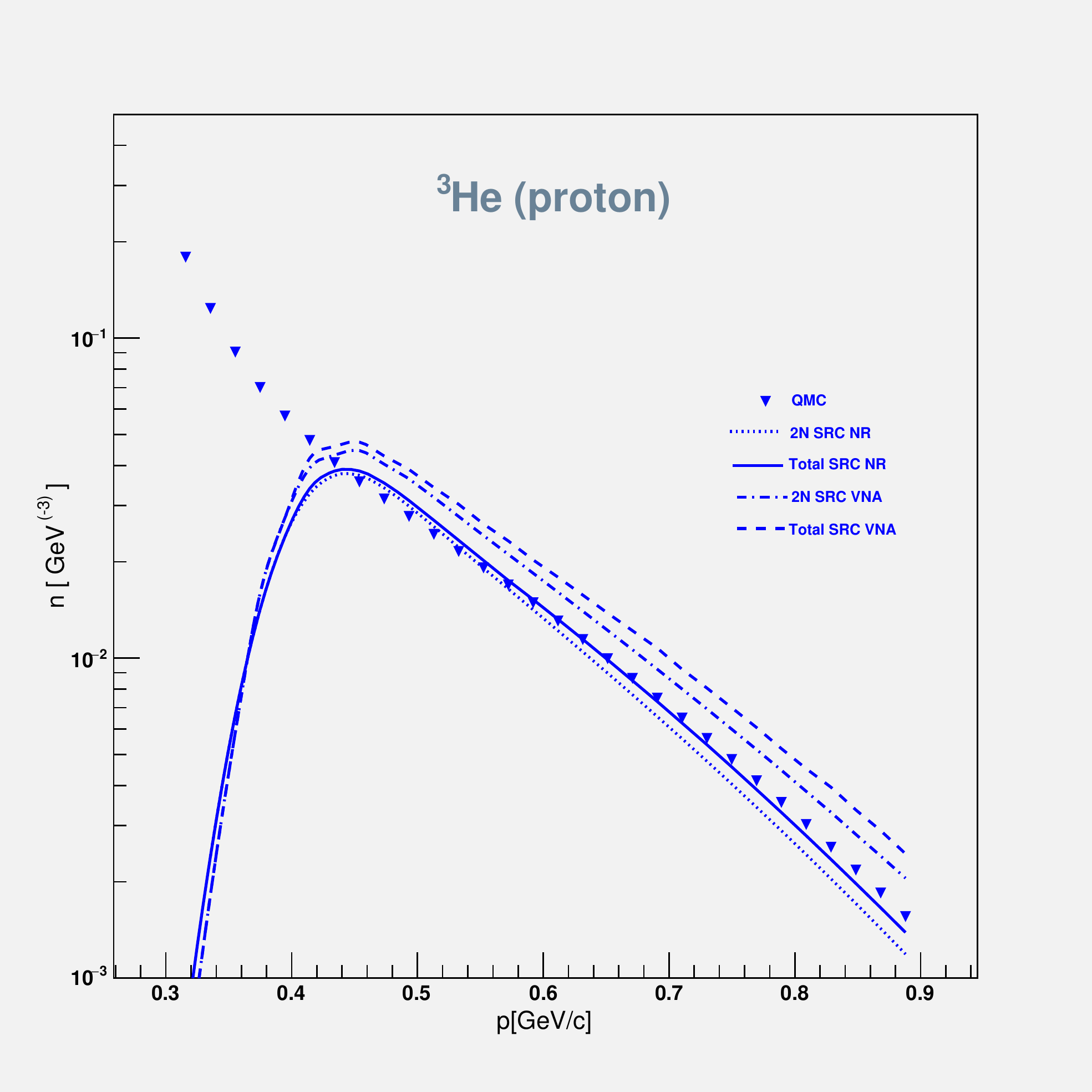}
\caption{SRC  momentum distribution  for proton in $^{3}$He, in VNA and  in the nonrelativistic limit of the VNA, the predictions in the nonrelativistic limit of the VNA are compared to QMC calculations}
\label{VNA_mdco_He3}
\end{figure}

\begin{figure}[H]
\centering\includegraphics[scale=0.50]{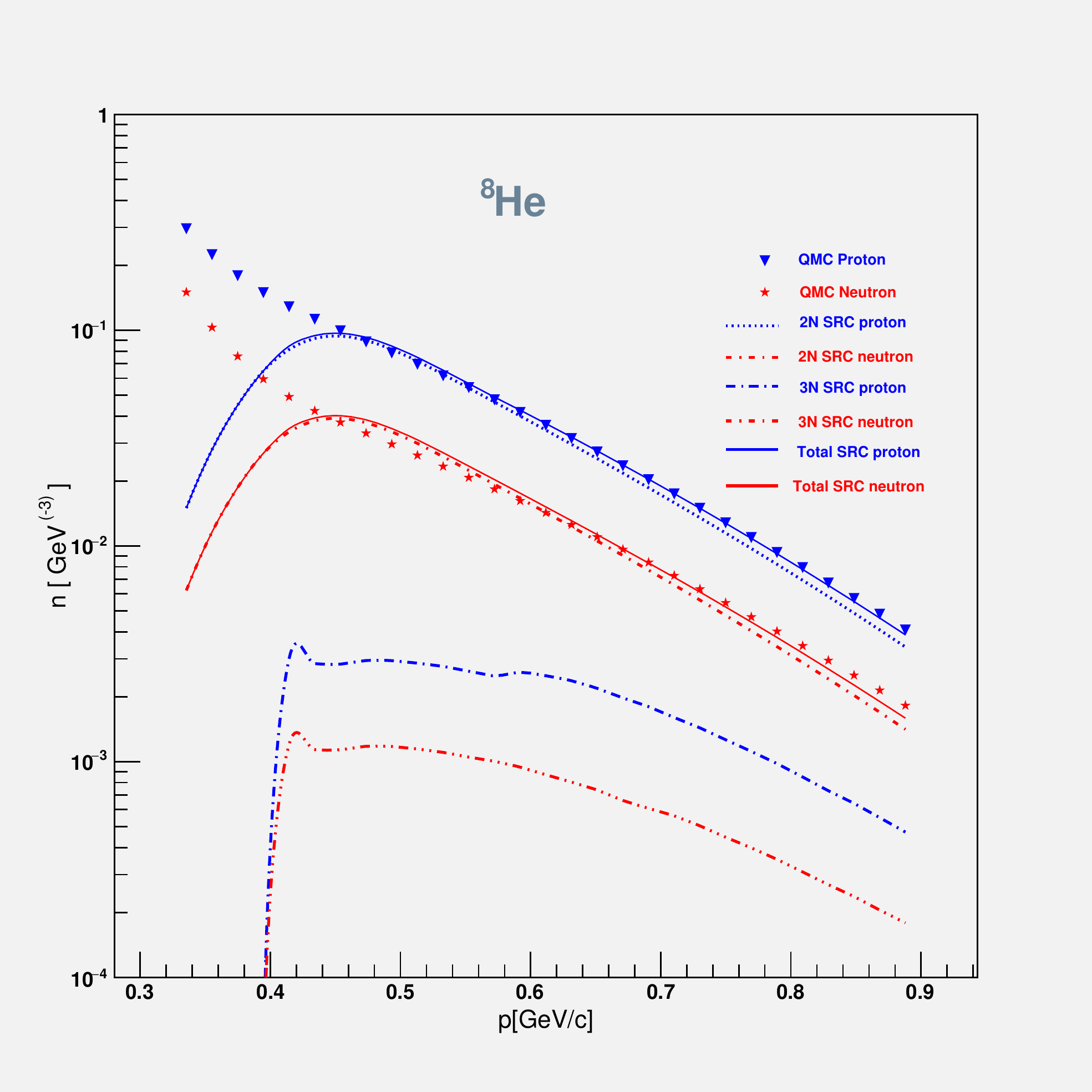}
\caption{SRC  momentum distribution  for $^{8}$He in the nonrelativistic limit of the VNA compared to QMC calculations}
\label{NR_mdco_He8}
\end{figure}

\begin{figure}[H]
\centering\includegraphics[scale=0.50]{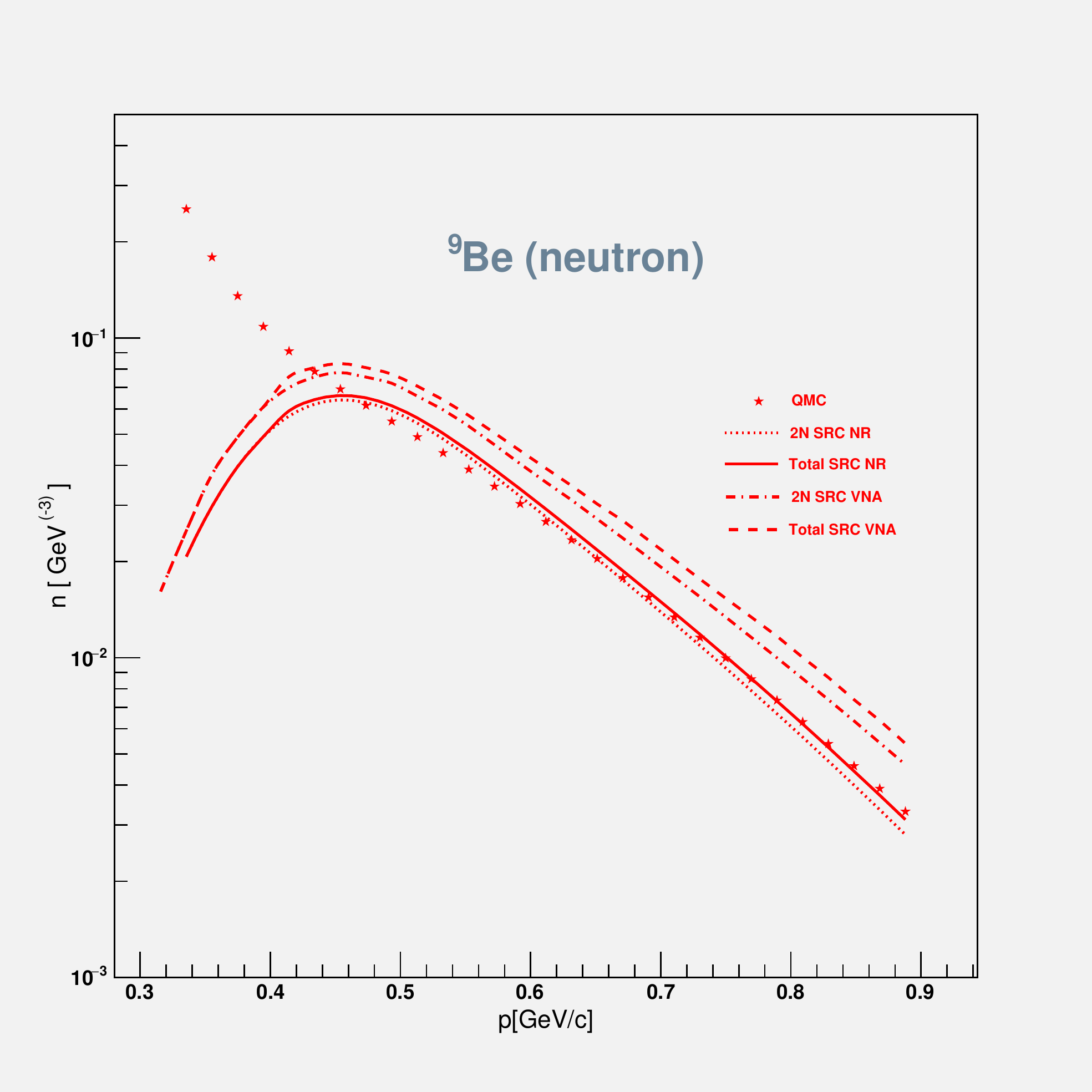}
\caption{SRC  momentum distribution  for neutron in  $^{9}$Be, in VNA and in the nonrelativistic limit of the VNA, the predictions in the nonrelativistic limit of the VNA are compared to QMC calculations}
\label{VNA_mdco_Be9}
 \end{figure}

 The  predictions for the SRC momentum distribution for proton and neutron  for light nuclei  (A $\leq $ 12), calculated with  Eqs. (\ref {SRC-pnmd2})  and with the SRC parameters given in Tables \ref  {table2NSRC} and \ref {table3NSRC},   are compared   with  the momentum distributions from \textit {ab initio} quantum Monte Carlo calculations (QMC) \cite {VMCpc}, the comparison is shown in 
  Figs. \ref {VNA_mdco_He3} , \ref {NR_mdco_He8}, and \ref {VNA_mdco_Be9}  for $^{3}$He, $^{8}$He, and $^{9}$Be respectively. 
   A good agreement, up to $\pm$ 15 $\%$,  is found between  the predictions and the QMC calculations for momenta in the range 450-1000 MeV/c, both for proton and neutron. The results show  that the SRC high momentum nuclear  spectral function models developed in the present  dissertation,  describe reasonably well the high momentum spectral function and the high momentum distributions for light  nuclei in the nonrelativistic limit of the VN approximation. 
 The values of  momentum distribution for proton in $^{8}$He are higher that for neutron in Fig.\ref {NR_mdco_He8}. This result is expected since the relative fraction factor for proton,
 $x_{p}$, is smaller than the same factor for neutron, $x_{n}$, in a neutron-rich nucleus as $^{8}$He. Thus, since there are more neutrons than protons,  the probability for a proton to be in SRC to a neutron is much higher than the  probability for a neutron to be in SRC to a proton. 
   
\begin{figure}[H]
\centering\includegraphics[scale=0.50]{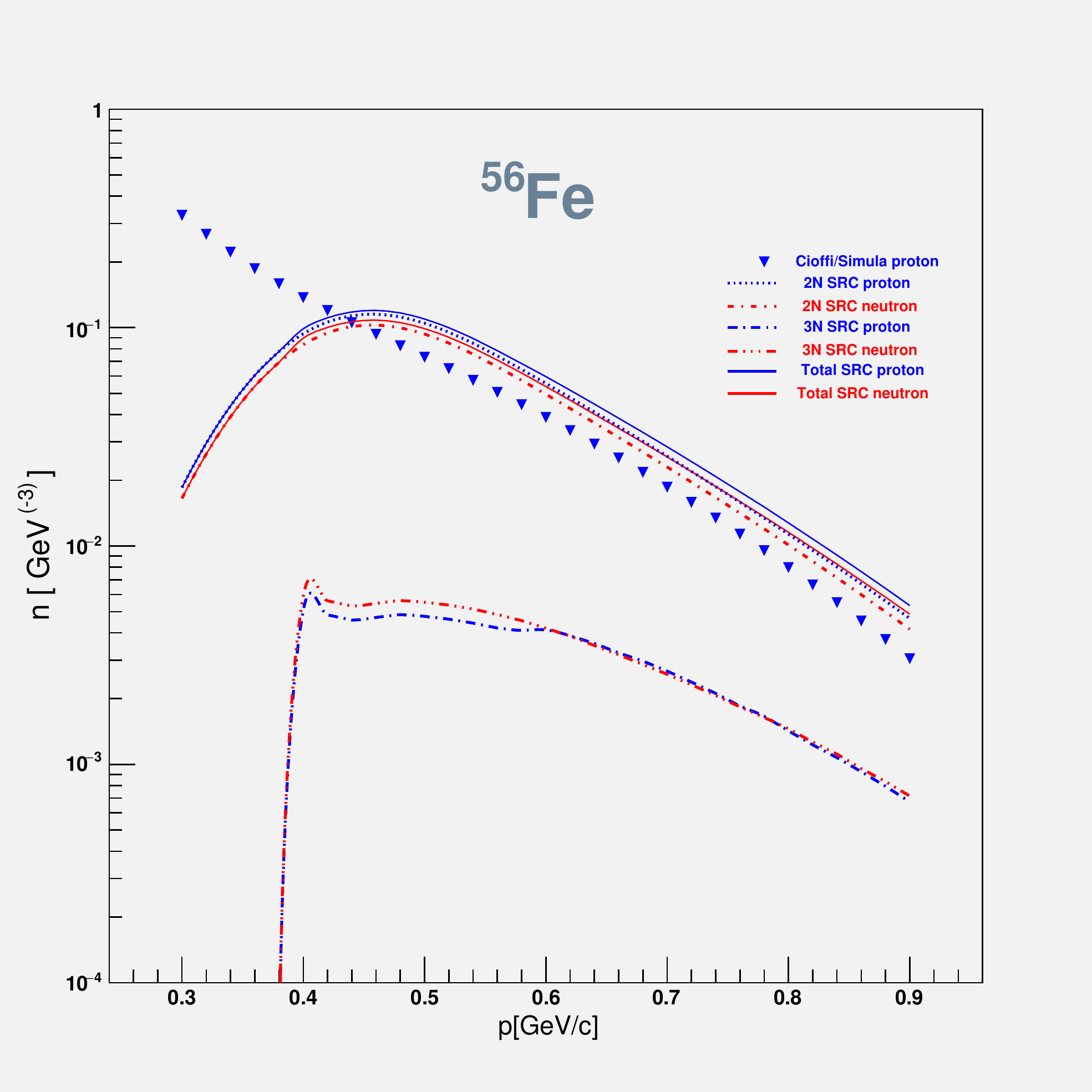}
\caption{SRC  momentum distribution  for $^{56}$Fe in the nonrelativistic limit of the  VNA compared to 2N SRC momentum distribution for proton in Ciofi/Simula  \cite {CiofiSimula}}
\label{NR_mdco_Fe56}
\end{figure}

The nonrelativistic   predictions of momentum distribution, calculated with  Eqs. (\ref {SRC-pnmd2})  and with the SRC parameters given in the Tables \ref  {table2NSRC} and \ref {table3NSRC},  for  $^{56}$Fe is depicted in the Fig. \ref  {NR_mdco_Fe56}. The  predictions   are compared to the 2N SRC momentum distribution Ciofi/Simula results  for proton given in \cite {CiofiSimula}. The predictions have the same slope and greater values  as compared to the Ciofi/Simula results for proton. The result  is explained by considering that the predictions include the 3N SRC contributions as well as the corrections represented by the proton and neutron relative fraction factor $x_{p/n}$  (\ref  {2N-fraction}),  contributions and  corrections that are not present in the  Ciofi/Simula results. 

The  predictions for the SRC  momentum distribution for proton and neutron in the VN approximation are compared to the  momentum distribution in the nonrelativistic limit of the VNA for  $^{208}$Pb.    The results are  calculated with  the  Eqs. (\ref {SRC-pnmd2})  and with the SRC parameters given in the Tables \ref  {table2NSRC} and \ref {table3NSRC}.   The VNA momentum distributions are as expected between 30 $\%$  and 80 $\%$  greater than the momentum distributions for the non relativistic case, the difference being greater for higher momentum. Similar results are  obtained for the   momentum distribution  for proton in $^{3}$He (Fig. \ref {VNA_mdco_He3}), and for neutron in $^{9}$Be (Fig. \ref  {VNA_mdco_Be9}). 

\begin{figure}[H]
\centering\includegraphics[scale=0.42]{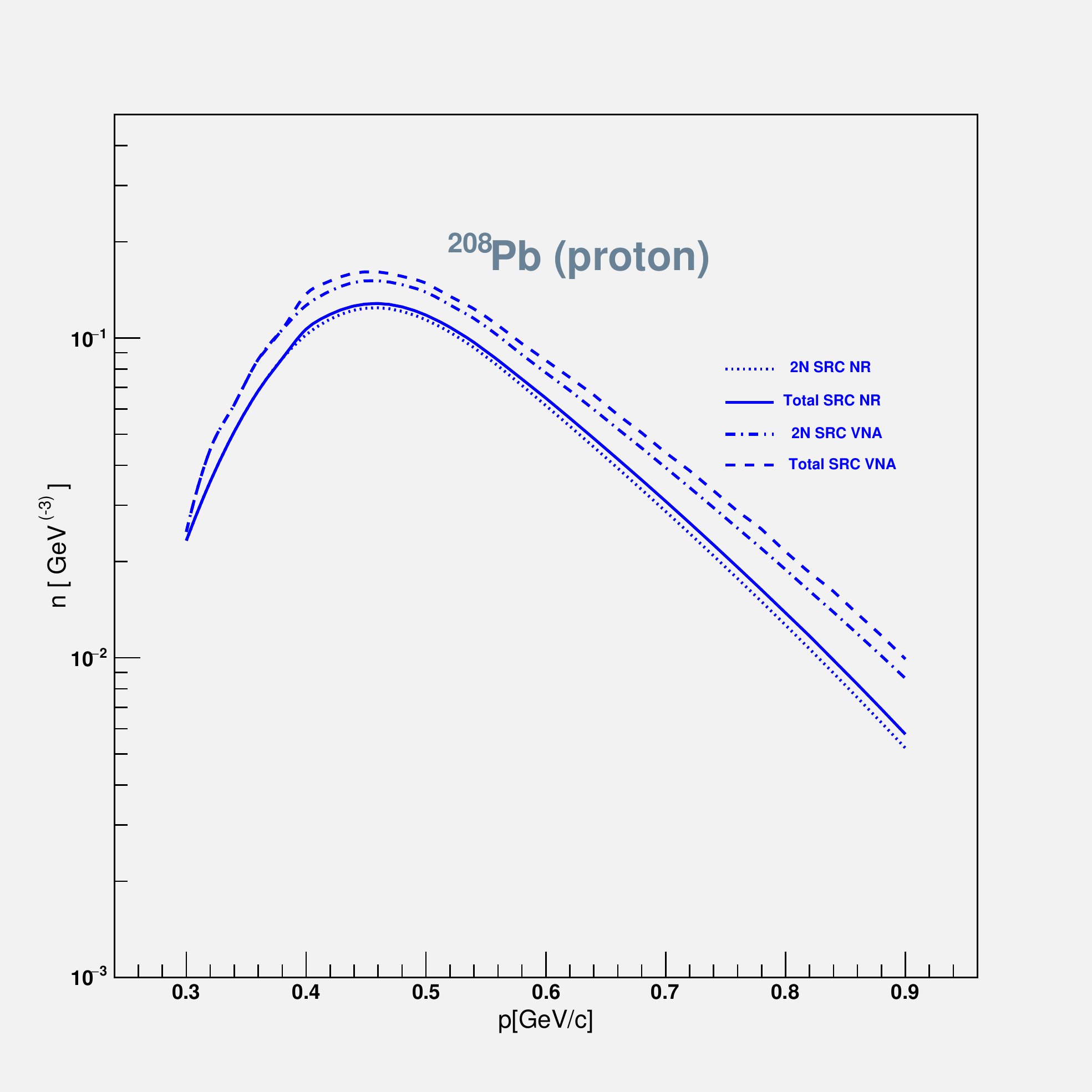}
\caption{SRC  momentum distribution  for $^{208}$Pb in VN approximation compared to predictions in the nonrelativistic limit of the VNA }
\label{NR_mdco_Pb208}
\end{figure}

 
 
 
 
 
 
\section{Parametrization\; of\;  SRC\; and\; mean\; field\; momentum\; distribution } 
\label{sec:ParMd}
 The parametrization results for the mean field and the SRC momentum distributions are given in the present section. 
The parametrization of the nucleon    momentum distribution for light (A$\le$12) and heavy nuclei (16 $\le$A$\le$208) are obtained  for  the mean field momenta in the range \; (0 $ \le$ p $ \le$ 1.5 fm$^{-1}$)  and for  the SRC momenta in the range  (1.5 $ \le$ p $ \le$ 4.5 fm$^{-1}$). 

 The  nonrelativistic mean field momentum distribution n$_{MF}$ for each nucleus is calculated with the following expressions 
\begin {align}
\mathrm {n_{MF}(p})  & = \mathrm {\frac {1-4\pi\int^1_0  p^2 [n^{p/n}_{A,2N}(p) +n^{p/n}_{A,3N}(p) ]d p}  {4\pi\int^1_0 k^2 n_0(p) } n_0(p) } 
\label {mf_md}
\end {align}  
so that the total (MF plus SRC) momentum distribution, $ \mathrm {  n_{total}}$,  is normalized to unity, namely
\begin {align}
 4\pi\int \mathrm { p^2 n_{total}(p) d p  = 4\pi\int ^1_0p^2 [n_{MF}(p) + n^{p/n}_{A,2N}(p) +n^{p/n}_{A,3N}(p)  ] d p} & =  1.0 
 \label {md_norm}
\end {align} 

The  SRC momentum distributions $n^{p/n}_{A,2N}$  and $n^{p/n}_{A,3N}$  are calculated  by Eq.(\ref {SRC-pnmd2}), with the  SRC nuclear spectral functions in the VN approximation and VN approximation in the nonrelativistic limit. 
The mean field  momentum distributions n$_0$ for light nuclei  are obtained from the QMC results given in Ref. \cite {VMCpc}; for $^{16}$O and  $^{40}$Ca from reference  \cite {CiofiSimula}; and  for $^{27}$Al, $^{56}$Fe, and $^{208}$Pb from the mean field wave function  given for Zverev\cite {MSprivate2015}

The equations for the parametrization of the mean field and the SRC momentum distributions  are those given by Ciofi and Simula  in Ref. \cite {CiofiSimula}.
The parametrization of the mean-field momentum distribution, in the range 0.0 $\le$ p $\le$ 1.50 fm $^{-1}$ for light nuclei (A $\le$ 12),   are obtained with  the following equations 
\begin {align}
n_{MF}(\mathrm {p})  = \frac {A^{(0)}_{1}  \frac {e^{-B^{(0)}_{1 } \mathrm {p}^2}}{(1+C^{(0)}_1 \mathrm {p}^2)^2} + A^{(0)}_{2}  \frac {e^{-B^{(0)}_{2} \mathrm {p}^2}}{(1+C^{(0)}_2 \mathrm {p}^2)^2}  } {4\pi},
\label {E:n0a}
\end {align}  
and  for heavy nuclei (16 $\le$A$\le$208) with
\begin {align}
n_{MF}(\mathrm {p})  = \frac {A^{(0)}  e^{-B^{(0)}\mathrm {p}^2} (1+C^{(0)} \mathrm {p}^2 + D^{(0)}\mathrm {p}^4 +E^{(0)} \mathrm {p}^6 + F^{(0)} \mathrm {p}^8)}  {4\pi},
\label {E:n0b}
\end {align}    
where  ${4\pi} \int \mathrm {dp{p}^2} n_{MF}(\mathrm  p)=1 $. 
For the parametrization of the SRC momentum distributions, in the range 1.5 $\le$ p $\le$ 4.50 fm $^{-1}$, the equations are  
\begin {align}
n_{SRC}(\mathrm {p})  =  \frac {A^{(1)}  e^{-B^{(1)} \mathrm {p}^2}+ C^{(1)}  e^{-D^{(1)} \mathrm {p}^2}+E^{(1)} e^{-F^{(1)}\mathrm {p}^2}  }{4\pi}
\label {E:n1}
\end {align}  
where  ${4\pi} \int \mathrm {dp{p}^2} n_{SRC}(\mathrm  p) =1 $.  
 
 \subsection{Parametrization\; of\;  the \; SRC\; and\; the\; mean\; field\; momentum \; distribution\; for the \; VN \;approximation\; in \; the\; nonrelativistic\; limit  } 

The parametrization of the momentum distribution for nuclei with $3\le$A$\le$208 calculated with the nuclear spectral functions in the  nonrelativistic limit of the VN approximation is obtained with Eqs. (\ref {E:n0a})  to ( \ref {E:n1})  for the  mean-field and the SRC momentum distributions respectively.  The corresponding parameters for the mean field and the SRC momentum distributions are  given in Tables \ref {tableMF_NR_light} to \ref {tableSRC_NR_heavy}, for light and heavy nuclei respectively.

 \begin{table}[H]
\caption {Parameters for the  non-relativistic mean-field  momentum distribution  for light nuclei A $\le$12   } 
\centering
\renewcommand {\arraystretch}{1.0}
\begin{tabular}{|c || c | c | c| c| c |c| }
\hline
Nucleus &  $A^{(0)}_1(\mathrm {fm}^{3})$ & $B^{(0)}_1(\mathrm {fm}^{2})$ & $C^{(0)}_1(\mathrm {fm}^{2})$ & $A^{(0)}_2(\mathrm {fm}^{3})$ & $B^{(0)}_2(\mathrm {fm}^{2})$ & $C^{(0)}_2(\mathrm {fm}^{2})$ \\ \hline
$^{3}$He (p) &  36.20 &1.07 &5.09 &0.36& 5.22&0.17\\ 
$^{3}$He (n) &  21.60 &0.49&5.13 &5.54& 4.80& \\ 
 $^{4}$He &  9.78 &1.04&1.38 &0.67& 5.20& \\ 
 $^{6}$He (p) &  3.06 &4.50&0.68 &8.60& 1.14&1.23 \\ 
 $^{6}$He(n) &  27.68 &1.19&3.29 &-20.91& 7.08&6.50 \\ 
 $^{8}$He(p)&  12.14 &0.99&1.72 &-1.80& 8.65& \\ 
 $^{8}$He(n)& 16.45 &1.78&1.21 &-13.58&7.79 &3.07\\ 
 $^{6}$Li& 14.04 &1.34&1.20 &-4.30&13.20 & \\ 
$^{7}$Li (p) & 11.11 &1.33 &1.13&-2.70& 8.58&\\  
$^{7}$Li (n) &  10.08 &1.60&0.71 &-3.59& 9.45& \\ 
 $^{8}$Li (p) & 8.17&1.72 &0.46 &-1.55& 8.37&\\  
$^{8}$Li (n) & 14.17 &1.23&1.37 &-10.48& 7.97& \\ 
 $^{9}$Li (p) &  7.88 &1.34 &0.70&-1.80& 7.65&\\  
$^{9}$Li (n) &  7.68 &2.09&0.094&-4.92& 10.82& \\ 
$^{8}$Be(p)& 7.98 &1.34&0.70 &-1.64& 7.50& \\ 
 $^{9}$Be(p)&  8.17 &1.44&0.63 &-1.92&6.57 & \\ 
 $^{9}$Be(n)& 7.99 &1.89&0.27&-3.21&8.64 & \\ 
$^{10}$Be(p)& 7.08 &2.32&-0.026 &-2.02& 10.37& \\ 
 $^{10}$Be(n)&  8.36 &2.01&0.20 &-5.41&8.89 & \\ 
 $^{12}$C& 6.70 &1.26&0.45&-3.41&4.90 & \\  \hline
 \end{tabular}
\label{tableMF_NR_light}
\end{table}

\begin{table}[H]
\caption{ Parameters for the non-relativistic mean-field momentum distribution  for heavy nuclei  16 $\le$ A $\le$ 208 } 
\centering
\renewcommand {\arraystretch}{1.0}
\begin{tabular}{|c || c|  c | c |c |c| c| }
\hline
Nucleus &  $A^{(0)}(\mathrm {fm}^{3})$ & $B^{(0)}(\mathrm {fm}^{2})$ & $C^{(0)}(\mathrm {fm}^{2})$ & $D^{(0)}(\mathrm {fm}^{4})$ & $E^{(0)}(\mathrm {fm}^{6})$ & $F^{(0)}(\mathrm {fm}^{8})$ \\ \hline 
$^{16}$O(p)&  2.93&3.01&6.60&& &\\ 
 $^{27}$Al (p)&  1.83&3.10&7.00&3.51& &\\ 
 $^{27}$Al (n)& 2.30 &3.24&3.75&6.18& &\\ 
 $^{40}$Ca(p)&  3.24&3.60&0.00&11.10& &\\ 
 $^{56}$Fe (p)&  3.03&3.84&-0.17&11.42& 3.63&\\ 
  $^{56}$Fe (n)& 2.65 &3.80&2.60&1.89&12.18 &\\ 
  $^{208}$Pb (p)&  2.59&5.55&1.08&49.22& -121.16&210.69\\ 
 $^{208}$Pb (n)& 1.89 &3.61&5.66&-9.36&24.30 &\\ \hline
 \end{tabular}
\label{tableMF_NR_heavy}
\end{table} 

\begin{figure}[H]
\centering\includegraphics[scale=0.70]{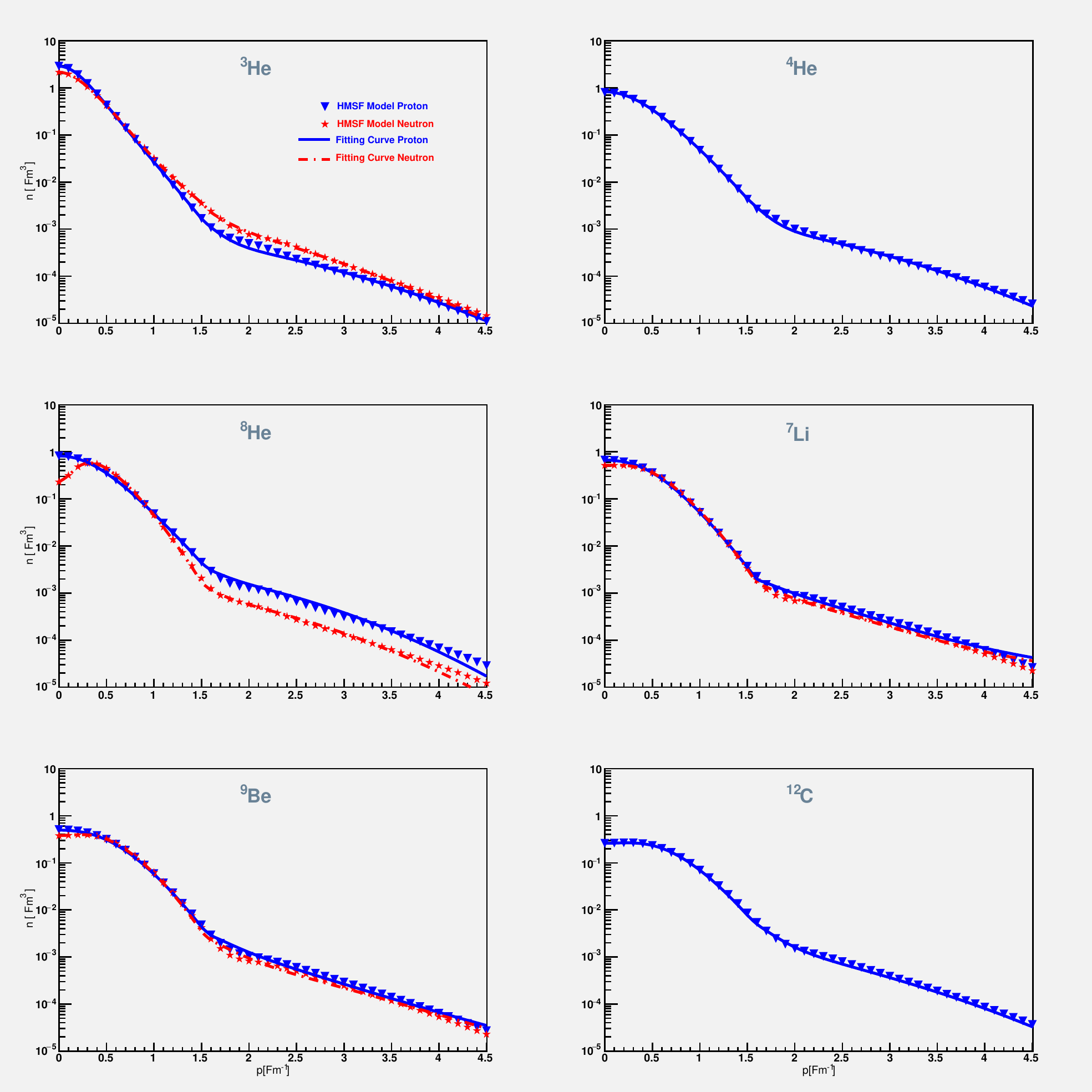}
\caption{ Momentum distribution  for light nuclei calculated with the SRC nuclear spectral functions in the nonrelativistic limit of   the  VNA, compared to the curve fitting obtained with the mean field parameters in Table \ref  {tableMF_NR_light} and SRC parameters in Table \ref {tableSRC_NR_light} }
\label{Fig_checkfit_NR_mdLigth}
\end{figure}

\begin{figure}[H]
\centering\includegraphics[scale=0.70]{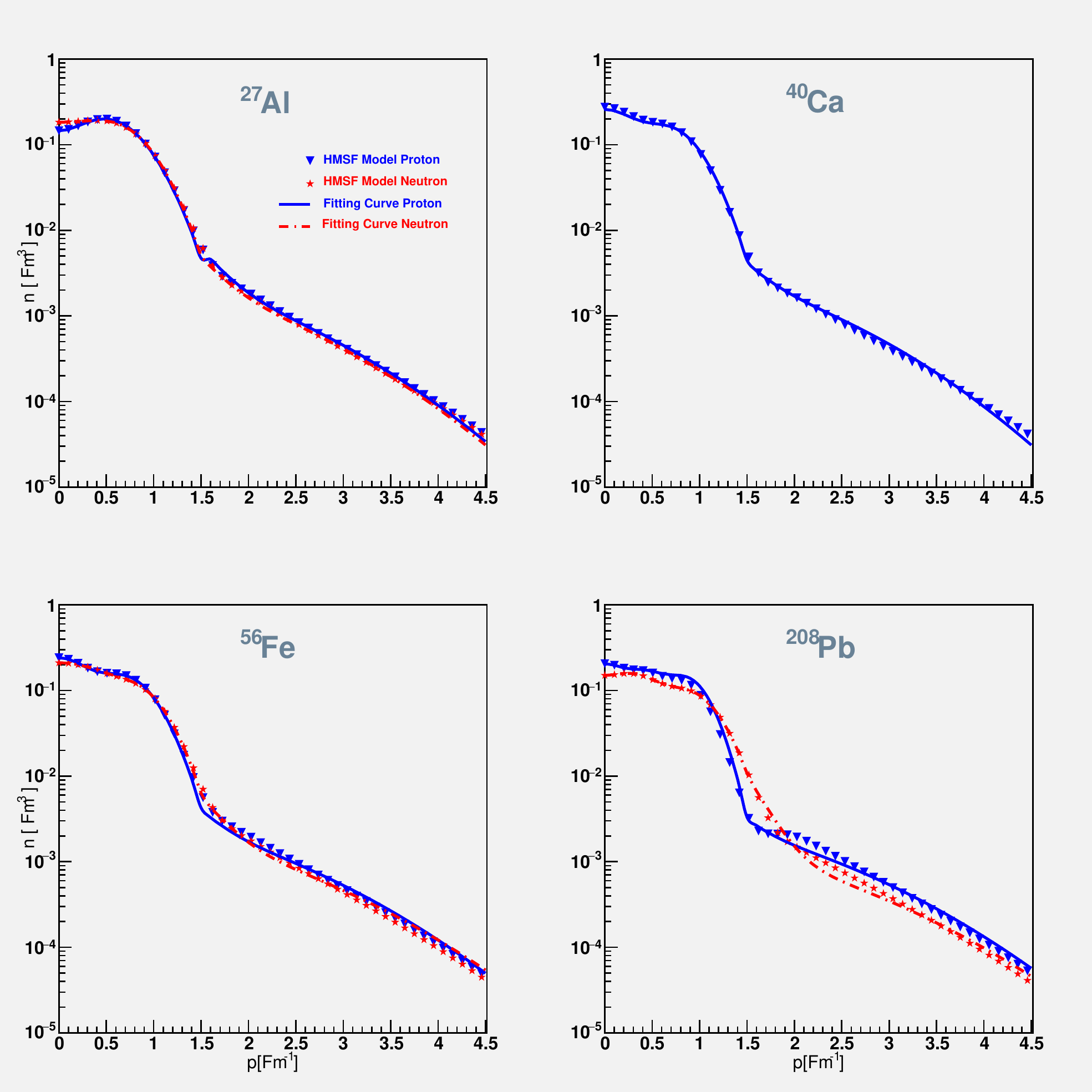}
\caption{Momentum distribution  for heavy nuclei calculated with the SRC nuclear spectral functions  in the nonrelativistic limit of   the  VNA, compared to the curve fitting obtained with the mean field parameters in Table \ref  {tableMF_NR_heavy} and SRC parameters in Table \ref {tableSRC_NR_heavy}}
\label{Fig_checkfit_NR_mdHeavy}
\end{figure}

The momentum distributions   calculated with the SRC nuclear spectral functions in the nonrelativistic limit of the  VN approximation,  compared to the curve fitting obtained with the parameters in Table \ref  {tableMF_NR_light} to  Table \ref {tableSRC_NR_heavy}, are  presented in  Figs. \ref {Fig_checkfit_NR_mdLigth} and  \ref {Fig_checkfit_NR_mdHeavy}, for some light and heavy nuclei respectively. The agreement between the curve fitted with the  parameters and those from numerical calculations is, in general, good for all the range of momentum considered. 
\label{sec:ParMdNRA}

\begin{table}[H]
\caption{Parameters for the  SRC  momentum distribution in the  nonrelativistic limit of the VNA for light nuclei A $\le$12   } 
\centering
\renewcommand {\arraystretch}{1.0}
\begin{tabular}{|c || c | c | c |c |c |c |}
\hline
Nucleus &   $A^{(1)}(\mathrm {fm}^{3})$ & $B^{(1)}(\mathrm {fm}^{2})$ & $C^{(1)}(\mathrm {fm}^{3})$ & $D^{(1)}(\mathrm {fm}^{2})$& $E^{(1)}(\mathrm {fm}^{3})$ & $F^{(1)}(\mathrm {fm}^{2})$   \\ \hline
$^{3}$He (p)&0.93 &1.87&0.009&0.22&0.0014& 0.18\\  
$^{3}$He (n) &0.94&1.64&0.026&0.34&0.006&0.19 \\  
$^{4}$He  &3.16&1.92&0.021&0.22&0.0023&0.19 \\   
$^{6}$He (p)  &0.35&0.95&0.015&0.19&& \\   
$^{6}$He (n)  &0.60&1.60&0.0114&0.20&& \\   
$^{8}$He (p)  &0.87&1.67&-0.040&0.34&0.091& 0.29\\   
$^{8}$He (n)  &0.50&1.83&0.02&0.27&& \\   
$^{6}$Li (p) & 3.00 &1.90&0.02&0.22&0.001& 0.30\\  
$^{7}$LI (p)& 0.20 &1.21&0.033&0.32&0.002&0.07 \\  
$^{7}$Li (n) & 0.26&1.32&0.024&0.30 &0.0013&0.058\\  
$^{8}$LI (p)& 0.24 &1.12&0.041&0.30&0.0013&0.072 \\  
$^{8}$Li (n) & 0.27&1.18&0.021&0.27&0.001&0.09 \\  
$^{9}$LI (p)& 0.26 &1.10&0.041&0.30&0.0017&0.07 \\  
$^{9}$Li (n) & 0.25&1.20&0.02&0.27&0.0012& 0.08\\  
$^{8}$Be  &0.36&1.07&0.015&0.22&0.0052&0.16 \\   
$^{9}$Be (p) & 0.31 &1.11&0.035&0.33&0.0045&0.12 \\ 
$^{9}$Be (n) & 0.29&1.12&0.019&0.30&0.0047&0.13\\  
$^{10}$Be (p) &0.31 &1.11&0.035&0.33&0.0045&0.12 \\  
$^{10}$Be (n) &0.28&1.13&0.018&0.31&0.0045&0.14\\  
$^{12}$ C  & 1.79&1.50&0.032&0.57&0.029&0.21\\  \hline
 \end{tabular}
\label{tableSRC_NR_light}
\end{table}

\begin{table}[H]
\caption{Parameters for  the SRC  momentum distribution in the  nonrelativistic limit of the VNA for  heavy  nuclei 16 $\le$ A $\le$ 208 } 
\centering
\renewcommand {\arraystretch}{1.00}
\begin{tabular}{|c || c | c | c| c | c| c  |}
\hline
Nucleus &  $A^{(1)}(\mathrm {fm}^{3})$ & $B^{(1)}(\mathrm {fm}^{2})$ & $C^{(1)}(\mathrm {fm}^{3})$ & $D^{(1)}(\mathrm {fm}^{2})$& $E^{(1)}(\mathrm {fm}^{3})$ & $F^{(1)}(\mathrm {fm}^{2})$   \\ \hline
$^{16}$O(p)  &0.48&1.33&0.057&0.25&-0.011&0.35 \\   
$^{27}$Al(p)  &0.80&1.25&0.045&0.23&& \\   
$^{27}$Al(n) &0.60 &1.25&0.041&0.23 &&\\ 
$^{40}$Ca(p)  &0.45&1.31&0.058&0.246&-0.011&0.35 \\   
$^{56}$Fe (p) &0.40&1.27&0.044&0.21&&\\  
$^{56}$Fe (n) &0.75 &1.20&0.032&0.19&& \\  
$^{208}$Pb (p) &0.36&1.45&0.041&0.20&&\\  
$^{208}$Pb (n)  & 3.12&1.48&0.0022&0.18&&\\  \hline
\end{tabular}
\label{tableSRC_NR_heavy}
\end{table}

 \subsection{The\; parametrization\; of\;  the \; SRC\; and\; the \; mean\; field\; momentum\; distribution\;in\; virtual\; nucleon \; approximation  } 
\label{sec:ParMdVNA}
 
The parametrization of the momentum distribution for nuclei with $3\le$A$\le$208 calculated with the nuclear spectral functions in the VN approximation is obtained with Eqs. (\ref {E:n0a}) to ( \ref {E:n1})  for the mean-field and the SRC momentum distributions respectively.  The corresponding parameters for the mean field and the SRC momentum distributions are  given in the Tables \ref {tableMF_VNA_light} to  \ref {tableSRC_VNA_heavy}, for light and heavy nuclei respectively.
 
 \begin{table}[H]
\caption{Parameters for  the mean-field  momentum distribution in VNA for light nuclei A $\le$12   } 
\centering
\renewcommand {\arraystretch}{1.00}
\begin{tabular}{|c || c | c | c| c| c |c| }
\hline
Nucleus &  $A^{(0)}_1(\mathrm {fm}^{3})$ & $B^{(0)}_1(\mathrm {fm}^{2})$ & $C^{(0)}_1(\mathrm {fm}^{2})$ & $A^{(0)}_2(\mathrm {fm}^{3})$ & $B^{(0)}_2(\mathrm {fm}^{2})$ & $C^{(0)}_2(\mathrm {fm}^{2})$ \\ \hline
$^{3}$He (p) &  36.20 &0.97 &5.09 &0.36& 6.22&0.17\\ 
$^{3}$He (n) &  21.60 &0.49&5.13 &5.54& 4.80& \\ 
 $^{4}$He &  9.78 &1.04&1.38 &0.67& 5.20& \\ 
 $^{6}$He (p) &  3.06 &4.50&0.68 &9.00& 1.14&1.23 \\ 
 $^{6}$He(n) &  27.68 &1.19&3.15 &-20.91& 7.08&6.50 \\ 
 $^{8}$He(p)&  12.14 &0.99&1.72 &-1.80& 8.65& \\ 
 $^{8}$He(n)& 16.45 &1.78&1.13 &-13.58&7.79 &3.07\\ 
 $^{6}$Li& 13.40&1.34&1.20 &-4.30&13.20 & \\ 
$^{7}$Li (p) & 11.11 &1.33 &1.13&-2.70& 8.58&\\  
$^{7}$Li (n) &  10.08 &1.60&0.71 &-3.59& 9.45& \\ 
 $^{8}$Li (p) & 8.17&1.62 &0.46 &-1.55& 8.37&\\  
$^{8}$Li (n) & 14.17 &1.23&1.37 &-10.48& 7.97& \\ 
 $^{9}$Li (p) &  7.88 &1.34 &0.70&-1.80& 7.65&\\  
$^{9}$Li (n) &  7.68 &2.09&0.094&-4.92& 10.82& \\ 
$^{8}$Be(p)& 7.98 &1.34&0.70 &-1.64& 7.50& \\ 
 $^{9}$Be(p)&  8.17 &1.44&0.63 &-1.92&6.57 & \\ 
 $^{9}$Be(n)& 7.99 &1.89&0.27&-3.21&8.64 & \\ 
$^{10}$Be(p)& 7.08 &2.32&-0.026 &-2.02& 10.37& \\ 
 $^{10}$Be(n)&  8.36 &2.01&0.20 &-5.41&8.89 & \\ 
 $^{12}$C& 6.70 &1.26&0.45&-3.41&4.90 & \\  \hline
 \end{tabular}
\label{tableMF_VNA_light}
\end{table}

\begin{table}[H]
\caption{  Parameters for  the mean-field  momentum distribution in VNA  for heavy nuclei  16 $\le$ A $\le$ 208 } 
\centering
\renewcommand {\arraystretch}{1.00}
\begin{tabular}{|c || c|  c | c |c |c| c| }
\hline
Nucleus &  $A^{(0)}(\mathrm {fm}^{3})$ & $B^{(0)}(\mathrm {fm}^{2})$ & $C^{(0)}(\mathrm {fm}^{2})$ & $D^{(0)}(\mathrm {fm}^{4})$ & $E^{(0)}(\mathrm {fm}^{6})$ & $F^{(0)}(\mathrm {fm}^{8})$ \\ \hline 
$^{16}$O(p)&  2.73&3.00&6.60&& &\\ 
 $^{27}$Al (p)&  1.65&3.10&7.00&6.00& &\\ 
 $^{27}$Al (n)& 2.30 &3.24&3.75&7.18& &\\ 
 $^{40}$Ca(p)&  3.27&3.55&0.07&10.41& &\\ 
 $^{56}$Fe (p)&  2.66&3.85&-0.20&11.46& 6.99&\\ 
  $^{56}$Fe (n)& 2.50 &3.81&2.55&2.05&12.20&\\ 
  $^{208}$Pb (p)&  2.15&5.20&1.16&44.33& -120.00&180.96\\ 
 $^{208}$Pb (n)& 1.76 &3.59&5.61&-9.47&24.12 &\\ \hline
 \end{tabular}
\label{tableMF_VNA_heavy}
\end{table}

\begin{table}[H]
\caption{Parameters for  the SRC  momentum distribution in VNA for light nuclei A $\le$12   } 
\centering
\renewcommand {\arraystretch}{1.00}
\begin{tabular}{|c || c | c | c |c |}
\hline
Nucleus &   $A^{(1)}(\mathrm {fm}^{3})$ & $B^{(1)}(\mathrm {fm}^{2})$ & $C^{(1)}(\mathrm {fm}^{3})$ & $D^{(1)}(\mathrm {fm}^{2})$   \\ \hline
$^{3}$He (p)&0.39&1.35&0.014&0.21 \\  
$^{3}$He (n) &0.10&0.59&0.015&0.19\\  
$^{4}$He  &0.20&0.83&0.027&0.20 \\   
$^{6}$He (p)  &0.24&0.95&0.030&0.20\\   
$^{6}$He (n)  &0.17&1.10&0.020&0.21\\   
$^{8}$He (p)  &0.24&0.85&0.030&0.19\\   
$^{8}$He (n)  &0.14&1.15&0.018&0.22\\   
$^{6}$Li (p) & 0.19 &0.83&0.027&0.20 \\  
$^{7}$LI (p)& 0.19&0.90&0.028&0.20\\  
$^{7}$Li (n) & 0.12&0.77&0.020&0.19\\  
$^{8}$LI (p)& 0.19 &0.90&0.028&0.19\\  
$^{8}$Li (n) & 0.14&0.73&0.020&0.19\\  
$^{9}$LI (p)& 0.21&0.77&0.028&0.19\\  
$^{9}$Li (n) & 0.13&0.76&0.017&0.18 \\  
$^{8}$Be  &0.21&0.80&0.030&0.20\\   
$^{9}$Be (p) & 0.21 &0.77&0.028&0.19\\ 
$^{9}$Be (n) & 0.14&0.75&0.019&0.175\\  
$^{10}$Be (p) &0.15&0.83&0.026&0.18\\  
$^{10}$Be (n) &0.13&0.78&0.018&0.17\\  
$^{12}$ C  & 0.33&0.73&0.025&0.16\\  \hline
 \end{tabular}
\label{tableSRC_VNA_light}
\end{table}
\begin{table}[H]
\caption{Parameters for the SRC  momentum distribution in VNA for heavy  nuclei 16 $\le$ A $\le$ 208     } 
\centering
\renewcommand {\arraystretch}{1.00}
\begin{tabular}{|c || c | c | c| c | c| c  |}
\hline
Nucleus &  $A^{(1)}(\mathrm {fm}^{3})$ & $B^{(1)}(\mathrm {fm}^{2})$ & $C^{(1)}(\mathrm {fm}^{3})$ & $D^{(1)}(\mathrm {fm}^{2})$& $E^{(1)}(\mathrm {fm}^{3})$ & $F^{(1)}(\mathrm {fm}^{2})$   \\ \hline
$^{16}$O(p)  &0.20&0.96&-0.028&0.22&0.069&0.21\\   
$^{27}$Al(p)  &0.52&1.11&0.047&0.20&-0.035& 0.89\\   
$^{27}$Al(n) &0.52 &1.11&0.043&0.20 &-0.020&0.73\\ 
$^{40}$Ca(p)  &0.21&0.90&-0.027&0.24&0.069&0.21\\   
$^{56}$Fe (p) &0.24&0.86&-0.025&0.23&0.069&0.20\\  
$^{56}$Fe (n) &0.42 &1.04&0.048&0.20&-0.05&1.00 \\  
$^{208}$Pb (p) &0.23&0.90&-0.025&0.25&0.071&0.20\\  
$^{208}$Pb (n)  & 1.46&1.18&0.026&0.16&&\\  \hline
\end{tabular}
\label{tableSRC_VNA_heavy}
\end{table}

\begin{figure}[H]
\centering\includegraphics[scale=0.75]{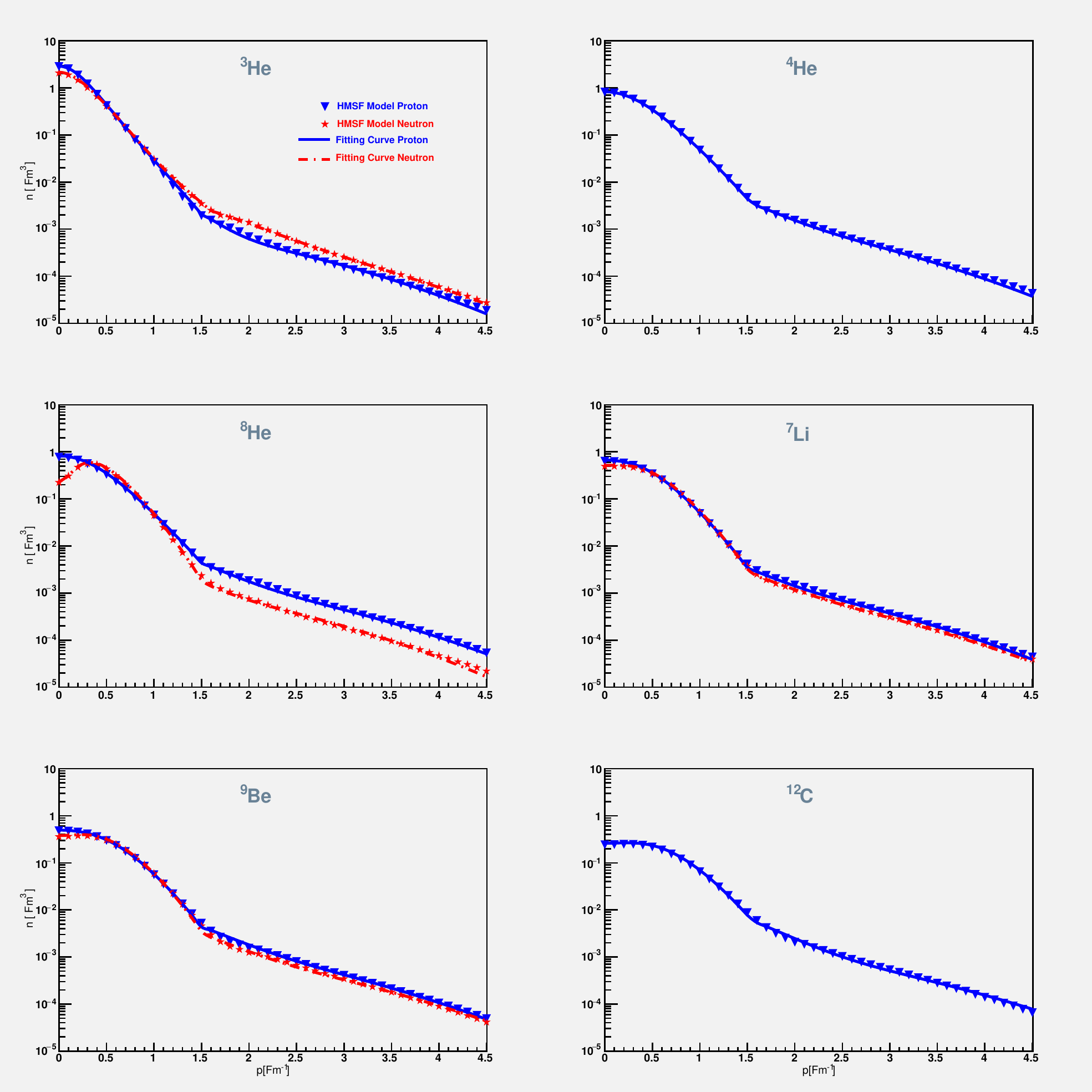}
\caption{Momentum distribution  for light nuclei calculated with the SRC nuclear spectral functions for VNA, compared to the curve fitting obtained with the mean field parameters in Table \ref  {tableMF_VNA_light} and the SRC parameters in Table \ref {tableSRC_VNA_light}}
\label{Fig_checkfit_VNA_mdLigth}
\end{figure}

\begin{figure}[H]
\centering\includegraphics[scale=0.65]{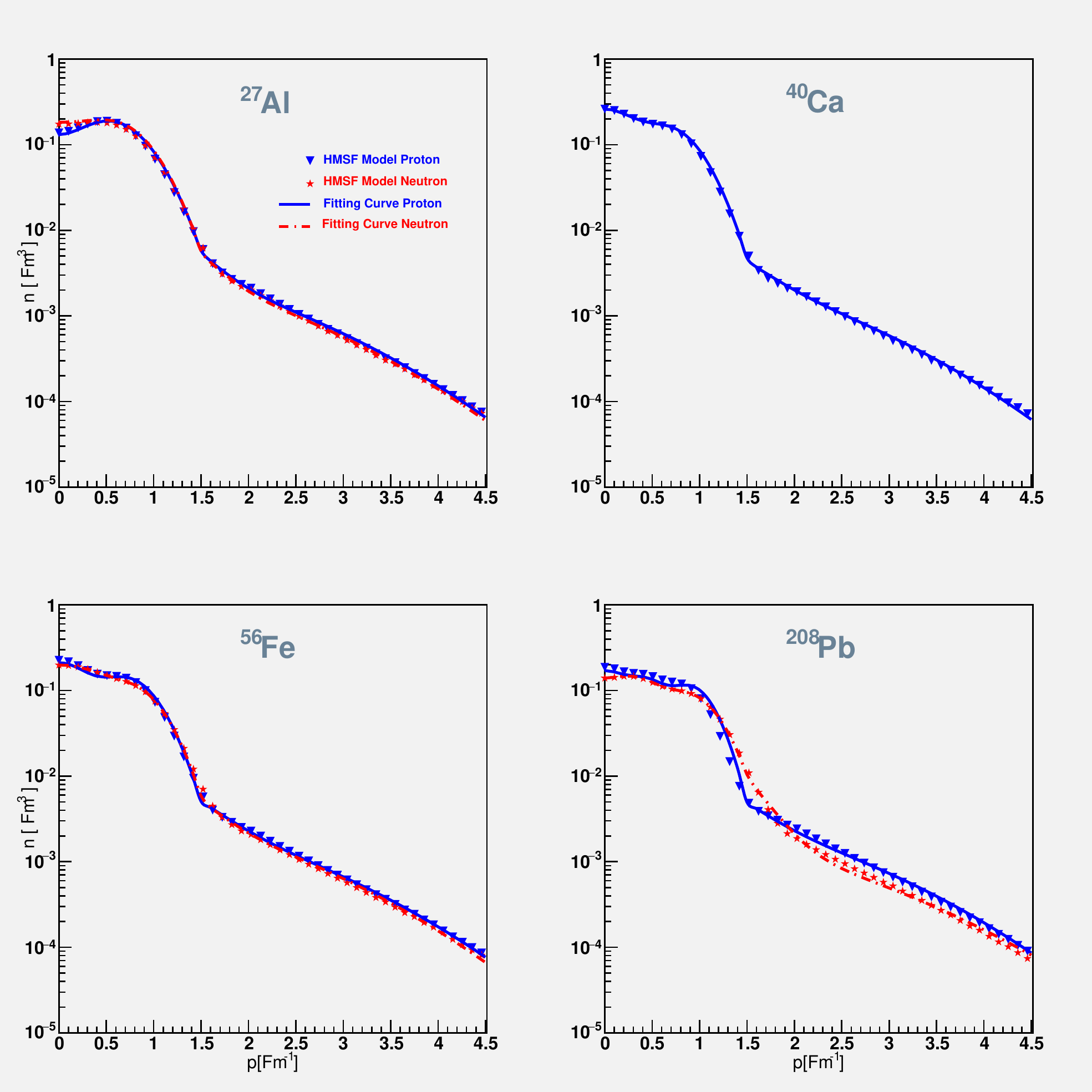}
\caption{Momentum distribution  for heavy nuclei calculated with the SRC nuclear spectral functions for VNA,  compared to the curve fitting obtained with the mean field parameters in Table \ref  {tableMF_VNA_heavy} and the SRC parameters in Table \ref {tableSRC_VNA_heavy}}
\label{Fig_checkfit_VNA_mdHeavy}
\end{figure}

 The momentum distribution   calculated with the SRC nuclear spectral functions in  the  VN approximation,  compared to the curve fitting obtained with the parameters in Table \ref  {tableMF_VNA_light} to  Table \ref {tableSRC_VNA_heavy}, is  presented in  Figs. \ref {Fig_checkfit_VNA_mdLigth} and  \ref {Fig_checkfit_VNA_mdHeavy}, for some light and heavy nuclei respectively. The agreement between the curve fitted with the  parameters and those from numerical calculations is, in general, good for all the range of momentum considered. 
 
\section{Summary}
\label{sec:Ch4Summ}

The numerical estimates of the 2N and 3N SRC nuclear spectral functions in the LF and VN approximations, as well as  the numerical estimates and the parametrization of momentum distribution of the bound nucleon,   both for VN and VN in the nonrelativistic limit approximations, are obtained  in the present  chapter. The numerical estimates for the LF SRC density matrices are also calculated.  The nuclear spectral functions and the LF SRC density matrices are calculated with the computational models developed in chapter 3. 
	
Before calculating the numerical estimates of the nuclear spectral functions, the range of validity and set of parameters used  for the  numerical estimates are briefly   discussed. 	
	
 The D and S-waves deuteron momentum distribution ($n_{d}$) are defined as the result from a  wave function modeled with  the Argone V18 (AV18) two-nucleon potential, that is specially suitable for the spectral function model since fits both $pp$ and $np$ data \cite {AV1895}.  The numerical estimates for the deuteron LF density matrix are also included in the present  chapter. 
 
The numerical estimates for  the 2N and 3N SRC momentum distribution for proton and neutron  are calculated by integration of the nuclear spectral functions obtained in the computational models of chapter 3. These momentum distributions are compared  for light nuclei (A $\le 12$)   with  the \textit{ab initio} quantum Monte Carlo calculations given in \cite {VMCpc}, and for some heavy nuclei (A $> 12$)  with  the calculations for proton included in\cite {CiofiSimula}. 
 
The mean-field and SRC momentum distributions in VNA and VN in the  nonrelativistic limit are parametrized  for a wide range of nuclei ($3\le$A$\le$208), resulting in a set of curve fitting parameters useful for further calculations.

\chapter{Conclusions}


For momenta below the Fermi momentum $\mathrm k_F \sim$ 250 MeV/$c$, the nuclear shell model, derived from the mean field picture of the nucleus,   has been successful in correctly  predicting all nuclear magic numbers, as well as in describing  a large amount of nuclear data. The main assumption of the nuclear mean field  model is that the nucleons are independent particles moving in an average or mean   field  generated by  the remaining $\;(A-1)\;$ nucleons in the nucleus. As a result each nucleon is independent of the exact instantaneous position of all other nucleons.  The nuclear shell model is valid for  nucleons with long range ($\ge$ 2 fm) mutual separations.

The nuclear shell model however is found to break down  for inter nucleon distances smaller  
than $2r_N$, where $r_N\approx 0.85$ fm is the radius of the nucleon, for which  two nucleons start to overlap and the notion of the mean field become invalid. Their dynamics  are mainly defined by the NN interaction at short distances  which is dominated by the tensor interaction~($\sim 0.8-1.2$ fm)  and repulsive core ($\le 0.5-0.7$ fm). Such  configurations  are generally referred to as 2N Short Range Correlations (SRCs).  The domain of multi-nucleon SRCs is characterized by relativistic momenta of the probed  nucleon. Another high energy nuclear phenomena for  which the nuclear shell model is  not valid, is the  modification of  quark distribution  of a bound nucleon in the nucleus, as compared to that of  a free nucleon, the so called EMC effect. 

In 2011, the strong correlation between the EMC effect and the 2N SRC in the nucleus was observed through  high energy experiments (well above $\mathrm k_F$). Since the SRC structure in  the nucleus implies high momentum bound nucleons, the experiments indicate  that the EMC effect is only due to the high momentum component of the nuclear wave function, so that the possible modification of the parton distributions in nucleons in the nucleus occur only in nucleons belonging to SRCs.  

A further understanding  of   the dynamics involved  in the    above described  areas of high energy nuclear physics require theoretical models to predict the high  momenta and binding  energies of bound nucleons in the nuclei  that  describe 2N and 3N SRC in a consistent way. Such models are described  as nuclear spectral functions that  define the joint probability   of finding a nucleon in the nucleus  with momentum $\mathrm{p}$  and removal (binding) energy $E_m$. 

The  main motivation of  the research presented in the present dissertation  was therefore  to develop a self consistent theoretical model for  the calculation of  nuclear spectral functions in the domain of  the 2N and 3N short range correlations. For momenta above 400 MeV/c, a relativistic  multi-nucleon short range correlation model of the  spectral function  was   obtained. This model describes   the high momentum and high missing energy of   two and three   nucleons in short range correlations (2N and 3N SRC), for  symmetric and asymmetric nuclei. 

The main assumption for the nuclear spectral function  models developed in the present  dissertation  is the dominance of NN SRC in the nuclear dynamics for internal  momenta \; \; $\mathrm {p\gtrsim k_{src} > k_F}$, where $\mathrm { k_{src} }$ is the relative momentum threshold at which a NN system with such relative momentum can be considered in the short-range correlation. The  next major assumption is the dominance of the isosinglet $pn$ component in the NN SRC. It was  also assumed that the single proton or neutron  momentum distributions in the 2N SRC domain are inversely  proportional to their relative fractions in  nuclei. The inverse proportionality of the high momentum component  to the relative fraction of  proton or neutron is  important for asymmetric nuclei and they need to be included in the modeling of nuclear spectral functions in the 2N  SRC region.  An additional benefit of considering the inverse   proportionality is that distinct nuclear spectral function  models for neutron and for proton were obtained in the present dissertation. 

Considering the dominance of NN SRC described above, a main development  in the present   dissertation was  that   the contribution of 3N SRCs to the nuclear spectral function was modeled assuming  that such correlations are generated by two sequential 2N short range correlations. Hence,  the  phenomenological knowledge of the properties of 2N SRCs was sufficient to  calculate both the 2N and 3N  SRC contributions to the nuclear spectral function. This assumption added to  the strong dominance of $pn$ SRCs in nucleus, made it possible to use the momentum distribution of the deuteron as a fundamental component of the developed nuclear spectral function models.

A main focus  in the developed models  was to  treat the relativistic effects which are important for the bound nucleon momenta   exceeding characteristic Fermi momentum, $k_{F}$, in the nucleus.  The relativistic effects in the present  dissertation  were treated based on the effective Feynman diagrammatic approach, in which  the  Lorentz covariant amplitudes are reduced  to the nuclear spectral functions, in a process of calculation designed  to trace the relativistic effects entering in such   functions.

One of the main ambiguities related to the treatment of relativistic effects considered in the present dissertation was to account for the vacuum fluctuations  ($Z$ graphs) which ultimately alter the  definition of the spectral function as a probability of finding a bound nucleon in the nucleus with the given momentum and  removal energy. Two suitable  approaches were  employed to  deal with the relativistic  vacuum fluctuations:  the virtual nucleon and the  light-front approximations. In the virtual nucleon approximation the vacuum fluctuations were neglected, while in the light-front approximation  the vacuum fluctuations are kinematically suppressed. 

The amplitudes for the covariant Feynman diagrams for bound nucleons in the mean field of the nucleus, and in 2N and 3N SRCs, were defined by applying  the effective Feynman diagrammatic rules. The covariant  amplitudes were consistently transformed into nuclear spectral functions    to represent  bound nucleons in the mean field of the nucleus, and in 2N and 3N SRCs.  The obtained mean field models are valid for  momenta below the Fermi momentum, $\mathrm k_F$, and were constructed  by using a nonrelativistic approach  to estimate the ground state wave functions. The models for 2N and 3N SRCs are valid  for momenta above 400 MeV. It is  therefore expected that the nuclear spectral function  models developed in the present dissertation  should be valid for the analysis of phenomena important to understand the internal dynamics and structure of a wide  range  of  atomic nuclei. 

The computational models for 2N and 3N SRC  high momentum  nuclear  spectral functions in VN and LF approximations, as well as in  the non-relativistic limit of the VN approximation,  were  obtained in the present  dissertation   by integration of the delta functions in the mathematical models described above.    

The computational models are  a set of computable integrals  that represent the high momentum nuclear spectral functions,  as well as equations for    momenta and  cosine of angles between momentum vectors,  and for auxiliary variables.  Restrictions in the values of momenta and  cosine of angles were also defined to ensure that the results are physically valid. The computational models are ready to  be    programed in   a computer  to obtain  numerical estimates of nuclear spectral functions, density matrices and momentum distributions.  

By applying the developed computational models, the  numerical estimates of 2N and 3N SRC nuclear spectral functions, as well as  the numerical estimates and the parametrization of momentum distributions of the bound nucleon,   both for VN and VN in the nonrelativistic limit approximations, were  obtained  in the present dissertation. 	

The  predictions for the SRC momentum distribution for proton and neutron  for light nuclei  (A $\leq $ 12), calculated with  the computational models   were  compared with the  momentum distributions from \textit {ab initio} quantum Monte Carlo calculations.
 A good agreement, up to $\pm$ 15 $\%$,  was  found between  the predictions and the QMC calculations for momenta in the range 450-1000 MeV/c, both for proton and neutron. The results showed that the SRC high momentum nuclear  spectral function models developed in the present dissertation,  describe reasonably well the high momentum spectral function and the high momentum distributions for the light nuclei in the nonrelativistic limit of the VN approximation.
 
The mean-field and SRC momentum distributions in VNA and VN in the  nonrelativistic limit were parametrized  for a wide range of nuclei ($3\le$A$\le$208), resulting in a set of  fitting parameters useful for further calculations. 
\section {Application of the multi-nucleon short range correlation model for the nuclear spectral functions }

Some of the experiments for which the multi-nucleon short range correlation model for the nuclear spectral functions  developed in the present dissertation may be applied for their analysis are the following.

\subsection { Experiments at Jefferson Lab Hall A}

\begin{itemize}
 \item { E08-14: $x > 2$ Short range correlations: Three-nucleon short range correlations studies in Inclusive scattering for $0.8 < Q^2 < 2.8 \mathrm {(Gev/c)}^2$   }
 \item { E07-006 : Short Range Correlation: Studying short-range correlations in nuclei at the repulsive core limit via the triple coincidence $(e,e'pN)$ reaction.  }
 \item { E01-015 : Short Range Correlation Experiment:  Studying  the internal small distance structure of nuclei via the   triple coincidence $(e,e'p+N)$ measurement.   }
\item { E97-111 : Systematic Probe of Short-Range Correlations via 4He(e,e'p)3He:  The experiment consists of making measurements of the cross section for the reaction 4He(e,e'p)3H.}
\item { E89-004 : Selected studies of the $^{3}$He and $^{4}$He Nuclei Through Electrodisintegration at High Momentum Transfer}
 \end{itemize}

\subsection { Experiments at the BooNE collaboration}

The experiments of interest at the BooNE (Booster Neutrino Experiment)  collaboration are the neutrino-nucleus scattering which has the goal of understanding neutrino oscillations. The experiments are described in reference \cite {Boone2011} .

\subsection { Experiments at the NUTEV  collaboration}

The experiments of interest at the NUTEV  collaboration are  the neutrino-nuclei scatterings on  which  a discrepancy have been found between measurements of $\sin^2 \theta_w$ ($ \theta_w$ is the weak  mixing angle)  involving free particles and those involving bound nucleons in the nuclear medium, the so-called  NuTev anomaly \cite {Nutev, mNutev, GPZ02}. A possible  explanation to the experimental results is  the presence of more energetics protons in neutron rich, large A,  asymmetric nuclei which implies that u-quarks are more modified than d-quarks, resulting in  a negative correction for the experimental value of  $\sin^2 \theta_W $ for bound nucleons \cite {mNutev}. The application of the models of the nuclear spectra functions may increase   the understanding  of  the medium modification of the parton distribution functions (PDFs) as well as the realistic treatment of SRCs involved in such experiments.







\newpage
\addcontentsline{toc}{chapter}{VITA}
\begin{center}
  VITA \\
  ~\\
  OSWALDO ARTILES \\
  ~\\
\end{center}
\vspace{-4mm}

\bgroup
\def\arraystretch{1}
\begin{table}[h!]
  \begin{tabular}{ l l }
    ~~~~~~1964-1970~~~~~~~~~~~~~~~~~~~~~~~~~& Electrical Engineering \\
    ~& Universidad Central de Venezuela \\
    ~& Caracas, Venezuela \\
    ~&~ \\
    ~~~~~~1976-1978~~~~~~~~~~~~~~~~~~~~~~~~~~& M.Sc., Power Systems Analysis \\
    ~& The University of Manchester, \\
    ~&  Institute of Science and   Technology (UMIST), \\
    ~& Manchester, England \\
    ~&~ \\
    ~~~~~~1996-1997~~~~~~~~~~~~~~~~~~~~~~~~~~& Specialist in Finance\\
    ~& Instituto de Estudios Superiores de Administraci\'{o}n (IESA), \\
    ~& Caracas, Venezuela \\
    ~&~ \\
    ~~~~~~1972 - 1999 & Electrical Engineer and Manager \\
    ~& C.V.G. Electrificaci\'{o}n del Caroni (EDELCA)\\
    ~& Caracas, Venezuela \\
    ~&~ \\
    ~~~~~~1999 - 2003 & President  \\
    ~& C.V.G. Electrificaci\'{o}n del Caroni (EDELCA)\\
    ~& Caracas, Venezuela \\
    ~&~ \\
    ~~~~~~2005 - 2007 & B.Sc., Physics \\
    ~& University of Miami\\
    ~& Miami, Florida, USA.\\
    ~&~ \\
    ~~~~~~2009 - 2012 & M.Sc., Physics \\
    ~& Florida International University \\
    ~& Miami, Florida, USA.\\
    ~&~ \\
    ~~~~~~2012 - 2017 & Ph.D., Physics \\
    ~& Florida International University \\
    ~& Miami, Florida,USA.\\\\
   ~~~~~~1979 - present  & Life Senior Member of the Institute of Electrical and Electronic\\
  ~& Engineers (IEEE)\\\\
   ~~~~~~2012 - present  & Member of the American Physical Society (APS)
  \end{tabular}
\end{table}

\newpage

SELECTED PUBLICATIONS AND PRESENTATIONS \\
\begin{enumerate}
\item Oswaldo Artiles and Misak Sargsian,
  {\sl Multinucleon short-range correlation model for nuclear spectral functions:  Theoretical framework},
  Physical Review C{\bf 88}, 044604 (2013)\\
\item Oswaldo Artiles and Misak Sargsian,
  {\sl Multinucleon short-range correlation model for nuclear spectral functions:  Numerical estimates},
  in preparation\\
\item Oswaldo Artiles ,
  {\sl Multinucleon short-range correlation model for nuclear spectral functions},
    talk presented at APS April meeting,
  Washington,DC,  January 2015. \\
\end{enumerate}


\doublespacing


\begin{thebibliography}{99}
\singlespacing


\bibitem{Rut11} 
E. Rutherford, "The Scattering of $\alpha$ and $\beta$ Particles by Matter and the Structure of Atoms", Phil. Mag. $\boldsymbol{21}$, 669 (1911).

\bibitem{Rut19} 
E. Rutherford, "Collision of $\alpha$ Particles With Light Atoms, part IV. An Anomalous Effect in Nitrogen", Phil. Mag. $\boldsymbol{37}$, 581 (1919).

\bibitem{Rut20} 
E. Rutherford, "Bakerian Lecture: Nuclear Constitution of Atoms", Proc. Roy. Soc.  $\boldsymbol{A97}$, 374(1920).

\bibitem{Moot29} 
N. F. Mott, "The Scattering of Fast Electrons by Atomic Nuclei", Proc. Roy. Soc. A $\boldsymbol{124}$, 425(1929).

\bibitem{Chad32} 
J. Chadwick,  "The Existence of a Neutron", Proc. Roy. Soc. A $\boldsymbol{136}$, 692 (1932).

\bibitem{CuJ32} 
I. Curie-Joliot and  F. Joliot-Curie, "The Emission of High energy Photons from Hydrogenous Substances Irradiated with Very Penetrating Alpha Rays", Compt. Rend. $\boldsymbol{194}$, 273 (1932).

\bibitem{He32} 
W. Heisenberg , "On the Structure of Atomic Nuclei. I. ",  Z.Phys. $\boldsymbol{77}$,1 (1932).

\bibitem{Yu35} 
H. Yukawa, "On the Interaction of Elementary Particles", Proc. Phys. Math. Soc. Japan $\boldsymbol{17}$, 48 (1935).

\bibitem{LMOP47} 
C.M.G. Lates,  H. Muirhead, G.P.S. Occhialini, and C.F. Powell, "Processes  Involving Charged  Mesons", Nature $\boldsymbol{159}$, 694 (1947).

\bibitem{Dirac49} 
P.A.M. Dirac, "Forms of Relativistic Dynamics", Rev. Mod. Phys.  $\boldsymbol{21}$, 392 (1949).

\bibitem{RO50} 
M.Rosenbluth, "High Energy Elastic Scattering of Electrons on Protons", Phys. Rev. $\boldsymbol{79}$, 615 (1950).

\bibitem{SB51}
E.E. Salpeter, and H.A. Bethe , "A Relativistic Equation for Bound-State Problems", Phys. Rev. $\boldsymbol{84}$, 1232 (1951).


\bibitem {YM54} 
C.N. Yang and R.L. Mills, "Conservation of Isotopic Spin and Isotopic Gauge Invariance", Phys. Rev. $\boldsymbol{96}$, 191 (1954).

\bibitem{MJ55} 
M.G. Mayer and J.H.D. Jensen,  \textit{ Elementary Theory of  Nuclear Shell Structure},  (John Wiley and Sons, New York, 1955).

\bibitem {HO58} 
R. Hofstadter, F. Bumiller, and M.R. Yearian, "Electromagnetic Structure of the Proton and Neutron", Rev. Mod. Phys. $\boldsymbol{30}$, 482 (1958).

\bibitem{GE61}
M. Gell-Mann, \textit{The Eightfold Way: A Theory of Strong Interaction Symmetry},  (Washington DC, U.S. Dept. of Energy, 1961).

\bibitem{NE61}
Y. Ne'eman,  "Derivation of strong interactions from a gauge invariance", Nuc. Phys. $\boldsymbol{26}$, 222 (1961).

\bibitem{HA63} 
L.N. Hand, D.G. Miller, and R. Wilson, "Electric and Magnetic Form Factors of the Nucleon", Rev. Mod. Phys. $\boldsymbol{35}$, 335 (1963).

\bibitem{ST63} 
A. de-Shalit and I. Talmi, \textit{ Nuclear Shell Theory}, ( Academic Press, New York, 1963).

\bibitem{GE64}
M. Gell-Mann, "A Schematic Model of Baryons and Mesons", Phys. Lett. $\boldsymbol{8}$, 214 (1964).

\bibitem{ZW64a}
G. Zweig, "An SU3 model for strong interaction symmetry and its breaking", Version 1, CERN-8182/Th.401 (Jan.1964), unpublished.

\bibitem{ZW64b}
G. Zweig, "An SU3 model for strong interaction symmetry and its breaking", Version 2, CERN-8182/Th.412 (Feb.1964), unpublished.

\bibitem{DW64} 
S.D. Drell, and J.D. Walecka, "Electrodynamic processes with nuclear targets", Ann.Phys. $\boldsymbol{28}$, 18 (1964).

\bibitem{MOTT}
N.F. Mott and H.S.W. Massey, \textit{ The Theory of Atomic Collisions}, (Oxford University Press, New York, 1965).


\bibitem{Weinberg} 
  S.~Weinberg,  "Dynamics at infinite momentum'', Phys.\ Rev.\  {\bf 150}, 1313 (1966).  
  
\bibitem{BP69} 
J.D. Bjorken and E.A. Paschos, "Inelastic Electron-Proton and $\gamma$ Proton Scattering and the Structure of the Nucleon", Phys. Rev. $\boldsymbol{185}$, 1975 (1969).

\bibitem{Gribov}
V.~N.~Gribov, "Interaction of gamma quanta and electrons with nuclei at high-energies",  Sov.\ Phys.\  JETP {\bf 30}, 709 (1970).

\bibitem{MSW71}  
  E. J. Moniz, I. Sick,  R. R. Whitney, J. R. Ficenec, R. D. Kephart, and W. P. Trower
  "Nuclear Fermi Momenta from Quasielastic Electron Scattering",
  Phys.\ Rev.\ Lett.\  {\bf 26}, 445 (1971).  

\bibitem{RF72} 
R. Feynman, \textit{ Photon-Hadron Interactions}, ( Addisson-Wesley,Massachusetts, 1972).

 \bibitem{Bertocchi}
 L.~Bertocchi, "Graphs and glauber", Nuovo Cimento A {\bf 11}, 45  (1972).
 
 \bibitem {FG73} 
 H. Fritzsch, M. Gell-Mann,  and  H. Leutwyler, "Advantages of the color octet gluon picture", Phys. Lett. B $\boldsymbol{47}$, 365 (1973).

\bibitem{GR74} 
F. Gross, "New theory of nuclear forces. Relativistic origin of the repulsive core", Phys. Rev. D $\boldsymbol{10}$ , 223 (1974).

\bibitem{PB75} 
M.A. Preston  and R.K. Bhaduri, \textit{ Structure of the Nucleus},  (Addison-Wesley Publishing Company, Reading, 1975).

\bibitem{CL79} 
F. E. Close,  \textit{ An Introduction  to Quarks and Partons},  (Academic Press,London, 1979).

\bibitem{BO79} 
A. Bodek et al., "Experimental studies of the neutron and proton electromagnetic structure functions", Phys. Rev. D $\boldsymbol{20}$, 1471 (1979).



\bibitem{FS81} 
  L.~L.~Frankfurt and M.~I.~Strikman,
  "High-Energy Phenomena, Short Range Nuclear Structure and QCD,''
  Phys.\ Rept.\  {\bf 76}, 215 (1981).
  
\bibitem{Gross:1982} 
  F.~Gross, "The Relativistic Few Body Problem. 1. Two-Body Equations,''
  Phys.\ Rev.\ C {\bf 26}, 2203 (1982).

\bibitem{Au83} 
J. J. Aubert {\it et. al.}, "The ratio of the nucleon structure functions $F^N_2$ for iron and deuterium", Phys. Lett. B  $\boldsymbol{123}$, 275 (1983).

\bibitem{DF83} 
Taber De Forest, "Off-shell electron-nucleon cross sections: The impulse approximation", Nucl.  Phys. A $\boldsymbol{392}$, 232 (1983).

\bibitem{HM84} 
 M. Hanzel and A. D. Martin, \textit{Quarks and Leptons: An Introductory Course in Modern Particle Physics}, (John Wiley and Sons, New York, 1984).


\bibitem{FS87}  
  L.~L.~Frankfurt and M.~I.~Strikman,
  "On the Normalization of Nucleus Spectral Function and the EMC Effect",
  Phys.\ Lett.\ B {\bf 183}, 254 (1987).  
  

\bibitem{As88} 
J. Ashman {\it et. al.}, (European Muon Collaboration), "Measurement of the ratios of deep inelastic muon-nucleus cross sections on various nuclei compared to deuterium", Phys. Lett. B  $\boldsymbol{202}$, 603 (1988).

\bibitem{FS88}
  L.~L.~Frankfurt and M.~I.~Strikman,
  $"$Hard Nuclear Processes and Microscopic Nuclear Structure$"$,
  Phys.\ Rept.\  {\bf 160}, 235 (1988).




\bibitem{RW91}
R.B. Wiringa, "Variational calculations of few-body nuclei",  Phys. Rev. C $\boldsymbol{43}$,1585 (1991).

\bibitem{Ciofi1991} 
  C.~Ciofi degli Atti, S.~Simula, L.~L.~Frankfurt and M.~I.~Strikman,
 $"$Two nucleon correlations and the structure of the nucleon spectral function at high values of momentum and removal energy$"$, Phys.\ Rev.\ C {\bf 44}, R7(R) (1991).
  
 \bibitem{FS92} 
  L.~Frankfurt and M.~Strikman,
  $"$Short range correlations in nuclei as seen in hard nuclear reactions and light cone dynamics$"$,
 in  \textit {Modern topics in electron scattering},   edited by B.~Frois and I.~Sick 
 (World Scientific, Singapore, 1991), pp. 645-694. 

\bibitem{CDL92} 
C. Ciofi degli Atti, B.D. Day, and S. Liuti, "Quasielastic and inelastic inclusive electron scattering by nuclear systems: Nucleon momentum distributions, spectral functions, and off-shell effects", Phys. Rev. C $\boldsymbol{46}$, 1045 (1992).

 \bibitem{FGross} 
 F.~Gross, J.~W.~Van Orden and K.~Holinde,
 $"$Relativistic one boson exchange model for the nucleon-nucleon interaction$"$,
  Phys.\ Rev.\ C {\bf 45}, 2094 (1992).
  
  

\bibitem{FSDS93} 
  L.~L.~Frankfurt, M.~I.~Strikman, D.~B.~Day and M.~Sargsian,
   $"$Evidence for short range correlations from high Q$^2$ (e, e-prime) reactions$"$,
  Phys.\ Rev.\ C {\bf 48}, 2451 (1993)

\bibitem{IT93}  
I. Talmi, \textit {Simple Models of Complex Nuclei},  (Hardwood Academic Publishers GmbH, Switzerland, 1993).

\bibitem{AV1895} 
R.B. Wiringa, V.G.J. Stoks and R.Schiavilla, "Accurate nucleon-nucleon potential with charge-independence breaking",  Phys. Rev. C $\boldsymbol{51}$, 38 (1995).


 \bibitem{CiofiSimula}
  C.~Ciofi degli Atti and S.~Simula,
  $"$Realistic model of the nucleon spectral function in few and many nucleon systems$"$,
  Phys.\ Rev.\ C {\bf 53}, 1689 (1996).

\bibitem{gea}
  L.~L.~Frankfurt, M.~M.~Sargsian and M.~I.~Strikman,
  $"$Feynman graphs and generalized eikonal approach to high-energy knockout processes$"$,
  Phys.\ Rev.\ C {\bf 56}, 1124 (1997)
  
\bibitem{BRDH91} 
L.M. Brown, M. Riordan, M. Dresden, and L. Hodddeson in:  \textit{The Rise of the Standard Model: Particle Physics in the 1960s and 1970s}, eds. L.M. Brown, M. Riordan, M. Dresden, and L. Hodddeson,  (Cambridge University  Press, Cambridge, 1997) p3.

\bibitem{Sa97} 
N. Samios in:  \textit{ The Rise of the Standard Model: Particle Physics in the 1960s and 1970s}, eds. L.M. Brown, M. Riordan, M. Dresden, and L. Hodddeson,  (Cambridge University  Press, Cambridge, 1997) p525.

\bibitem{Fr97} 
J. Friedman in:  \textit{The Rise of the Standard Model: Particle Physics in the 1960s and 1970s}, eds. L.M. Brown, M. Riordan, M. Dresden, and L. Hodddeson,  (Cambridge University  Press, Cambridge, 1997) p566.

\bibitem{Bj97} 
J.Bjorken in:  \textit{The Rise of the Standard Model: Particle Physics in the 1960s and 1970s}, eds. L.M. Brown, M. Riordan, M. Dresden, and L. Hodddeson,  (Cambridge University  Press, Cambridge, 1997) p589.

\bibitem{MI97}
G.A. Miller, "Light front treatment of nuclei: Formalism and simple applications", Phys. Rev. C $\boldsymbol{56}$, 2789 (1997).

\bibitem{PPC97} 
B. S. Pudliner, V. R. Pandharipande, J.Carlson, S.C. Pieper, and R. B. Wiringa, "Quantum Monte Carlo calculations of nuclei with $A<7$", Phys. Rev. C $\boldsymbol{56}$, 1720(1997).

\bibitem{FZC97}
M.S. Fayache, L. Zamick, and B. Castel, "The nuclear tensor interaction", Phys. Rep. $\boldsymbol{290}$, 201 (1997).

\bibitem{geaproceed}
  L.~Frankfurt, M.~Strikman, G.~Piller and M.~Sargsian,
  $"$Feynman diagram approach to high-energy scattering from lightest nuclei$"$,
  Nucl.\ Phys.\ A {\bf 631}, 502 (1998).
  
  \bibitem{Brodsky:1997de} 
  S.~J.~Brodsky, H.~C.~Pauli and S.~S.~Pinsky,
 $"$Quantum chromodynamics and other field theories on the light cone$"$,
  Phys.\ Rept.\  {\bf 301}, 299 (1998).

\bibitem{AH98}
A. Harindranath, "An Introduction to Light-Front Dynamics for Pedestrians", arXiv:hep-ph: 9612244v2  (1998).


\bibitem{Aclander1999} 
  J.~L.~S.~Aclander {\it et. al.},
  $"$The large momentum transfer reaction $C-12(p,2p + n)$ as a new method for measuring short range NN correlations in nuclei$"$, Phys.\ Lett.\ B {\bf 453}, 211 (1999).

\bibitem{WPC00} 
R. B. Wiringa ,  S.C. Pieper, J.Carlson, and V. R. Pandharipande, "Quantum Monte Carlo calculations of A = 8 nuclei ", Phys. Rev. C $\boldsymbol{62}$, 014001 (2000).


\bibitem{GM00}   G.~A.~Miller,
  $"$Light front quantization: A Technique for relativistic and realistic nuclear physics$"$,
  Prog.\ Part.\ Nucl.\ Phys.\  {\bf 45}, 83 (2000)
  
\bibitem{VMC01}
S.~C.~Pieper and R.~B.~Wiringa, "Quantum  Monte  Carlo Calculations  of  Light  Nuclei", 
 Ann.\ Rev.\ Nucl.\ Part.\ Sci.\  {\bf 51}, 53 (2001). 

 \bibitem{ms01} 
 M.~M.~Sargsian,
$"$Selected topics in high energy semiexclusive electronuclear reactions$"$,
 Int.\ J.\ Mod.\ Phys.\ E {\bf 10}, 405 (2001)
  
 \bibitem{VMCpc} 
 R.~B.~Wiringa,   http://www.phy.anl.gov/theory/research/momenta/. 
 
 \bibitem{SSS02} 
M.M. Sargsian,  S. Simula, and M.I. Strikman,  "Neutron structure function and inclusive deep inelastic scattering from $^3$H  and $^3$He at large Bjorken $x$",  Phys. Rev. C $\boldsymbol{66}$, 024001 (2002). 
 

\bibitem{PVW02}
S.C. Pieper, K.Varga, and  R. B. Wiringa, "Quantum Monte Carlo calculations of $A = 9,10$ nuclei", Phys. Rev. C $\boldsymbol{66}$, 044310 (2002).


\bibitem{CERN02} 
The LEP collaborators, "A Combination of Preliminary Electroweak Measurements and Constraints on the Standard Model", Report No. CERN-EP/2001-98, arXiv:hep-ex:0112021v2 (2002). 

\bibitem{GPZ02} 
G.P. Zeller {\it et. al.} (NuTeVCollaboration),   "Precise Determination of Electroweak Parameters in Neutrino-Nucleon Scattering", Phys. Rev. Lett. $\boldsymbol{88}$, 091802 (2002).

\bibitem{Yaron2002} 
  I.~Yaron, J.~Alster, L.~Frankfurt, E.~Piasetzky, M.~Sargsian and M.~Strikman,
  $"$Investigation of the high momentum component of nuclear wave function using hard quasielastic $A(p,2p)X$ reactions$"$,
  Phys.\ Rev.\ C {\bf 66}, 024601 (2002).
  
 \bibitem{MMSA03}    
 M. M. Sargsian {\it et. al.}, $"$Hadrons in the Nuclear Medium$"$,
 J. Phys.\ Phys.\ G\  {\bf 29}, R1 (2003)

 \bibitem{PN2003}    
P. R. Norton, 
$"$The EMC effect$"$,
Rep. Prog.\ Phys.\  {\bf 66}, 1253 (2003)

\bibitem{Nutev}    
G.~P.~Zeller {\it et. al.} (NuTeVCollaboration), 
$"$A Precise determination of electroweak parameters in neutrino nucleon scattering$"$,
 Phys.\ Rev.\ Lett.\  {\bf 90}, 239902 (2003) 
  
\bibitem{Kim1}  K.~S.~Egiyan {\it et. al.}  (CLAS Collaboration),
  $"$Observation of nuclear scaling in the A(e, e-prime) reaction at x(B) greater than 1$"$,
  Phys.\ Rev.\ C {\bf 68}, 014313 (2003).
  
\bibitem{Tang} A. Tang {\it et. al.}, "$n-p$ Short-Range Correlations from $(p,2p+n)$, Phys Rev. Lett. {\bf 90}, 042301 (2003).  

\bibitem{RS04}
P. Ring and P. Schuck, \textit{The Nuclear Many Body Problem} ( Springer-Verlag, New York, 2004).

\bibitem{eheppn1} 
  M.~M.~Sargsian, T.~V.~Abrahamyan, M.~I.~Strikman and L.~L.~Frankfurt,
  $"$Exclusive electro-disintegration of He-3 at high Q$^2$. I. Generalized eikonal approximation$"$,
  Phys.\ Rev.\ C {\bf 71}, 044614 (2005). 

\bibitem{eheppn2} 
  M.~M.~Sargsian, T.~V.~Abrahamyan, M.~I.~Strikman and L.~L.~Frankfurt,
  $"$Exclusive electro-disintegration of He-3 at high Q$^2$. II. Decay function formalism$"$,
  Phys.\ Rev.\ C {\bf 71}, 044615 (2005). 

\bibitem{AW05}
 G.B. Arfken and H.J. Weber, \textit{Mathematical Methods for Physicists}, Sixth edition 
 (Elsevier Academic Press, Burlington,  2005).

\bibitem{Kim2} 
 K.~S.~Egiyan {\it et. al.}  (CLAS Collaboration),
 $"$Measurement of 2- and 3-nucleon short range correlation probabilities in nuclei$"$,
 Phys.\ Rev.\ Lett.\  {\bf 96}, 082501 (2006).

\bibitem{isosrc} 
 E.~Piasetzky, M.~Sargsian, L.~Frankfurt, M.~Strikman and J.~W.~Watson,
 $"$Evidence for the strong dominance of proton-neutron correlations in nuclei$"$,
 Phys.\ Rev.\ Lett.\  {\bf 97}, 162504 (2006).



\bibitem{RSRB07} 
  R.~Schiavilla, R.~B.~Wiringa, S.~C.~Pieper and J.~Carlson,
  $"$Tensor Forces and the Ground-State Structure of Nuclei$"$,
  Phys.\ Rev.\ Lett.\  {\bf 98}, 132501 (2007).
  
  \bibitem{Shneor} R. Shneor et al. (Jefferson Lab Hall A Collaboration)
$"$Investigation of Proton-Proton Short-Range Correlations via the C12(e,e?pp) Reaction$"$,
Phys. Rev. Lett. {\bf 99}, 072501 (2007). 

\bibitem{srcrev} 
  L.~Frankfurt, M.~Sargsian and M.~Strikman,
  $"$Recent observation of short range nucleon correlations in nuclei and their implications for the structure of nuclei and neutron stars$"$,
  Int.\ J.\ Mod.\ Phys.\ A {\bf 23}, 2991 (2008).

\bibitem{FSS08A} 
L. Frankfurt, M. Sargsian  M. Strikman, "Future directions for probing two and three nucleon short-range correlations at high energies", AIP.Conf.Proc.$\boldsymbol{1056}$, 322 (2008).

\bibitem{Eip3} 
  R.~Subedi, R.~Shneor, P.~Monaghan, B.~D.~Anderson, K.~Aniol, J.~Annand, J.~Arrington and H.~Benaoum {\it et al.},  
  $"$Probing Cold Dense Nuclear Matter$"$,
  Science {\bf 320}, 1476 (2008)
 
\bibitem{Se09}
J. Seely {\it et. al.},  "New Measurements of the European Muon Collaboration Effect in Very Light Nuclei", Phys. Rev. Lett. $\boldsymbol{103}$, 202301 (2009).

\bibitem{HGP10}  
D.V. Higinbotham, J.Gomez and E. PiasetzKy, "Nuclear Scaling and the EMC Effect",  arXiv:hep-ph: 1003.4497v2  (2010).

\bibitem{noredepn} 
M.~M.~Sargsian,$"$Large Q$^2$ Electrodisintegration of the Deuteron in Virtual Nucleon Approximation$"$,   Phys.\ Rev.\ C {\bf 82}, 014612 (2010)
  
\bibitem{MS1} 
M.M. Sargsian, private communication, 2011. 

\bibitem{Boone2011} 
R. Tayloe, Indiana U. APS-DPF 2011 Providence, RI, 8/11, http://www-boone.fnal.gov

\bibitem{Boeglin11}  
  W.~U.~Boeglin {\it et. al.}   [Hall A Collaboration],  
  $"$Probing the high momentum component of the deuteron at high Q$^2$ $"$,
  Phys.\ Rev.\ Lett.\  {\bf 107}, 262501 (2011)
  
\bibitem{LW1} 
L.B. Weinstein, E. Piasetzhy, D.W. Higinbotham, J.Gomez, O. Hen, and R. Shneor,  
$"$Short Range Correlations and the EMC Effect$"$,
Phys. Rev. Lett. $\boldsymbol{106}$, 052301 (2011).

\bibitem{edepnx}
W.~Cosyn and M.~Sargsian, $"$Final-state interactions in semi-inclusive deep inelastic scattering off the Deuteron$"$, Phys.\ Rev.\ C {\bf 84}, 014601 (2011).

\bibitem{AHR12} 
J. Arrington, D.V. Higinbotham, G. Rosner, and  M. Sargsian,
$"$Hard probes of short-range nucleon-nucleon correlations$"$,
 Prog. Part. Nucl. Phys.$\boldsymbol{67}$,898 (2012).

\bibitem{LW2}  O.~Hen, E.~Piasetzky and L.~B.~Weinstein,
  $"$New data strengthen the connection between Short Range Correlations and the EMC effect$"$,
  Phys.\ Rev.\ C {\bf 85}, 047301 (2012)
 
\bibitem{PDG12a}
S.J. Beringer {\it et. al.} (Particle Data Group, Structure Functions),  Phys. Rev. D$\boldsymbol{6}$, 010001(2012)

\bibitem{PDG12b}
S.J. Beringer {\it et. al.} (Particle Data Group, Electro-weak model) ,  Phys. Rev. D$\boldsymbol{6}$, 010001(2012).

\bibitem{Fomin2011} 
 N.~Fomin, J.~Arrington, R.~Asaturyan, F.~Benmokhtar, W.~Boeglin, P.~Bosted, A.~Bruell and M.~H.~S.~Bukhari,
 $"$New measurements of high-momentum nucleons and short-range structures in nuclei$"$,
 Phys.\ Rev.\ Lett.\  {\bf 108}, 092502 (2012). 

\bibitem{proa2} 
  M.~McGauley and M.~M.~Sargsian,
  $"$Off-Fermi Shell Nucleons in Superdense Asymmetric Nuclear Matter$"$,
  arXiv:1102.3973v3 (2012).

\bibitem{Jan2012} 
  M.~Vanhalst, J.~Ryckebusch and W.~Cosyn,
  $"$Quantifying short-range correlations in nuclei$"$,
  Phys.\ Rev.\ C {\bf 86}, 044619 (2012).
  
  \bibitem{srcprog} 
  J.~Arrington, D.~W.~Higinbotham, G.~Rosner and M.~Sargsian,
  $"$Hard probes of short-range nucleon-nucleon correlations$"$,
  Prog.\ Part.\ Nucl.\ Phys.\  {\bf 67}, 898 (2012).

\bibitem{MS0912} 
M. Sargsian, 
$"$The EMC effect and short-range correlation$"$,
AIP Conf.Proc. {\bf 1560}, 480 (2013).
 
    
\bibitem{HHM13} 
  O. Hen, D. W. Higinbotham, G. A.Miller, E. Piasetzhy and L. B. Weinstein
  $"$The EMC effect and high momentum nucleons in nuclei$"$,
    Int.\ J. \ Mod. \ Phys. \ E {\bf 22}, 1330017 (2013)

\bibitem{Alvioli2012} 
  M.~Alvioli, C.~Ciofi degli Atti, L.~P.~Kaptari, C.~B.~Mezzetti and H.~Morita,
 $"$Nucleon momentum distributions, their spin-isospin dependence and short-range correlations$"$,
  Phys.\ Rev.\ C {\bf 87}, 034603 (2013).
 
\bibitem{Alvioli2013} 
  M.~Alvioli, C.~Ciofi Degli Atti, L.~P.~Kaptari, C.~B.~Mezzetti and H.~Morita,
  $"$Universality of nucleon-nucleon short-range correlations and nucleon momentum distributions$"$,
  Int.\ J.\ Mod.\ Phys.\ E {\bf 22}, 1330021 (2013).

\bibitem{mNutev} M. Sargsian,  "Neutrino interactions in the nuclear environment", APS April Meeting 2013, April 13-16, 2013 (unpublished) 

\bibitem{newprops} 
  M.~M.~Sargsian,
 $"$New properties of the high-momentum distribution of nucleons in asymmetric nuclei$"$,
  Phys.\ Rev.\ C {\bf 89}, no. 3, 034305 (2014).
  
  \bibitem{CMS14} 
  W. Cosyn, W. Melnitchouk and M.~M.~Sargsian,
 $"$Final-state interactions in inclusive deep-inelastic scattering from the deuteron$"$,
  Phys.\ Rev.\ C {\bf 89}, 014612 (2014).
 
  \bibitem{MMSNN14} 
  M.~M.~Sargsian,
 $"$Nucleon-Nucleon Interactions at Short Distances$"$,
  arXiv:nucl-th:1403.0678v1 (2014).
  
  \bibitem{Boeglinproposal} 
  W.~U.~Boeglin {\it et. al.},
  $"$Deuteron Electro-Disintegration at Very High Missing Momenta$"$,
  arXiv:nucl-ex:1410.6770v1 (2014).

\bibitem{Rios2013}   
A.~Rios, A.~Polls and W.~H.~Dickhoff,
$"$Density and isospin asymmetry dependence of high-momentum components$"$,
  Phys.\ Rev.\ C {\bf 89},  044303 (2014).
  
  \bibitem{PDG14b}
  S.J. Beringer {\it et. al.}, (Particle Data Group, Quantum Chromodynamics)  Chin. Phys. C,$\boldsymbol{86}$, 090001(2014).

\bibitem{Jansen:2014qxa} 
  G.~R.~Jansen, J.~Engel, G.~Hagen, P.~Navratil and A.~Signoracci,
  $"$Ab-initio coupled-cluster effective interactions for the shell model: Application to neutron-rich oxygen and carbon isotopes$"$,
  Phys.\ Rev.\ Lett.\  {\bf 113}, 142502 (2014).

\bibitem{Eip4} 
  O.~Hen, M.~Sargsian, L.~B.~Weinstein, E.~Piasetzky, H.~Hakobyan, D.~W.~Higinbotham, M.~Braverman and W.~K.~Brooks,
  $"$Momentum sharing in imbalanced Fermi systems$"$, 
  {\it et al.},Science {\bf 346}, 614 (2014)
  
  \bibitem{Wir2014} 
 R.~B.~Wiringa, R.~Schiavilla, S.~C.~Pieper and J.~Carlson, 
 $"$Nucleon and nucleon-pair momentum distributions in $A \le 12$ nuclei$"$,
 Phys.\ Rev.\ C {\bf 89}, no. 2, 024305 (2014). 
 
 \bibitem{Korover}I. Korover et al. (Jefferson Lab Hall A Collaboration), 
 "Probing the Repulsive Core of the Nucleon-Nucleon Interaction via the He4(e,e?pN) Triple-Coincidence Reaction",
Phys. Rev. Lett. {\bf 113}, 022501 (2014).

\bibitem{Colle:2013nna} 
 C.~Colle, W.~Cosyn, J.~Ryckebusch and M.~Vanhalst,
$"$Factorization of exclusive electron-induced two-nucleon knockout$"$,
   Phys.\ Rev.\ C {\bf 89}, 024603 (2014).

\bibitem{MSprivate2015} 
M. Sargsian, 
$"$Zverev mean field wave function$"$,
private communication. 

 \bibitem{Jan2014} 
 J.~Ryckebusch, W.~Cosyn and M.~Vanhalst,
 $"$Stylized features of single-nucleon momentum distributions$"$,
 J.\ Phys.\ G {\bf 42}, 055104 (2015).
  
\bibitem{FSS15} 
  A.~J.~Freese, M.~M.~Sargsian and M.~I.~Strikman,  
  $"$Probing superfast quarks in nuclei through dijet production at the LHC$"$,
  Eur.\ Phys.\ J.\ C {\bf 75},  534 (2015).
  
  \bibitem{FS15}
  A.~J.~Freese and M.~M.~Sargsian,
  $"$QCD Evolution of Superfast Quarks$"$,
  arXiv:1511.06044v1 (2015).
  
  \bibitem{BS15}   
  W.~Boeglin and M.~Sargsian,  
  $"$Modern Studies of the Deuteron: from the Lab Frame to the Light Front$"$,
  Int.\ J.\ Mod.\ Phys.\ E {\bf 24}, no. 03, 1530003 (2015)
 
\bibitem{Ciofi2015} 
  C.~Ciofi~degli~Atti,
  $"$`In-medium short-range dynamics of nucleons: Recent theoretical and experimental advances$"$,
  Phys.\ Rept.\  {\bf 590}, 1 (2015).
  
\bibitem{Neff2015} 
  T.~Neff, H.~Feldmeier and W.~Horiuchi,
  $"$Short-range correlations in nuclei with similarity renormalization group transformations$"$,
  Phys.\ Rev.\ C {\bf 92},  024003 (2015)
  
  \bibitem{Cosyn2015} 
  C.~Colle, O.~Hen, W.~Cosyn, I.~Korover, E.~Piasetzky, J.~Ryckebusch and L.~B.~Weinstein,
  $"$Extracting the mass dependence and quantum numbers of short-range correlated pairs from $A(e,ep)$ and $A(e,epp)$ scattering$"$,
  Phys.\ Rev.\ C {\bf 92}, 024604 (2015).
  
  \bibitem{contact} 
  O.~Hen, L.~B.~Weinstein, E.~Piasetzky, G.~A.~Miller, M.~M.~Sargsian and Y.~Sagi,
  $"$Correlated fermions in nuclei and ultracold atomic gases$"$,
  Phys.\ Rev.\ C {\bf 92},  045205 (2015).
  
  \bibitem{Salme} 
  S.~Scopetta, A.~Del Dotto, L. Kaptari, E. Pace, M. Rinaldi and G. Salm,   $"$A Light-Front Approach to the $^{3}$He Spectral Function$"$,
  Few Body Syst.\  {\bf 56}, 425 (2015).
  
  \bibitem{OrDouglas} 
  D.~W.~Higinbotham and O.~Hen,
  $"$Comment on Measurement of Two- and Three-Nucleon Short-Range Correlation Probabilities in Nuclei$"$,
  Phys.\ Rev.\ Lett.\  {\bf 114}, 169201 (2015).
  
   




\end{thebibliography}
\end{document}